# Advanced Discrete-Time Control Methods for Industrial Applications

By

Arash KHATAMIANFAR

A THESIS IN FULFILMENT OF THE REQUIREMENT FOR THE DEGREE OF DOCTOR OF PHILOSOPHY

SCHOOL OF ELECTRICAL ENGINEERING AND TELECOMMUNICATIONS

FACULTY OF ENGINEERING

July 2015

# Preface

This thesis was submitted as fulfilment of the requirements for the Doctor of Philosophy at the School of Electrical Engineering and Telecommunications at the Faculty of Engineering, the University of New South Wales (UNSW), Australia. This Ph.D. study was partly sponsored by the Australian Research Council (ARC). ARC is a commonwealth entity and an independent institution. It advises the federal government of Australia on research matters, manages the National Competitive Grants Program (NCGP), a significant component of Australia's investment in research and development, and has responsibility for Excellence in Research for Australia (ERA). The subject of the thesis is development of advanced discrete-time control methods for two industrial processes, including wind power dispatch control with battery energy storage system (BESS) and overhead crane motion control. The work has been carried out from March 2011 to April 2014 under the supervision of Professor Andrey V. Savkin from UNSW.

The University of New South Wales, April 2015
Arash KHATAMIANFAR

# Abstract

In this thesis, we focus on developing advanced control methods for two industrial systems in discrete-time with the aim of enhancing their performance in delivering the control objectives as well as considering the practical aspects of the designs such as the nature of the industrial process, control configurations, and implementation.

In the first part, the problem of dispatching wind power into the electricity network using a battery energy storage system (BESS) is addressed. To manage the amount of energy sold to the electricity market, a novel control scheme is developed based on discrete-time model predictive control to ensure the optimal operation of the BESS in the presence of practical system constraints. The control scheme follows a decision policy to sell more energy at peak demand times and store it at off-peak periods in compliance with the Australian National Electricity Market rules. The performance of the control system is assessed under different scenarios using actual wind farm and electricity price data in the simulation environment.

The second part of this thesis deals with the modeling and control of overhead crane systems for high-performance automatic operation. To be able to achieve high-speed load transportation with high precision in load positioning as well as minimizing load swings, a new modeling approach is developed based on independent joint control strategy which considers the system actuators as the main plant. The nonlinearities of the overhead crane dynamics are then treated as disturbances acting on each actuator. The resulting model enables us to estimate the unknown parameters of the system including coulomb friction constants thanks to its decoupled and linear-in-parameter form. To suppress load swings, a novel load swing control is designed based on passivity-based control. Two discrete-time controllers are then developed based on model predictive control and state feedback control to track the reference trajectories in conjunction with a feedforward control to compensate for the disturbances using computed torque control and a novel disturbance observer. The practical results on an experimental overhead crane setup demonstrate the high performance of the designed control systems.

# Content

# List of Figures

# List of Tables

# Chapter 1

# Introduction

## 1.1 Background and Motivation

When it comes to design and implementation of a control system for a specific industrial application, many practical aspects of the process should be taken into account to be able to improve their performance in real-world operation. The first line of the consideration in the design is the controller platform to which the control algorithm is implemented. They are mostly built using digital computers and microprocessors which work in discrete-time domain known as computer-controlled systems. In addition, some particular processes are naturally sample-data control problems and operated in discrete time. Therefore, there is a great benefit to be able to design the control system directly in discrete-time domain. To name a few, the issues regarding to incompatibilities of continuous-time control systems with the controller platform, such as sampling time and quantization issues, can be easily avoided. Moreover, ordinary differential equations in continuous-time systems are integrated by approximating them by difference equations, whereas discrete-time control systems are based on difference equations. Furthermore, many advance control algorithms and system identification techniques have been developed in discrete-time domain since they can be described in terms of difference equations and easily solved using numerical methods. Also, computer-controlled systems make it very easy to include logic statements and sophisticated calculations in the control law. Hence, in this thesis, we focus on developing advanced control methods for two industrial systems in discrete-time with the aim of enhancing their performance in delivering the control objectives and requirements as well as considering the practicality of the designs naming the nature of the industrial process, control configurations, and implementation.

The problem of harnessing more energy from intermittent renewable resources, like wind power, using energy storage systems (ESSs) is considered as the first topic in this thesis. The aim is to be able to increase the economic viability of wind farm integrated with a battery energy storage system (BESS) in the competitive electrify market as the price changes with a fixed period (inherently sampled-data control problem). As an example of its significance, in Canada, a new Wind R&D Park was commissioned in April 2013 with a combined total generating capacity of 10 MW that will be able to demonstrate the benefit of energy storage systems under various scenarios including time shifting, power smoothing and voltage control, as reported in the 2013 international energy agency wind annual report [1]. This project was awarded 12.0 million CAD (8.2 million EUR; 11.3 million USD) from the government of Canada's Clean Energy Fund, as well as a 12.0 million CAD (8.2 million EUR; 11.3 million USD) loan from the government of Prince Edward Island. The loan will be repaid from the sale of power produced by the Wind R&D Parkwith.

The second topic is the intriguing problem of controlling overhead crane motion for high performance automatic load transportation as one of the complex mechanical systems in industry with many practical issues for full automation using computer-controlled system. According to the most recent market research report on overhead crane manufacturing in the United States industry conducted by Supplier Relations US, LLC., by 2018, the US overhead traveling crane, hoist, and monorail system manufacturing industry demand will grow around eight percent [2]. This shows that research and development (R&D) in this field will also grow for more advanced overhead crane systems. For instance, at Konecranes' R&D sector, as one of the leading companies in overhead crane industry, the researchers are working on smart solutions to simplifying difficult maneuvers, eliminating load swing, and helping position loads in predefined locations [3].[1]

[1] The first one and a half years of the PhD candidacy was dedicated to the first part of this thesis (wind power dispatch control with BESS), and the remaining two and a half years was committed on the second part of the thesis (overhead crane control) as part of the studies on advanced discrete-time control for industrial applications.

## 1.2 Objective and Focus

In this thesis, the primary focus is on the application of advanced discrete-time control systems for two industrial applications with the aim of meeting the requirements for high performance control operation. Thus, the process of controlling wind power dispatch with a BESS is first discussed and then, the overhead crane control operation is considered.

### 1.2.1 Control of Wind Power Dispatch with BESS

Recent developments in wind energy generation, both technically and economically, have led to significant rise in deployment of this renewable energy source for electricity production cycle in many developed countries [1], [4]. However, as the penetration level of wind energy into the electricity networks grows significantly, it is important to consider the problems and challenges facing its integration to the electricity grid. From among these challenges, the intermittent nature of wind power and occasional large fluctuations due to stochastic behaviour of weather conditions need to be managed in order to prevent some undesirable and potentially destructive impacts on the stability of the electricity grids [5]. This intermittency in the wind power generation reduces its capability to compete with conventional power plants in the regions with deregulated energy market where the energy price is determined based on supply and demand.

One technically feasible solution to mitigate these problems is the integration of an energy storage system (ESS) with the wind farms. Such a solution can provide added value through greater reliability, improved power quality, energy availability, and overall reduced energy generation cost, although it is currently an expensive one [6]. A battery energy storage system (BESS) has been shown to be a suitable choice among different ESSs' technologies for integration with wind farms to achieve maximum benefit [7], [8], and that is why BESS technology is chosen in this study. In addition, Integration of BESS with a wind farm makes it possible to control its combined generated power in the grid-connected mode similar to conventional power plants. Thus, in the countries, like Australia, where the electricity price is determined by market driven electricity supply and demand, a BESS enables wind energy to be stored at off-peak demand times when the price of electricity is relatively low. This stored energy can be sold simultaneously with the generated wind power at peak demand times when the electricity price is significantly high. This is typically referred to as *time shifting*. The policies and strategies used to design such dispatch control schemes are highly dependent on the electricity market rules of the region where the power is sold as well as the constraints of the applied BESS. However, little attention has been paid to improve the controllability of a wind farm with a BESS in the light of control systems considering electricity price and optimal use of the BESS for the purpose of trading within a competitive electricity market.

Therefore, considering the fact that the electricity price changes every five minutes in Australia– as the country where the wind farm under study is located –which makes the process to operate naturally in discrete-time, the main objective is set to the design of an intelligent control system for wind farm dispatching using a BESS in time shifting application. Hence, a discrete-time control system to achieve the above-mentioned objective is developed in this thesis which is comprised of three key parts: A decision-making system based on fuzzy logic for preparing an online reference tracking power signal based on electricity price and peak/off-peak demand periods of the day; a discrete-time controller to determine the optimum amount of charging/discharging power for the BESS and follow the reference power signal designed based on model predictive control (MPC), and a fuzzy logic controller (FLC) to update the reference power signal based on the BESS conditions and wind power availability.

MPC algorithms are now widely used in industry since their first application in chemical process industry and then gained popularity in different types of industrial control problems [9]. Recently, they are reported to be used in power and renewable energy systems as well [10], [11]. The main advantage of MPC algorithms that makes them so practical is that the process constraints can be explicitly taken into account in the controller design. In our case, the BESS energy capacity and rated power are the main constraints that should be considered for its optimal operation. The FLC is also applied in the proposed control system to update the generated reference power signal to facilitate the operation of the MPC for the tracking performance objective using the BESS charging/discharging conditions. FLCs are applied in many control system applications, particularly as a role of coordination and complementary controller [12]. Moreover, it has been shown that FLC can be used for both continuous-time and discrete-time applications due to their inherent nature of processing IF-THEN rules.

The effectiveness of the proposed discrete-time control system is examined under different scenarios of selling power using the actual wind farm and electricity price data to show the potential of our proposed control scheme for the Australian national electricity market (NEM) in terms of key performance index and earning comparison from selling the power.

#### 1.2.1.1 Previous and Related Works

Over the last couple of years, two main research topics have been the center of attraction in the field of integrating the ESSs with wind power: Smoothing the fluctuations of the generated power for its secure connection to the electricity grid, and making wind farms more dispatchable like conventional power plants according to the electricity market variations. Different techniques have been suggested and developed in the area of power smoothing and power quality which vary from simple schemes of charging and discharging the BESS as the wind power output goes beyond a minimum or maximum threshold as in [13] and [14], to much more sophisticated control algorithms. For the latter case, the authors of [15] applied an optimal control method on the linear model of the lead–acid battery to smooth the generated wind power and make it dispatchable on an hourly basis. They used one-hour ahead average forecasted wind power as the reference power for dispatching. In [16], an open-loop MPC scheme was proposed to find the optimal wind power output integrated with a BESS that meets the requirements for low-fluctuated power output. They used a new prediction model for wind speed and direction to reduce the wind power intermittency which was improved later in [17] and [18]. Similar technique was suggested by [19] for the purpose of frequency control of grid-conned wind farm with BESS using MPC. A dual-layer control strategy for a BESS to mitigate the wind farm power output fluctuations was proposed in [20] which consists of a fluctuation mitigation control layer and a power allocation control layer. The first layer uses a flexible first-order low-pass filter with an optimization of time constant to calculate the power for the BESS so that the combined wind farm and BESS power output meets fluctuation mitigation requirement. The second layer optimizes power allocation among the battery units of the BESS using a mixed-integer quadratic programming model.

The other topic closely related to power smoothing for intermittent power sources like wind power is to find the optimal size of the ESS such that the overall cost of the generated power can be reduced [5], [21]. For instance, a constraint-based monotonic charge/discharge strategy for multiple batteries of a BESS was proposed in [22] to determine the optimal capacity of each individual battery satisfying a set of given operating constraints for the purpose of smoothing the intermittent wind power. The authors in [23] used similar approach as in [20] to reduce the wind power fluctuations with a BESS using two-time scale coordination control. The BESS capacity and power rating for the wind farm is then estimated to meet the two-time-scale maximal power fluctuation restrictions for the combined output of the wind farm and the BESS. Most recently, a scheme to minimize the capacity of BESS in a distributed configuration using MPC and wind power prediction is developed in [24].

In the case of dispatchability of wind farms in grid-conned mode, several efforts have been made to make wind power more dispatchable using ESSs in relation to the electricity market. For example, a dynamic programming algorithm was employed in [25] to determine the optimal wind energy exchange with the electricity market for a specified scheduling period using an ESS, taking into account the transmission constraints with emphasizing on the impact of the ESS sizing and weather forecasting accuracy on system operation and economics. The writers of [26] performed an economic and technical analysis for hourly energy management of a wind farm with three different ESSs through detecting peak and off-peak electricity consumption periods. These time periods are identified via an optimization software developed by the same authors [27]. In another work, the profitability of a wind power plant integrated with a BESS was examined from the supply chain perspective considering price volatility in the electricity market [28]. In [29], an iterative optimization technique for scheduling wind power was applied based on an hourly electricity tariff prediction with a dual BESS structure. And finally, a methodology based on dynamic programming algorithm was proposed in [30] to determine the hourly-profile energy delivery of the combined wind power and a BESS that fits the generation forecast and the BESS features and complies with electricity market requirements with economic feasibility analysis of the methodology.

### 1.2.2 Control of Overhead Crane Motion

The problem of controlling the overhead crane motion for full automatic load transportation has drawn more attention in the last couple of years in the control engineering research community due to its complex nonlinear dynamics and its pivotal role in transportation industry, especially in heavy machinery industry. The most important factors in load transportation are time efficiency and accurate load positioning. In other words, the load should be transported as fast as possible from initial location to the destination with high accuracy in the final point. However, the higher the speed of the overhead crane motion, the larger the load swing which not only poses danger to the surrounding objects but also could damage the overhead crane itself due to exerting massive load force. The reason is that in the complex dynamics of the overhead crane, the number of control inputs is less than the number of control variables. This means that there are three control inputs for motions along the main three coordinates in *XYZ* plain (traveling along *X* axis, traversing along *Y* axis, and hoisting along *Z* axis), but no direct control input exists for swing angles dynamics. That is why overhead crane systems are classified as underactuated systems with swing dynamics as unactuated dynamics and the rest as actuated dynamics. Moreover, to avoid obstacle and increase time efficiency, it is common practice to hoist the load as the overhead crane accelerates, but this load lifting during acceleration intensifies the swings if it is conducted with high speed. As a result, many would avoid load hoisting during acceleration for the sake of safety, which slows down the entire operation.

In the manual operation of the overhead crane, an expert operator (with the help of a second person as the ground guide) controls the overhead crane along a typical anti-swing trajectory that consists of three motion zones: An accelerating zone, a constant-velocity zone, and a decelerating zone. In the accelerating zone, the overhead crane is initially accelerated to a normal velocity with zero load swing and the load is hoisted up if necessary. This process allows a certain level of load swing until the normal velocity is reached. Then, in the constant-velocity zone, the overhead crane is controlled such that they move at the normal velocity with zero load swing. Finally, in the decelerating zone, the overhead crane is decelerated to a complete stop with zero load swing and the load is hoisted down if necessary. This process also allows a certain level of load swing until it reaches the final point.

To be able to achieve a high-performance control in the overhead crane operation in fully automated fashion, the control system should be designed such that it can deliver high-speed load transportation with high accuracy in load positioning (for higher efficiency) as well as the ability to minimize load swings during the entire operation (for safety). Furthermore, it should provide the capability to handle high-speed load hoisting during accelerating zone. Therefore, taking into account these primary control objectives for the overhead crane system, a discrete-time control system is designed in this thesis that is composed of four main parts. The first part is a reference signal generator that provides reference trajectories similar to typical anti-swing trajectory performed by an expert crane operator considering all the physical constraints on the overhead crane actuators and workspace. The second part is the load swing control that modifies the reference traveling and traversing accelerations enabling robust load swing suppression. The third part is the main discrete-time controller, which calculates the final control inputs to perform trajectory tracking. This discrete-time controller is designed using MPC and state feedback approaches. And finally, a feedforward control action as the forth part to compensate for the disturbances and uncertainties and improve load positioning accuracy and robustness using the idea of computer torque control [31]. The foundation of the proposed control system is the so-called *independent joint control strategy* adopted from the field of robot manipulator control. In this control strategy, the process actuators are considered as the main plant to be controlled, and all the nonlinearities caused by the coupling effects between the mechanical structure of the process and the actuators are treated as disturbances acting on each actuator [31], [32]. The main advantage of applying this idea is the simplicity in the design of the controller without compromising the effectiveness of the control performance. The proposed discrete-time control systems are implemented on a laboratory-sized overhead crane setup and extensive tests are carried out to demonstrate the effectiveness of the designed discrete-time control systems in delivering the high-performance control for the automated overhead crane operation.

#### 1.2.2.1 Previous and Related Works

Over the past couple of decades, extensive research has been conducted on controlling the overhead crane motion intended to act similarly to what an expert human operator can provide[2]. The early works in the 80s and 90s were mostly focused on two-dimensional (2D) overhead crane due to less complexity in its dynamics compared to the three-dimensional (3D) form using linear control theory which involves traveling and hoisting motions (some papers referred to 2D overhead crane as the one moves only along *X* and *Y* axes with no hoisting). For instance, a minimum-time control problem was solved in [33] for swing-free velocity profiles of a crane under the constraint of zero load swing at the start and end of acceleration, which seems to be the earliest work reported in this area. In [34], a feedback control based on the swing dynamics of the load was proposed, and in [35], root locus method was used to design the feedback control law for an overhead crane. Later on, the first attempts to derive the full nonlinear dynamics of the overhead crane were made in [36] and [37]. The equations of motion of the overhead crane in [36] is derived based on spherical coordinates, whereas in [37], a new swing angle definition was proposed in Cartesian coordinates which results in having a set of equations of motion equivalent to those of a three-link flexible robot manipulator having the first flexible mode [38]. However, both works used simplified linear models with linear feedback control to control the overhead crane motion. The former applied linearization around the equilibrium point, and the latter simplified the nonlinear dynamics assuming that the hoisting rope is varying slowly and the trolley acceleration is much smaller than the gravitational acceleration to obtain the simplified linear models.

After that, many other works used similar linearized models to control the overhead crane with different linear and nonlinear control algorithms including damping the linearized system by an observer-based controller and applying a dynamic inversion procedure in order to assure a predetermined oscillation-free polynomial motion law for the payload in [39], using constraint MPC on the linearized model obtained by sub-space identification algorithm in [40], and applying a discrete-time integral sliding mode control on the non-minimal linear model of the overhead crane in [41]. However, many nonlinear control techniques have also been applied using the overhead crane equations of motions to tackle the nonlinearity of the overhead crane dynamics. For instance, the authors in [42] used additional nonlinear feedback terms with a PD controller to increase the coupling between gantry and payload and improve the transient behavior of the overhead carne. This category of controllers is known as energy-based controller which has been developed and used in some recent works as well [43]–[45].

[2] Since the focus of this thesis is on the overhead crane only, the literature on other types of crane was not considered.

The linear-in-parameter form of the overhead crane nonlinear dynamics makes it possible to use adaptive control algorithms to reduce the effect of parameter uncertainty such as those reported in [46] and [47] which used passivity-based adaptive control as a known adaptive technique in robot manipulators. Some other nonlinear techniques such as partial feedback linearization [48], full feedback linearization using the swing angular rate as well as the swing angle using the spherical-coordinates model [49], gain scheduling [50], and nonlinear MPC [51], [52] have also been used on overhead crane.

Model-free control algorithms, on the other hand, have been suggested to be used for overhead crane control to avoid dealing with the complex nonlinear dynamics, including the early work in [53] where a fuzzy logic controller was used for reducing the load swing with a simple PD controller for position control. In the paper [54], a sliding mode controller with fuzzy tuning for the sliding surface was used on the linearized model of the overhead crane as in [37] with constant hoisting rope length. A full fuzzy controller was developed in [55] with an adaptive algorithm to tune the free parameters of the control system with no load hoisting. In a recent work, the nonlinear dynamics of the overhead crane was modeled as a three-rule Takagi-Sugeno fuzzy model with a saturated input and then a state-feedback controller was designed based on the fuzzy model so that trajectories of the system that start from an ellipsoid will remain in it [56]. Moreover, A combination of PID controller with neural network compensation for fining the PID gains using standard weights training algorithms was proposed in [57]. More recently, some attempts have been made to apply visual-based feedback control using standard CCD (charge-coupled device) cameras to capture the dynamic movement of the overhead crane. For instance in [58], visual tracking is based on color histograms which involves comparison of the color in a model image with the color in a sequence of images to track a dynamic object in a 2D overhead crane. Similar approach proposed for a 3D overhead crane in [59] where visual tracking method involves comparison of the lightest or darkest points in the tracking or positioning area of a dynamic object and then computes the necessary trolley position and load swing in 3-D space. Both works then used an adaptive fuzzy sliding mode controller to control the overhead crane motion.

A major issue with the aforementioned works is that their control systems are designed for set-point control (following constant reference signals). Whereas, as explained before, typical anti-swing trajectory is mostly used in practice to reduce the transportation time, not to mention that some of the works mentioned before ignored the load hosting action. Supposedly, the first work on controlling the overhead crane which allows high-speed load hoisting was initially proposed by Lee in [60], where a trajectory tracking controller was designed based on Lyapunov stability theorem using the full nonlinear model he had initially introduced in [37]. The same author developed a different control approach in [61] with load-hoisting capability using Lyapunov stability theorem where load swing dynamics is coupled with the trolley motion by defining a linear PD-type sliding surface. Similar technique was adopted in [62] using sliding mode control. Also, a second-order sliding mode controller was developed for the 3D overhead crane in [63] and [64] considering load hoisting.

In addition to closed-loop control systems discussed above, the open-loop control of an overhead crane has also been suggested recently known as motion planning. The aim in this category is to find the reference trajectory such that it can provide the minimum-time motion with less swing angle while satisfying the physical constraints of the overhead crane. The pioneering work in this area was developed by Lee in [65] where the motion-planning problem is solved as a kinematic problem using swing dynamics and Lyapunov stability theorem for a 2D overhead crane. The improved version of this motion-planning scheme for 3D overhead cranes was proposed in [66]. Most recently, some efforts have also been made to suggest different motion-planning algorithm such as [67]–[70]. However, a major issue with these works is that the hoisting rope is assumed to be constant which is not the case in practice compared to [65] and [66], not to mention that those algorithms seem to be complicated and they were developed only for 2D overhead cranes.

## 1.3 Contributions

As the focus of this thesis is to demonstrate the ability of advanced discrete-time control systems for industrial applications with emphasis on two practical processes, naming the wind power dispatch control with BESS and overhead crane control, the primary contributions can be expressed separately as follows.

**Part I**: The challenges facing the higher penetration of wind power into the electricity production cycle was highlighted and the integration of BESS with wind farm was considered as a feasible solution to cope with these challenges. Based on the structure of wind power system with a BESS in grid-connected mode, a discrete-time model was suggested knowing that the power grid-connected mode dispatches with a fixed sampling rate determined by the electricity market operator of the region where the grid is located. As the aim is to increase the financial benefits for the wind farm from the sale of its generated power to the electricity market, a new control scheme was developed in time shifting application. The proposed control scheme was designed to manage the wind power dispatch taking into account the electricity market rules and dispatch operation (in our case Australian NEM), and the constraints on the BESS energy capacity and rated power. To achieve the control objective, the control system was designed in three parts. A decision-making system was developed based on fuzzy logic to generate online reference power signal using electricity dispatch price and time of the day information. A discrete-time controller based on MPC was designed to optimize the BESS charging/discharging process, perform reference tracking, and handle system constraints. And finally, a feedback fuzzy controller was applied to update the reference signal according to the BESS conditions and wind power availability.

The application of these known control algorithms together with an online reference power signal generator in designing the proposed control system is a novel insight to the problem of controlling wind farm power dispatch integrated with a BESS. The proposed control scheme realizes higher controllability of the wind farm power dispatch with the BESS in the electricity market in a stable and robust manner, not to mention its practicality for real-time operation. The effectiveness of the proposed control system was examined under different scenarios of selling power using the actual wind farm and electricity price data to show the potential of our proposed control scheme for the Australian NEM in terms of key performance index and earning comparison from selling the power.

**Part II**: The problem of high-performance load transportation using overhead crane system in automatic operation was considered in this part of the thesis. To be able to design an effective discrete-time control system, the application of independent joint control strategy was introduced for overhead crane to deal with complex nonlinearity of the overhead crane, and to our knowledge, this has not been reported in the literature so far (although this is a common method in robot manipulator control field). In this strategy, the process actuators are considered as the main plant to be controlled, and the nonlinear dynamics of the process are modelled as disturbances acting on the actuators. Thus, the overall control system design is significantly simplified without compromising the performance of the control operation as one of the primary contributions of this work, which is a great advantage. Moreover, the resulting dynamic model enabled us to develop a system identification procedure to determine the main physical parameters of the overhead crane and its actuators with a high precision. It should be mentioned that by using this approach, we were able to identify the coulomb friction constants as another contribution, which are one of the main factors in reducing the overhead crane position accuracy if not compensated. The coulomb friction forces were used as part of pre-known load disturbances in addition to the highly nonlinear dynamics to be compensated in the proposed discrete-time control system. Very few works have mentioned the negative impacts of friction forces in the load positioning accuracy for high-speed control operation of the overhead crane [62]. In addition, the resulting model has less system order compared to the original nonlinear model since swing dynamics are separated from trolley and hoisting dynamics, and consequently, it could be easily transformed into discrete-time form due to its linear-in-parameter from.

To be able to have high-performance control operation for the overhead crane, naming high-speed load transportation with accurate position control and as minimum load swing as possible, the main control requirements were investigated in details. Based on those requirements, the overall structure of the discrete-time control system was established which consists of four main parts. A reference signal generator, as the first part, provides reference trajectories for traveling, traversing, and hoisting motions using the desired accelerations. The second part is a new load swing control designed to suppress load swings robustly by modifying the reference traveling and traversing acceleration. To compensate the effects of load disturbance, a feedforward control was designed using the idea of computed torque control in the third part. A discrete-time controller as the main and final part of the proposed control system was designed based on the MPC and state feedback approach to calculate control input voltages for the overhead crane actuators such that it can follow the reference trajectory with high performance. A trajectory planning was also developed in conjunction with reference signal generator and load swing control which guarantees the satisfaction of the physical constraints of the overhead crane and actuators as well as maintaining the minimum-time control operation.

The major achievement of the proposed discrete-time control system for the overhead crane is that it can deliver high-performance control operation with much less complexity in terms of implementation and control configuration compared to the existing methods in the literature. This is an important factor when it comes to applicability of the control system and easy understating of the controller settings for the operator in practice. Furthermore, it allows high-speed load hoisting during acceleration of the overhead crane without deteriorating load swings that improves time efficiency. The proposed discrete-time control system can be applied in both 2D and 3D overhead cranes for either set-point tracking or trajectory tracking. It can even be used for other underactuated systems.

An extensive number of tests and experiments were carried out to verify the performance of the designed discrete-time control systems in several scenarios. Realization of any control system in real-time is always one of the most challenging tasks in the implementation phase. In that regard, all of the proposed control systems were constructed via MATLAB® software and SIMULINK® environment[3]. The generic block-diagram form of the control systems makes it easy to change the settings and run the tests repeatedly without any interruption as another important contribution of this thesis. The obtained results also showed that the proposed control systems are robust against massive changes in the overhead crane load mass due to the inclusion of load mass as part of load disturbances. In addition, a new disturbance observer was designed that can estimate the amount of load disturbance without the need to know the value of the load mass which is a great advantage in improving the robustness of the control operation.

[3] MATLAB and SIMULINK are registered trademarks of The MathWorks, Inc.

## 1.4 Thesis Outline

Since two industrial processes are the focus of this thesis, the outline is divided into two parts in addition to the first chapter, where each part comprises of two chapters presenting the modeling procedure and control system design for each industrial process.

**Chapter 2** covers the modeling process of the wind power integrated with a BESS. It starts with an overview of the wind power and its promising role in harnessing more energy from wind renewable source. It continues with the application of ESSs as a way to deal with the intermittency nature of wind power that limits its competitiveness against conventional power plants. After that, a summary of the Australian national energy market (NEM) is presented, followed by discrete-time model proposed for the wind power with a BESS in grid-connected mode, and a brief discussion at the end.

**Chapter 3** includes the details of the discrete-time control system design for the wind power dispatch with BESS. The control objectives and requirements are first specified for economic viability of wind power sale with a BESS in the electricity market. Next, each part of the proposed control system is described beginning with an introduction to fuzzy logic systems since the decision-making system for generating reference power signal is designed based on fuzzy logic, which is discussed after the fuzzy systems overview. Following that, an overview and basic formulation of model-based predictive control is provided. The design procedure of the discrete-time controller for wind power dispatch with BESS using MPC is given after the MPC overview. The chapter continues with the design of the fuzzy logic controller for updating the reference power signal. Then, the simulation results using the actual wind power and electricity price data obtained from Australian energy market operator (AEMO) database are given, and the chapter is finished with a discussion.

**Chapter 4** provides a comprehensive modeling procedure for the overhead crane system starting with an overview of overhead cranes in transportation industry and then the derivation of the overhead crane equations of motion and actuator dynamics. Following that, the application of independent joint control strategy in modeling of the overhead crane is discussed. Next, the parameter identification procedure is explained and the results of the identification are given. At the end, the discrete-time representation of the obtained model of the overhead crane is provided followed by a short conclusion.

**Chapter 5** deals with the design of each part of the proposed discrete-time control system for the overhead crane. After discussing the control objectives and requirements for high-performance operation of the overhead crane, the overall configuration of the discrete-time control system is established. Then, the details of load swing control based on passivity-based control and $\mathcal{L}_2$ stability theorem along with swing angle observer is given. Next, the trajectory planning for typical anti-swing motion of the overhead crane is explained. Following that, the formulation of the MPC and state feedback control for discrete-time controller is provided along with the design of disturbance observer using state estimation error. Finally, the results of practical tests and validation of the performance of the discrete-time control system is provided, and the chapter is finished with a brief discussion.

**Chapter 6** concludes the thesis by discussing the achievements of this thesis and potential future works.

## 1.5 Author's Publications

The following papers were the outcome of the work conducted in this thesis by the author during the PhD candidacy.

**Peer-Reviewed Conference Publications:**

A. Khatamianfar, M. Khalid, A. V. Savkin, and V. G. Agelidis, "Wind Power Dispatch Control with Battery Energy Storage using Model Predictive Control," in *Proc. 2012 IEEE Multi-Conf. Syst. Control, (MSC2012),* Dubrovnik, Croatia, 2012, pp. 733–738.

A. Khatamianfar and A. V. Savkin, "A new tracking control approach for 3D overhead crane systems using model predictive control," in *Proc. 2014 European Control Conf., (ECC2014),* Strasbourg, France, 2014, pp. 796–801.

A. Khatamianfar and A. V. Savkin, "A new discrete time approach to anti-swing tracking control of overhead cranes," in *Proc. 2014 IEEE Multi-Conf. Syst. Control, (MSC2014),* Nice, France 2014, pp. 790–795.

A. Khatamianfar, "A New Approach to Overhead Crane Parameter Estimation and Friction Modeling," in *Proc. 2014 IEEE Int. Conf. Syst., Man, Cybern., (SMC2014)*, San Diego, CA, USA 2014, pp. 2581–2586.

A. Khatamianfar, "Discrete-Time Servo Control of Overhead Cranes with Robust Load Swing Damping," in *Proc. 2015 European Control Conf., (ECC2015)*, Linz, Austria, 2015, pp. 1106–1113.

**Peer-Reviewed Journal Publications:**

A. Khatamianfar, M. Khalid, A. V. Savkin, and V. G. Agelidis, "Improving Wind Farm Dispatch in the Australian Electricity Market With Battery Energy Storage Using Model Predictive Control," *IEEE Trans. Sustain. Energy*, vol. 4, no. 3, pp. 745–755, Jul. 2013.

A. Khatamianfar and M. Bagheri "Application of Independent Joint Control Strategy for Discrete-Time Servo Control of Overhead Cranes," *J. Model., Simul. Electr., Electron. Eng. (MSEEE)*, vol. 1, no. 2, pp. 81–88, Mar. 2015.

# **Part I:** Discrete-Time Control of Wind Power Dispatch Using BESS

# Chapter 2

# Modeling of Wind Power Dispatch with BESS

In this chapter, a dynamic model is presented for the integration of wind power with a battery energy storage system (BESS). An overview on the wind power as one of the major renewable energy sources and the role of energy storage systems (ESSs) in harnessing more energy from wind for participation in electricity markets is given in Section 2.1, along with different applications of ESSs in power industry. A summary of the Australian national energy market operator for which the control system should operate in compliance is also given in Section 2.1. Then, the discrete-time dynamic model for wind power dispatch with a BESS is derived in Section 2.2, followed by a short conclusion at the end in Section 2.3.

## 2.1 Overview of Wind Power and BESS

Wind power has gained a significant role in the cycle of electricity power generation as one of the most prominent renewable energy resources due to the maturity in wind turbine technology and economically viable in recent years, not to mention that its penetration level in the capacity of the power production is increasing worldwide. As an example, Denmark has the largest share of electricity generation from wind farms with more than 20 percent of its annual national demand in 2010 [71] which has grown up to 32 percent in 2013 [1]. In addition, other leading countries in this field have been adding to their installed capacity of wind power generation. As can be seen in Fig. 2.1–1, the total capacity of wind power generation in the international energy agency (IEA) wind member countries has increased from less than five giga-watts (GW) in 1995 to more than 268.8 GW in 2013 [1].

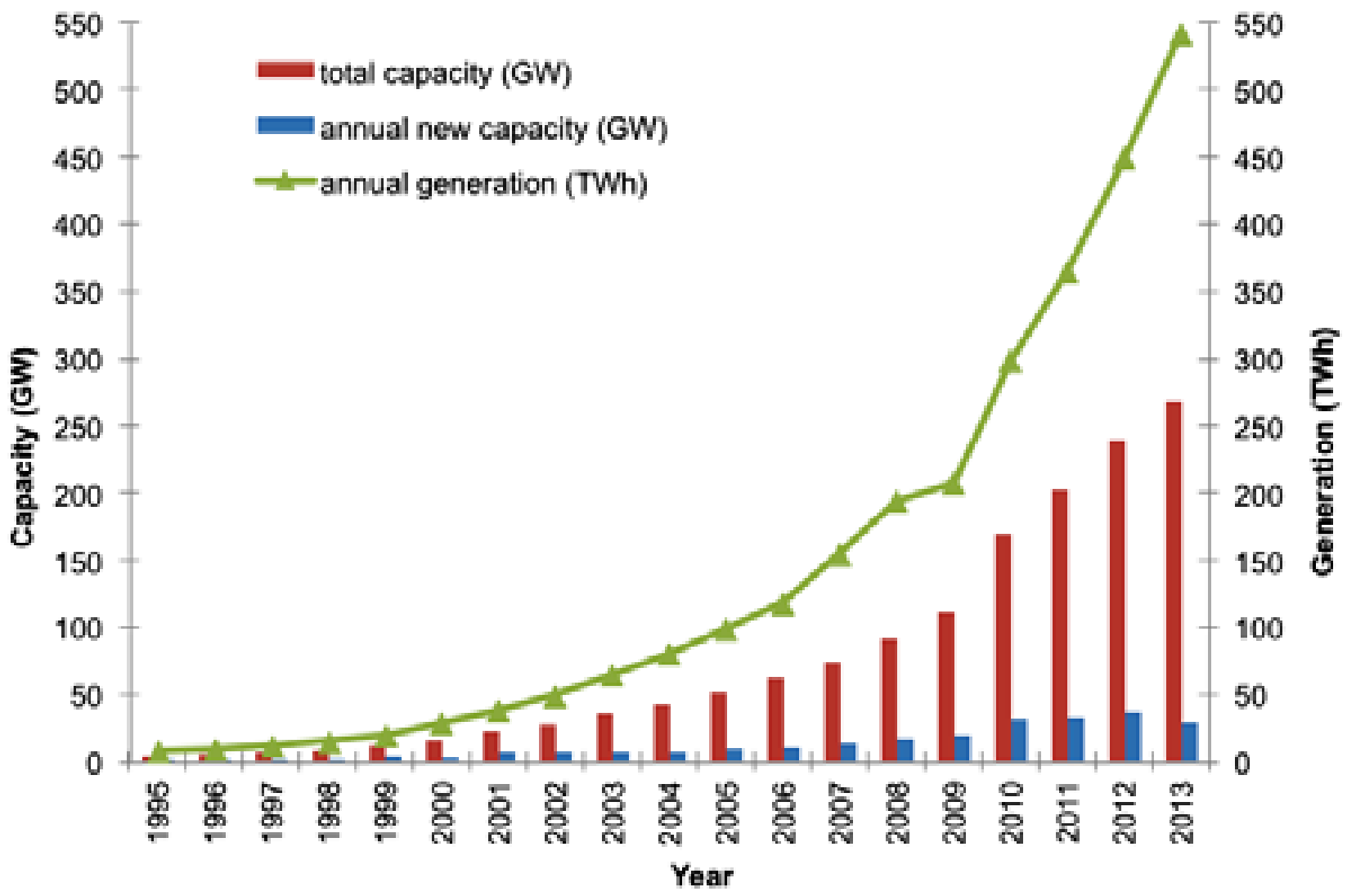

Fig. 2.1–1. Annual new capacity (net), cumulative capacity, and electricity generation for IEA wind member countries, 1995–2013 with China first represented in 2010 [1].

In spite of all the benefits and growth in the generation of wind power as an important renewable energy source, the intermittent and stochastic nature of wind speed make it difficult to have an accurate estimation of the amount of power that a wind farm can produce to meet the energy demand. In addition, occasional large fluctuations in the generated power could have destructive impacts on the appliances using wind energy. If the electricity grid to which the wind farm is connected is not strong enough, the fluctuations of generated power could lead to network failure and power outage.

One feasible solution to mitigate these problems is the integration of energy storage systems (ESSs) with wind power. ESSs can store electrical energy into different forms of energy and convert them back to electrical energy when needed. There are several types of ESSs depending on the technology used to store energy, which are mostly chemical and mechanical energies with some examples shown in Fig. 2.1–2. Popular types of ESSs are categorized as the following[4],

[4] Further details on different ESSs technologies can be found in [72]–[75].

– Battery energy storage systems (BESSs) such as Lead-Acid battery (Pb-Acid), Sodium-Sulphur battery (NaS), Lithium-Ion battery (Li-Ion), and Nickel-Cadmium battery.
– Flow batteries.
– Fuel cells
– Flywheels
– Compressed air energy storage (CAES)
– Pumped hydro storage (PHS)
– Superconducting magnetic energy storage (SMES)
– Super capacitors

Between the above-mentioned energy storage technologies, BESS have shown to be more applicable with intermittent renewable energy source like wind and solar power [7], [8].

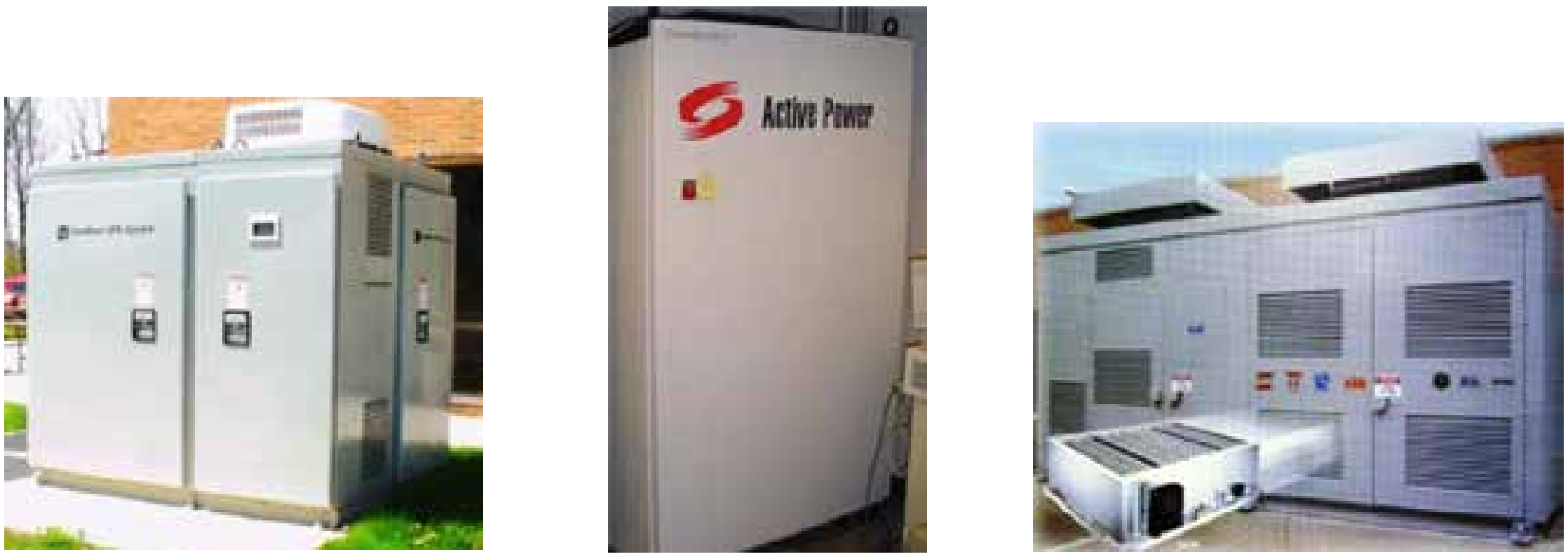

(a) (b) (c)

Fig. 2.1–2. Energy storage sysems. (a) S&C PureWave Lead Acid BESS, (b) MGE/UPS Active Power flywheel, (c) ABB/NGK NaS BESS [76].

The main applications of the ESSs with intermitted power generation sources like wind power are power smoothing, power quality, voltage control, and time shifting. In power smoothing, the fluctuations in the generated power can be reduced using ESSs that gives more reliability and stability to the intermittent power source in grid-connection mode as well as the ability to participate in the electricity market in accordance with the electricity grid rules as illustrated in Fig. 2.1–3. The stored energy in ESSs can also be used as a backup power in case of unexpected utility disruption, wind turbine outage, or insufficient wind speed. Thus, the ESS can continuously supply power to the load and avoid penalties imposed by the grid operator for the failure of the utility to comply with its power production obligation (similar to the function of uninterruptable power supply (UPS) in a larger scale). The reactive power and voltage of the generated wind power can rapidly be changed by means of power electronics interfaces into the desired amount known as voltage control using ESSs.

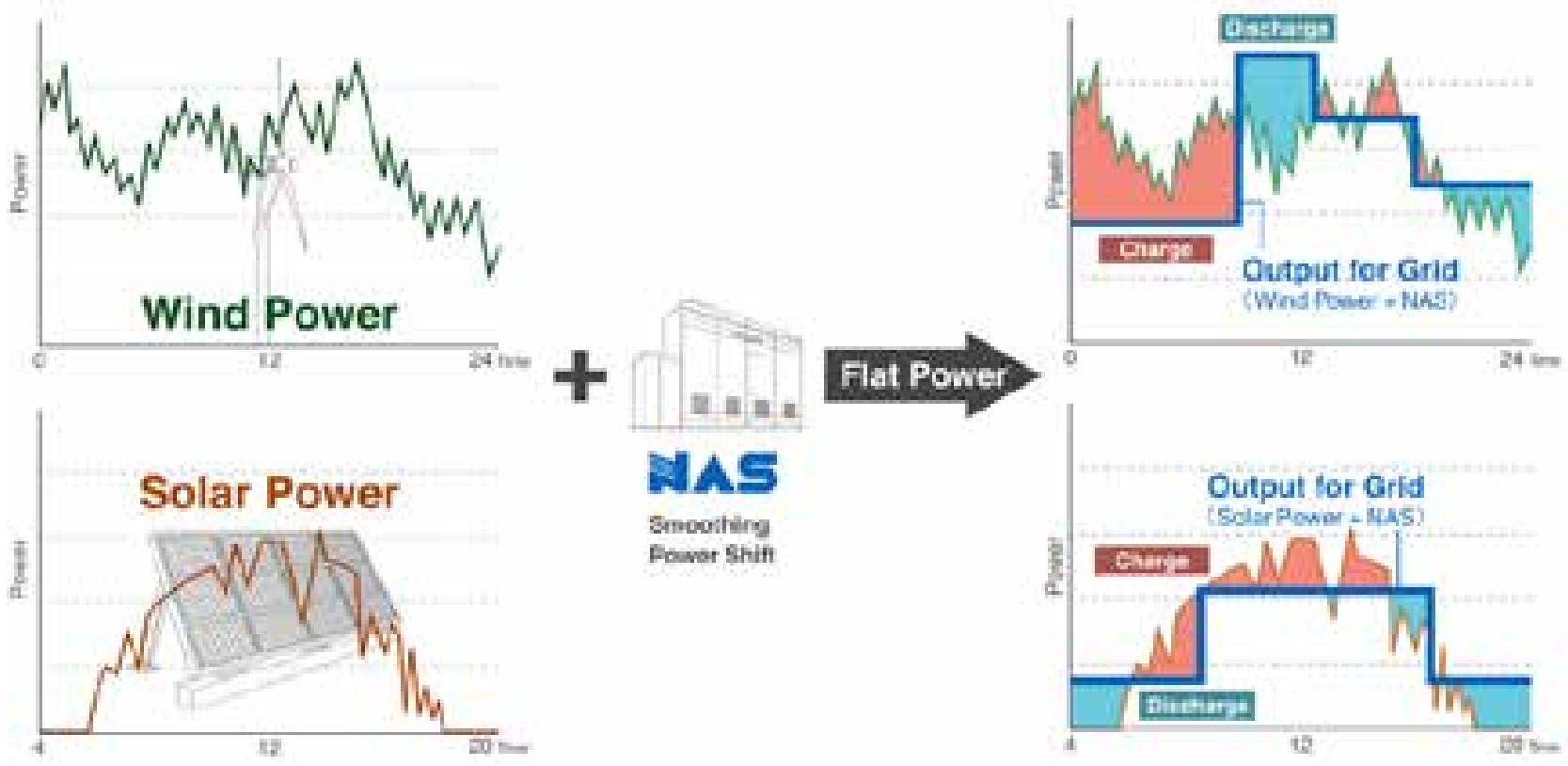

Fig. 2.1–3. Application of BESS (NaS BESS in this example) in power smoothing for intermittent renewable energy recourses like wind power and solar power [77].

In time shifting application (also known as load levelling or peak shaving), as the main focus in the first part of this thesis, the ESSs can provide the capability to store energy in off-peak periods when the energy price is low and then discharge that stored energy during peak demand for electricity consumption. This will add to the value of the stored energy both for the utility and for the customers since they can reduce their consumption from the grid. Time shifting has become applicable as a result of competitive electricity markets in countries where restructuring in their electricity industry has taken place [78], [79]. However, time shifting should be in compliance with the electricity market rules where the power is being sold. In our case, it is the Australian energy market operator (AEMO) which is responsible for managing the national electricity market (NEM) in Australia. A brief summary of the AEMO operation is given in the following section. Further details on AEMO can be found in [80]–[82].

### 2.1.1 An Introduction to the Australian NEM Operation

In the Australian NEM, a trading day is the time period from 04:01 a.m. to 04:00 a.m. of the next day. Each trading interval represents a half-hourly period which is divided into six 5-minute dispatch intervals. Therefore, a trading day is comprised of 48 trading intervals and 288 dispatch intervals, subsequently. The Australian NEM is operated as follows. Generators with a power capacity greater than 30 mega-watts (MW) are required to submit their offers for each trading interval in 10 price bands with an increasing order. These price bands correspond to 10 incremental energy quantities that generators are willing to sell. These offers have to be received by the AEMO one day ahead. On the trading day, AEMO runs an optimization program every five minutes to determine which generators to be dispatched and to meet demand based on their offers and some technical constraints. Therefore, the generators with the lowest price offers are allocated to dispatch first, but the dispatch price is the price of the most expensive generator dispatched on that 5-minute interval.

However, the actual price paid to the generators for their metered generation is the average of six dispatch prices for each half-hourly trading interval which is called spot price or regional reference price (RRP). Fig. 2.1–4 illustrates the dispatch procedure in the first half-hour of the trading day as an example taken from AEMO website.

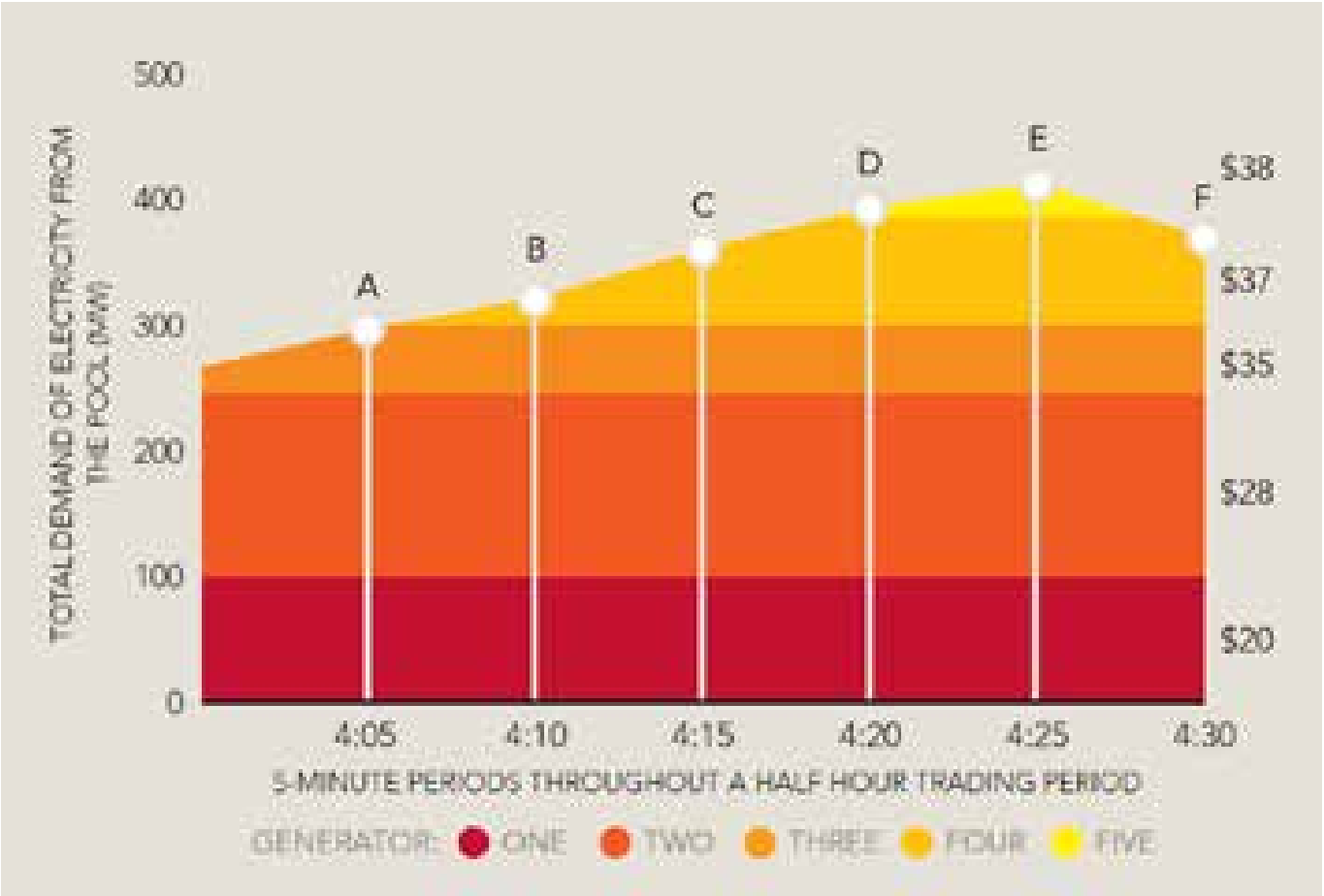

Fig. 2.1–4. AEMO 5-minute dispatch process for from 04:01 a.m. to 04:30 a.m. with five generators having different capacities (left axis) and bid offers (right axis), and six dispatch prices from A to F.

As can be seen in Fig. 2.1–4, the generators with cheaper offers are dispatched first (the red and dark orange horizontal bands), and more expensive ones (light orange through to yellow) are utilized to cover the peak load. Thus, the base generator that generates a flat 100 MW during the first half an hour with the bid offer of $20 per MW-hour (MWh) will be paid the spot price of $37 per MWh for each mega-watt of energy supplied which is the average of dispatch prices A to F as shown in Fig. 2.1–4, i.e., $37×100/2 = $1,850 for that half-hour. Generators have the opportunity to rebid their offers based on the latest changes in the market. This means that they can adjust the energy quantities by shifting them between different price bands without changing the price band levels just before each 5-minute dispatch.

## 2.2 Discrete-Time Model for Wind Power Integrated with BESS

The structure of the wind farm plus BESS in a grid-connected mode is illustrated in Fig. 2.2–1. In this structure, the amount of power sent to the grid ($P_g$), is managed by a control system using the generated wind power ($P_w$) and the stored energy in the BESS ($P_c$). The purpose of the BESS is to store excess energy from wind power in charging mode or add the required amount of energy to the wind power in discharging mode when needed.

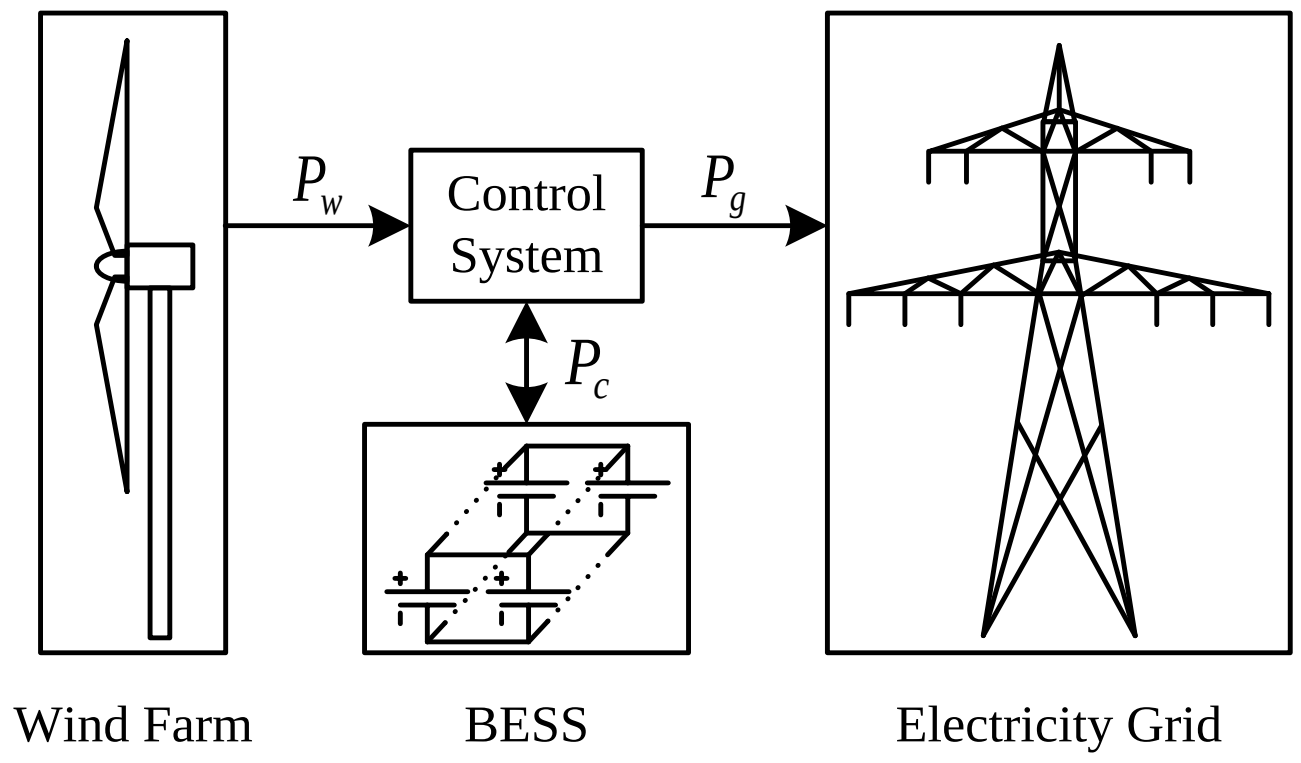

Fig. 2.2–1. Schematic diagram of the grid-connected wind farm plus BESS.

According to the 5-minute dispatch procedure in AEMO, the following discrete-time dynamic equations are considered to model the behavior of the wind power plus BESS in a grid-connected mode as shown in the system structure in Fig. 2.2–1, [83], [84],

$$\begin{aligned} P_g(k+1) &= P_c(k) + P_w(k), \\ E_b(k+1) &= E_b(k) - t_d P_c(k), \end{aligned} \tag{2.2–1}$$

where $P_g(k)$ is the generated power output sent to the grid; $E_b(k)$ is the available battery energy at time step $k$; $P_c(k)$ is the power control command which is the amount of the BESS charging power from wind (negative value) or discharging power (positive value) added to the wind power output; $P_w(k)$ is the real wind power, and $t_d$ is conversion coefficient (MW to MWh for each five minutes), i.e., $t_d = 5/60 = 1/12$ . This value is defined according to the AEMO power dispatch process performed for each five minutes. Therefore, each $k$ step is 5 min ($t = kT_s$, $k = 0, 1, 2, \ldots$, and $T_s = 5$ is the sampling time in minutes). Subsequently, the state space representation of the discrete-time dynamic equations of the wind power plus BESS can be written as follows,

$$\begin{aligned} \boldsymbol{x}(k+1) &= A\boldsymbol{x}(k) + B_1 u(k) + B_2 r(k), \\ y(k) &= C\boldsymbol{x}(k), \end{aligned} \tag{2.2–2}$$

where $\boldsymbol{x}(k)$ is defined as state vector; $u(k)$ is the control input; $r(k)$ is the second input which is not under control; $y(k)$ is the output; $A$ is the system matrix; $B_1$ is the control input matrix; $B_2$ is the second-input matrix, and $C$ is the output matrix, all given as the below,

$$\begin{aligned} &\boldsymbol{x}(k) = [x_1(k) \quad x_2(k)]^T = [P_g(k) \quad E_b(k)]^T, \\ &u(k) = P_c(k), \quad r(k) = P_w(k), \quad y(k) = P_g(k), \\ &A = \begin{bmatrix} 0 & 0 \\ 0 & 1 \end{bmatrix}, \quad B_1 = \begin{bmatrix} 1 \\ -t_d \end{bmatrix}, \quad B_2 = \begin{bmatrix} 1 \\ 0 \end{bmatrix}, \quad C = [1 \quad 0]. \end{aligned} \tag{2.2–3}$$

## 2.3 Discussion and Conclusion

The discrete-time model for wind power dispatch integrated with a BESS is provided in this chapter. The significance of the wind power contribution in the overall energy production around the world and the utilization of ESSs alongside the intermittent renewable energy sources like wind power were described. The procedure of how AEMO dispatches power and determines the electricity price was explained since the control system design should be in compliance with the electricity market rule of the region where the power is sold.

# Chapter 3

# Wind Power Dispatch Control Using BESS

The details of the discrete-time control system design for dispatching wind power with BESS is presented in this Chapter. At first, the overall objective and requirements in controlling wind power with BESS in the electricity market is established in Section 3.1. The structure of the designed discrete-time control system with its components is elaborated in Section 3.2. Section 3.3 provides a description on fuzzy logic systems and control as part of the overall discrete-time control system, followed by the design of the reference power signal generator using fuzzy decision-making system in Section 3.4. Then, an introduction to MPC is provided in 3.5 followed by the design of the discrete-time controller for wind power with BESS using MPC in Section 3.6. The design of a fuzzy logic controller for updating the reference power signal is given in Section 3.7. Section 3.8 covers the results of simulating the discrete-time control system for wind power integrated with a BESS under different scenarios of selling the generated power using the actual data of the wind power and electricity price. Finally, the chapter ends with a discussion in Section 3.9.

## 3.1 Control Objectives and Requirements

Based on the overall objective expressed in Chapter 1, The main goal of the first part of this thesis is to develop a new control scheme for making wind power generation more controllable using a BESS in time shifting application in the Australian NEM. To achieve this goal, the problems facing the management of the generated wind power integrated with BESS in the electricity market with aim of increasing the competitiveness and profitability of wind farms should be considered. Therefore, the first requirement for the control system is to provide a reference power signal that is able to change dynamically with the status of the electricity price and peak/off-peak periods of power consumption. The designed controller should then be able to calculate the proper amount of BESS charging or discharging power such that it enables the control system to produce power as close as possible to the reference power signal and inject it to the electricity grid. This should be conducted by taking into account the BESS constraints and the available wind energy. Having such a control system can provide the opportunity for the wind farm owners to trade within the competitive electricity market and increase their earning from the sale of the controlled power. It should be noted that the control scheme should be in compliance with the market rules where the power is being sold.

## 3.2 Control Configuration

### 3.2.1 Decision-Making System

For any closed-loop control system, the reference signal or desired output should be determined according to the desired requirements and control objectives of the system. As mentioned before, in time sifting application for wind power integrated with BESS, the control system is required to follow a reference power signal that reflects the changes in the electricity price and peak/off-peak periods during the day. Thus, a decision-making system should be designed for generating the reference tracking signal online which is a function of electricity price variations and time intervals during the day.

The basic idea behind designing such a decision-making system is quite straightforward. To increase the earning received from the electricity market, wind energy should be stored in the BESS during low prices and time periods at which they are usually supposed to have low demand for electricity consumption. This stored energy would be more financially valuable to be discharged in addition to wind energy generated at peak times when the electricity price is significantly high. Such an increase in the power generation at high prices and peak times up to its maximum power capacity and vice versa is allowable due to the rebidding opportunity provided by the AEMO before the actual dispatch time, as explained in Section 2.1.1. This means that the real dispatch price signal determined by the AEMO just before each 5-min dispatch can be used with time periods as inputs to the decision-making system to generate the proper reference power signal. Therefore, we use fuzzy logic to design the proposed decision-making system since they are well-known as a suitable and applicable option for this purpose. More details about fuzzy logic systems and control are provided in Section 3.3.

### 3.2.2 Control System Structure

After defining the discrete-time state space model for wind power with BESS in Section 2.2 and establishing control objectives for controlling the generated power in the grid-connected mode, the overall control system structure is given in Fig. 3.2–1. As can be seen, the proposed structure for wind power control with BESS consists of three main blocks. Reference signal generator is responsible for supplying reference power trajectory profile online with 5-minite dispatch price and time of the day (peak/off-peak periods) as inputs using the decision-making system which is designed via fuzzy logic system. However, as it will be explained in Section 3.7, this reference power should be updated in accordance with the current state of charge (SoC) of the BESS to prevent large tracking error. This correction is performed using a fuzzy logic controller (FLC) that generates the correction power using the battery SoC and its rate of change [84].

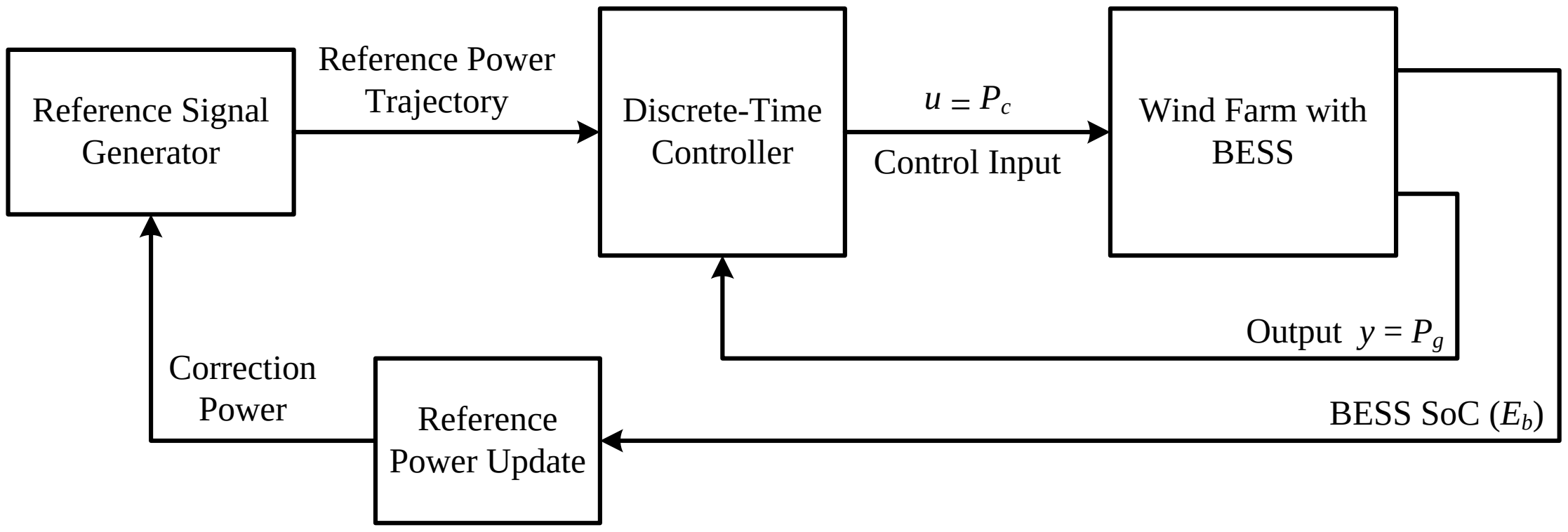

Fig. 3.2–1. The structure of the control system for wind power with BESS.

The main control algorithm is implemented in the discrete-time controller to find the proper charging/discharging power of the BESS as the control input. It is designed based on the discrete-time model obtained in (2.2–2) and (2.2–3). Therefore, considering the main objective and control requirements described earlier, the discrete-time controller is designed using model predictive control (MPC) in order to deliver the desired control action in managing the generated power from wind farm and BESS. The main advantage of MPC compared to conventional controllers in terms of practicality is that the ability to handle process constraints since they are included in the formulation of MPC. It should be mentioned that in the wind farm integrated with a BESS, the main constrains having a great impact on system behavior are the maximum BESS energy and the power capacity. It is important that the control system maintains the BESS constrains within their permitted range to have an optimal operation of the BESS as well as avoiding damages that are caused by overcharging or depleting the battery. Other benefits of the MPC include online optimization and being able to be implemented in any digital computer. In the following sections, the details of designing each part of the control system structure for wind power dispatch with BESS are given.

## 3.3 Fuzzy Logic Systems

### 3.3.1 Overview of Fuzzy Logic Systems

Since the introduction of fuzzy theory by Prof. L. A. Zadeh in 1965, the application of fuzzy logic and control has greatly developed in different fields of study [85], [86]. In a simple language, fuzzy systems are knowledge-based or rule-based systems where the linguistic rules, known as IF-THEN rules, and inference logic are utilized to create a mapping from some input variables to some output variables. The basic configuration of a fuzzy system commonly used in engineering field is shown in Fig. 3.3–1 which comprises of four main parts: fuzzifier, defuzzifier, fuzzy rule base, and fuzzy inference engine. The fuzzifier transforms the real-valued (crisp) input variables into their corresponding fuzzy sets. The defuzzifier, on the other hand, transforms the fuzzy sets into the real-valued output variables. The fuzzy rule base represents the collection of fuzzy IF-THEN rules. The fuzzy inference engine combines these fuzzy rules into a mapping from input fuzzy sets to the output fuzzy sets based on the fuzzy logic principles. One of the benefits of fuzzy logic systems is that they can be designed for both continuous-time and discrete-time systems. In addition, they can be easily implemented in digital computers with fast response time. In the next section, more details on the fuzzy logic systems and their operation are provided.

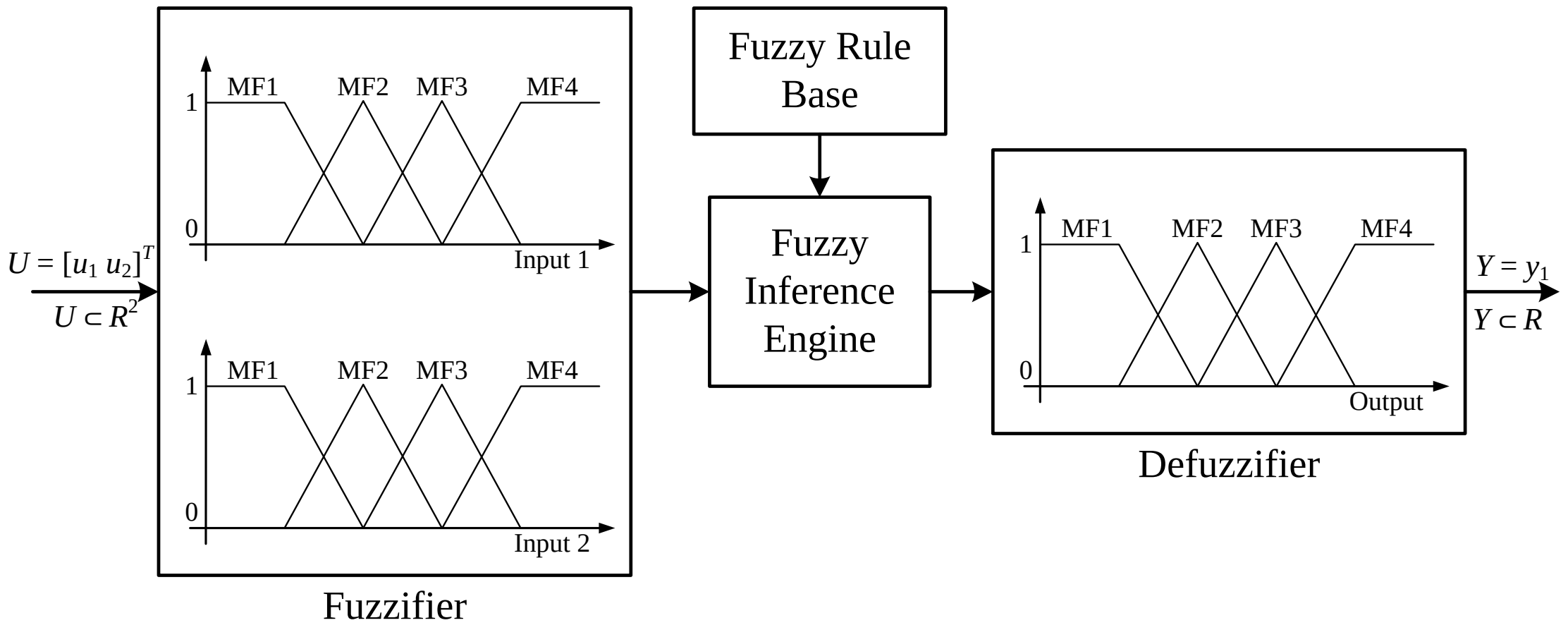

Fig. 3.3–1. The basic structure of a fuzzy logic system with fuzzification and defuzzification (two-input single-output system with inputs and output belong to the real numbers set, i.e., $U \in R^2$, $Y \in R$).

## 3.3.2 Basics of a Fuzzy System Operation

The major components of a fuzzy system are fuzzy sets, membership functions (MFs), fuzzy logic operations, and IF-THEN rules. A fuzzy set is a set defined for a variable or *universe of discourse* that describes the degree of membership of each value of the variable in the set. The degree of membership is determined by a membership function which varies from zero (not belong to the set) to one (full membership to the set) and can have different geometric forms. The most common MFs are trapezoidal, Gaussian distribution function, sigmoid curve, piecewise linear function, and triangle function (as special form of trapezoidal function). Mathematically speaking, a fuzzy set is an extension of a classical set. If $U$ is the universe of discourse defined as a subset of real number set, i.e., $U \subset R$, and its elements are denoted by $x$, then a fuzzy set $A$ in $X$ is defined as a set of ordered pairs as follows,

$$A = \{x, \mu_A(x) \mid x \in U \subseteq R\},$$
$$\mu_A(x): U \rightarrow [0,1], \qquad (3.3\text{–}1)$$

where $\mu_A(x)$ is called the membership function of $x$ in $A$ that maps each element of $U$ to a membership value between the closed interval of (0, 1). The process of mapping a real-valued input variable to a fuzzy set with its membership degree is called *fuzzification*. For example, the electricity price, as the universe of discourse and a linguistic variable, can be described by three fuzzy sets: Low, Medium, and High. Each of these sets is represented by a MF covering a specific range of price for electricity as shown in Fig. 3.3–2. Due to different human interpretation in characterizing the fuzzy variables, fuzzy sets can have intersection, i.e., any element in the universe of discourse can belong to two neighboring MFs. As can be seen in the example of Fig. 3.3–2, the electricity price of $p$ = 16 $/MWh belongs to the Medium fuzzy set with membership degree/value of 0.8 ($\mu_M$ $(p)$ = 0.8) as well as the Low fuzzy set with membership degree/value of 0.2 ($\mu_L$ $(p)$ = 0.2). Later on, we will see that this results in activation of more than one IF-THEN rule in the process of fuzzy inference, i.e., more than one rule has true value.

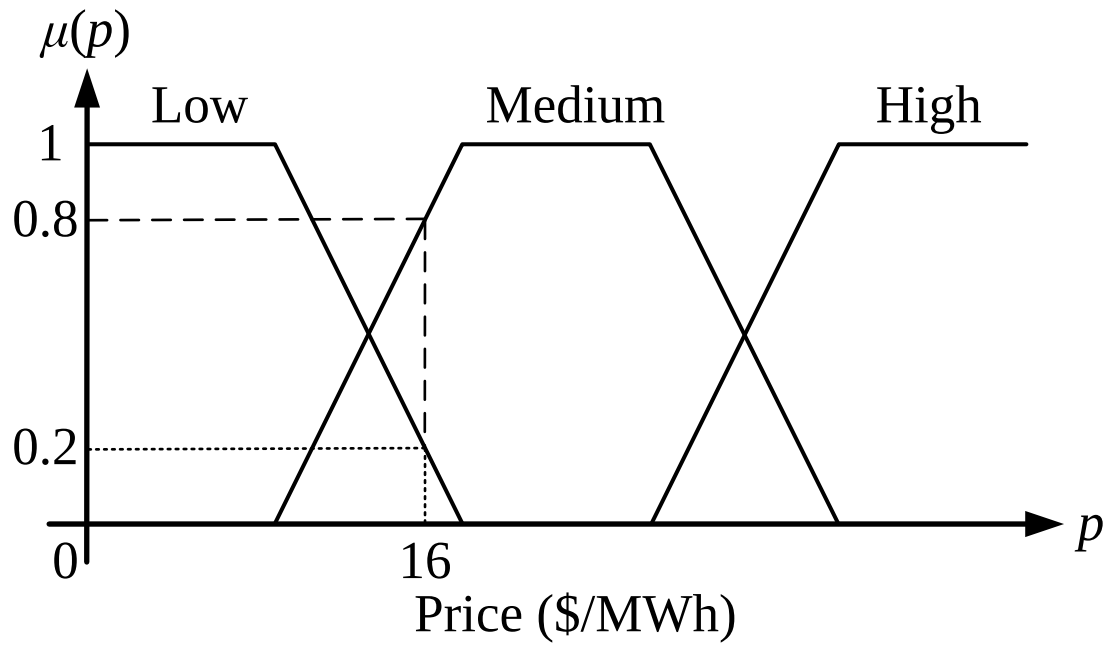

Fig. 3.3–2. The fuzzy sets for electricity price with three MFs: Low, Medium, and High.

The basic fuzzy operations are similar to the standard classical set operations and Boolean logic including intersect/conjunction (AND), union/disjunction (OR), and complement (NOT) operators. The difference is that the “truth” value of any statement in fuzzy logic is a matter of degree defined as a real number in the interval of [0, 1] rather than distinct values of zero and one. Therefore, assuming that $A$ and $B$ are fuzzy sets defined in the same universe of discourse $U$, the basic operations on fuzzy set are defined as follows,

– $A$ and $B$ are equal if and only if $\mu_A(x) = \mu_B(x)$ for all $x \in U$.
– The complement of $A$ is a fuzzy set $\bar{A}$ in $U$ whose MF is given by

$$\mu_{\bar{A}}(x) = 1 - \mu_A(x). \tag{3.3–2}$$

– The intersection of $A$ and $B$ is a fuzzy set in $U$ denoted by $A \cap B$ whose MF is given by

$$\mu_{A \cap B}(x) = \min\,[\mu_A(x), \mu_B(x)]. \tag{3.3–3}$$

– The union of $A$ and $B$ is a fuzzy set in $U$ denoted by $A \cup B$ whose MF is given by

$$\mu_{A \cup B}(x) = \max\,[\mu_A(x), \mu_B(x)]. \tag{3.3–4}$$

where min[.] and max[.] are minimum and maximum functions, respectively. These operations can be customized in a way to vary the gain on the function so that it can be very restrictive or very permissive. Moreover, these fuzzy operations can be performed on two fuzzy sets from two different variables.

The fuzzy IF-THEN rules are at the heart of a fuzzy system. These IF-THEN rules are simple IF-THEN statements obtained from human experts or based on domain knowledge, and they are expressed in the following form,

IF < *Fuzzy Proposition*1 >, THEN < *Fuzzy Proposition*2 >, (3.3–5)

where the IF-part of the rule is called the *antecedent* or premise, and the THEN-part of the rule is called the *consequent* or conclusion. The fuzzy proposition can be a single proposition or a compound proposition. A single fuzzy proposition is a single statement as in the form of < *x in A* > where $x$ is the linguistic variable and $A$ is a linguistic value of $x$ given by the fuzzy set $A$ defined in the physical domain of $x$. A compound fuzzy proposition is a composition of single fuzzy propositions using the connectives AND, OR, and NOT which represent fuzzy intersection, fuzzy union, and fuzzy complement, respectively, also known as *fuzzy relation*. For example, if $p$ is the dispatch price for the electricity, $t$ is the time of the day, and $y_r$ is the reference power signal, then the following can be considered as an IF-THEN rule with single and compound fuzzy propositions,

$$\text{IF } p \text{ is } L \text{ AND } t \text{ is } EP, \text{ THEN } y_r \text{ is } M, \tag{3.3–6}$$

where L, EP, and M denote fuzzy sets Low for dispatch price, Evening Peak for time of the day, and Medium for reference power signal, respectively. In general, the input to an IF-THEN rule is the current value of the input variables (dispatch price $p$ and time of the day $t$ in (3.3–6)) and the output is a fuzzy set obtained by using the result of the IF-part and the THEN-part (Medium reference power in (3.3–6)). The process of calculating the final value for the reference power signal ($y_r$) from the fuzzy set obtained from IF-THEN rule is known as *defuzzification* which will be explained later in this section.

Interpreting an IF-THEN rule involves determining the MF and membership degree of fuzzy relations in IF-part and THEN-parts. This is done in three steps. First, the crisp value of the inputs should be fuzzified to determine the corresponding fuzzy sets and membership degrees. Then, IF-part or antecedent should be evaluated by applying any necessary fuzzy operators on the fuzzy relation based on (3.3–2)–(3.3–4) to find the degree to which the fuzzy relation is true. Finally, the truth value of the THEN-part or consequent should be obtained using the result of the IF-part evaluation known as *implication*. There are several implication methods in the literature, but the most widely used one in the fuzzy systems and control is the Mamdani Implication which is defined as below [85].

***Def. 1.*** *Assume that fuzzy proposition1 ($FP_1$) and Fuzzy proposition2 ($FP_2$) in* (3.3–5) *are fuzzy relations defined in* $U = U_1 \times U_2 \times \ldots \times U_n \subset R^n$ *and* $V \subset R$, *respectively (multi-input single-output system), and* $\mathbf{x} = [x_1\ x_2\ \ldots\ x_n]^T$ *and y are linguistic variables in U and V, respectively.*[5] *The IF-THEM rule given in* (3.3–5) *is interpreted as a fuzzy relation $Q_{MM}$ in U×V with the following MF or fuzzy set,*

$$\mu_{Q_{MM}}(\mathbf{x},y) = \min\,[\mu_{FP_1}(\mathbf{x}),\, \mu_{FP_2}(y)], \tag{3.3–7}$$

[5] Bold notation is used to represent vector-space variables. It should also be noted that multi-input multi output fuzzy systems could be decomposed into a collection of single output systems.

*where* $\mu_{FP_1}(\mathbf{x})$ *and* $\mu_{FP_2}(y)$ *are the fuzzy sets obtained for the IF-part and THEN-part using fuzzy operations and* $\mu_{QMM}(\mathbf{x},y)$ *is the fuzzy set associated with the IF-THEN rule*.

Now that the major components of a fuzzy system are explained, we can move on to the procedure of fuzzy inference and defuzzification in generating the output value from the inputs in a fuzzy system. The basic structure of the fuzzy inference process is shown in Fig. 3.3–3. As mentioned before, fuzzy inference is the process of formulating the mapping from a given input to an output (or multiple inputs to multiple outputs) using fuzzy logic. The process of fuzzy inference involves all of the components that are described before including fuzzy sets, MFs, fuzzy logical operations, and IF-THEN rules evaluation. In a simple language, fuzzy inference process comprises of five steps:

1. Fuzzification of the input variables to find the corresponding fuzzy sets and their membership degrees ($\mu_A(x)$).
2. Application of fuzzy operators in the IF-part and THEN-part of each activated rule to find the fuzzy set of each fuzzy relation ($\mu_{FP_1}(\mathbf{x})$ and $\mu_{FP_2}(y)$).
3. Application of implication method (3.3–7) on each activated rule to find the fuzzy set associated with those IF-THEN rules ($\mu_{QMM}(\mathbf{x},y)$).
4. Combination/Aggregation of the results of the implications to find a fuzzy set for the output variable ($\mu_{A_g}(z)$).
5. Defuzzification of the resulting output fuzzy set to assign a real value to the output variable.

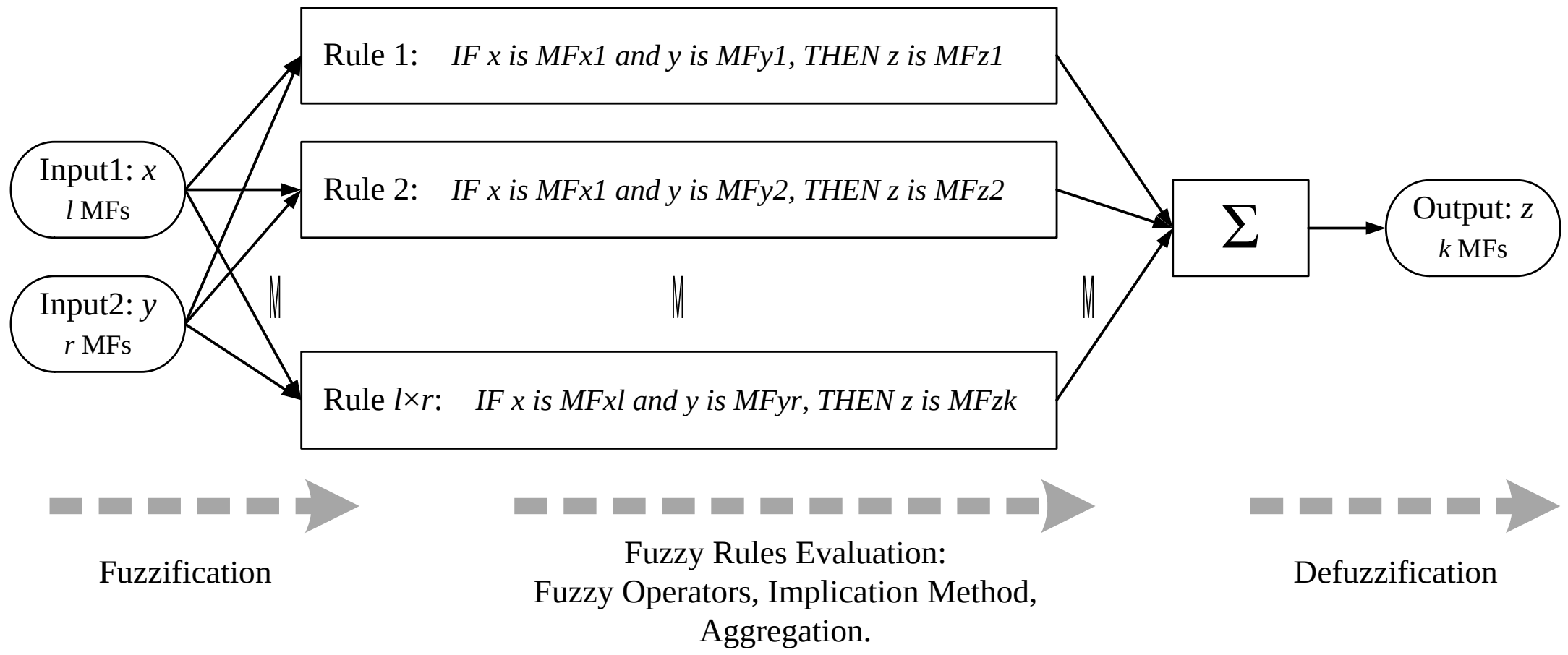

Fig. 3.3–3. The basic structure of the fuzzy inference system for a two-input one-output system.

The first three steps have already been explained earlier and only steps four and five need to be elaborated. Because it is always possible to have more than one IF-THEN rule as true or activated after the fuzzificaton process for specific values of inputs (refer to Fig. 3.3–2), the results of the implications on the activated rules must be combined in some logical manner to find a proper fuzzy set for the output. *Aggregation* is the process by which the fuzzy sets that represent the outputs of each rule are combined into a single fuzzy set. The input of the aggregation process is the list of truncated fuzzy sets returned by the implication process for each rule. The output of the aggregation process is one fuzzy set for each output variable. Since the aggregation methods are always commutative, the order in which the rules are executed is not important. The common aggregation method is the use of maximum function in combining the fuzzy sets coming from implication process which is illustrated in Fig. 3.3–4.

As mentioned before, defuzification is the process of assigning a single value to the output from the resulting aggregated fuzzy set. However, the final fuzzy set covers a range of output values. The most popular defuzzification method is the *centroid* method, which returns the center of the area under the MF curve representing the final fuzzy set as the proper value for the output. This method is known as the center of gravity defuzzifier (CoG). If the MF representing the final fuzzy set form aggregation is $\mu_{A_g}(z)$, the final output value using centroid method is obtained as follows,

$$z = \text{CoG} = \frac{\int_a^b z\mu_{A_g}(z)dz}{\int_a^b \mu_{A_g}(z)dz}, \tag{3.3–8}$$

where *a* and *b* are the output range covered by $\mu_{A_g}(z)$.

The summary of the fuzzy inference process is depicted graphically in Fig. 3.3–4 for the same example of choosing the reference power signal ($y_r$) as the output uisng electricity dispatch price ($p$) and time of the day ($t$) as inputs. As can be seen, based on the crisp values of dispatch price ($p$ = 16 \$/MWh similar to Fig. 3.3–2) and time of the day ($t$ = 16:00 hours), two IF-THEN rules form the list of fuzzy rule base are activated (the full set of IF-THEN rules and fuzzy sets for reference signal generator will be given in Section 3.4). Since the fuzzy relations in both IF-parts are defined using AND, the minimum function is used as the fuzzy operator, and subsequently the implication method produces the truncated fuzzy sets for each IF-THEN rules, $\mu_{QMM1}(p,t)$ and $\mu_{QMM2}(p,t)$, respectively, using the Medium and High fuzzy sets defined for the output, $\mu_M(y_r)$ and $\mu_H(y_r)$, respectively. The aggregation process combines the truncated fuzzy sets using their maximum values as shown in Fig. 3.3–4 resulting in $\mu_{Ag}(y_r)$. And finally, the defuzzification process using CoG method in (3.3–8) generates the crisp value for the reference power signal as $y_r$ = 103.45 MW.

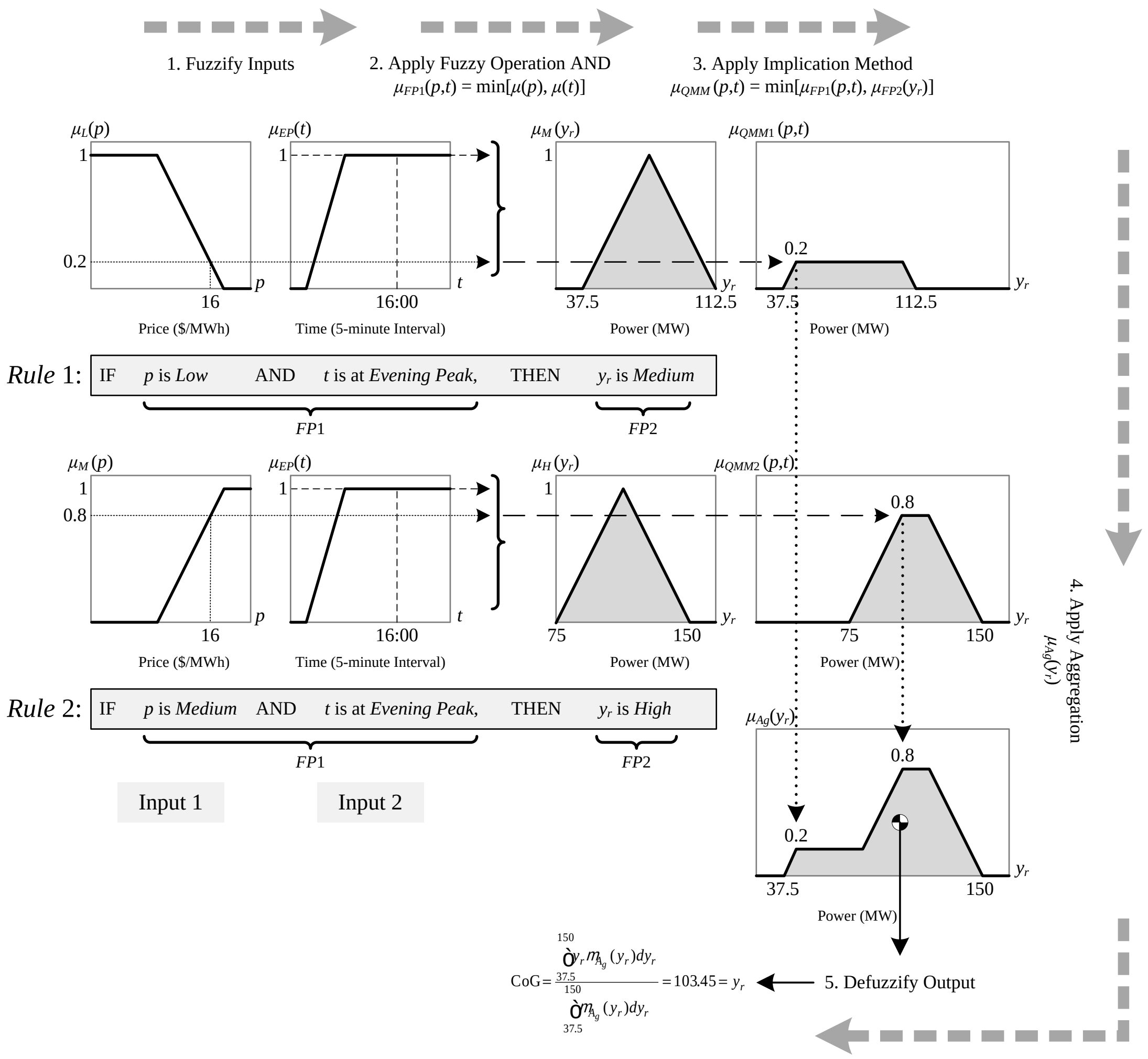

Fig. 3.3–4. The diagram of the fuzzy inference process for a fuzzy system with two inputs (dispatch price and time of the day), one output, and two activated IF-THEN rules based on the values of the inputs.

## 3.4 Reference Signal Generation Using Fuzzy System

As we explained in Section 3.2.1, the decision-making system based on fuzzy logic is designed for generating a tracking reference power signal online which is a function of electricity price variations and time intervals during the day. The reference signal generator is given as follows,

$$y_r(k) = \text{FD}\ (p(k), t(k)), \tag{3.4–1}$$

where $y_r(k)$ is the reference power signal, FD(.) is the fuzzy decision-making system, $p(k)$ and $t(k)$ are the electricity price in dispatch interval $k$ and time of the day as an integer at each time step, respectively (one trading day is divided into 288 of 5-minute intervals where 4:00 am is zero and 4:00 am of the next day is 288). Based on the information given in the New South Wales (NSW) government resources and energy website [87], energy cost periods, including electricity, are given for weekdays and weekends/public holidays for both summer and winter, as shown in Fig. 3.4–1 (although the electricity consumption is different in summer and winter, but the peak/shoulder/off-peak periods are the same for both seasons in NSW region based on [87]). These periods are mostly considered by all electricity retailers in NSW region as “time-of-use-pricing.” This method of pricing persuades costumers to use smart meters in order to coordinate their consumption with these periods and reduce their energy costs. On the other hand, the retailers charge customers more during these periods. According to this fact, the reference power signal should be determined separately for weekdays and weekends/public holidays. Therefore, two different fuzzy sets for time of the day and two sets of fuzzy decision-making rules are designed in such a way that on weekdays, the first fuzzy rules with the corresponding time MFs are used and on weekends/public holidays, it switches to the second one. The fuzzy rule bases are defined according to (3.3–5) which are summarized in Table 3.4–1 and Table 3.4–2 for weekdays and weekends/public holidays, respectively, and they comply with the following form,

$$\begin{gathered}\text{IF } p(k) \text{ is } MF_i \text{ AND } t(k) \text{ is } MF_j, \text{ THEN } y_r(k) \text{ is } MF_l\,, \\ \textit{for } i = 1, 2, 3,\ \ j = 1, 2,.., 6, (\textit{or } j = 1, 2, 3) \textit{ and } l = 1, 2,\dots 5.\end{gathered} \tag{3.4–2}$$

where the abbreviations in Table 3.4–1 and Table 3.4–2 denote the fuzzy sets for dispatch price and time of the day which are shown in Fig. 3.4–2, and they are as follows: MOP is morning off-peak, MP is morning peak, DS is day shoulder, EP is evening peak, ES is evening shoulder, NOP is night off-peak, L, M, and H are low, medium, and high, respectively, VL is very low, and VH is very high.

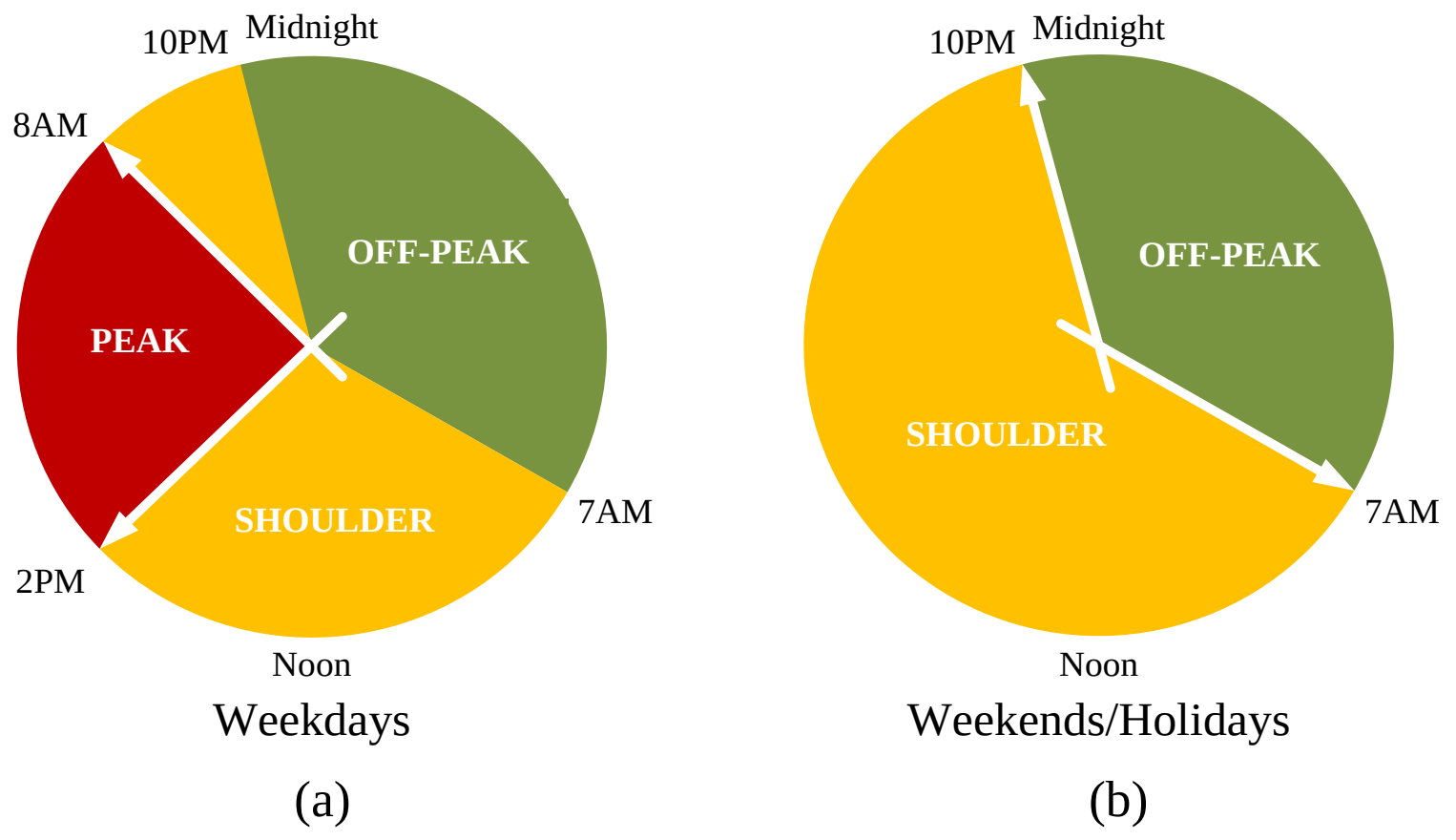

Fig. 3.4–1. Three energy cost periods in NSW region, (a) Weekdays energy cost periods, (b) Weekends/holidays energy cost periods [87].

Table 3.4–1. Fuzzy Decision-Making Rules (Weekdays)

| $y_r(k)$ | | $t(k)$ | | | | |
|---|---|---|---|---|---|---|
| | | MOP | DS | EP | ES | NOP |
| $p(k)$ | L | VL | L | M | L | VL |
| | M | L | M | H | M | L |
| | H | M | H | VH | H | M |

Table 3.4–2. Fuzzy Decision-Making Rules (Weekends/Public Holidays)

| $y_r(k)$ | | $t(k)$ | | |
|---|---|---|---|---|
| | | MOP | DS | NOP |
| $p(k)$ | L | VL | L | VL |
| | M | L | M | L |
| | H | M | H | M |

The dispatch price range shown in Fig. 3.4–2(a) is chosen based on annual average RRP and peak price available at AEMO database [88]. The bound for each MF is determined according to some experts' viewpoint in the Australian NEM and also some useful points given in [80]. Different bounds for the time of the day MFs are selected based on the residential electricity consumption periods for weekdays in Fig. 3.4–2($b_1$) and for weekends/public holidays in Fig. 3.4–2($b_2$) [87]. These intervals can be updated readily if the correlation between the wholesale market and the residential pricing is shown to be inaccurate. The range of reference power signal in Fig. 3.4–2(c) is selected according to the maximum capacity of the Woolnorth wind farm (where the actual data was collected for simulations) which is 140 MW, and the MF bounds are chosen based on some wind farm experts. It is also tested that the maximum value of the reference power signal will be exactly 140 MW if the maximum range is 150 MW based on the centroid defuzzification method, defined fuzzy rules, and the input MFs as shown in Fig. 3.4–2 (c). For fuzzy rule definition, the trial-and-error approach and the idea of time shifting application are applied to get the best results [85], [86]. The reason for choosing trapezoidal MFs is that the calculation of the output value using the centroid defuzzification method in (3.3–8) will be simple and fast as it deals with the integration of polynomial functions if trapezoidal MFs are used. This is considered as an advantage for easy implementation in any processor and having fast response which is important for online control applications [85], [86].

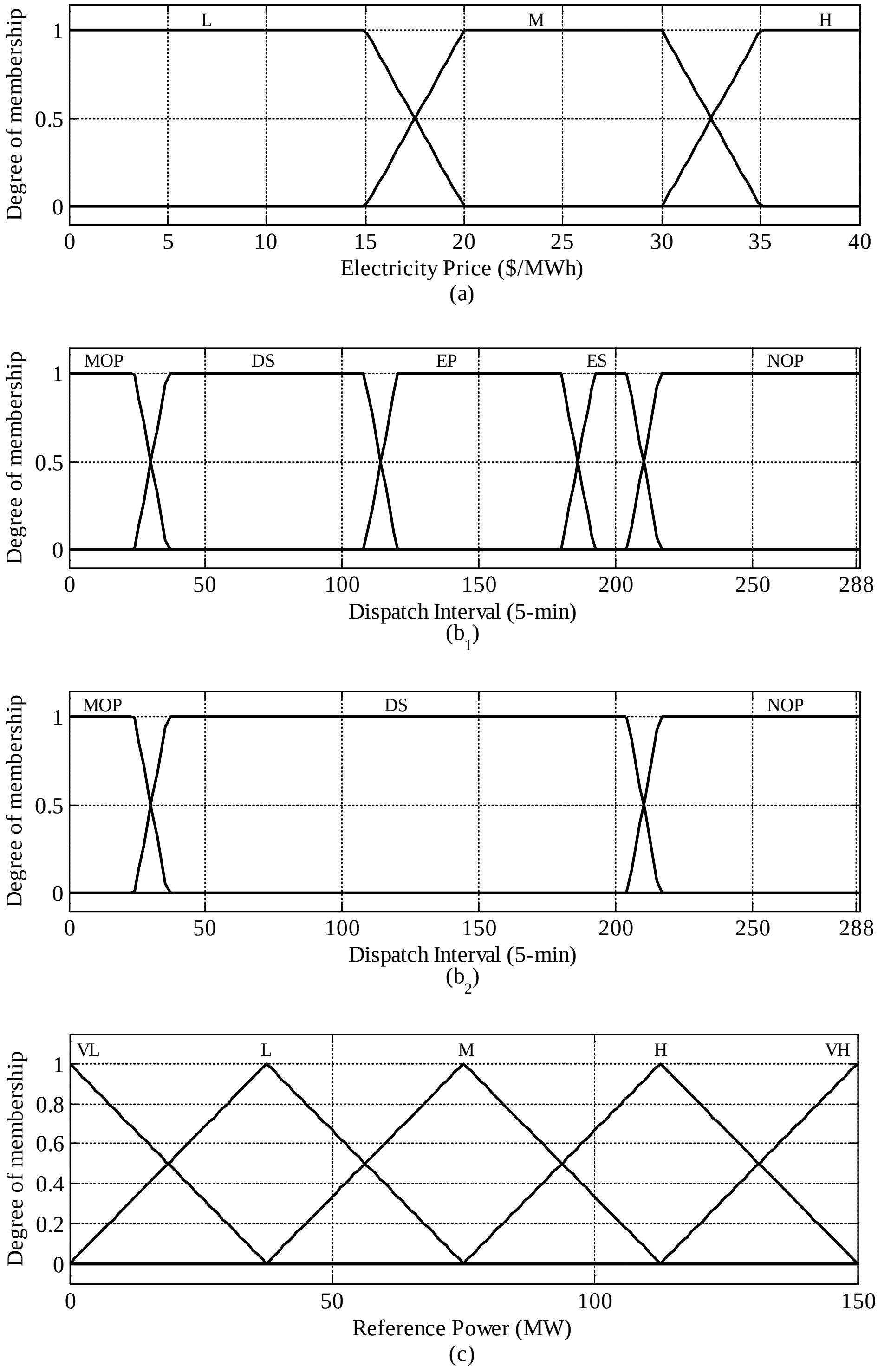

Fig. 3.4–2. Fuzzy sets and MFs for two inputs and the output of the fuzzy decision-making system. (a) Price MFs. ($b_1$) Time MFs for weekdays, ($b_2$) Time MFs for weekends/public holidays, (c) Reference power signal MFs.

# 3.5 Model Predictive Control

## 3.5.1 Overview of MPC

The MPC is actually based on the solution of an online optimal control problem where a receding horizon approach is applied in such a way that for any current state vector $\boldsymbol{x}(k)$ at time step $k$, an optimal control problem is solved over finite future intervals taking into account the current and future constraints on the control input, output, and the states. The MPC algorithm calculates a sequence of manipulated variables (control inputs) in order to optimize the future behavior of the control system. The first value of this optimal sequence is applied to the process. The procedure is then repeated at time $k + 1$ using the current measurements. Fig. 3.5–1 illustrates the concept of the predictive controller at time step $k$ tracking a constant reference signal after operating for some sampling instances.

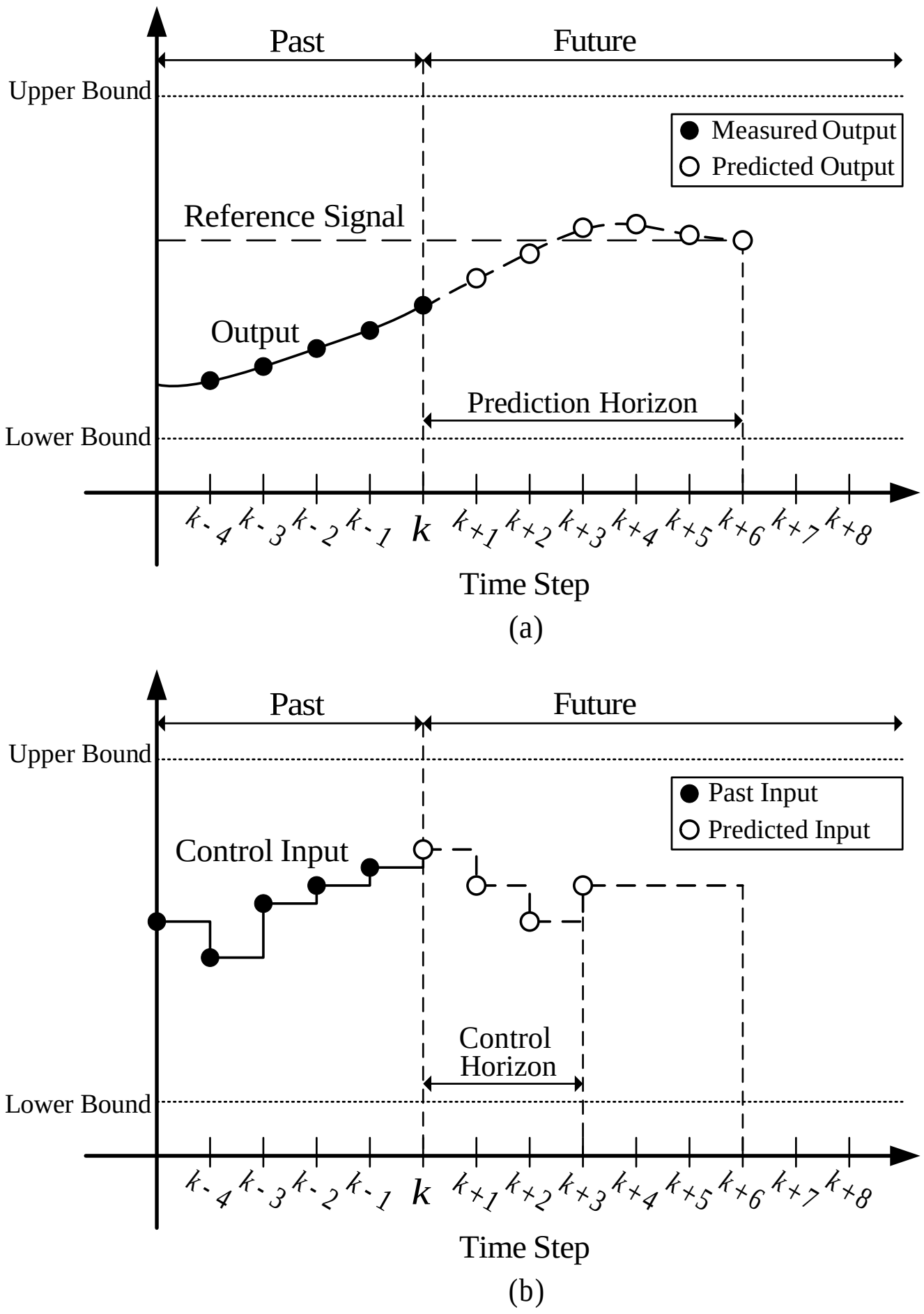

Fig. 3.5–1. The basic idea of MPC. (a) The past and future behavior of the output, (b) The past and the future values of the control signal.

The main advantage of MPC compared to conventional controllers is that the process constraints can be explicitly taken into account in the controller design, naming input voltage constraints of the crane actuators and overhead crane workspace limits. Discrete-time nature of MPC also provides easy implementation using digital computers, and not to mention, it offers online optimization which is quite useful in real-time control applications. In addition, feedforward disturbance compensation can be easily integrated into MPC formulation [89], [90].

### 3.5.2 Basic Formulation

Consider the nominal discrete-time linear time-invariant system as follows,

$$\begin{aligned} \boldsymbol{x}(k+1) &= A\boldsymbol{x}(k) + B\boldsymbol{u}(k) + W\boldsymbol{d}(k), \qquad \boldsymbol{x}(0) = \boldsymbol{x}_0, \\ \boldsymbol{y}(k) &= C\boldsymbol{x}(k), \end{aligned} \tag{3.5–1}$$

where $\boldsymbol{x} \in R^n$ is the state vector; $\boldsymbol{u} \in R^p$ is the control input; $\boldsymbol{y} \in R^m$ is the measured/controlled output; $\boldsymbol{d} \in R^p$ is the input disturbance; the $n \times n$ matrix $A$, $n \times p$ matrix $B$, $n \times p$ matrix $W$, and $m \times n$ matrix $C$ are system matrix, control input matrix, input disturbance matrix, and output matrix, respectively; $\boldsymbol{x}_0$ is the initial conditions, and $k$ is the time step defined as $t = kT_s$ for $k = 0, 1, 2, \ldots$, with $T_s$ as the sampling time. The procedure of calculating the optimum control input at current time step $k$ is conducted in two phases: prediction and optimization.

In prediction phase, the current value of the system states $\boldsymbol{x}(k)$ or measured output $\boldsymbol{y}(k)$ (solid circle at time step $k$ in Fig. 3.5–1(a)), the dynamic model ((3.5–1)), and the previous value of the control input $\boldsymbol{u}(k-1)$ are required in order to make predictions of the future values of the output (empty circles in Fig. 3.5–1(a)). These predictions are made over a finite horizon known as prediction horizon. The future values of the reference signal (a fixed set-point or a predetermined trajectory), any measured/known disturbances, and system constraints are also needed to be calculated over the same prediction horizon. Since not all the system states are available, a state observer is often used to estimate the values of the states from the current and previous output measurements, previous control inputs and the dynamic model of the system. Then, the state estimates can be used to obtain the predicted outputs.

Once the predictions are made, the current and future values of the control input is computed in the sense of optimal control problem over a finite horizon known as control horizon in optimization phase. The optimum control inputs are determined by minimizing a cost function, normally defined in a quadratic form, that penalizes deviations of the predicted outputs from future values of the reference trajectory and the changes in the control input $\Delta\boldsymbol{u}$ rather than the control input itself, i.e., $\Delta\boldsymbol{u}(k) = \boldsymbol{u}(k) - \boldsymbol{u}(k-1)$. The main reason is that this formulation coincides with that used in the majority of the predictive control literature (as we will use in Part II for controlling the overhead crane), but it should be noted that the input constraints have to be written in terms of $\Delta\boldsymbol{u}$ in case of using this formulation. However, we can still use similar formulation to penalize $\boldsymbol{u}(k)$ in the cost function if needed (as we will use in Part I for controlling wind power dispatch with BESS). Therefore, the cost function is defined as follows [89],

$$V(k) = \sum_{i=1}^{H_p} \left\| \hat{\boldsymbol{y}}(k+i\,|\,k) - \boldsymbol{y}_{ref}(k+i\,|\,k) \right\|^2_{Q(i)} + \sum_{i=0}^{H_u-1} \left\| \Delta\hat{\boldsymbol{u}}(k+i\,|\,k) \right\|^2_{R(i)}, \tag{3.5–2}$$

or

$$V(k) = \sum_{i=1}^{H_p} \left\| \hat{\boldsymbol{y}}(k+i\,|\,k) - \boldsymbol{y}_{ref}(k+i\,|\,k) \right\|^2_{Q(i)} + \sum_{i=0}^{H_u-1} \left\| \hat{\boldsymbol{u}}(k+i\,|\,k) \right\|^2_{R(i)}, \tag{3.5–3}$$

subject to system equations, state, input, and output constrains as below,

$$\begin{aligned} &\boldsymbol{x}_{min} \le \boldsymbol{x}(k) \le \boldsymbol{x}_{max} && k = 1, 2, \ldots, H_p, \\ &\boldsymbol{y}_{min} \le \boldsymbol{y}(k) \le \boldsymbol{y}_{max} && k = 1, 2, \ldots, H_p, \\ &\boldsymbol{u}_{min} \le \boldsymbol{u}(k) \le \boldsymbol{u}_{max} && k = 0, 1, \ldots, H_u - 1, \end{aligned} \tag{3.5–4}$$

where $H_p$ is prediction horizon; $H_u$ is control horizon with $H_p \geq H_u$ (the control horizon does not necessarily need to be equal to the prediction horizon); $\hat{\boldsymbol{y}}(k+i\,|\,k)$ is prediction of output at time $k+i$ made at time $k$; $\boldsymbol{y}_{ref}(k+i\,|\,k)$ is future values of reference or desired trajectories available at time $k$; $\hat{\boldsymbol{u}}(k+i\,|\,k)$ and $\Delta\hat{\boldsymbol{u}}(k+i\,|\,k)$ are the prediction of future control input and its changes made at time $k$, respectively, i.e., $\Delta\hat{\boldsymbol{u}}(k+i\,|\,k) =$

$\hat{\mathbf{u}}(k + i \mid k) - \hat{\mathbf{u}}(k + i - 1 \mid k)$; $\mathbf{x}_{min}$ and $\mathbf{x}_{max}$ are the vectors of lower and upper bounds on the states, respectively; $\mathbf{y}_{min}$ and $\mathbf{y}_{max}$ are the vectors of lower and upper bounds on the output, respectively; $\mathbf{u}_{min}$ and $\mathbf{u}_{max}$ are the vectors of lower and upper bounds on the input, respectively; $Q(i)$ and $R(i)$ are square diagonal weighting matrices for tracking error and control input changes, respectively. Here, $\|x\|_Q^2$ is a notation for quadratic form $x^TQx$ as the square of a "weighted norm" since we have $x^TQx = \| Q^{½}x \|^2$. It is also assume that if $H_u < H_p$, the future control inputs after control horizon remain unchanged as shown in Fig. 3.5–1(b), i.e., $\hat{\mathbf{u}}(k + i \mid k) = \hat{\mathbf{u}}(k + H_u - 1 \mid k)$ or $\Delta\hat{\mathbf{u}}(k + i \mid k) = 0$ for $H_u \leq i \leq H_p - 1$.

It should be noted that MPC would work better if the trajectory of the controlled outputs were available in advance (that means designing the MPC for tracking control rather than set-pint control). The future values of the reference trajectory can then be assumed to be equal to its current value at time step $k$ during the prediction horizon. If the reference trajectory is available in advance, their predesigned values can be used in the prediction horizon unless the control algorithm requires that the reference trajectory is being updated during the control operation. In that case, the former approach is mostly applied. The similar approach is utilized for future values of disturbances unless a disturbance model is available that can be used to generate the predicted values of the disturbances. Again, similar to reference trajectory predictions, if the disturbance model cannot give a proper estimate of the future values because some changes in the control system during the operation, using the current value of the estimated disturbances during the prediction horizon will be more effective.

Now, the cost function given in (3.5–2) can be written in a matrix form as below,

$$\begin{aligned} V(k) &= \| Y(k) - Y_{ref}(k) \|_Q^2 + \| \Delta U(k) \|_R^2 \\ &= (Y(k) - Y_{ref}(k))^T Q (Y(k) - Y_{ref}(k)) + \Delta U(k)^T R \Delta U(k), \end{aligned} \tag{3.5–5}$$

or

$$\begin{aligned} V(k) &= \| Y(k) - Y_{ref}(k) \|_Q^2 + \| U(k) \|_R^2 \\ &= (Y(k) - Y_{ref}(k))^T Q (Y(k) - Y_{ref}(k)) + U(k)^T R U(k), \end{aligned} \tag{3.5–6}$$

where $Y(k) = [\hat{\boldsymbol{y}}(k+1|\,k)\ \ \hat{\boldsymbol{y}}(k+2|\,k)\ \ \ldots\ \ \hat{\boldsymbol{y}}(k+H_p\,|\,k)]^T$; $Y_{ref}(k) = [\boldsymbol{y}_{ref}(k+1|\,k)\ \ \boldsymbol{y}_{ref}(k+2|\,k)\ \ \ldots\ \ \boldsymbol{y}_{ref}(k+H_p\,|\,k)]^T$; $\Delta U(k) = [\Delta\hat{\boldsymbol{u}}(k\,|\,k)\ \Delta\hat{\boldsymbol{u}}(k+1|\,k)\ \ldots\ \Delta\hat{\boldsymbol{u}}(k+H_u-1|\,k)]^T$; $U(k) = [\hat{\boldsymbol{u}}(k\,|\,k)\ \hat{\boldsymbol{u}}(k+1|\,k)\ \ldots\ \hat{\boldsymbol{u}}(k+H_u-1|\,k)]^T$; $Q = BlockDiag\ \{Q(1), Q(2), \ldots\ Q(H_p)\}$[6], and $R = BlockDiag\{R(0), R(1), \ldots\ R(H_u-1)\}$.

The minimization problem in (3.5–2) is a quadratic programming problem (QP problem) considering that the system constraints in (3.5–4) are defined as linear inequalities. Depending on whether the cost function $V$ penalizes $\Delta\boldsymbol{u}(k)$ or $\boldsymbol{u}(k)$, the future values of all the constraints should be converted into the constraint on $\Delta\boldsymbol{u}(k)$ or $\boldsymbol{u}(k)$ since the main variable of minimization is the control input. The QP problem can be solved using available QP algorithms such as *active set method* or *interior point method* [89], [90]. Thus, based on the receding horizon approach, the optimal control input ($\boldsymbol{u}(k)_{opt}$) is obtained by solving the constraint optimization problem as follows,

$$\Delta U(k)_{opt} = \arg \min_{\substack{\Delta U(k) \\ \Omega_\Delta \Delta U(k) \le \omega_\Delta}} \left\| Y(k) - Y_r(k) \right\|_Q^2 + \left\| \Delta U(k) \right\|_R^2, \tag{3.5–7}$$

or

$$U(k)_{opt} = \arg \min_{\substack{U(k) \\ \Omega U(k) \le \omega}} \left\| Y(k) - Y_r(k) \right\|_Q^2 + \left\| U(k) \right\|_R^2, \tag{3.5–8}$$

where $\Omega_\Delta$, $\Omega$, $\omega_\Delta$, and $\omega$ are the matrices and vectors for all inequality constraints on control input changes (with subscript $\Delta$) and control input itself to be specified. Finally, if using (3.5–2), the optimal control input to the plant at time step $k$ is obtained by taking discrete integration from the first element in the sequence of optimal control input changes $\Delta U(k)_{opt}$ in (3.5–7) as follows

$$\begin{aligned} \Delta U(k)_{opt} &= [\Delta\boldsymbol{u}(k)_{opt}\ \ \Delta\boldsymbol{u}(k+1)_{opt} \ldots \Delta\boldsymbol{u}(k+H_u-1)_{opt}]^T, \\ \boldsymbol{u}(k)_{opt} &= \Delta\boldsymbol{u}(k)_{opt} + \boldsymbol{u}(k-1). \end{aligned} \tag{3.5–9}$$

Otherwise, the first element of $U(k)_{opt}$ in (3.5–8) will be the optimal control input ($\boldsymbol{u}(k)_{opt}$) to be applied to the plant at time step $k$, if (3.5–3) is used.

[6] *BlockDiag*{.} denotes block diagonal matrix.

A major problem that can occur with constraint optimization in predictive control is that the problem may become infeasible. This can happen because of an unexpected large disturbance that can make it impossible for the plant to be kept within the specified constraints. In addition, huge uncertainty in the model used to make predictions in MPC can contribute to the infeasibility due to different behavior between the real plant and the model towards disturbances and control input. Therefore, the robust feasibility is closely connected with the robust stability in MPC. As a result, it is essential to have a strategy on either how to deal with the possibility of infeasibility by having some back-up plan for computing the control signal, or avoid facing an infeasible problem at a time step. Various approaches to this issue have been suggested including,

- Avoid hard constraints on the output,
- Actively manage the constraint definition at each time step known as constraints softening,
- Actively manage the horizons at each time step,
- Use non-standard optimization algorithms,
- Try to have a good system modeling approach to reduce uncertainties,
- Try to compensate disturbances by having some disturbance model to estimate them,
- Outputting the same control input calculated in the previous step, or better that that, use the second element in the previous sequence of optimum control inputs successfully computed, i.e., $\hat{\boldsymbol{u}}(k+1|\, k)$.

The latter one is the back-up plan used in many practical optimization algorithm like those used in MATLAB® software[7] (we will use this approach in case of infeasibility of the MPC in our control system design). Furthermore, there is a trade-off between the choice of prediction/control horizon and the control accuracy. In case of having feedforward compensation (as in overhead crane control in Part II) or an uncontrolled input to the system (as in wind power with BESS in Part I), the common practice is that not to choose a long prediction horizon since it increases the prediction error for measured disturbances or uncontrolled input.

[7] MATLAB is registered trademark of The MathWorks, Inc.

However, other possible methods for having a robust MPC controller have been proposed. It is possible to use state feedback control to stabilize the predictions and then use optimization over the control horizon $H_u$ to modify the baseline predictions. This means that the predictions of the future input values is given by $\hat{\boldsymbol{u}}(k + 1| k) = -K\hat{\boldsymbol{x}}(k + 1| k) + \hat{\boldsymbol{u}}_p(i)$ where $\hat{\boldsymbol{u}}_p(i) = 0$ for $i \geq H_u$ and the values of the $\hat{\boldsymbol{u}}_p(i)$ for $i$=0, 1, …, $H_u - 1$ are chosen by the optimizer. It can be readily shown that $\Delta\hat{\boldsymbol{u}}_p(i) = \hat{\boldsymbol{u}}_p(i) - \boldsymbol{u}_p(i-1)$, and hence $\Delta\hat{\boldsymbol{u}}_p(i)$ is linearly related to $\Delta\hat{\boldsymbol{u}}(k+i|k)$. This yields the system constrains written in terms of $\Delta\hat{\boldsymbol{u}}_p(i)$ remain as linear inequalities and we still have a QP problem to solve. This use of stabilized predictions was first introduced in [91] and [92] with the emphasis on the use of state feedback to obtain deadbeat or finite impulse response (FIR) behavior prior to the use of optimization. However, a remaining question is how to select the stabilizing state feedback gain $K$. For systems with state dimension less than 5, it is possible to use pole placement technique, but for higher order systems, as in our case, that cannot work properly because of the difficulty of knowing what closed-loop pole locations are reasonably attainable. So, the only practical alternatives are to obtain $K$ by solving an LQR, or possibly an $H_\infty$ problem [89]. For robustness, however, we only need to show that the optimization will be feasible at each step. This can be done by using the ideas of "maximal output admissible sets" or "robust admissible and invariant sets" [93]–[95]. The other alternative to have a robust MPC is to make state feedback gain $K$ as the decision variable for the optimizer to choose at each time step. This can be done by using the LMI approach which leads to an LMI optimization problem rather than QP problem [96]. Similar proposal is given in [97] which is the combination of stabilized prediction and LMI approach. A recent approach known as tube MPC was proposed in [98] in which it uses an independent nominal model of the system, and employs a feedback system to ensure the actual state converges to the nominal state.

## 3.6 MPC Formulation for Wind Power with BESS

According to the control system structure for wind power dispatch with BESS described in Section 3.2, MPC is utilized to design the discrete-time controller for its ability in online optimization, constraint handling, easy implementation and the discrete-time nature of the system. As mentioned in Section 3.5.2, MPC can be formulated to penalize the control input rather than control input changes. Therefore, let us recall the discrete-time state-space model we obtained for wind power with BESS in Section 4.62.2 given as below,

$$\begin{aligned} \boldsymbol{x}(k+1) &= A\boldsymbol{x}(k) + B_1 u(k) + B_2 r(k), \\ y(k) &= C\boldsymbol{x}(k), \end{aligned} \tag{3.6–1}$$

$$\begin{gathered} \boldsymbol{x}(k) = [x_1(k) \quad x_2(k)]^T = [P_g(k) \quad E_b(k)]^T, \\ u(k) = P_c(k), \quad r(k) = P_w(k), \quad y(k) = P_g(k), \\ A = \begin{bmatrix} 0 & 0 \\ 0 & 1 \end{bmatrix}, \quad B_1 = \begin{bmatrix} 1 \\ t_d \end{bmatrix}, \quad B_2 = \begin{bmatrix} 1 \\ 0 \end{bmatrix}, \quad C = [1 \quad 0], \end{gathered} \tag{3.6–2}$$

and the cost function defined in Section 3.5.2 for MPC that penalizes trajectory tracking error and control input in (3.5–3) subject to system equations, control input, and BESS energy constraints as follows,

$$V(k) = \sum_{i=1}^{H_p} \left\| \hat{y}(k+i\,|\,k) - \hat{y}_r(k+i\,|\,k) \right\|_{Q(i)}^2 + \sum_{i=0}^{H_u-1} \left\| \hat{u}(k+i\,|\,k) \right\|_{R(i)}^2, \tag{3.6–3}$$

$$\begin{aligned} x_{2min} \le x_2(k) \le x_{2max} \qquad & k = 1, 2, \ldots, H_p, \\ -u_{max} \le u(k) \le u_{max} \qquad & k = 0, 1, \ldots, H_u - 1, \end{aligned} \tag{3.6–4}$$

where $\hat{y}_r(k)$ is the updated reference power signal (which is updated using a fuzzy controller that will be explained in the next section); $u_{max}$ is the maximum charging/discharging rated power capacity of the BESS; $x_{2min}$ and $x_{2max}$ are the minimum and maximum SoC of the BESS, respectively, i.e., the BESS should not be charged over $x_{2max}$ or discharged below $x_{2min}$.

Now, the cost function given in (3.6–3) can be written in a matrix form as below,

$$V(k) = \| Y(k) - Y_r(k) \|_Q^2 + \| U(k) \|_R^2$$
$$= (Y(k) - Y_r(k))^T Q (Y(k) - Y_r(k)) + U(k)^T R U(k), \quad (3.6–5)$$

where $Y(k) = [\hat{y}(k+1|\ k)\ \ \hat{y}(k+2|\ k)\ \ \dots\ \ \hat{y}(k+H_p\ |\ k)]^T$ is the vector of output predictions; $Y_r(k) = [\hat{y}_r(k+1|\ k)\ \ \hat{y}_r(k+2|\ k)\ \ \dots\ \ \hat{y}_r(k+H_p\ |\ k)]^T$ is the vector of future values for reference power signal; $U(k) = [\hat{u}(k\ |\ k)\ \hat{u}(k+1|\ k)\ \dots\ \hat{u}(k+H_u-1|\ k)]^T$ is the vector of control input; $Q = Diag\{Q(1), Q(2), \dots Q(H_p)\}$, and $R = Diag\{R(0), R(1), \dots R(H_u-1)\}$ are diagonal weighting matrices for tracking error $Q(k) = q_y$, and control input $R(k) = t_d^2 r_u$, respectively (The term $t_d$ is considered in the control input weight $R(k)$ to achieve the desired tracking with the minimum BESS energy consumption at each optimization step).

The minimization of the cost function $V$ in (3.6–5) requires the predictions of the output and reference power signal up to horizon $H_p$. Thus, the wind power with BESS model given in (3.6–1) can be used to calculate $\hat{y}(k+i\ |\ k)$ as based on the measurement of the states $\boldsymbol{x}(k)$ at time $k$ as follows,

$$\hat{y}(k+1|k) = C\hat{x}(k+1|k) = CA\boldsymbol{x}(k) + CB_1\hat{u}(k|k) + CB_2\hat{r}(k|k), \quad (3.6–6)$$

$$\hat{y}(k+2|k) = CA^2\boldsymbol{x}(k) + CAB_1\hat{u}(k|k) + CAB_2\hat{r}(k|k) + CB_1\hat{u}(k+1|k)$$
$$+ CB_2\hat{r}(k+1|k), \quad (3.6–7)$$

$$\vdots$$

$$\hat{y}(k+H_p|k) = CA^{H_p}\boldsymbol{x}(k) + CA^{H_p-1}B_1\hat{u}(k|k) + \dots + CB_1\hat{u}(k+H_p-1|k)$$
$$+ CA^{H_p-1}B_2\hat{r}(k|k) + \dots + CB_2\hat{r}(k+H_p-1|k), \quad (3.6–8)$$

where $\hat{r}(k+i\ |\ k)$ is the prediction of wind power at $k+i$ made at time step $k$. Based on the general assumption that the control input will remain constant after control horizon $H_c$, i.e., $\hat{u}(k+i\ |\ k) = \hat{u}(k+H_u-1|\ k)$ for $H_u \le i \le H_p - 1$, (3.6–6)–(3.6–8) are rewritten as follows,

$$\hat{y}(k+1|k) = C\hat{x}(k+1|k) = CA\boldsymbol{x}(k) + CB_1\hat{u}(k|k) + CB_2\hat{r}(k|k), \quad (3.6–9)$$

$$\hat{y}(k+2|k) = CA^2\boldsymbol{x}(k) + CAB_1\hat{u}(k|k) + CAB_2\hat{r}(k|k) + CB_1\hat{u}(k+1|k)$$
$$+ CB_2\hat{r}(k+1|k), \quad (3.6–10)$$

$$\vdots$$

$$\begin{aligned}\hat{y}(k+H_u \mid k) = {} & CA^{H_u}\boldsymbol{x}(k) + CA^{H_u-1}B_1\hat{u}(k \mid k) + \dots + CB_1\hat{u}(k+H_u-1 \mid k) \\ & + CA^{H_u-1}B_2\hat{r}(k \mid k) + \dots + CB_2\hat{r}(k+H_u-1 \mid k),\end{aligned} \tag{3.6–11}$$

$$\begin{aligned}\hat{y}(k+H_u+1 \mid k) = {} & CA^{H_u+1}\boldsymbol{x}(k) + CA^{H_u}B_1\hat{u}(k \mid k) \\ & + \dots + CA^2B_1\hat{u}(k+H_u-2 \mid k) \\ & + (CAB_1 + CB_1)\hat{u}(k+H_u-1 \mid k) \\ & + CA^{H_u}B_2\hat{r}(k \mid k) + \dots + CB_2\hat{r}(k+H_u \mid k),\end{aligned} \tag{3.6–12}$$

$$\vdots$$

$$\begin{aligned}\hat{y}(k+H_p \mid k) = {} & CA^{H_p}\boldsymbol{x}(k) + CA^{H_p-1}B_1\hat{u}(k \mid k) \\ & + \dots + CA^{H_p-H_u+1}B_1\hat{u}(k+H_u-2 \mid k) \\ & + (CA^{H_p-H_u} \dots + CAB_1 + CB_1)\hat{u}(k+H_u-1 \mid k) \\ & + CA^{H_p-1}B_2\hat{r}(k \mid k) + \dots + CB_2\hat{r}(k+H_p-1 \mid k).\end{aligned} \tag{3.6–13}$$

These predictions can be written in matrix form as the following,

$$Y(k) = \Psi\boldsymbol{x}(k) + \Theta_1 U(k) + \Theta_2 R_w(k), \tag{3.6–14}$$

where $R_w(k) = [\hat{r}(k \mid k) \;\; \hat{r}(k+1 \mid k) \; \dots \; \hat{r}(k+H_p-1 \mid k)]^T$, and matrices $\Psi$, $\Theta_1$, and $\Theta_2$ are obtained using (3.6–9)–(3.6–13) as below,

$$\Psi = \begin{bmatrix} CA \\ CA^2 \\ \vdots \\ CA^{H_p} \end{bmatrix}, \tag{3.6–15}$$

$$\Theta_1 = \begin{bmatrix} CB_1 & \underline{0} & \cdots & \underline{0} & \underline{0} \\ CAB_1 & CB_1 & \ddots & \vdots & \vdots \\ \vdots & \vdots & \ddots & \ddots & \vdots \\ \vdots & \vdots & \ddots & \ddots & \underline{0} \\ CA^{H_u-1}B_1 & CA^{H_u-2}B_1 & \cdots & \cdots & CB_1 \\ CA^{H_u}B_1 & CA^{H_u-1}B_1 & \cdots & CA^2B_1 & CAB_1+CB_1 \\ \vdots & \vdots & \ddots & \vdots & \vdots \\ CA^{H_p-1}B_1 & CA^{H_p-2}B_1 & \cdots & CA^{H_p-H_u+1}B_1 & \sum_{i=0}^{H_p-H_u} CA^iB_1 \end{bmatrix}, \quad (3.6\text{–}16)$$

$$\Theta_2 = \begin{bmatrix} CB_2 & \underline{0} & \cdots & \underline{0} \\ CAB_2 & CB_2 & \ddots & \vdots \\ \vdots & \vdots & \ddots & \underline{0} \\ CA^{H_p-1}B_2 & CA^{H_p-2}B_2 & \cdots & CB_2 \end{bmatrix}, \quad (3.6\text{–}17)$$

where $\underline{0}$ is a zero matrix with proper size.

The wind power data which is available up to time step $k$ is used for the predicted values during the prediction horizon (i.e., $\hat{r}(k + i|\, k) = \hat{r}(k \mid k)$ for $0 \le i \le H_p - 1$). This is a common practice in predictive control as the case for feedforward compensation of a disturbance that its future values are considered to remain constant for the whole prediction horizon [89]. The same condition is also assumed for the updated reference power signal $\hat{y}_r\,(k)$. It should be mentioned that the performance of the control system would be improved if some model for wind power prediction were available. However, if the predictive controller can respond well enough in the sense of stability and tracking performance, there would be no need to use such a prediction system.

As mentioned before, the system constraints given in (3.6–4) should also be translated into linear inequalities in terms of $\hat{u}(k + i \mid k)$ that should hold for the entire prediction and control horizon. The direct constraint on the control input in (3.6–4) can be readily extended for the entire control horizon as follows,

$$-u_{max} \le \hat{u}(k\,|\,k) \le u_{max} \quad \Rightarrow \quad \begin{array}{l} \hat{u}(k\,|\,k) \le u_{max} \\ \hat{u}(k\,|\,k) \ge -u_{max} \end{array}, \quad (3.6\text{–}18)$$

which can be written into two separate inequalities if the lower bound of (3.6–18) is inverted as below,

$$\begin{aligned} \hat{u}(k\,|\,k) &\le u_{max}, \\ -\hat{u}(k\,|\,k) &\le u_{max}. \end{aligned} \qquad (3.6\text{–}19)$$

Repeating this for $\hat{u}(k + i \mid k)$ up to $i = H_u - 1$ leads to the following control input constraints with the $2H_u \times H_u$ matrix $\Omega_1$ and the $2H_u \times 1$ vector $U_m$,

$$\Omega_1 U(k) \le U_m, \qquad (3.6\text{–}20)$$

$$\Omega_1 = \begin{bmatrix} I \\ -I \end{bmatrix}, \qquad (3.6\text{–}21)$$

$$U_m = \begin{bmatrix} u_{max} \\ u_{max} \\ \vdots \\ u_{max} \end{bmatrix}, \qquad (3.6\text{–}22)$$

where $I$ is a $H_u \times H_u$ identity matrix. The second constraint in (3.6–4), which represents the BESS energy capacity, can also be separated into two inequalities in the same way as in (3.6–19) in terms of the future values of state vector $\hat{\boldsymbol{x}}(k + i \mid k)$ as follows,

$$\begin{aligned} [0 \quad 1]\hat{\boldsymbol{x}}(k+i\,|\,k) &\le x_{2max}, \\ -[0 \quad 1]\hat{\boldsymbol{x}}(k+i\,|\,k) &\le -x_{2min}, \qquad i = 1, 2, \ldots, H_p, \end{aligned} \qquad (3.6\text{–}23)$$

which can then be written in matrix form with the $2H_p \times 2H_p$ matrix $\Omega_2$, $2H_p \times 2H_p$ matrix $I_2$, $H_p \times 1$ vector $X(k)$, and the $2H_p \times 1$ vector $X_m$ as below,

$$\Omega_2 X(k) \le X_m, \qquad (3.6\text{–}24)$$

$$\Omega_2 = \begin{bmatrix} I_2 \\ -I_2 \end{bmatrix}, \qquad (3.6\text{–}25)$$

$$X(k) = \begin{bmatrix} \hat{\boldsymbol{x}}(k+1\,|\,k) \\ \hat{\boldsymbol{x}}(k+2\,|\,k) \\ \vdots \\ \hat{\boldsymbol{x}}(k+H_p\,|\,k) \end{bmatrix} \qquad (3.6\text{–}26)$$

$$X_m = \begin{bmatrix} x_{2max} \\ x_{2max} \\ \vdots \\ x_{2max} \\ -x_{2min} \\ -x_{2min} \\ \vdots \\ -x_{2min} \end{bmatrix}, \qquad (3.6\text{–}27)$$

$$I_2 = \begin{bmatrix} 0 & 1 & 0 & \cdots & 0 \\ 0 & 0 & 1 & \cdots & 0 \\ \vdots & \vdots & \ddots & \ddots & \vdots \\ 0 & 0 & \cdots & 0 & 1 \end{bmatrix}. \qquad (3.6\text{–}28)$$

The prediction of the state variables $X(k)$ can be obtained in terms of the future values of the control input $U(k)$ is the same fashion as for the future values of the output $Y(k)$ given in (3.6–9)–(3.6–13) and (3.6–14) using the system model in (3.6–1) as the following,

$$X(k) = \Psi' x(k) + \Theta'_1 U(k) + \Theta'_2 R_w(k), \qquad (3.6\text{–}29)$$

where

$$\Psi' = \begin{bmatrix} A \\ A^2 \\ \vdots \\ A^{H_p} \end{bmatrix}, \qquad (3.6\text{–}30)$$

$$\Theta'_1 = \begin{bmatrix} B_1 & \underline{0} & \cdots & \underline{0} & \underline{0} \\ AB_1 & B_1 & \ddots & \vdots & \vdots \\ \vdots & \vdots & \ddots & \ddots & \vdots \\ \vdots & \vdots & \ddots & \ddots & \underline{0} \\ A^{H_u-1}B_1 & A^{H_u-2}B_1 & \cdots & \cdots & B_1 \\ A^{H_u}B_1 & A^{H_u-1}B_1 & \cdots & A^2B_1 & AB_1 + CB_1 \\ \vdots & \vdots & \ddots & \vdots & \vdots \\ A^{H_p-1}B_1 & A^{H_p-2}B_1 & \cdots & A^{H_p-H_u+1}B_1 & \sum_{i=0}^{H_p-H_u} A^i B_1 \end{bmatrix}, \tag{3.6–31}$$

$$\Theta'_2 = \begin{bmatrix} B_2 & \underline{0} & \cdots & \underline{0} \\ AB_2 & B_2 & \ddots & \vdots \\ \vdots & \vdots & \ddots & \underline{0} \\ A^{H_p-1}B_2 & A^{H_p-2}B_2 & \cdots & B_2 \end{bmatrix}. \tag{3.6–32}$$

Thus, by using (3.6–29), the constraints in (3.6–24) can be given in terms of $U(k)$ and then combined with (3.6–20) as one set of linear constraints on control input as follows,

$$\Omega \Delta U(k) \le \omega, \tag{3.6–33}$$

where

$$\Omega = \begin{bmatrix} \Omega_1 \\ \Omega_2 \Theta'_1 \end{bmatrix}, \qquad \omega = \begin{bmatrix} U_m \\ X_m - \Omega_2(\Psi' x(k) + \Theta'_2 R_w(k)) \end{bmatrix} \tag{3.6–34}$$

Having all the constraints written in terms of control input, the optimal BESS charging/discharging power ($\boldsymbol{u}(k)_{opt}$) is obtained based on the receding horizon strategy in the sense of MPC for wind power integrated with BESS by minimizing the quadratic cost function defined in (3.6–5) as follows,

$$U(k)_{opt} = \arg \min_{\substack{U(k) \\ \Omega U(k) \le \omega}} \left\| Y(k) - Y_r(k) \right\|_Q^2 + \left\| U(k) \right\|_R^2, \tag{3.6–35}$$

$$U(k)_{opt} = [u(k)_{opt} \;\; u(k+1)_{opt} \;\; \ldots \;\; u(k+H_u-1)_{opt}]^T,$$

$$u(k)_{opt} = P_c(k)_{opt}. \tag{3.6–36}$$

## 3.7 Reference Power Update Using Fuzzy Logic Controller

As mentioned before, the optimization problem defined in (3.6–5) penalizes the deviations between the future values of the reference power signal, which is updated through adding a feedback signal to the reference signal generator ($\hat{y}_r(k)$), and the power output predictions, and also minimizes the amount of energy charged in or discharged from the BESS. To generate this correction power as a feedback signal, a fuzzy logic controller (FLC) is designed based on the basics of fuzzy systems explained in Section 3.3.2. The main reason for applying FLC to update the reference power signal is that the effect of wind power variations is not considered in the generation of the reference power signal. In this case, the MPC alone can only maintain the tracking error as small as possible and keeping the constraints within their limits even if the electricity price is high but wind power is insufficient. This causes the BESS to discharge completely and therefore, makes the MPC to put more weight on the minimization of BESS discharging against the tracking error to prevent BESS depletion. Thus, the MPC loses the tracking performance but maintains the system within its constraints. If this situation or the opposite case continues, i.e., the BESS is kept in depletion or overcharge for a long time, the BESS could be damaged and its lifetime would be significantly reduced.

In this regard, one solution is to get short-term wind power prediction methods (as in [17]) involved in the decision-making system design for generating the reference signal. This may alleviate the tracking problem to some extent, but it makes the decision-making system design very complicated, and due to the uncertainties associated with weather prediction methods, the tracking performance problem might not be resolved. Therefore, another alternative is to use a feedback from the battery's SoC in order to coordinate the reference signal with the current available wind power and battery's SoC. Hence, for increasing the lifetime of the BESS and maintaining good tracking performance, the FLC has been employed to intelligently update the reference power signal as illustrated in the overall control system block diagram for wind power dispatch with BESS in Fig. 3.7–1.

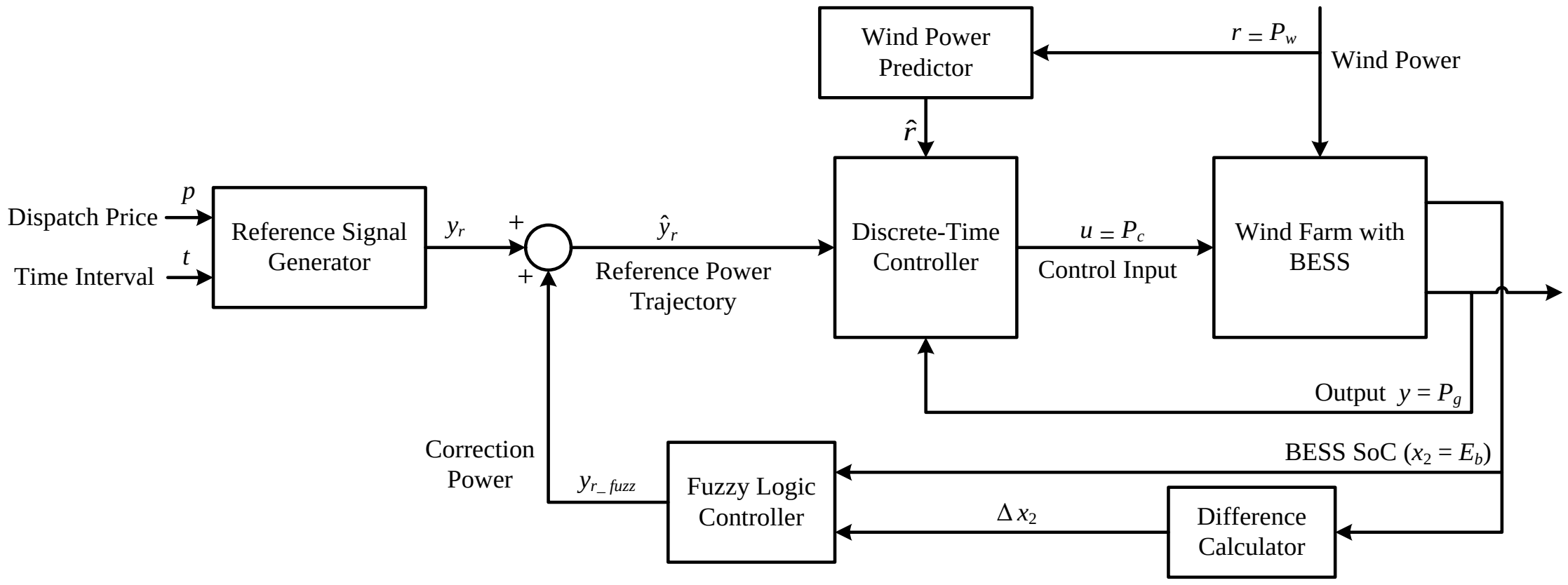

Fig. 3.7–1. The block diagram of the discrete-time control system for wind power dispatch with BESS.

It can be seen from Fig. 3.7–1 that the FLC uses the current battery's SoC and the speed of charging/discharging ($\Delta x_2$) as inputs and adjusts the reference power signal to prevent overcharge or depletion of the BESS. The proposed FLC can be described as follows,

$$y_{r_fuzz}(k) = K_o \text{FC}(x_2(k), \Delta x_2(k)), \tag{3.7–1}$$

where $y_{r_fuzz}(k)$ is the correction value added to the reference signal in MW, FC(.) is the fuzzy controller, $\Delta x_2(k)$ is the rate of change of the battery's SoC, and $K_o$ is the FC(.) output scaling factor. The fuzzy rules are shown in Table 3.7–1 for which the following rule base is used,

$$\begin{gathered}\text{IF } x_2(k) \text{ is } MF_i \text{ AND } \Delta x_2(k) \text{ is } MF_j, \text{THEN } y_{r_fuzz}(k) \text{ is } MF_l, \\ for\ i = 1, 2, 3,\ \ j = 1, 2, 3,\ \ and\ \ l = 1, 2, \ldots 7.\end{gathered} \tag{3.7–2}$$

where the abbreviations in Table 3.7–1 denote the fuzzy sets for battery's SoC and its rate of change, and the reference power signal which are shown in Fig. 3.7–2. These fuzzy set abbreviations are as follows: L is low, M is medium, H is high, Z is zero, P is positive, N is negative, and HP, MP, HN, and MN are high positive, medium positive, high negative, and medium negative, respectively.

As can be seen in Fig. 3.7–2(a), the range of the battery's SoC is selected based on the maximum capacity of the BESS used in this study (480MWh) which is explained in Section 3.8.2. The range for the rate of change of the battery's SoC shown in Fig. 3.7–2(b) is obtained using system model in (3.6–1) and the fact that rate of change of a variable in discrete-time can be calculated using forward difference as follows,

$$\Delta x_2(k) = x_2(k+1) - x_2(k) = t_d u(k), \tag{3.7–3}$$

which shows the relation between the rate of change of the battery's SoC and the control input. As the charging/discharging power of the battery is bounded to $u_{max}$ = 80MW in each 5 minutes (based on the selected type of the BESS which will be explained in Section 3.8.2), the range of the battery's SoC variation rate is obtained using (3.7–3), system constraints in (3.6–4) and the value of $t_d$ = 1/12 as follows,

$$-6.67 = -t_d u_{\max} \le \Delta x_2(k) \le t_d u_{\max} = 6.67. \tag{3.7–4}$$

The range for the FLC output shown in Fig. 3.7–2(c) is the normalized power in MW which is a common choice when designing FLC using scaling factor based on a trial-and-error approach [85], [86]. The same procedure as for the fuzzy decision-making system explained Section 3.3.2 is used for the FLC design including rule definition, MFs type and bound selection to get the best results.

Table 3.7–1. Fuzzy Logic Controller Rules

| $y_{r_fuzz}(k)$ | | $\Delta x_2(k)$ | | |
|---|---|---|---|---|
| | | N | Z | P |
| $x_2(k)$ | L | HN | MN | N |
| | M | N | Z | P |
| | H | P | MP | HP |

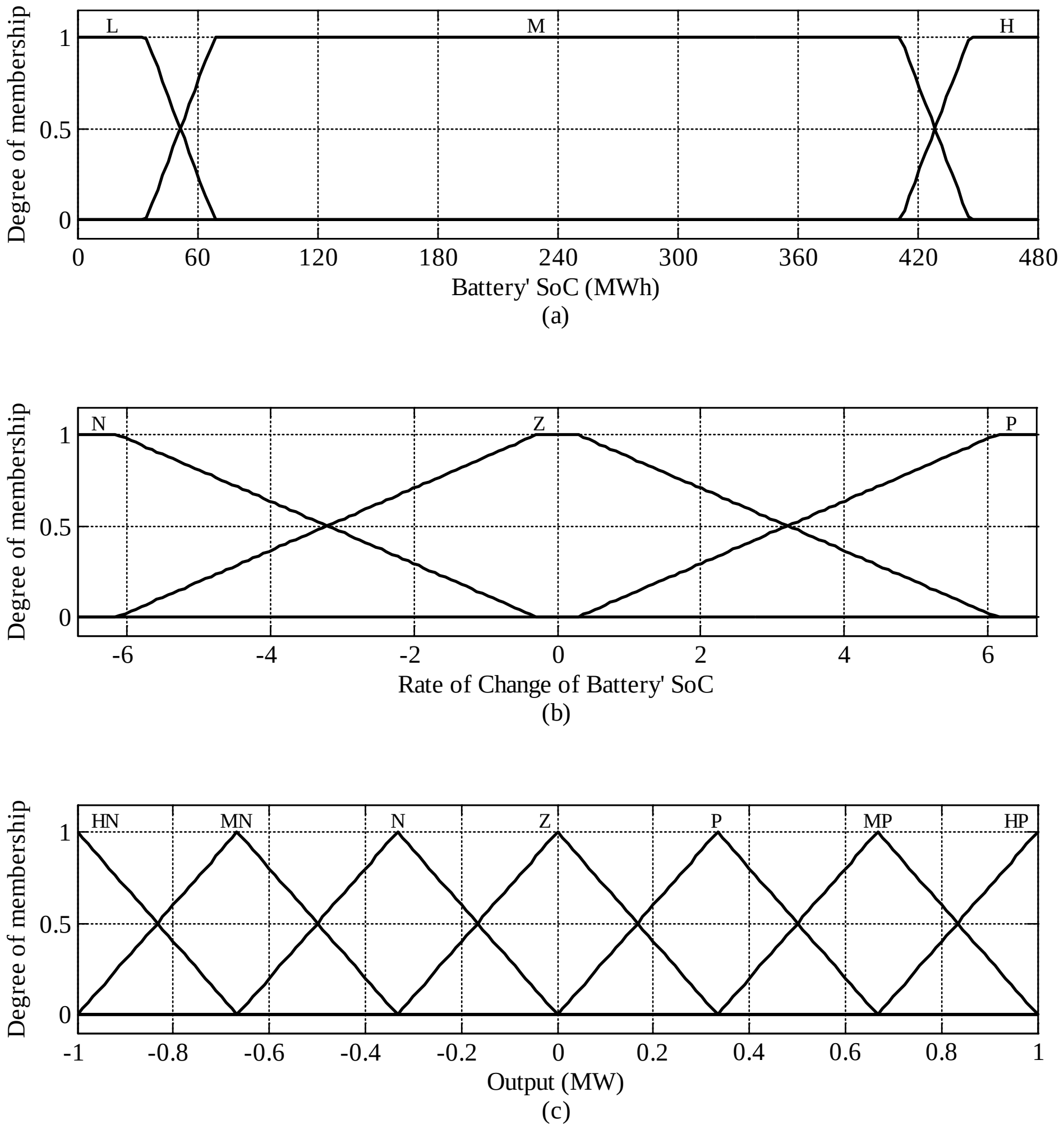

Fig. 3.7–2. Fuzzy sets and MFs for two inputs and the output of the FLC. (a) Battery's SoC MFs, (b) Variation of battery's SoC MFs, (c) Output MFs that determine the correction power added to the reference signal to be updated in normalized MW.

## 3.8 Simulation Results

### 3.8.1 Database

In order to verify the proposed control system, extensive simulations for assessing different scenarios are carried out based on the actual wind farm and dispatch price data using MATLAB® software. The data used in this study are obtained from the AEMO database [88] including Woolnorth wind farm power generation data, which is located in Tasmania, Australia with the maximum generation capacity of 140 MW, and the dispatch electricity price data of the NSW electricity market for the corresponding dates of wind power generation. These data are available in 5-minute resolution and they are properly filtered to eliminate outliers which may exist in the data. The data covers two time periods including a successive 6-month period from June 2010 to November 2010, and discrete sequence of 6 days in each month with a 5-day time distance for a one-year period from June 2010 to May 2011 started from the first day in each month, i.e., days 1, 6, 11, 16, 21, and 26. It is assumed that within 5 minutes the wind power fluctuation is smoothed and so there is no need to consider ramp rate constraint in the control system design. The reason for selecting electricity price data from a different state is that the NSW electricity market is more dynamic compared with the state of Tasmania.

### 3.8.2 BESS Type Selection

From among different BESSs suitable to be used with wind power generation, the sodium–sulphur (NaS) battery technology is shown to be more promising for integration with wind farms in comparison with other types due to its high efficiency (89%), high energy capacity, and long life span at 100% depth of discharge (DoD) up to 2500 cycles [7], [8], [76]. Thus, an 80-MW NaS battery with the energy capacity of 480 MWh is assumed for this simulation based on the available information from the manufacturer of NaS battery [77], i.e., the battery's power rating constraint $u_{max}$ is 80MW ($-80 \le u(t) \le 80$) and its energy rating constraint is $0 \le x_2(t) \le 480$, i.e., $x_{2min} = 0$ and $x_{2min} = 480$ MWh, as in (3.6–4). It is obvious that the more the BESS capacity is selected, the higher the capital cost would be. Therefore, a trade-off between the storage capacity and the cost of the BESS should be made depending on the application.

### 3.8.3 Results

The first set of simulations is carried out for one day in each month with a fixed interval successively for 12 months (June 2010 to May 2011). The simulations are then repeated for another day in each month. Dates are selected as follows, first days of each month, then days 6 of each month, 11, 16, 21, and 26. In other word, it is hypothetically assumed that the wind farm (integrated with a BESS) sells its generated power for 12 days in one year with monthly resolution. Thus, choosing six different days within a month, we would have six different yearly scenarios for simulation. This could give an approximately appropriate view for yearly operation of the proposed control system. After that, for a full period of 6 months the simulation is carried out to compare with the discrete-period simulations (June 2010 to November 2010). Finally, the total earning from the sale of the combined net power (i.e., the wind power with the BESS) and wind power only are calculated for all the scenarios based on the half-hourly spot prices and they are compared with each other. For simplicity, the persistent method for wind power prediction in the MPC algorithm is used with control horizon to be the same as prediction horizon and to be equal to three ($H_c = H_p = 3$). The values for MPC weights ($Q(k) = q_y$ and $R(k) = t_d^2 r_u$) and FLC output scaling factor ($K_o$) are 45, 1.67, and 1.285, respectively, in all simulations.

Simulation results for the first days of each month are shown in Fig. 3.8–1 and Fig. 3.8–2, followed by the wind power output profile in Fig. 3.8–1(a) as some examples. Because of the discontinuity of the data, some huge spikes in the graphed wind power profile appear, like on March 1, 2011. It can be seen from Fig. 3.8–1(b) that during February 1, 2011, the electricity price reached its maximum value that can be bidden by the market participants in the Australian NEM, which is set by the AEMO to be $12,500/MWh. This price is called value of lost load. Therefore, the zoomed-in view of the electricity price variations for the first days of each month is illustrated in Fig. 3.8–1(c). The load-duration curve for the corresponding time period (i.e., first days of each month) in the NSW region is shown in Fig. 3.8–1(d) which is obtained from demand data available in [88]. It should be noted that in the case of negative prices, the reference generator would output zero reference power signal to the control system. Therefore, wind power can be stored in the BESS in the negative price case unlike thermal power plants as, not very often, they have to dispatch power and pay to the market. This happens as thermal power plants normally operate 24 hours, 7 days a

week, and it would not be economical for them to stop their operation just for a short period when the price is negative.

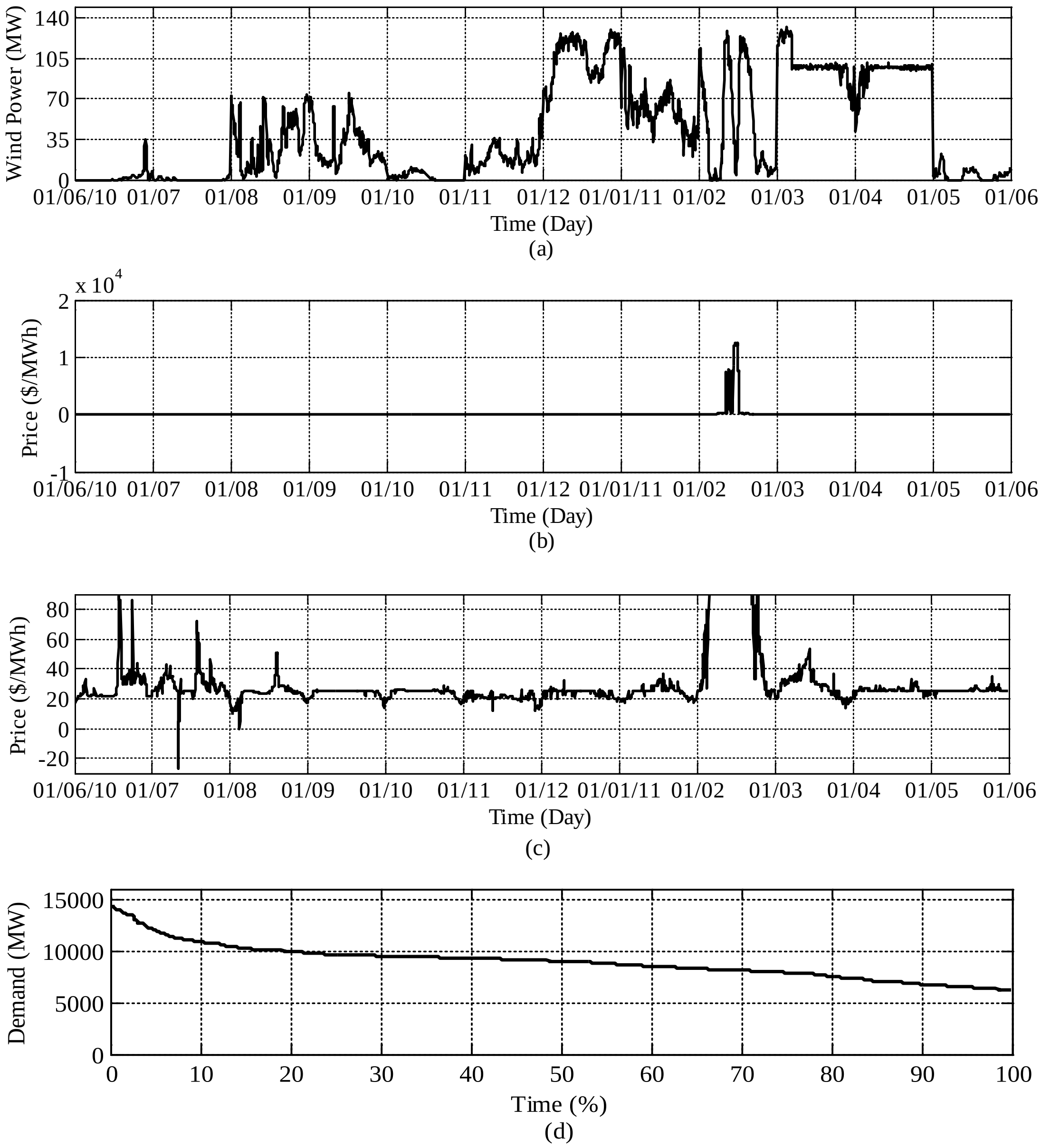

Fig. 3.8–1. Actual wind power and dispatch price data on one day in month basis from 1 June 2010 to 1 May 2011. (a) Generated wind power data of Woolnorth wind farm, (b) Dispatch prices of the NSW electricity market, (c) Zoomed-in dispatch prices, (d) Load-duration curve of the NSW region demand for the corresponding dates.

The reference power signal generated by the decision-making system and the updated one are shown in Fig. 3.8–2(a) and (b), respectively. It can be observed that our decision-making system generates reference power signal according to the electricity price variation and peak/off peak times in each day. Since the peak/off-peak periods are considered to be the same for both summer and winter according to NSW government resources and energy website (see Section 3.4 and [87]), it would make sense that the generated reference power signal exhibits some periodic pattern throughout the year. Besides, at times when the BESS is about to fully discharge (mostly in the first half) or overcharge (mostly in the second half) in Fig. 3.8–2(c), the reference power is intelligently updated by the FLC in Fig. 3.8–2(b) to match itself with the wind power availability and BESS limitations. By employing the FLC, the battery's SoC is almost maintained between 10% and 90%, as illustrated in Fig. 3.8–2(c). In this case, the lifetime would be increased by 4500 cycles rather than 2500 cycles for 100% DoD in a full charging and discharging cycle based on the NaS battery's manufacturer information[8] [77].

The behaviour of battery's SoC in Fig. 3.8–2(c) can be better described when seen along with the corresponding wind power and electricity price profiles given in Fig. 3.8–1(a) and Fig. 3.8–1(c), respectively. It can be clearly seen that in the first half of the battery's SoC in Fig. 3.8–2(c), due to lack of wind power and frequent high electricity prices, the control system had to use the available BESS energy to generate as much power as it can to follow the reference power signal. However, once the battery's SoC reaches 10%, the reference power signal is updated to its minimum value to avoid further BESS discharge and protect the battery. In the second half, though, there are plenty wind power available as seen in Fig. 3.8–1(a), whereas the electricity price is relatively low, except a massive spike between February 1st and March 1st as shown in Fig. 3.8–1(b). Thus, it is completely natural for the control system to charge BESS for most of the time except for that specific period when the control system tries to generate as much power as possible due to huge electricity price rise.

[8] It should be noted that in this type of application, the BESS is faced to partial charging/discharging cycles that makes it difficult to count the number of full cycles for estimating the remaining life span of the BESS.

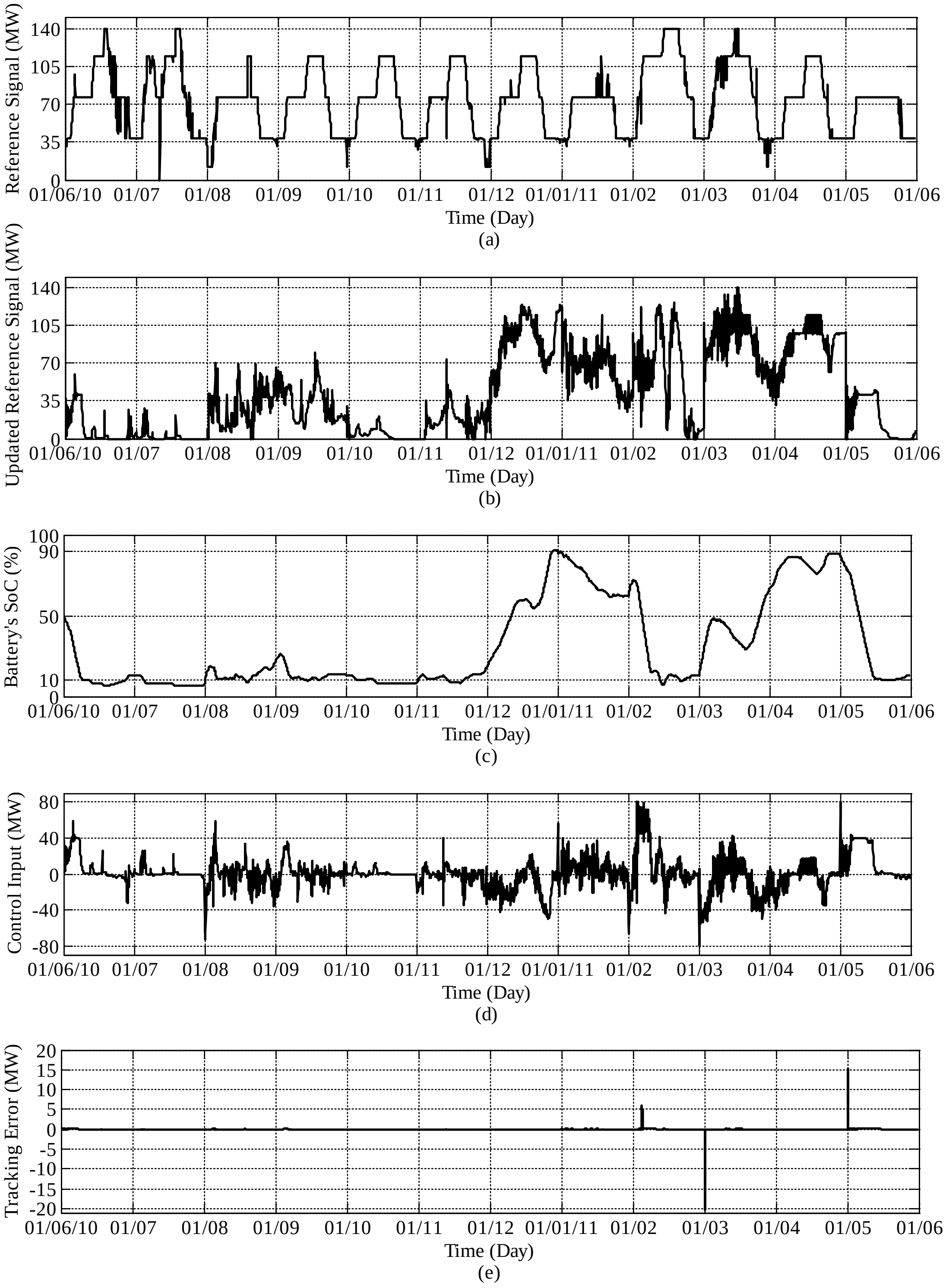

Fig. 3.8–2. Simulation results of the proposed control system. (a) Original reference power signal, (b) Updated reference power signal, (c) Battery's SoC, (d) Control input or charging/discharging power of the battery, (e) Tracking power error.

Also, the MPC action keeps controlling the input signal in Fig. 3.8–2(d) within its rated power limit. As the NaS battery has a fast response from 0% to 100% of its rated power within 10 seconds [77], it can provide continuous 80-MW power charging/discharging during each 5 minutes. Finally, Fig. 3.8–2(e) shows the tracking error . It can be seen that the total tracking performance is quite good with hardly ever having errors larger than 5 MW, except two large errors happened at the spikes of the wind power data due to discontinuity on March 1 and May 1, 2011 as in Fig. 3.8–2(a). As a result of attaching wind power profiles for 12 discontinuous days in a year, it is possible to have such huge jumps from one day to another. However, as the control input has an upper and lower bound (80 MW), in the transition from March 1 to April 1 and May 1 to June 1, 2011, the MPC calculates the maximum allowable control input in Fig. 3.8–2(d) for forcing the system to generate the power demanded by the updated reference power signal. But, the system constraint prevents more than 80-MW charging or discharging power command to the battery for protecting it. Thus, it is obvious that the tracking is failed temporarily for just that dispatch interval. In fact, this shows that the proposed control system could work practically to maintain all system constraints for system safety.

After keeping the wind power dispatch under control with the BESS, our goal is to indicate the ability to increase the earning from the sale of the generated power using our proposed control system. Therefore, we repeated the simulations for the same dates mentioned earlier and also for the consecutive 6 months. The total earning from selling wind power only and the controlled one for all scenarios are calculated based on the Australian NEM operation explained in Section 2.1.1, which can be described by the following formula for each scenario,

$$\text{Total Earning} = \sum_{l=1}^{n} \left[ \sum_{j=0}^{47} \left( \left( \frac{1}{6} \sum_{i=0}^{5} p_l(6j+i) \right) \left( \sum_{i=0}^{5} P_{g_l}(6j+i) \right) \right) \right] \tag{3.8–1}$$

where $p_l$ is the dispatch price for day $l$; $P_{gl}$ is the generated power for day $l$; $i$ is a counter for six 5-minute power dispatch; j is a counter for each half-hour throughout a trading day, and $n$ is total number of days in each scenario.

Furthermore, we consider two modes for the initial charge of the BESS, one with full charge and one with 50% SoC. For comparing the effectiveness of our proposed control system under different simulation scenarios in tracking performance, we use the key performance index (KPI) introduced by [15] with the acceptable tracking errors up to ±3 MW as follows,

$$\mathrm{KPI} = \sum N_x \left| e_x \right|, \tag{3.8–2}$$

which adds the unacceptable power deviations greater than 3MW. Here $N_x$ denotes the number of unacceptable errors that occurs in each simulation scenario. The results of the KPI and the earning comparison are all provided in Table 3.8–1. It can be seen from Table 3.8–1 that the number of unacceptable tracking errors does not exceed 10 units is all scenarios (i.e., $N_x \leq 10$). This indicates the high performance of our proposed control system in improving wind power dispatch for long-term operation with 5-minute sampling rate in simulations, even with different initial conditions for battery's SoC.

Table 3.8–1. Performance and Total Earning from Power Sale Comparison for Different Scenarios

| Scenarios | KPI (50% SoC as initial condition) | KPI (100% SoC as initial condition) | Total Earning (AU$) Wind Only | Total Earning (AU$) MPC+FLC (50% SoC as initial condition) | Total Earning (AU$) MPC+FLC (100% SoC as initial condition) |
|---|---|---|---|---|---|
| 1st Day | 46.4283 ($N_x$ = 4) | 205.5859 ($N_x$ = 10) | 122,512 | 1,374,156 | 1,800,448 |
| 6th Days | 72.7851 ($N_x$ = 4) | 95.7511 ($N_x$ = 5) | 4,730,410 | 5,059,460 | 5,541,390 |
| 11th Days | 54.1024 ($N_x$ = 2) | 41.5057 ($N_x$ = 1) | 10,985,692 | 10,655,008 | 11,064,871 |
| 16th Days | 8.2408 ($N_x$ = 1) | 69.1632 ($N_x$ = 3) | 3,733,510 | 4,127,488 | 4,545,592 |
| 21st Days | 7.2963 ($N_x$ = 1) | 183.7023 ($N_x$ = 8) | 580,760 | 966,118 | 1,415,974 |
| 26th Days | 92.7878 ($N_x$ = 4) | 99.5753 ($N_x$ = 5) | 4,948,924 | 5,642,914 | 6,043,558 |
| 6 Months | 28.3374 ($N_x$ = 3) | 187.4950 ($N_x$ = 9) | 71,296,902 | 73,387,087 | 73,837,238 |

In addition, for all the scenarios, the total earning from the sale of power is greater when applying the proposed control system compared with selling wind power alone, except for the 11th days in each month using 50% SoC as initial condition. This is an exceptional happening as a result of having high wind power generation through mostly all the 11th days in each month, and on the contrary, the electricity price is low. Thus, having half-charged BESS at the beginning, the controls system will command the BESS to be charged from wind power and consequently less power is sold. While starting with full-charged BESS, the FLC increases the reference signal power to protect the BESS from being overcharged. This leads to selling more power compared to wind power alone. Therefore, it is clear that the total earning would be higher when starting with full-charged BESS than that with half-charged BESS.

## 3.9 Discussion and Conclusion

The detailed procedure of designing the discrete-time control system for dispatching wind power in the grid using the BESS with the aim of increasing financial benefits from the sale of power in the Australian NEM was presented in this chapter. After expressing the control requirements, the overall control system structure was proposed which consists of three main parts including a fuzzy decision-making system to generate the reference power signal using the online electricity price and time of the day, a discrete-time controller designed using MPC to provide a suitable trajectory tracking as well as maintaining battery's energy and rated power constraints within their permissible range, and a fuzzy logic controller to update the reference power signal to adapt the reference power signal to the wind power availability and BESS conditions. The designed discrete-time control system was evaluated using actual data for wind power and electricity 5-minute dispatch price under different simulation scenarios for selling the generate power to the market.

Preliminary indications from the obtained results suggest that the inclusion of the BESS with the proposed control system not only improves reliability, availability, and dispatch of the wind farm but also offers the potential to increase the generated income through higher earnings from the electricity market. However, such increased income generation needs to be properly assessed against the increased capital cost of the BESS and the implications on the economic viability of the solution. The reason is that not only our proposed solution increases the operating earning but also increases capital expenditure. Therefore, proper financial methods such as net present value (NPV) and return on investment (ROI) for any given project need to be used to assess the overall economic benefit. These assessments can be done provided that the lifetime of the battery is predicted under the operating conditions forced by the application. As the battery in the “time shifting” application faces irregular operating conditions such as partial state-of-charge cycling and different times between full charging, lifetime prediction is a difficult task to do, although it is essential for verifying economic benefits and lifecycle cost study. However, based on the work carried out in [99], a mathematical model for predicting a battery’s lifetime is derived for lead–acid batteries which is called “weighted Ah throughput” (Ah stands for ampere-hour). In that paper, it is mentioned that in order to find such a model for a different battery technology, some technical data are needed which can only be provided by the battery manufacturer (such as open circuit voltage at full charge, effective internal resistance, normalized reference current for current factor, acid stratification factor, etc.).

# **Part II:** Discrete-Time Control of Overhead Crane System

# Chapter 4

# Modeling of Overhead Crane

In this chapter, the overhead crane dynamic modeling is presented starting with an overview of the history of overhead crane in Section 4.1. Equations of motion for both 3D and 2D overhead cranes are derived in Section 4.2 with actuator description in Section 4.3. Following that, the application of independent joint modeling approach on overhead crane is discussed in Section 4.4. Section 4.5 covers the proposed procedure of model parameter identification with practical validation results. The derivation of discrete-time form of the overhead crane model obtained from independent joint modeling is presented in Section 4.6. Finally, a brief conclusion is given in Section 4.7.

## 4.1 Overview of Overhead Crane

The first overhead crane was built in Germany by Ludwig Stuckenholz AG (now Demag Cranes & Components GmbH) which was the first company in the world to mass-produce steam-powered cranes in mid 19$^{th}$ century. In 1876, Sampson Moore in England manufactured the first electric overhead crane. Fig. 4.1–1 shows a steam-powered overhead crane made in Germany in 1875 [100].

Unlike construction cranes, overhead cranes are typically used for either manufacturing or maintenance applications, mostly in heavy machinery industries, where efficiency and downtime are critical factors for the materials to be transported to different stages of manufacturing process until the finished product leaves the factory. There are many industries using overhead crane in their manufacturing process including, but not limited to, metal refinement industry (Fig. 4.1–2), paper mill industry (Fig. 4.1–3), automobile industry and many other manufacturing industries.

Overhead crane is one of the crane types used to move heavy and bulky loads through overhead space in a facility, warehouse or factory instead of through aisles or on the floor. It is also referred as industrial crane, bridge crane and overhead traveling crane. It can move the load in a three-dimensional (3D) Cartesian space and they have high lifting capacities for load movement. An overhead crane consists of three main parts: A parallel runways (rail), a traveling bridge (trolley/cart) spanning the gap between runways on a girder, and a hoist (lifting component of a crane) that travels along the bridge. Fig. 4.1–4 shows different parts of a 3D overhead crane in more details. If the bridge is rigidly supported on two or more legs running on a fixed rail at ground level, the crane is called gantry crane. Overhead cranes are normally directed by an expert operator either manually, with a wired pendant station or wireless control. The required forces for moving the load are mainly provided by electric or pneumatic-powered motors.

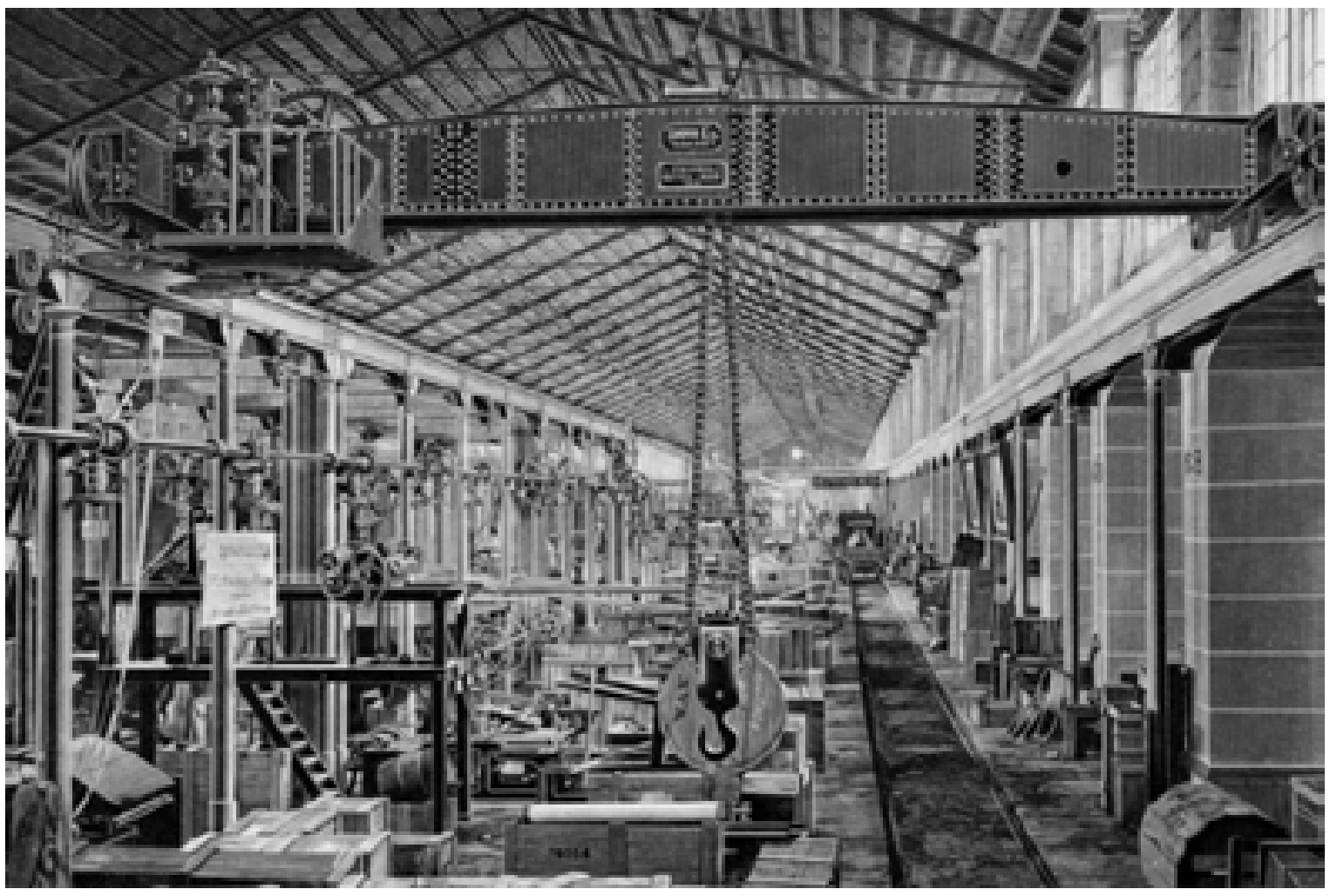

Fig. 4.1–1. A steam-powered overhead crane produced by Stuckenholz AG, Germany in 1875 [100].

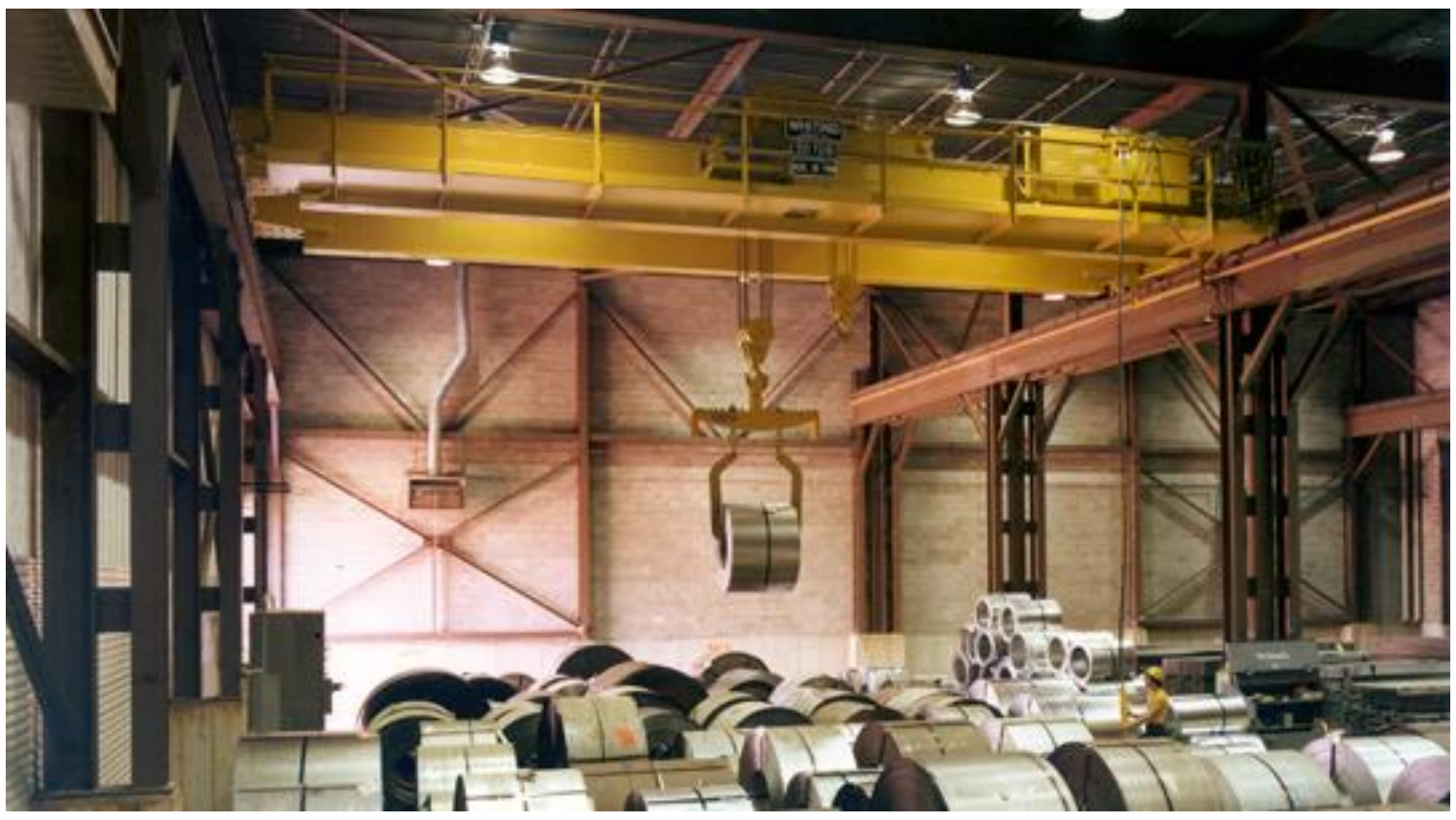

Fig. 4.1–2. Steel coil handling by an overhead crane in a steel refinement factory.

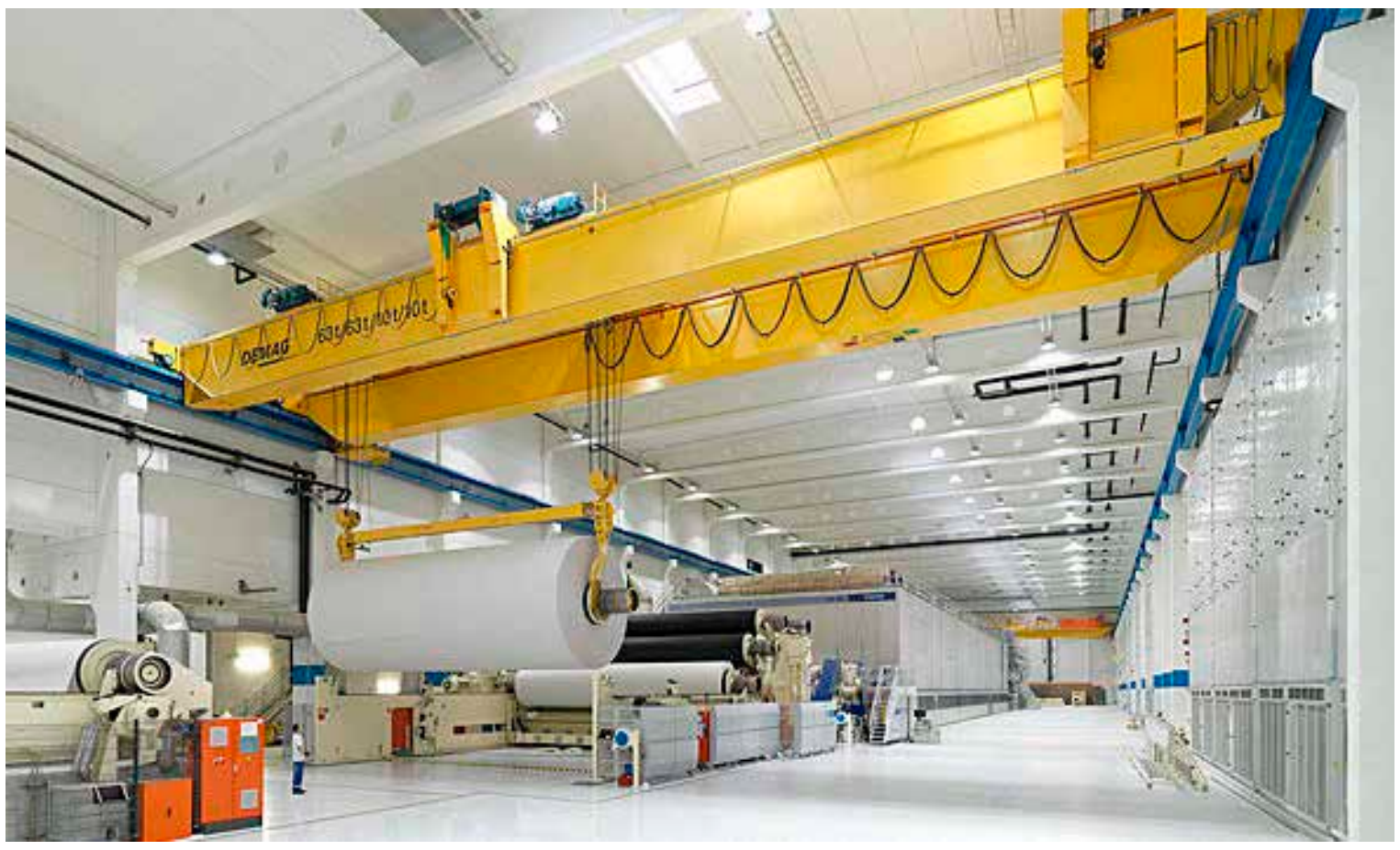

Fig. 4.1–3. Paper roll carried by an overhead crane in paper mill factory.

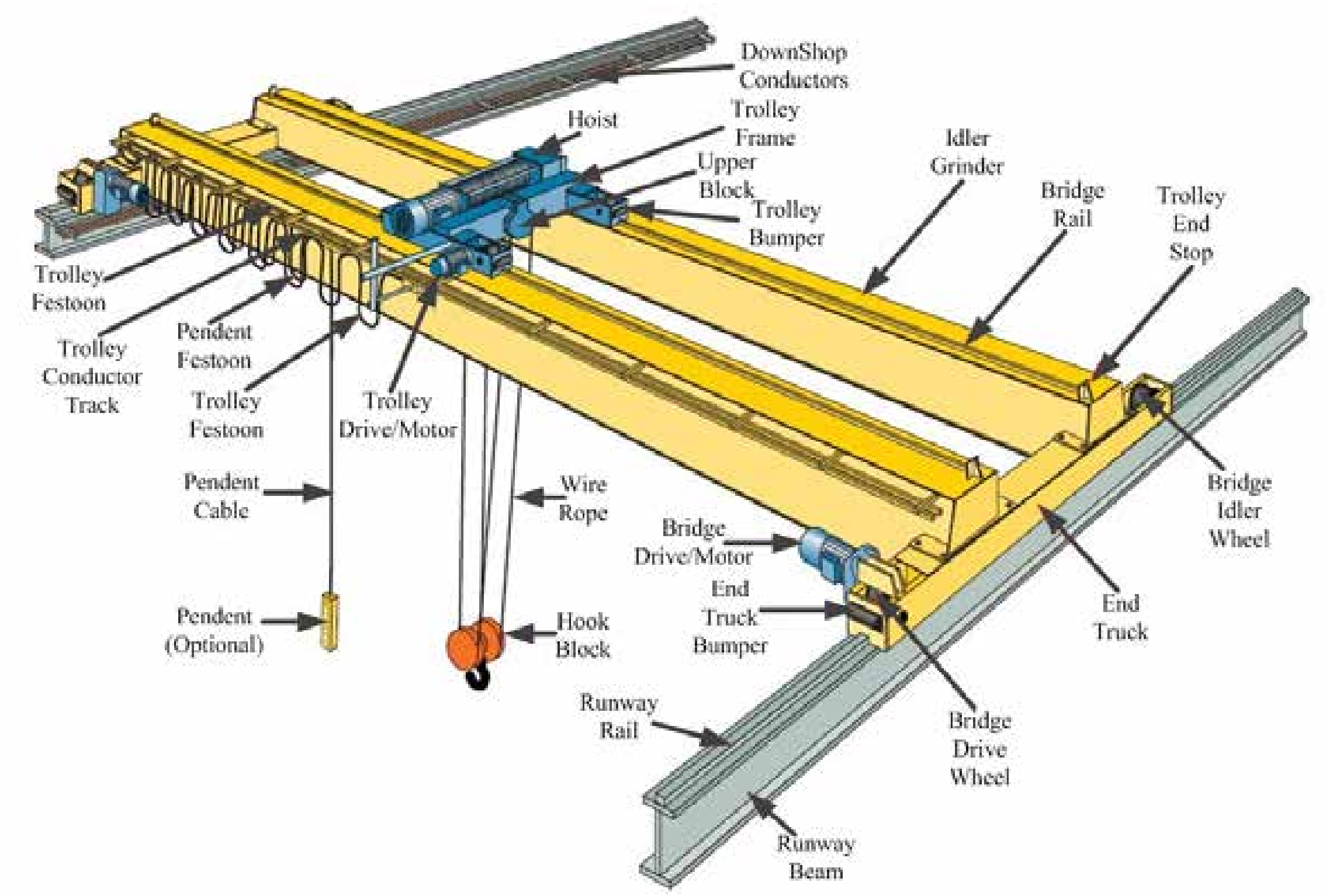

Fig. 4.1–4. Different parts of a 3D overhead crane.

# 4.2 Dynamic Model of Overhead Crane

## 4.2.1 Definition of Generalized Coordinates

The coordinate systems of a three-dimensional (3D) overhead crane and its load are illustrated in Fig. 4.2–1. The reference coordinate system is *XYZ* where the final position of the load is measured with respect to this coordinate system, and $X_TY_TZ_T$ is the trolley coordinate system, which is fixed on the trolley, and it is in parallel with the reference coordinate system. The position of the trolley in reference coordinate system is (*x*, *y*, 0). The trolley motion along *X* or $X_T$ direction (rail) is called traveling and its motion along *Y* or $Y_T$ direction (girder) is called traversing. Lifting the load up and down in *Z* or $Z_T$ direction is called hoisting. $\theta_l$ is the swing angle of the load in an arbitrary direction which can be separated into two components [37]: $\theta_x$ which is defined as swing angle along *X* direction (projection of $\theta_l$ on *XZ* plane) and $\theta_l$ defined as swing angle along *Y* direction (projection of $\theta_l$ on *XY* plane). Using these two swing angle definitions, the position of the load in reference coordinate system, i.e., ($x_m$, $y_m$, $z_m$), is obtained using the translation along the vector $[x \;\; y \;\; 0]^T$ as follows,

$$
\begin{aligned}
x_m &= x + lS_{q_x}C_{q_y}, \\
y_m &= y + lS_{q_y}, \\
z_m &= - lC_{q_x}C_{q_y},
\end{aligned}
\qquad (4.2\text{–}1)
$$

where $S_\theta$ and $C_\theta$ denote sin$q$ and cos$\theta$, respectively, and *l* is the length of the rope connecting the load to the hoist which is mounted on the trolley. In order to derive the dynamic model of the overhead crane and describe its motion, the generalized coordinates, $\boldsymbol{q} = [q_1 \;\; q_2 \;\; q_3 \;\; q_4 \;\; q_5]^T$,[9] are defined as *x* (position of trolley in *X*-axis direction), *y* (position of trolley in *Y*-axis direction), *l* (hoisting rope length), $\theta_x$ and $\theta_y$ (swing angles along *X* and *Y* directions, respectively). Thus, a 3D overhead crane has five degrees of freedom (5-DOF). Moreover, the following assumptions will be considered in the modeling procedure,

- Load mass is known and considered as a point mass.

[9] Lowercase bold italic font will be used to denote vector variable throughout this text.

- Mass and stiffness of the hoisting rope are neglected.
- Values of $x$, $y$, $l$, $\theta_x$ and $\theta_y$ are measurable.
- Connection between hoist and trolley is frictionless

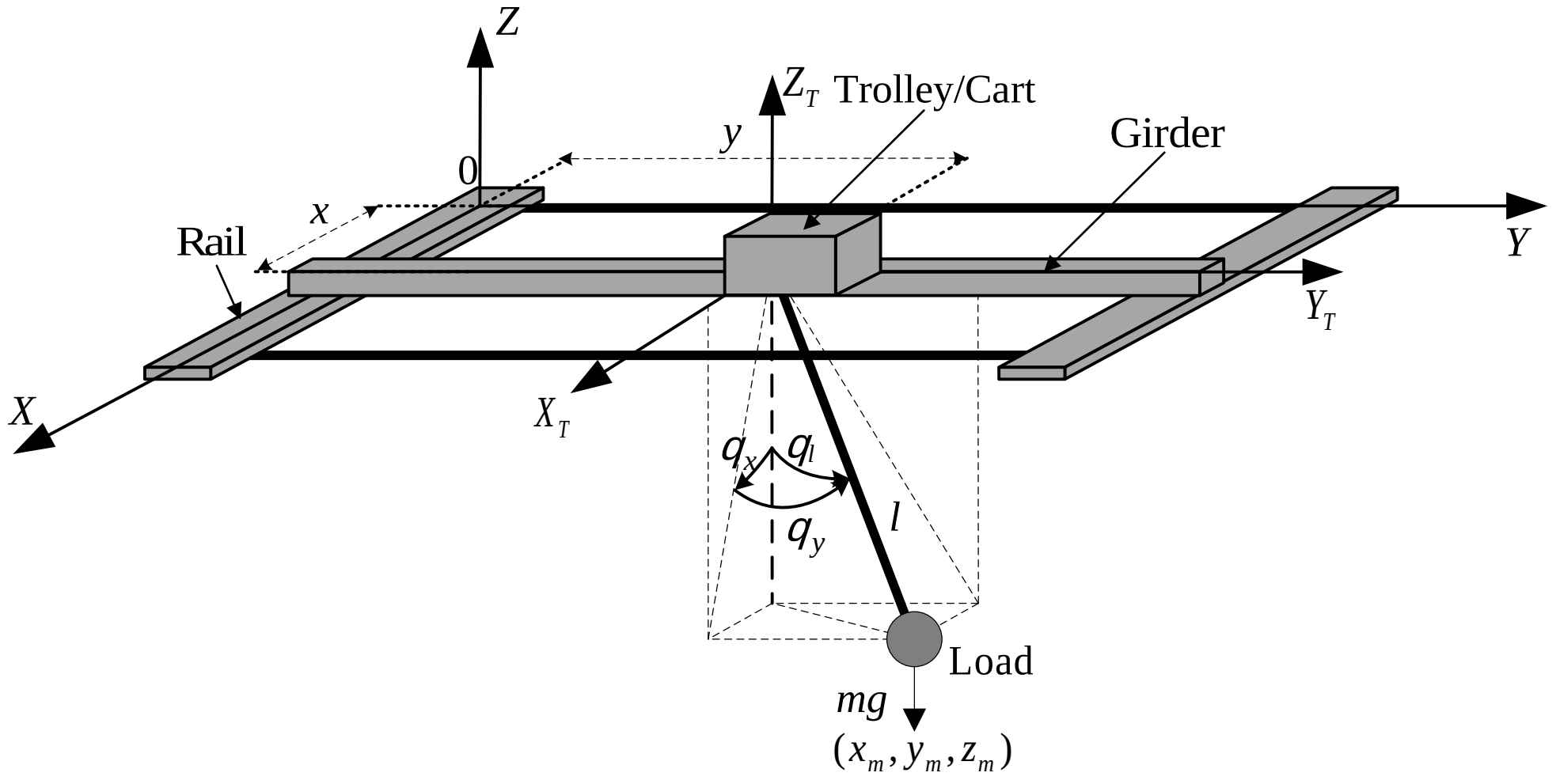

Fig. 4.2–1. Schematic structure and coordinate systems for a 3D overhead crane.

### 4.2.2 Equations of Motion for Overhead Crane

In this section, a set of coupled second-order ordinary differential equations that describe the time evolution of overhead crane system is derived using the method called Euler-Lagrange equations of motion which provides a formulation equivalent to those derived using Newton’s second law. To do this, we need to first find the kinetic and potential energies ($K$ and $P$, respectively) of the overhead crane in terms of the generalized coordinates we defined previously, i.e., $\boldsymbol{q} = [q_1 \;\; q_2 \;\; q_3 \;\; q_4 \;\; q_5]^T = [x \;\; y \;\; l \;\; \theta_x \;\; \theta_y]^T$, and then compute the equations of motion of overhead crane according to the Euler-Lagrange equations given as follows [32],

$$\frac{d}{dt}\frac{\partial L}{\partial \dot{q}_k} - \frac{\partial L}{\partial q_k} = f_k, \qquad for \quad k = 1, \ldots, 5, \tag{4.2–2}$$

where $f_k$ is the (generalized) force associated with $q_k$, $\dot{q}_k$ is the time-derivative of $q_k$,[10] and $L$ is the Lagrangian of the system defined as below,

[10] Throughout this text, the time-derivatives of a scalar $x$ is denoted by $\dot{x} = dx/dt$ and $\ddot{x} = d^2x/dt^2$.

$$L = K - P, \tag{4.2–3}$$

Since viscous damping effects mostly incorporates to the motion of mechanical systems when they accelerate, an extra term is added to the Euler-Lagrange equations in (4.2–2) to cover those velocity-related frictional forces as follows,

$$\frac{d}{dt}\frac{\partial L}{\partial \dot{q}_k} - \frac{\partial L}{\partial q_k} + \frac{\partial F_r}{\partial \dot{q}_k} = f_k, \qquad for \quad k = 1, \ldots, 5, \tag{4.2–4}$$

where $F_r$ is the Rayleigh dissipation function.

Kinetic energy of the overhead crane consists of kinetic energy of the overhead part and kinetic energy of the load. The kinetic and potential energies of the overhead crane are given as the following,

$$K = \frac{1}{2}(m_x \dot{x}^2 + m_y \dot{y}^2 + m_x \dot{l}^2) + \frac{1}{2} m \| v_m \|^2, \tag{4.2–5}$$

$$P = mgl(1 - C_{q_x} C_{q_y}), \tag{4.2–6}$$

where $m_x$, $m_y$, and $m_l$ are the traveling ($x$), traversing ($y$), and hoisting ($l$) components of the overhead crane mass, respectively, which each contains the equivalent masses of rotating parts such as motors and their drive trains; $m$ is the load mass; $g$ is the gravitational acceleration, and $v_m$ is the linear velocity vector of the load in reference coordinate system, i.e., $v_m = [\dot{x}_m \ \ \dot{y}_m \ \ \dot{z}_m]^T$, where its magnitude is obtained by taking time derivative from (4.2–1) as below [11],

$$\begin{aligned} \| v_m \|^2 &= \dot{x}_m^2 + \dot{y}_m^2 + \dot{z}_m^2 \\ &= (\dot{x} + \dot{l} S_{q_x} C_{q_y} + l C_{q_x} C_{q_y} \dot{q}_x - l S_{q_x} S_{q_y} \dot{q}_y)^2 \\ &\quad + (\dot{y} + \dot{l} S_{q_y} + l C_{q_y} \dot{q}_y)^2 \\ &\quad + (-\dot{l} C_{q_x} C_{q_y} + l S_{q_x} C_{q_y} \dot{q}_x + l C_{q_x} S_{q_y} \dot{q}_y)^2, \end{aligned} \tag{4.2–7}$$

which is simplified as follows,

[11] Throughout this text, the notation $\| v \|$ is used as the 2-norm or Euclidean norm for the magnitude of a vector.

$$\begin{aligned} \| v_m \|^2 = {} & \dot{x}^2 + \dot{y}^2 + \dot{l}^2 + l^2 C_{\theta_y}^2 \dot{\theta}_x^2 + l^2 \dot{\theta}_y^2 \\ & + 2(\dot{l} S_{\theta_x} C_{\theta_y} + l C_{\theta_x} C_{\theta_y} \dot{\theta}_x - l S_{\theta_x} S_{\theta_y} \dot{\theta}_y) \dot{x} \\ & + 2(\dot{l} S_{\theta_y} + l C_{\theta_y} \dot{\theta}_y) \dot{y}. \end{aligned} \tag{4.2–8}$$

Having found the kinetic and potential energies, the Lagrangian $L$ and Rayleigh dissipation function $F_r$ are given as below,

$$L = \frac{1}{2}(m_x \dot{x}^2 + m_y \dot{y}^2 + m_x \dot{l}^2) + \frac{1}{2} m \| v_m \|^2 + mgl(1 - C_{\theta_x} C_{\theta_y}), \tag{4.2–9}$$

$$F_r = \frac{1}{2}(D_x \dot{x}^2 + D_y \dot{y}^2 + D_x \dot{l}^2), \tag{4.2–10}$$

where $D_x$, $D_y$, and $D_l$ denote viscous damping coefficients associated with $x$, $y$, and $l$ motions, respectively. Substituting (4.2–8), (4.2–9), and (4.2–10) into Euler-Lagrange equations in (4.2–4) for each $q_k$ results in the following equations of motion for a 3D overhead crane,

$$\begin{aligned} & (m_x + m)\ddot{x} + ml C_{\theta_x} C_{\theta_y} \ddot{\theta}_x - ml S_{\theta_x} S_{\theta_y} \ddot{\theta}_y + m S_{\theta_x} C_{\theta_y} \ddot{l} + D_x \dot{x} + 2m C_{\theta_x} C_{\theta_y} \dot{l} \dot{\theta}_x \\ & - 2m S_{\theta_x} S_{\theta_y} \dot{l} \dot{\theta}_y - 2ml C_{\theta_x} S_{\theta_y} \dot{\theta}_x \dot{\theta}_y - ml S_{\theta_x} C_{\theta_y} \dot{\theta}_x^2 - ml S_{\theta_x} C_{\theta_y} \dot{\theta}_y^2 = f_x, \end{aligned} \tag{4.2–11}$$

$$\begin{aligned} & (m_y + m)\ddot{y} + ml C_{\theta_y} \ddot{\theta}_y + m S_{\theta_y} \ddot{l} + D_y \dot{y} \\ & + 2m C_{\theta_y} \dot{l} \dot{\theta}_y - ml S_{\theta_y} \dot{\theta}_y^2 = f_y, \end{aligned} \tag{4.2–12}$$

$$\begin{aligned} & (m_l + m)\ddot{l} + m S_{\theta_x} C_{\theta_y} \ddot{x} + m S_{\theta_y} \ddot{y} + D_l \dot{l} - ml C_{\theta_y}^2 \dot{\theta}_x^2 \\ & - ml \dot{\theta}_y^2 - mg C_{\theta_x} C_{\theta_y} = f_l, \end{aligned} \tag{4.2–13}$$

$$\begin{aligned} & ml^2 C_{\theta_y}^2 \ddot{\theta}_x + ml C_{\theta_x} C_{\theta_y} \ddot{x} + 2ml C_{\theta_y}^2 \dot{l} \dot{\theta}_x \\ & - 2ml^2 S_{\theta_y} C_{\theta_y} \dot{\theta}_x \dot{\theta}_y + mgl S_{\theta_x} C_{\theta_y} = 0, \end{aligned} \tag{4.2–14}$$

$$\begin{aligned} & ml^2 \ddot{\theta}_y + ml C_{\theta_y} \ddot{y} - ml S_{\theta_x} S_{\theta_y} \ddot{x} + 2ml \dot{l} \dot{\theta}_y \\ & + ml^2 C_{\theta_y} S_{\theta_y} \dot{\theta}_x^2 + mgl C_{\theta_x} S_{\theta_y} = 0, \end{aligned} \tag{4.2–15}$$

where $D_x$, $D_y$, and $D_l$ are the driving forces in $X$, $Y$, and $Z$ directions, respectively. The last two equations, i.e., (4.2–14) and (4.2–15), are called swing dynamics and as can be seen, there is no separate driving force for swing dynamics (right hand side of the equations are zero) which implies their unactuated behavior.

Using the obtained 3D overhead crane equations of motion, one can simply calculate the equations for a two-dimensional (2D) overhead crane (two motions in horizontal and vertical directions) by considering that there is no motion in either $X$ or $Y$ directions. Thus, the overhead crane variables and their time derivatives corresponding to that direction would be zero. Here, we assume no traversing motion, i.e. $y = \dot{y} = \ddot{y} = 0$ and $\theta_y = \dot{\theta}_y = \ddot{\theta}_y = 0$, and also the trolley is moved along $X$-axis, which then leads to the following equations of motion for a 2D overhead crane with traveling ($x$) and hoisting ($l$) motions with the schematic structure shown in Fig. 4.2–2,

$$(m_x + m)\ddot{x} + mlC_{\theta_x}\ddot{\theta}_x + mS_{\theta_x}\ddot{l} + D_x\dot{x} + 2mC_{\theta_x}\dot{l}\dot{\theta}_x - mlS_{\theta_x}\dot{\theta}_x^2 = f_x, \tag{4.2–16}$$

$$(m_l + m)\ddot{l} + mS_{\theta_x}\ddot{x} + D_l\dot{l} - ml\dot{\theta}_x^2 - mgC_{\theta_x} = f_l, \tag{4.2–17}$$

$$ml^2\ddot{\theta}_x + mlC_{\theta_x}\ddot{x} + 2ml\dot{l}\dot{\theta}_x + mglS_{\theta_x} = 0. \tag{4.2–18}$$

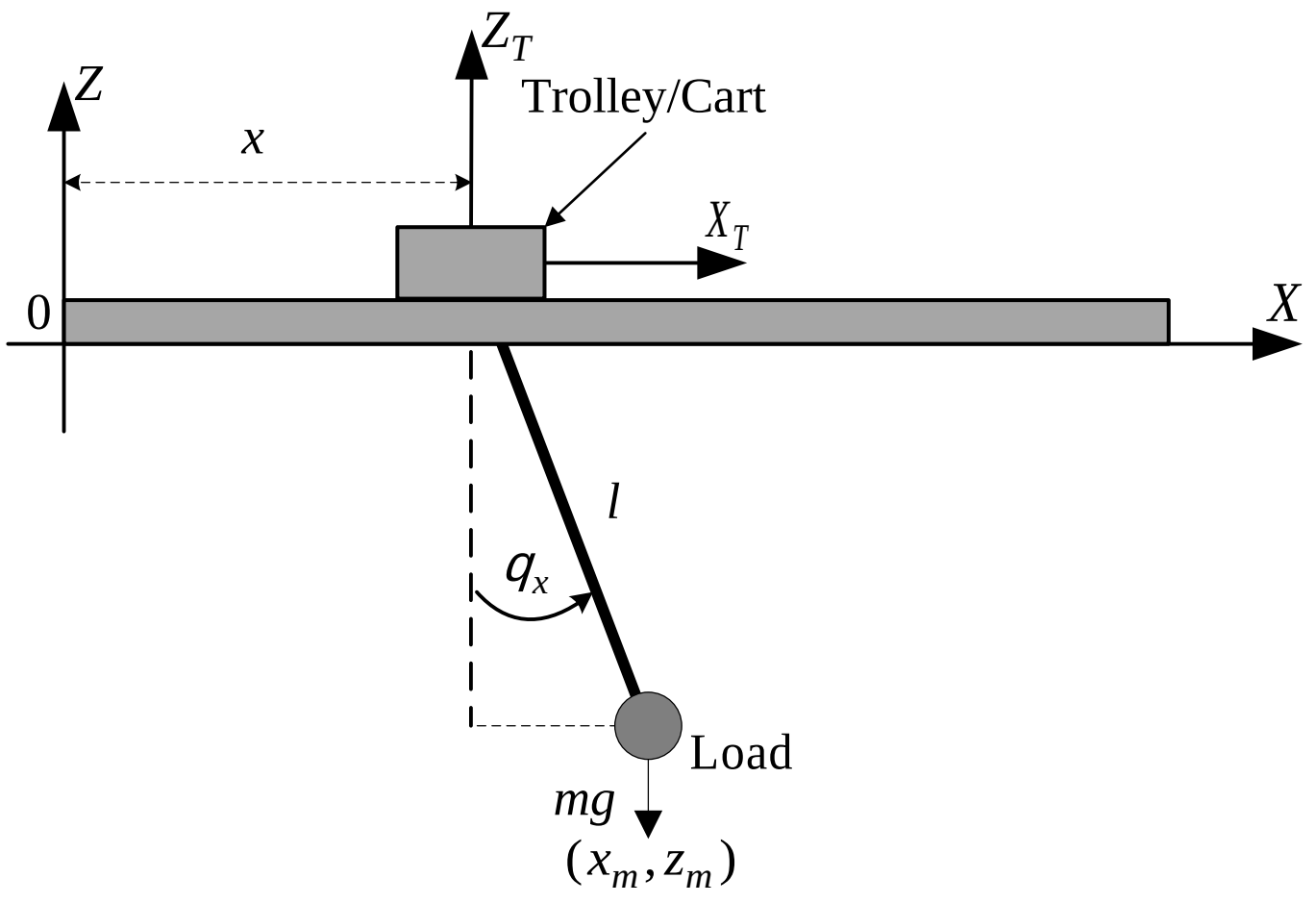

Fig. 4.2–2. Schematic structure and coordinate systems for a 2D overhead crane.

The obtained equations of motion for both 3D and 2D overhead crane can be simplified knowing that the load mass and hoisting rope length are always positive (i.e., $m > 0$ and $l > 0$), and both swing angles vary within the range of $\pm\pi/2$ (i.e., $|\theta_x| < \pi/2$, $|\theta_y| < \pi/2$, $|C_{\theta x}| > 0$, and $|C_{\theta y}| > 0$). Thus, the swing dynamics in (4.2–14) and (4.2–15) for 3D overhead crane can be divided by $mlC_{\theta y}$ and $ml$, respectively, leading to the following equations,

$$lC_{\theta_y}\ddot{\theta}_x = -C_{\theta_x}\ddot{x} - 2C_{\theta_y}\dot{l}\dot{\theta}_x + 2lS_{\theta_y}\dot{\theta}_x\dot{\theta}_y - gS_{\theta_x}, \tag{4.2–19}$$

$$l\ddot{\theta}_y = -C_{\theta_y}\ddot{y} + S_{\theta_x}S_{\theta_y}\ddot{x} - 2\dot{l}\dot{\theta}_y - lC_{\theta_y}S_{\theta_y}\dot{\theta}_x^2 - gC_{\theta_x}S_{\theta_y}. \tag{4.2–20}$$

Then, the terms $lC_{\theta y}\ddot{\theta}_x$ and $l\ddot{\theta}_y$ in traveling dynamics (4.2–11) and $l\ddot{\theta}_y$ in traversing dynamics (4.2–12) can be replaced by using (4.2–19) and (4.2–20). After further simplification using trigonometric identities, the overall simplified 3D overhead crane equations of motion are obtained as follows,

$$\begin{aligned}&(m_x + mS^2_{\theta_x}C^2_{\theta_y})\ddot{x} + mS_{\theta_x}S_{\theta_y}C_{\theta_y}\ddot{y} + mS_{\theta_x}C_{\theta_y}\ddot{l} + D_x\dot{x} - mlS_{\theta_x}C^3_{\theta_y}\dot{\theta}_x^2 \\ &- mlS_{\theta_x}C_{\theta_y}\dot{\theta}_y^2 - mgS_{\theta_x}C_{\theta_x}C^2_{\theta_y} = f_x,\end{aligned} \tag{4.2–21}$$

$$\begin{aligned}&(m_y + mS^2_{\theta_y})\ddot{y} + mC_{\theta_y}S_{\theta_x}S_{\theta_y}\ddot{x} + mS_{\theta_y}\ddot{l} + D_y\dot{y} - mlS_{\theta_y}C^2_{\theta_y}\dot{\theta}_x^2 \\ &- mlS_{\theta_y}\dot{\theta}_y^2 - mgC_{\theta_y}C_{\theta_x}S_{\theta_y} = f_y,\end{aligned} \tag{4.2–22}$$

$$(m_l + m)\ddot{l} + mS_{\theta_x}C_{\theta_y}\ddot{x} + mS_{\theta_y}\ddot{y} + D_l\dot{l} - mlC^2_{\theta_y}\dot{\theta}_x^2 - ml\dot{\theta}_y^2 - mgC_{\theta_x}C_{\theta_y} = f_l, \tag{4.2–23}$$

$$lC_{\theta_y}\ddot{\theta}_x + C_{\theta_x}\ddot{x} + 2C_{\theta_y}\dot{l}\dot{\theta}_x - 2lS_{\theta_y}\dot{\theta}_x\dot{\theta}_y + gS_{\theta_x} = 0, \tag{4.2–24}$$

$$l\ddot{\theta}_y + C_{\theta_y}\ddot{y} - S_{\theta_x}S_{\theta_y}\ddot{x} + 2\dot{l}\dot{\theta}_y + lC_{\theta_y}S_{\theta_y}\dot{\theta}_x^2 + gC_{\theta_x}S_{\theta_y} = 0. \tag{4.2–25}$$

Similarly, 2D overhead crane can be simplified if swing dynamics in (4.2–18) is divided by $ml$ and then, the term $l\ddot{\theta}_x$ in traveling dynamics (4.2–16) is replaced by simplified swing dynamics. This results in the following simplified 2D overhead crane equations of motion,

$$(m_x + mS^2_{\theta_x})\ddot{x} + mS_{\theta_x}\ddot{l} + D_x\dot{x} - mlS_{\theta}\dot{\theta}_x^2 - mgC_{\theta_x}S_{\theta_x} = f_x, \tag{4.2–26}$$

$$(m_l + m)\ddot{l} + mS_{\theta_x}\ddot{x} + D_l\dot{l} - ml\dot{\theta}_x^2 - mgC_{\theta_x} = f_l, \tag{4.2–27}$$

$$l\ddot{\theta}_x + C_{\theta_x}\ddot{x} + 2\dot{l}\dot{\theta}_x + gS_{\theta_x} = 0. \tag{4.2–28}$$

The reason for simplifying overhead crane equations of motion is to make traveling and traversing dynamics independent of swing angle accelerations since they are going to be incorporated into the design of the proposed control systems as nonlinear disturbances in addition to hoisting dynamics as will be explained in Chapter 5.

## 4.3 Actuator Dynamics

In many of today's overhead cranes, the required forces for traveling, traversing, and hoisting are commonly generated by electro-mechanical actuators such as permanent magnet (PM) DC motors with gearbox due to their high controllability and rated power [31], [32]. The PM DC motor dynamics consists of an electrical part and a mechanical part as shown in Fig. 4.3–1. The differential equation for the electrical part is given as follows,

$$L_m\dot{i}_a + R_m i_a + v_b = v_a, \qquad v_b = K_b\dot{\theta}_m, \tag{4.3–1}$$

where $L_m$ is the motor armature inductance; $R_m$ is the motor armature resistance; $i_a$ is the armature current; $v_b$ is the back electromotive force (EMF) voltage; $K_b$ is the back EMF constant; $\theta_m$ is the angular position of the motor before gearbox, and $v_a$ is the armature voltage being applied as the control input. In PM DC motor, the permanent magnet on the stator generates a constant flux. The torque on the rotor is then controlled by the armature current $i_a$ as below,

$$\tau_m = K_m i_a, \tag{4.3–2}$$

where $\tau_m$ is the motor torque and $K_m$ is the torque constant. The equation of motion for the mechanical part is given in terms of motor angle after gearbox is given as follows,

$$\frac{J_m}{r_g}\ddot{\theta}_g + \frac{B_m}{r_g}\dot{\theta}_g = \tau_m - r_g\tau_\ell - \tau_{cf}, \qquad (4.3\text{–}3)$$

where $\theta_g$, $\dot{\theta}_g$, and $\ddot{\theta}_g$ are the angular position, velocity, and acceleration of the motor after gearbox with gear reduction ratio $r_g$, respectively, (i.e., $\theta_g = r_g\,\theta_m$ for $0 < r_g < 1$); $J_m$ is the motor equivalent mass moment of inertia; $B_m$ is the equivalent viscous damping coefficient of the motor; $\tau_\ell$ is the load torque on the motor, and $\tau_{cf}$ is the total rotational coulomb friction including the friction caused by the interaction between the motor and its connected load. Since the mechanical time constant $L_m/R_m$ is frequently assumed to be much smaller than the mechanical time constant $J_m/B_m$, the inductive effects of motor winding can be ignored. This is a reasonable assumption for many electro-mechanical systems and leads to a reduced order model of the actuator dynamics [32]. Therefore, by using (4.3–1) in its steady state mode and (4.3–2), the electrical and mechanical dynamics can be combined together which results in the following dynamic equation for a geared PM DC motor with motor voltage $v_a$ as the control input and $\theta_g$ as the output,

$$\frac{J_m}{r_g}\ddot{\theta}_g + \frac{1}{r_g}\left(B_m + \frac{K_m K_b}{R_m}\right)\dot{\theta}_g = \frac{K_m}{R_m}v_a - r_g\tau_\ell - \tau_{cf}. \qquad (4.3\text{–}4)$$

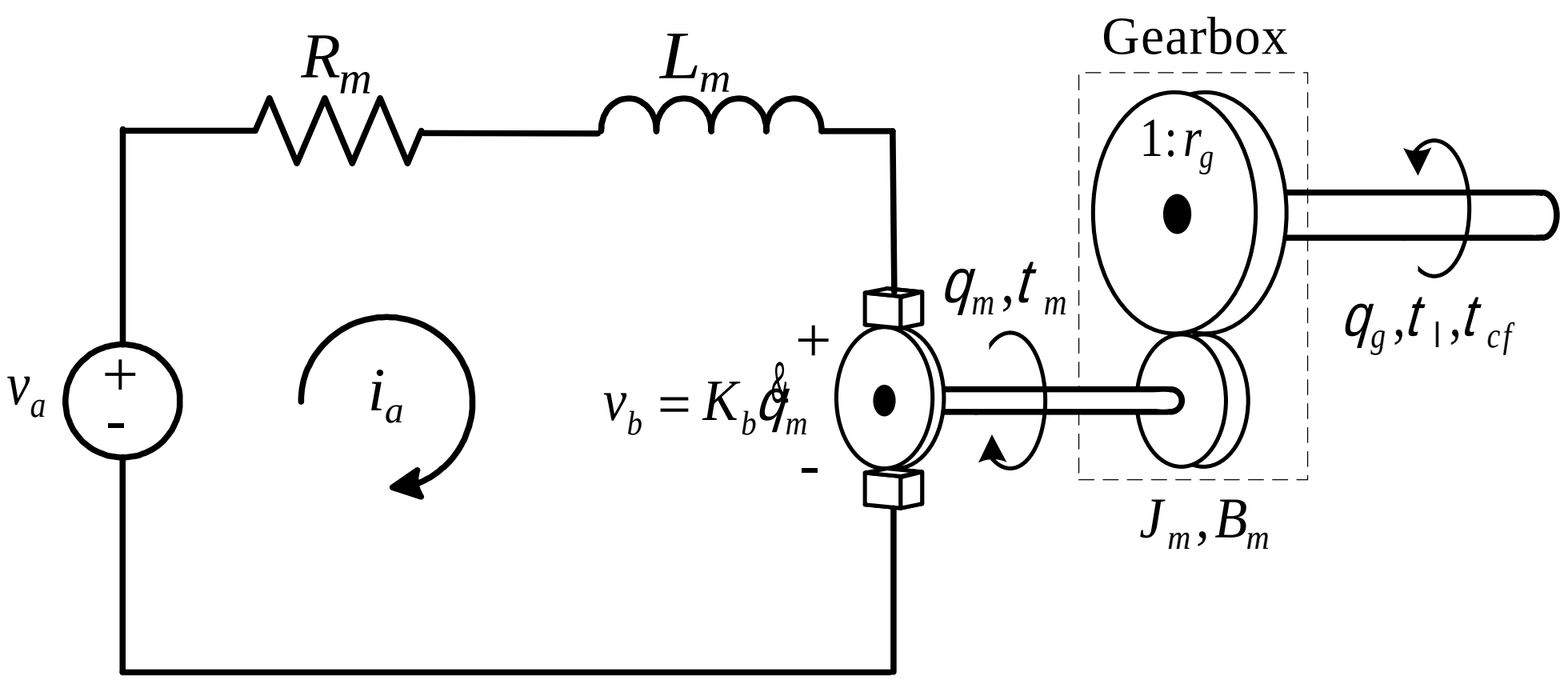

Fig. 4.3–1. Circuit diagram for an armature controlled PM DC motor.

## 4.4 Independent Joint Model

It should be noted that any control algorithm designed for an overhead crane should be converted in such a way that it can be applied to its actuators. Therefore, it is more convenient to consider the dynamics of the actuators in the overall system model [12]. Moreover, a system of pulleys and belts is often used to convert rotational motion of the motor to linear displacement $d$, i.e., $d = R_p\theta_g$ ($R_p$ is the radius of the pulley). Consequently, linear force becomes proportional to the torque as $f = \tau/R_p$ [31], [32]. Furthermore, in the case of a PM DC motor as the actuator for overhead crane, the load torque on each motor is generated by the overhead crane equations of motion. Moreover, the PM DC motor variables can be replaced by overhead crane variables, i.e., $\theta_{gx} = x/R_{px}$, $\theta_{gy} = x/R_{py}$, $\theta_{gl} = l/R_{pl}$, $t_{\ell x} = R_{px} f_x$, $t_{\ell y} = R_{py} f_y$, and $t_{\ell l} = R_{pl} f_l$. Therefore, by combining the simplified overhead crane equations of motion for traveling, traversing and hoisting obtained in (4.2–21), (4.2–22), and (4.2–23), respectively, and the actuator dynamics in (4.3–4) corresponding to $x$, $y$, and $l$ motions, we can rewrite the equations in terms of overhead crane variables as follows,

$$J_{ex}\ddot{x} + B_{ex}\dot{x} = K_{ex}v_{ax} - f_{dx} - f_{cfx}, \tag{4.4–1}$$

$$J_{ey}\ddot{y} + B_{ey}\dot{y} = K_{ey}v_{ay} - f_{dy} - f_{cfy}, \tag{4.4–2}$$

$$J_{el}\ddot{l} + B_{el}\dot{l} = K_{el}v_{al} - f_{dl} - f_{cfl}, \tag{4.4–3}$$

$$lC_{\theta_y}\ddot{\theta}_x + C_{\theta_x}\ddot{x} + 2C_{\theta_y}\dot{l}\dot{\theta}_x - 2lS_{\theta_y}\dot{\theta}_x\dot{\theta}_y + gS_{\theta_x} = 0, \tag{4.4–4}$$

$$l\ddot{\theta}_y + C_{\theta_y}\ddot{y} - S_{\theta_x}S_{\theta_y}\ddot{x} + 2\dot{l}\dot{\theta}_y + lC_{\theta_y}S_{\theta_y}\dot{\theta}_x^2 + gC_{\theta_x}S_{\theta_y} = 0, \tag{4.4–5}$$

where $J_{ex}$, $J_{ey}$, and $J_{el}$ are the total effective moment of inertia for the traveling, traversing, and hoisting motions, respectively, which include the effects of $m_x$ , $m_y$, and $m_l$ as well; $B_{ex}$, $B_{ey}$, and $B_{el}$ are the total damping effects of traveling, traversing, and hoisting motions, respectively, which include the effects of $D_x$, $Dy$, and $D_l$ as well; $f_{dx}$, $f_{dy}$, and $f_{dl}$ are the load effects of the overhead crane equations of motion on $X$, $Y$, and $Z$ directions, respectively, after gearbox; $f_{cfx}$, $f_{cfy}$, and $f_{cfl}$ are the coulomb friction forces acting on $x$, $y$, and $l$ motions, respectively[12], all are given in the following,

[12] Subscripts $x$, $y$, and $l$ refer to traveling, traversing, and hoisting motions, respectively.

$$J_{ex} = \frac{J_{mx}}{r_{gx}R_{px}} + r_{gx}R_{px}m_x, \qquad B_{ei} = \frac{B_{mi} + \frac{K_{mi}K_{bi}}{R_{mi}}}{r_{gi}R_{pi}} + r_{gi}R_{pi}D_i, \qquad K_{ei} = \frac{K_{mi}}{R_{mi}}, \qquad for \quad i = x, y, l, \tag{4.4–6}$$

$$\begin{aligned} f_{dx} &= r_{gx}R_{px}(f_x - m_x\ddot{x} - D_x\dot{x}), \\ &= r_{gx}R_{px}(mS_{\theta_x}^2 C_{\theta_y}^2 \ddot{x} + mS_{\theta_x} S_{\theta_y} C_{\theta_y}\ddot{y} + mS_{\theta_x} C_{\theta_y}\ddot{l} - mlS_{\theta_x} C_{\theta_y}^3 \dot{\theta}_x^2 \\ &\quad - mlS_{\theta_x} C_{\theta_y}\dot{\theta}_y^2 - mgS_{\theta_x} C_{\theta_x} C_{\theta_y}^2), \end{aligned} \tag{4.4–7}$$

$$\begin{aligned} f_{dy} &= r_{gy}R_{py}(f_y - m_y\ddot{y} - D_y\dot{y}), \\ &= r_{gy}R_{py}(mS_{\theta_y}^2 \ddot{y} + mC_{\theta_y} S_{\theta_x} S_{\theta_y}\ddot{x} + mS_{\theta_y}\ddot{l} - mlS_{\theta_y} C_{\theta_y}^2 \dot{\theta}_x^2 \\ &\quad - mlS_{\theta_y}\dot{\theta}_y^2 - mgC_{\theta_y} C_{\theta_x} S_{\theta_y}), \end{aligned} \tag{4.4–8}$$

$$\begin{aligned} f_{dl} &= r_{gl}R_{pl}(f_l - m_l\ddot{l} - D_l\dot{l}), \\ &= r_{gl}R_{pl}(m\ddot{l} + mS_{\theta_x} C_{\theta_y}\ddot{x} + mS_{\theta_y}\ddot{y} - mlC_{\theta_y}^2 \dot{\theta}_x^2 - ml\dot{\theta}_y^2 - mgC_{\theta_x} C_{\theta_y}), \end{aligned} \tag{4.4–9}$$

$$f_{cfi}(v_i) = \begin{cases} \alpha_{1i} & v_i > 0 \\ -\alpha_{2i} & v_i < 0 \end{cases}, \qquad for \quad i = x, y, l, \tag{4.4–10}$$

where $\alpha_{1i}$ and $\alpha_{2i}$ are the coulomb friction constants in positive and negative directions of motion with respect to the reference coordinate system (both are positive constants), and $v_i$ is the linear velocity, i.e., $v_i = di/dt$ for $i = x, y, l$.

The 2D overhead crane simplified dynamics in (4.2–26) and (4.2–27), respectively, can similarly be combined with actuator dynamics as below [101],

$$J_{ex}\ddot{x} + B_{ex}\dot{x} = K_{ex}v_{ax} - f_{dx} - f_{cfx}, \tag{4.4–11}$$

$$J_{el}\ddot{l} + B_{el}\dot{l} = K_{el}v_{al} - f_{dl} - f_{cfl}, \tag{4.4–12}$$

$$l\ddot{\theta}_x + C_{\theta_x}\ddot{x} + 2\dot{l}\dot{\theta}_x + gS_{\theta_x} = 0, \tag{4.4–13}$$

where $J_{ex}$, $J_{el}$, $B_{ex}$, $B_{el}$, $K_{ex}$, $K_{el}$, are defined similar as in (4.4–6); $f_{cfx}$ and $f_{cfl}$ defined similarly as in (4.4–10) for $x$ and $l$ motions, and $f_{dx}$ and $f_{dl}$ are given as below[13],

[13] Note that the unit of $f_{di}$ and $f_{cfi}$ in the obtained equations are N.m as in torque.

$$\begin{aligned} f_{dx} &= r_{gx} R_{px} ( f_x - m_x \ddot{x} - D_x \dot{x}), \\ &= r_{gx} R_{px} (m\ddot{x} + mlC_\theta \ddot{\theta} + mS_\theta \ddot{l} + 2mC_\theta \dot{l}\dot{\theta} - mlS_\theta \dot{\theta}^2 ), \end{aligned} \tag{4.4–14}$$

$$\begin{aligned} f_{dl} &= r_{gl} R_{pl} ( f_l - m_l \ddot{l} - D_l \dot{l}), \\ &= r_{gl} R_{pl} (m\ddot{l} + mS_\theta \ddot{x} - ml\dot{\theta}^2 - mgC_\theta). \end{aligned} \tag{4.4–15}$$

### 4.4.1 Remarks on the Dynamic Model

The main idea behind this modeling approach for overhead crane is inspired by independent joint control strategy which is a common control method in robot manipulator control field [31], [32]. In this method, the system actuators that are moving the joints are considered as the main process to be controlled. The coupling effects between joints mainly caused by nonlinear dynamics of the system are then modelled as disturbances acting on each actuator. This results in a decoupled dynamic model where the motion of each joint of the manipulator can be controlled by the corresponding actuator independently, not to mention that the control inputs are now the actual applied voltages to the motors which makes more sense in practice and that is why it is called independent joint model.

In the case of overhead crane, the traveling, traversing, and hoisting motors are the actuators to be considered as the main process. When their dynamics are combined with the overhead crane simplified equations of motion, it leads to two separate equations. The decoupled multi–input multi–output (MIMO) linear dynamic equations as derived in (4.4–1)–(4.4–3) and in (4.4–11) and (4.4–12) for 3D and 2D overhead cranes, respectively, and simplified swing dynamics obtained in (4.4–5) and (4.4–6) and in (4.4–13) for 3D and 2D overhead crane, respectively. The first set of equations can be used for tacking control purposes and swing dynamics can be used for load swing suppression. The effect of load swing on load positioning is reflected in the model with $f_{di}$ as nonlinear disturbances (for $i = x, y, l$), which can then be compensated using another technique used in robot manipulator control for disturbance rejection known as computed torque control [31] along with coulomb friction compensation which will be elaborated in Chapter 5.

In addition, not all the physical parameters of the actuator and overhead crane are provided by the manufacture and some are quite difficult to be measured manually such as the total moment of inertial and viscous damping coefficients due to inaccessibility of different mechanical parts already assembled on the overhead crane. The proposed model makes it much easier to make a good estimation of them using system identification techniques since they are combined together as linear coefficients in the decoupled equations which will be explained later in Section 4.5.

Moreover, coulomb friction effect ($f_{cfi}$) which is one of the significant load forces reducing the accuracy of load positioning ([62]) is added in the model as disturbance and its parameters can be identified alongside other unknown parameters of the system. The mass of the load is also included as part of disturbances, and therefore the uncertainty on the value of the load mass does not affect the model parameters accuracy.

Furthermore, the separation of tracking control and load swing damping, and the linear nature of the obtained dynamic model for overhead crane enable us to design high-performance control systems much simpler with less complexity due to developing and utilizing the independent join model for overhead crane. This is a great advantage when it comes to feasibility of implementation of any control system for an industrial process in practice.

## 4.5 Model Parameters Identification

Unless using model-free control systems like fuzzy control [85] or neural network [86], which do not need to have the system parameters, model-based control system designs require that the values of system parameters are identified. As mentioned before, the values of some parameters can be provided by the manufacturer but the rest has to be determined, either by manual measurements or by using system identification techniques. Besides, after performing several experiments, we understood that coulomb friction forces deteriorate the performance of load positioning especially at the beginning and the end of trajectories where motors are operated at low speeds. This phenomenon is a common type of velocity-dependent nonlinear characteristic in many mechanical systems. These effects are mostly ignored in the design of control systems since the mechanical systems are either operated at high speeds or it is compensated by using high gain controllers. However, for high-performance position control with constraints on the control input, the coulomb friction effects should be taken into account in the design of the control system for overhead crane.

It should be noted that nonlinear friction models have been widely discussed in the literature such as [102]–[104]. Based on the results of our initial experiments on the overhead crane setup, the friction model given in (4.4–10) appeared to be suitable for this study. The details of the overhead crane setup which was used in this research for implementation and verification of the proposed control systems will be described further in Section 5.9.

It can be seen from the proposed overhead crane model[14] obtained in (4.4–1)–(4.4–3) that most of the unknown parameters of the overhead crane and its actuators are combined in a linear form. In addition, the friction model in (4.4–10) can be written in a linear form using Sign function as below,

$$f_{cfi}(v_i) = b_{1i}\,\mathrm{sgn}(v_i) + b_{2i}, \quad for \quad i = x, y, l, \tag{4.5–1}$$

where $b_{1i} = (a_{1i} + a_{2i})/2$ and $b_{2i} = (a_{1i} - a_{2i})/2$.

The remaining questions here are firstly how to handle the effects of swing dynamics in (4.4–4) and (4.4–5), and secondly, the nonlinear terms in disturbances $f_{dx}$, $f_{dy}$, and $f_{dl}$ which depend on swing angles as can be seen in (4.4–7)–(4.4–9). The answer to the above questions lies in how the proposed parameter identification procedure is performed. Since we wanted to utilize linear recursive least squares (RLS) technique to determine the unknown parameters, it is obvious that the required regression model should be linear in terms of unknown parameters [105], [106]. According to this fact, the parameter identification is performed on each direction of motion separately. That means, the traveling motor is initially run to move the overhead carne only in $X$ direction with traversing and hoisting motors being off and no crane load is attached to the hoisting rope ($m$ = 0). This causes traversing and hoisting dynamics in (4.4–2) and (4.4–3) to be inactive. The required input/output data is then collected for traveling dynamics in (4.4–1) to be used for traveling parameter identification. In the second step, similar task is conducted for traversing dynamics with no load and no traveling and hoisting actions. Finally, data collection and parameter estimation is performed for hoisting dynamics in (4.4–3) with a known overhead crane load being lifted up and down without $x$ and $y$ motions.

[14] From this point forward, our focus is on the 3D overhead crane model obtained in Section 4.4 and 4.6 when we refer to the overhead crane model unless mentioned otherwise.

As a result, there will be no load swings in all steps which causes swing angles and their derivatives to be zero, i.e., $\theta_y = \dot{\theta}_y = \ddot{\theta}_y = 0$ and $\theta_x = \dot{\theta}_x = \ddot{\theta}_x = 0$, and also swing dynamics will have no effect on the overhead crane motion. In addition, all nonlinear terms in disturbances depending on swing angle will be cancelled. In other word, when the girder is moved alone in the first step we have $f_{dx} = 0$ and similarly in the second step we have $f_{dy} = 0$, and for hoisting action in the third step we have $f_{dl} = r_{gl}R_{pl}\,(m\ddot{l} - mg)$. Eventually, traveling, traversing, and hoisting dynamics in (4.4–1)–(4.4–3) are converted into three independent equations for construction of the regression models at each step as given below,

$$J_{ex}\dot{v}_x + B_{ex}v_x = K_{ex}v_{ax} - b_{1x}\,\mathrm{sgn}(v_x) - b_{2x}, \tag{4.5–2}$$

$$J_{ey}\dot{v}_y + B_{ey}v_y = K_{ey}v_{ay} - b_{1y}\,\mathrm{sgn}(v_y) - b_{2y}, \tag{4.5–3}$$

$$J_{elm}\dot{v}_l + B_{el}v_l = K_{el}v_{al} + M_l - b_{1l}\,\mathrm{sgn}(v_l) + b_{2l}, \tag{4.5–4}$$

where $J_{elm} = J_{el} + r_{gl}R_{pl}\,m$; $M_l = r_{gl}R_{pl}\,mg$, and $v_x$, $v_y$, and $v_l$ are the traveling, traversing, and hoisting velocities, respectively. It should be mentioned that the measurable variables are the trolley position in $XY$ plane, i.e., $x$ and $y$, the rope length $l$, and the input motor voltages $v_{ax}$, $v_{ay}$, and $v_{al}$. The reason for writing the above equations in terms of velocities is to have consistency between all variables in the equations since the friction is a function of velocity. Moreover, the parameters $m$, $g$, $r_{gi}$, $R_{pi}$, and $K_{mi}$ are considered to be known as they are mostly available from manufacturer datasheet, and therefore the main parameters to be determined are $J_{ei}$, $B_{ei}$, $a_{1i}$ and $a_{2i}$ for $i = x, y, l$.

Thus, the regression models for traveling ((4.5–5)–(4.5–7)), traversing ((4.5–8)–(4.5–10)), and hoisting ((4.5–11)–(4.5–13)) are obtained by applying the backward difference method to approximate derivatives as given in the following, (i.e., $v = (x(kT_s) - x((k-1)T_s))/T_s$, where $v$ is the velocity; $x$ is the position, and $T_s$ is the sampling time). However, it should be mentioned that in case of considerable measurement noise, more sophisticate discretization methods and robust regression would be required for better accuracy.

$$v_x(k) = a_{1x}v_x(k-1) + a_{2x}v_{ax}(k-1) + a_{3x}\,\mathrm{sgn}(v_x(k)) + a_{4x}, \tag{4.5–5}$$

$$\underbrace{v_x(k)}_{y_x(k)} = \underbrace{[v_x(k-1) \quad v_{ax}(k-1) \quad \mathrm{sgn}(v_x(k)) \quad 1]}_{\varphi_x^T(k)} \underbrace{\begin{bmatrix} a_{1x} \\ a_{2x} \\ a_{3x} \\ a_{4x} \end{bmatrix}}_{\hat{\theta}_x}, \tag{4.5–6}$$

$$y_x(k) = \varphi_x^T(k)\hat{\theta}_x, \tag{4.5–7}$$

$$v_y(k) = a_{1y}v_y(k-1) + a_{2y}v_{ay}(k-1) + a_{3y}\,\mathrm{sgn}(v_y(k)) + a_{4y}, \tag{4.5–8}$$

$$\underbrace{v_y(k)}_{y_y(k)} = \underbrace{[v_y(k-1) \quad v_{ay}(k-1) \quad \mathrm{sgn}(v_y(k)) \quad 1]}_{\varphi_y^T(k)} \underbrace{\begin{bmatrix} a_{1y} \\ a_{2y} \\ a_{3y} \\ a_{4y} \end{bmatrix}}_{\hat{\theta}_y}, \tag{4.5–9}$$

$$y_y(k) = \varphi_y^T(k)\hat{\theta}_y, \tag{4.5–10}$$

$$v_l(k) = a_{1l}v_l(k-1) + a_{2l}\underbrace{(v_{al}(k-1) + \frac{M_l}{K_{al}})}_{v'_{al}(k-1)} + a_{3l}\,\mathrm{sgn}(v_l(k)) + a_{4l}, \tag{4.5–11}$$

$$\underbrace{v_l(k)}_{y_l(k)} = \underbrace{[v_l(k-1) \quad v'_{al}(k-1) \quad \mathrm{sgn}(v_l(k)) \quad 1]}_{\varphi_l^T(k)} \underbrace{\begin{bmatrix} a_{1l} \\ a_{2l} \\ a_{3l} \\ a_{4l} \end{bmatrix}}_{\hat{\theta}_l}, \tag{4.5–12}$$

$$y_l(k) = \varphi_l^T(k)\hat{\theta}_l, \tag{4.5–13}$$

where $y_x(k)$, $y_y(k)$, and $y_l(k)$ are the measured outputs which are traveling, traversing, and hoisting velocities, respectively, at time step $k$ (i.e., $t = kT_s$ for $k = 1, 2, \ldots, N$, with $N$ is the total number of samples); $\varphi_x(k)$, $\varphi_y(k)$, and $\varphi_l(k)$ are the vectors of input/output data or regressors for traveling, traversing, and hoisting motions, respectively ($v'_{al}$ can be considered as hoisting input voltage with a DC offset), and $\hat{\theta}_x$, $\hat{\theta}_y$, and $\hat{\theta}_l$ are the vectors of estimated parameters for traveling, traversing, and hoisting motions, respectively, given as follows,

$$a_{1x} = \frac{J_{ex}}{J_{ex} + T_s B_{ex}}, \qquad a_{2x} = \frac{K_{ex} T_s}{J_{ex} + T_s B_{ex}},$$
$$a_{3x} = \frac{-b_{1x} T_s}{J_{ex} + T_s B_{ex}}, \qquad a_{4x} = \frac{-b_{2x} T_s}{J_{ex} + T_s B_{ex}}, \tag{4.5–14}$$

$$a_{1y} = \frac{J_{ey}}{J_{ey} + T_s B_{ey}}, \qquad a_{2y} = \frac{K_{ey} T_s}{J_{ey} + T_s B_{ey}},$$
$$a_{3y} = \frac{-b_{1y} T_s}{J_{ey} + T_s B_{ey}}, \qquad a_{4y} = \frac{-b_{2y} T_s}{J_{ey} + T_s B_{ey}}, \tag{4.5–15}$$

$$a_{1l} = \frac{J_{elm}}{J_{elm} + T_s B_{el}}, \qquad a_{2l} = \frac{K_{el} T_s}{J_{elm} + T_s B_{el}},$$
$$a_{3l} = \frac{-b_{1l} T_s}{J_{elm} + T_s B_{el}}, \qquad a_{4l} = \frac{-b_{2l} T_s}{J_{elm} + T_s B_{el}}. \tag{4.5–16}$$

Once the parameter identification is performed on the collected data from traveling, traversing, and hoisting motions separately, the unknown parameters of the overhead crane model are obtained as the following,

$$J_{ex} = \frac{K_{ex} T_s a_{1x}}{a_{2x}}, \qquad J_{ey} = \frac{K_{ey} T_s a_{1y}}{a_{2y}}, \qquad J_{el} = \frac{K_{el} T_s a_{1l}}{a_{2l}} - r_{gl} R_{pl} m, \tag{4.5–17}$$

$$B_{ex} = \frac{K_{ex}(1 - a_{1x})}{a_{2x}}, \qquad B_{ey} = \frac{K_{ey}(1 - a_{1y})}{a_{2y}}, \qquad B_{el} = \frac{K_{el}(1 - a_{1l})}{a_{2l}}, \tag{4.5–18}$$

$$b_{1i} = \frac{-K_{ei} a_{3i}}{a_{2i}}, \qquad b_{2i} = \frac{-K_{ei} a_{4i}}{a_{2i}}$$
$$\Rightarrow \quad a_{1i} = \frac{-K_{ei}(a_{3i} + a_{4i})}{a_{2i}}, \quad a_{2i} = \frac{-K_{ei}(a_{3i} - a_{4i})}{a_{2i}} \quad for \quad i = x, y, l. \tag{4.5–19}$$

## 4.5.1 Linear Recursive Least Squares

After building the regression model, linear least square technique can be used to estimate the system parameters by minimizing the squares of the differences between the actual output measurements and the estimated ones from the regression model [105], [106]. However, to increase the efficiency and accuracy of the estimated parameters, recursive least square technique can be implemented on the collected data. In this way, at each time step $k$, a correction term is added to the estimated parameters obtained at time step $k – 1$ using current measurements to update the estimations which is summarizes as below,

$$
\begin{aligned}
\hat{\theta}_i(k) &= \hat{\theta}_i(k-1) + \underbrace{P_i(k)\varphi_i(k)}_{C_i(k)}\varepsilon_i(k), \\
\varepsilon_i(k) &= y_i(k) - \varphi_i^T(k)\hat{\theta}_i(k-1), \\
P_i(k) &= P_i(k-1) - \frac{P_i(k-1)\varphi_i(k)\varphi_i^T(k)P_i(k-1)}{1+\varphi_i^T(k)P_i(k-1)\varphi_i(k)}, \quad for \quad i = x, y, l,
\end{aligned}
\tag{4.5–20}
$$

where $\varepsilon_i(k)$ is the estimation error and $C_i(k)$ is the estimator correction gain. This procedure is repeated for traveling, traversing and hoisting collected input/output data to obtain the model parameters in (4.5–17)–(4.5–19).

## 4.5.2 Practical Identification Results and Model Validation

The choice of the input test signal is an important factor in conducting successful parameter identification. To achieve that, a crucial condition is the so-called persistence excitation which is required for RLS to converge [105], [106]. In other words, the frequency content of the input test signal should contain both low and high frequencies to be able to excite most of the frequency modes of the system. Furthermore, the input voltage signal should be chosen such that the effects of coulomb friction force become noticeable in the output response. Thus, after studying different types of input voltage signals, the combination of sinusoidal voltage forms have shown to work well in terms of providing the above-mentioned requirements as shown in Fig. 4.5–1.

The practical results of identifying the parameters of traveling, traversing, and hoisting dynamics are provided in Fig. 4.5–2, Fig. 4.5–3, and Fig. 4.5–4, respectively. The estimated values of the identified parameters are also given in Table 4.5–1 with the values of the pre-known parameters. As can be seen from time history of the estimated parameters for traveling motion in Fig. 4.5–2, for instance, RLS identification ultimately converges to some constant values. Traversing and hoisting RLS identification similarly approach to their final values as illustrated in Fig. 4.5–3, and Fig. 4.5–4, respectively.

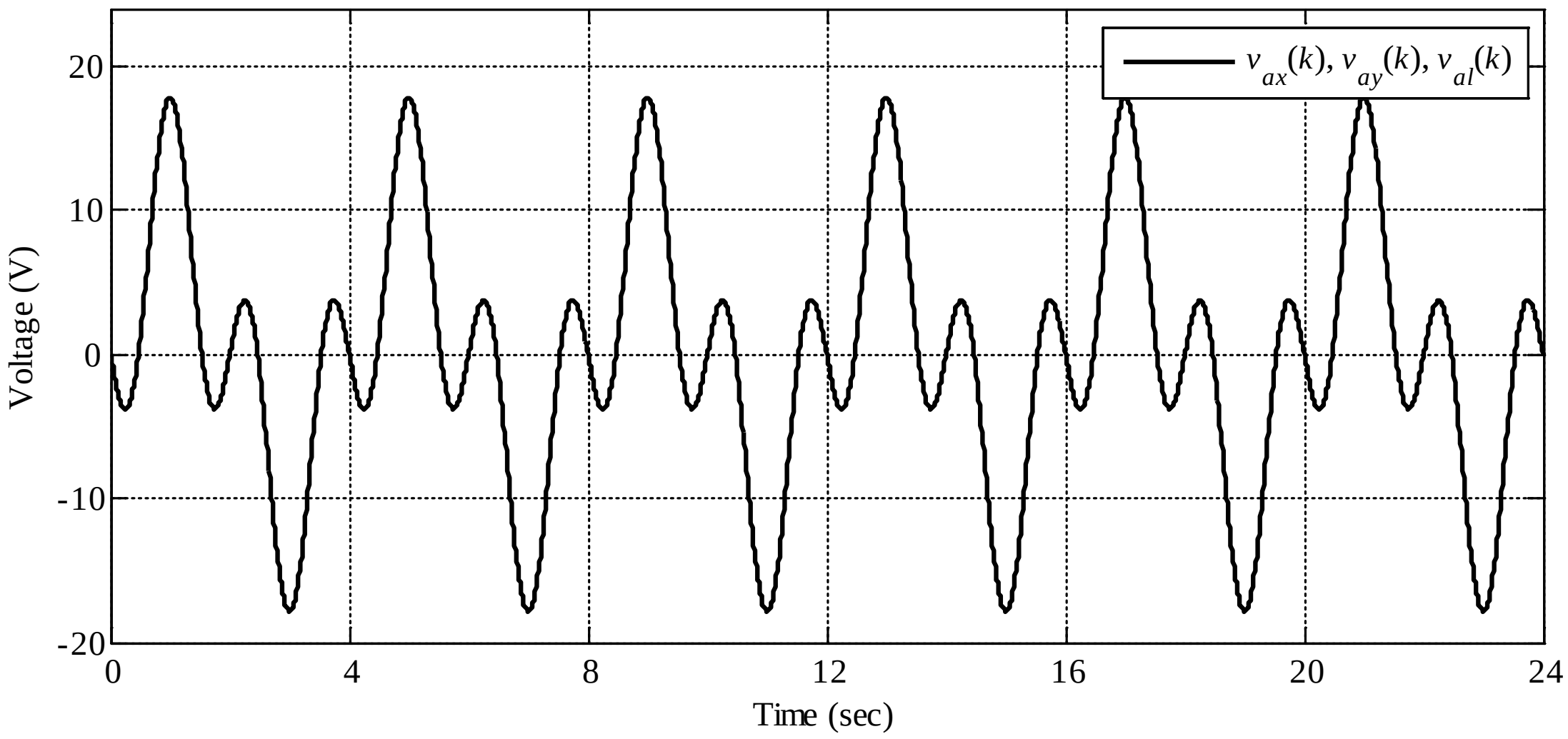

Fig. 4.5–1. Input test voltage signal for parameter identification applied to traveling, traversing, and hoisting actuators.

Table 4.5–1. Estimated Values of RLS Identification and Pre-Known Parameters

| *Parameters* | $J_{ei}$ (kg.m) | $B_{ei}$ (N.s) | $r_{gi}$ | $R_{pi}$ (m) | $K_{ei}$ (N.m/Amp.Ω) | $a_{1i}$ (N.m) | $a_{2i}$ (N.m) |
|---|---|---|---|---|---|---|---|
| Traveling | 75e−4 | 96.3e−3 | 13e−3 | 37.5e−3 | 14e−4 | 23e−4 | 21e−4 |
| Traversing | 40e−4 | 97.5e−3 | 13e−3 | 37.5e−3 | 14e−4 | 14e−4 | 11e−4 |
| Hoisting | 65e−4 | 24.55e−2 | 13e−3 | 13.5e−3 | 14e−4 | 13e−4 | 14e−4 |

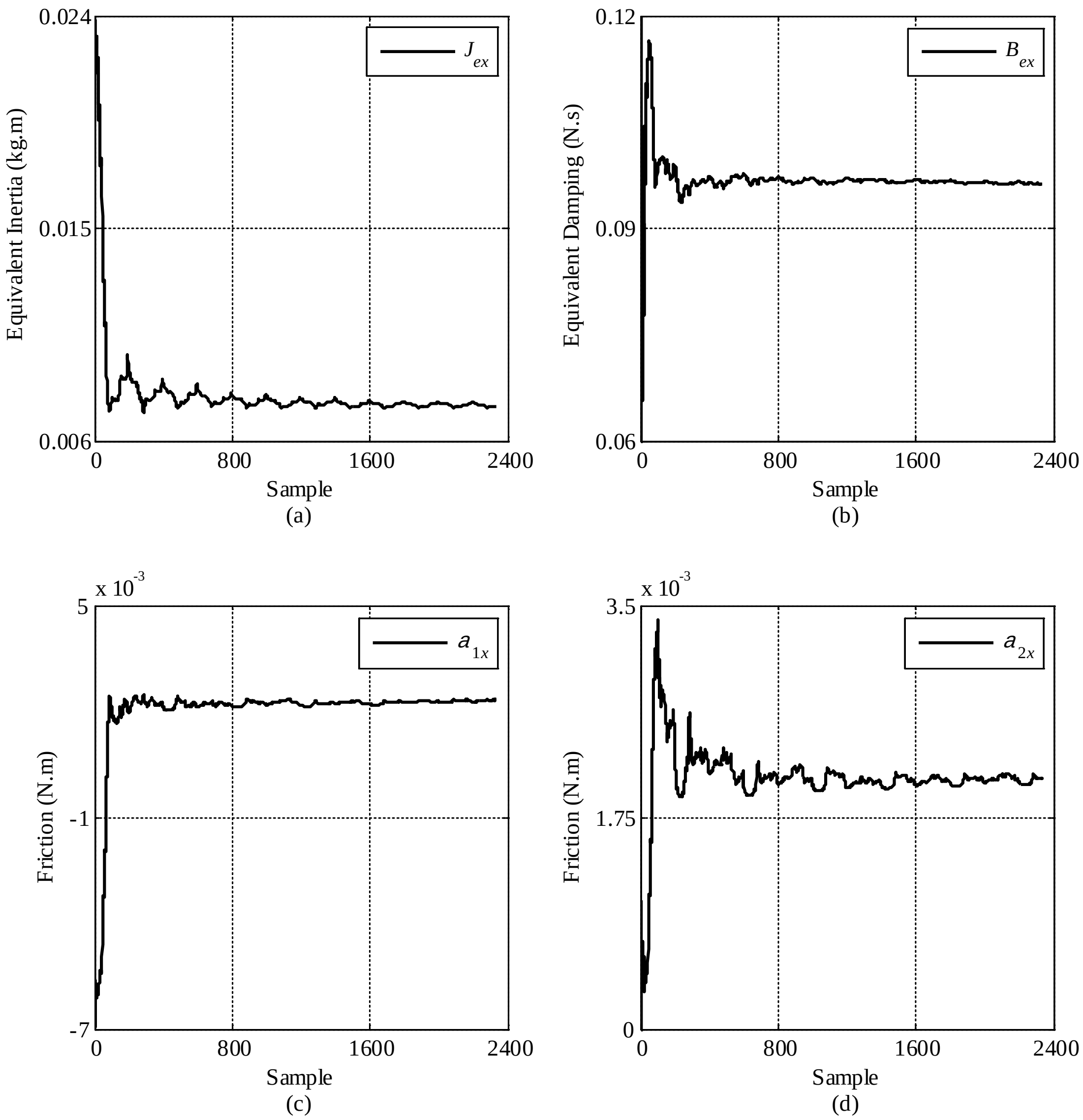

Fig. 4.5–2. Time history of traveling parameters estimation via RLS method, (a) Total effective moment of inertia, (b) Total effective viscous damping, (c) Coulomb friction constant in positive direction, (d) Coulomb friction constant in negative direction.

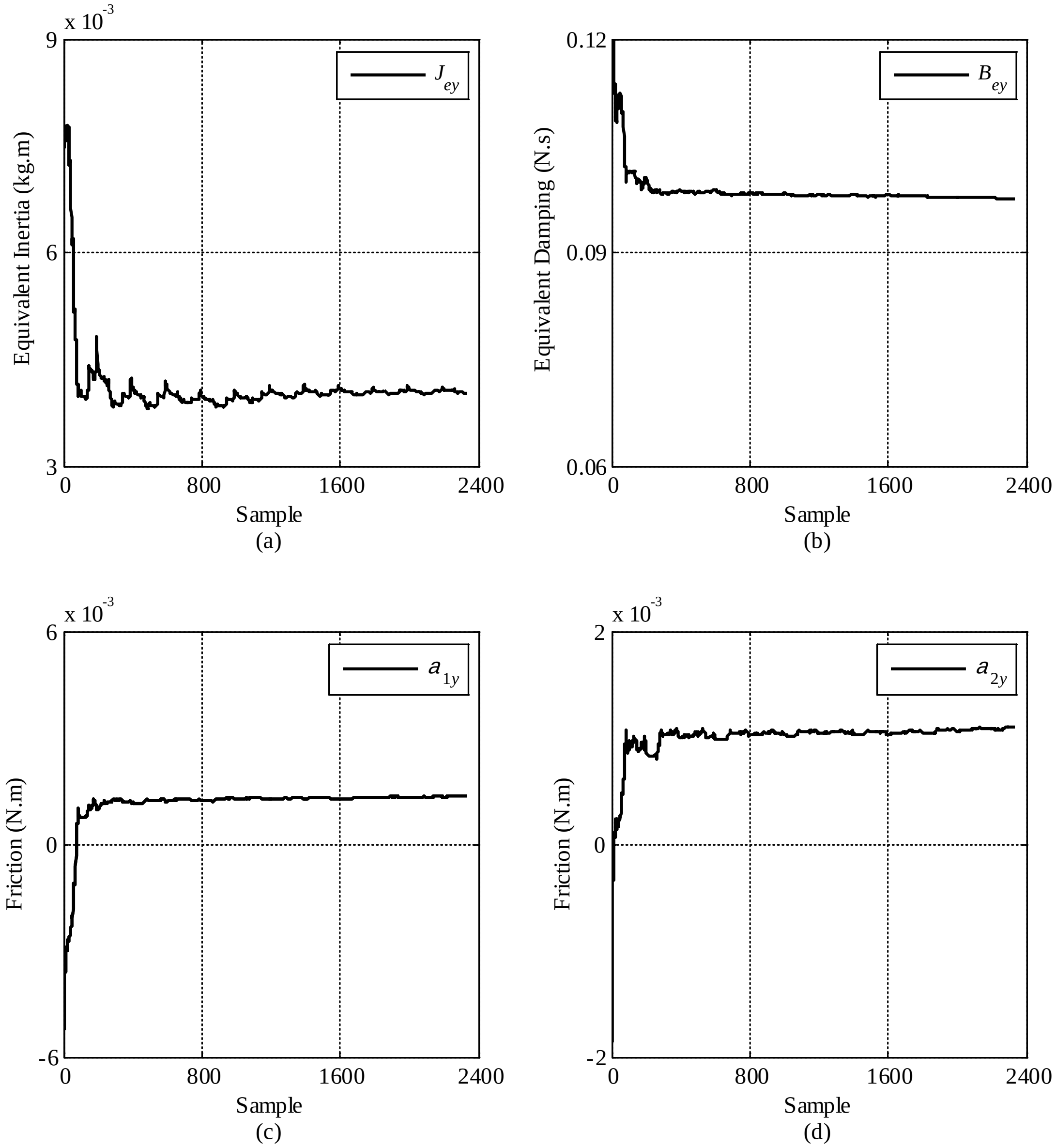

Fig. 4.5–3. Time history of traversing parameters estimation via RLS method. (a) Total effective moment of inertia, (b) Total effective viscous damping, (c) Coulomb friction constant in positive direction, (d) Coulomb friction constant in negative direction.

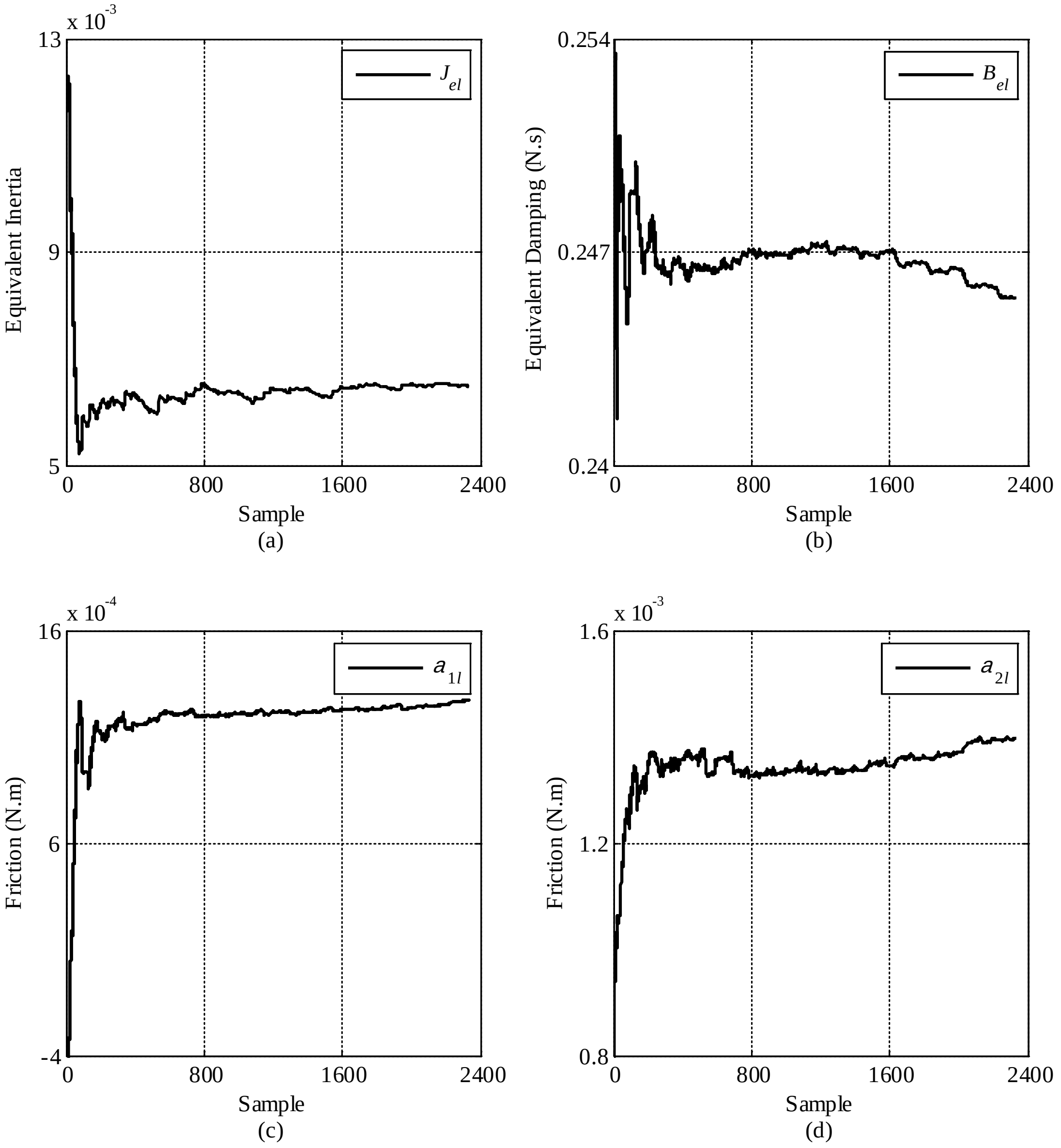

Fig. 4.5–4. Time history of hoisting parameters estimation via RLS method. (a) Total effective moment of inertia, (b) Total effective viscous damping, (c) Coulomb friction constant in positive direction, (d) Coulomb friction constant in negative direction.

More importantly, these results should be validated to show how accurately the identified model could represent the main plant behavior. To evaluate the results and demonstrate the accuracy of the identified model, the same input test signal is applied to the identified model. Then, the simulated position and velocity responses of the traveling, traversing, and hoisting models are compared with the actual responses separately, as pictured in Fig. 4.5–5, Fig. 4.5–6, and Fig. 4.5–7, respectively. It is interesting to observe how the coulomb friction force affects the responses, particularly in the position of the trolley in *X* direction (Fig. 4.5–5(b)) that causes it to fall behind the initial starting point as it continues moving back and forth. The effect of coulomb friction force is even stronger in traversing motion as depicted in Fig. 4.5–6(b). The hoisting rope length variations illustrate similar behavior as a result of coulomb friction force as can be seen in Fig. 4.5–7(b).

These results indicate the significance of including the coulomb friction effects in the dynamic model of the overhead crane. Moreover, Mean Square Error method (MSE) is used as a criterion to validate the precision of both velocity and position response comparison. The results are provided in Table 4.5–2 showing a high accuracy around the scale of $10^{-5}$. This is quite a promising outcome for the proposed identification approach in modeling the overhead crane with its actuators and coulomb friction force.

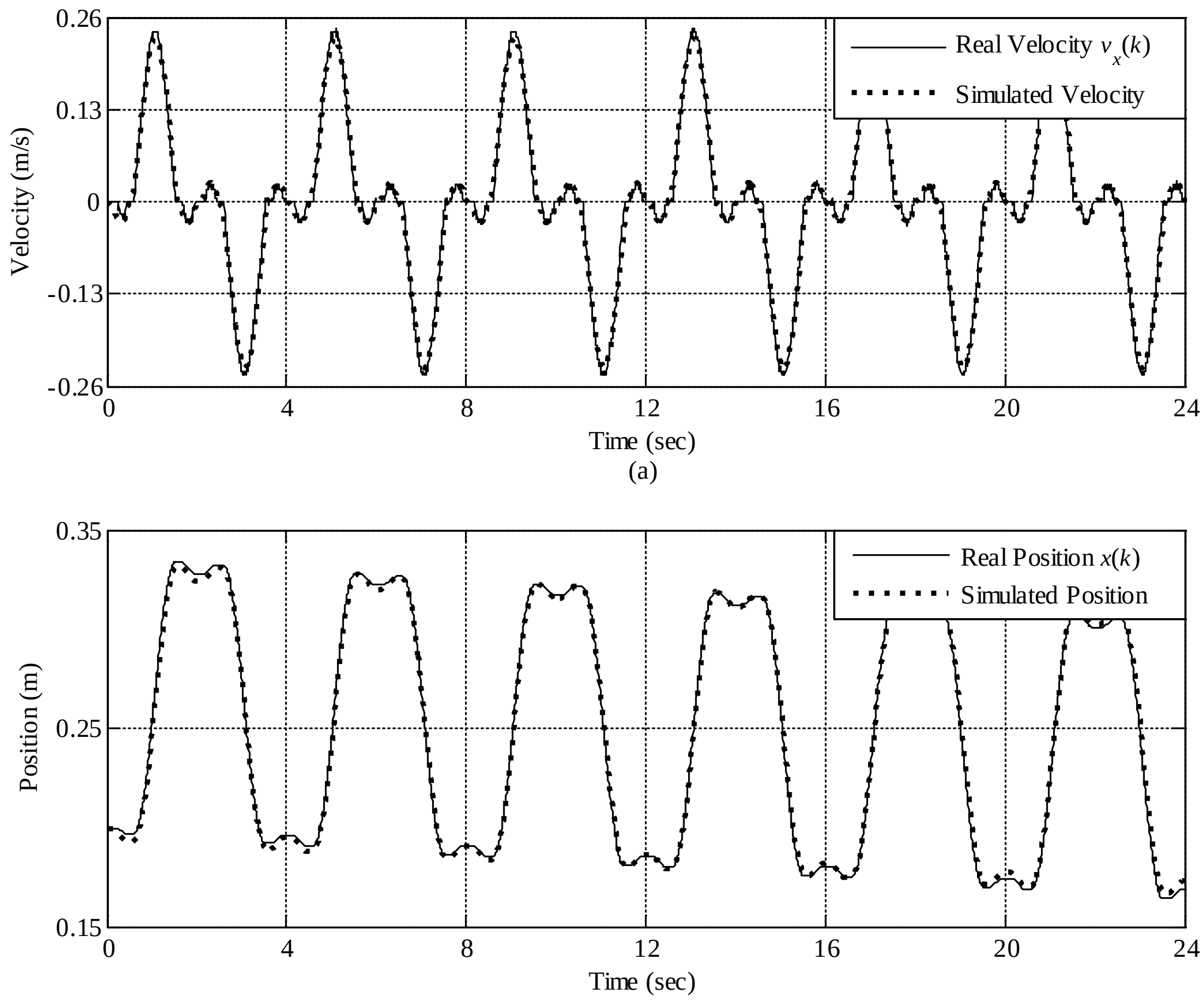

Fig. 4.5–5. The comparison between real and simulated responses for traveling motion. (a) Velocity responses, (b) Position responses.

Table 4.5–2. MSE Validation Criterion Results

| *MSE* | Traveling | Traversing | Hoisting |
|---|---|---|---|
| *Position* | 6.7004e−6 | 9.0855e−7 | 1.2911e−7 |
| *Velocity* | 7.8024e−5 | 4.1231e−5 | 4.3491e−6 |

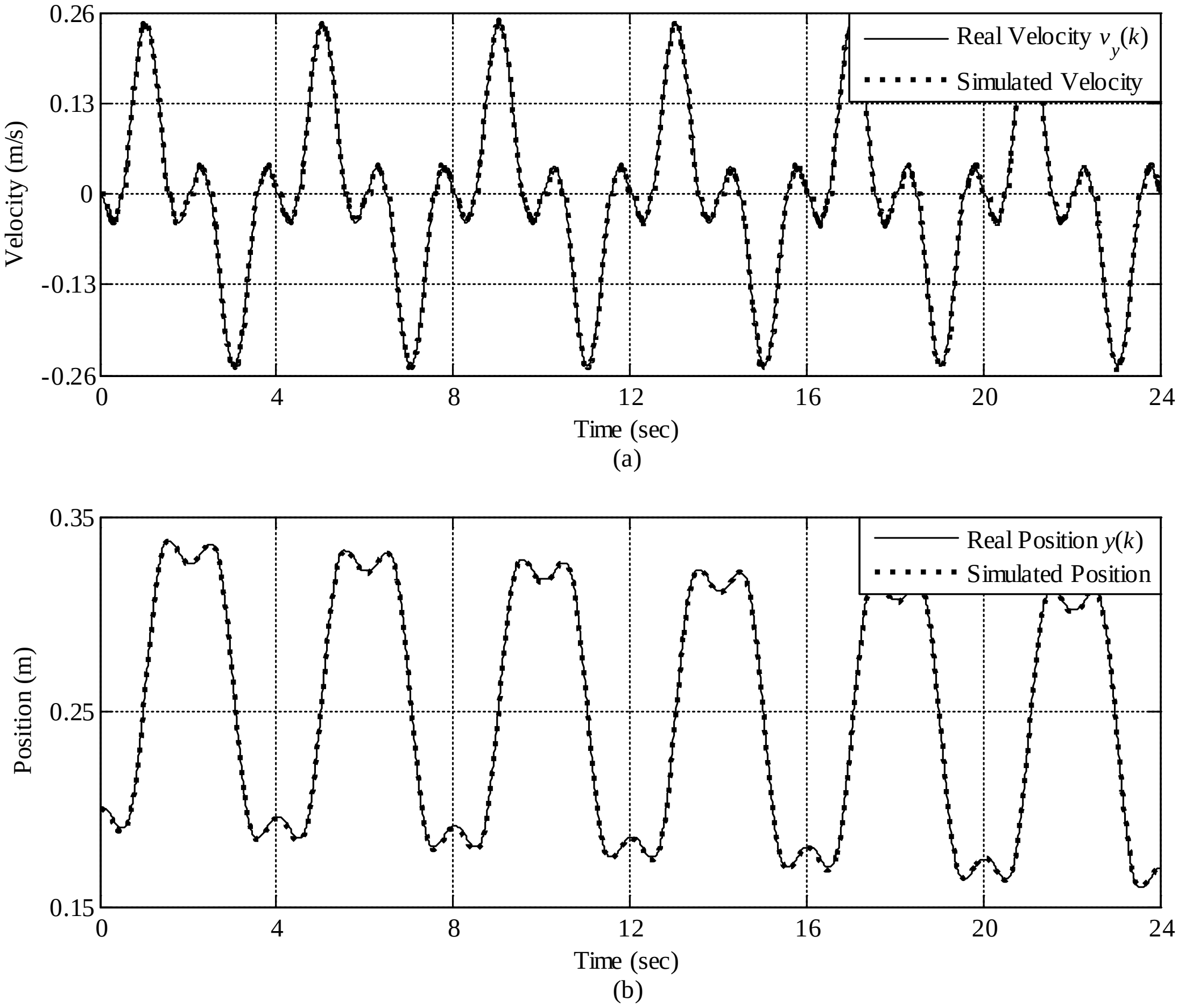

Fig. 4.5–6. The comparison between real and simulated responses for traversing motion. (a) Velocity responses, (b) Position responses.

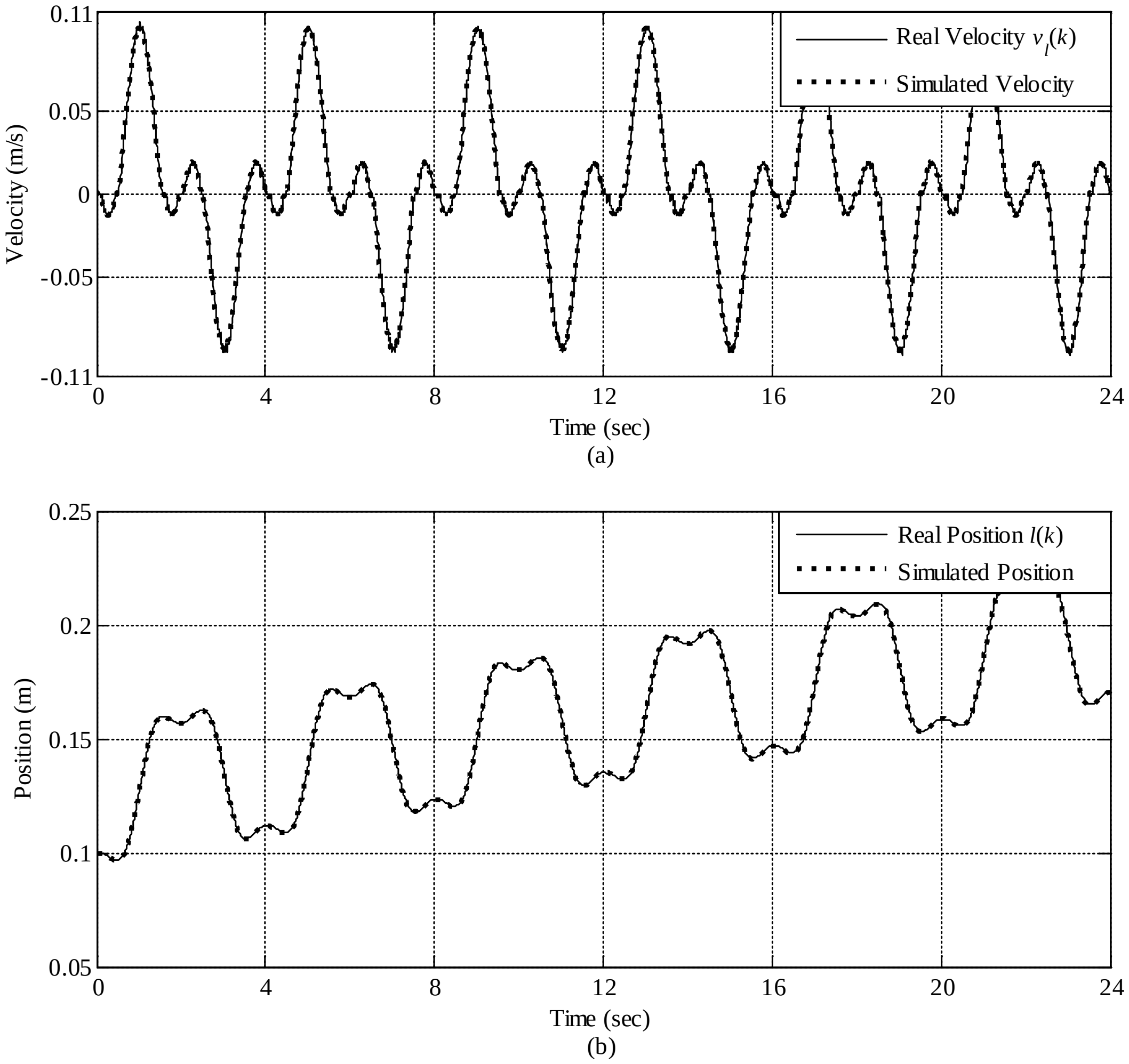

Fig. 4.5–7. The comparison between real and simulated responses for hoisting motion. (a) Velocity responses, (b) Position responses.

## 4.6 Discrete-Time State Space Model

Having obtained the overhead crane dynamic model using independent joint modeling, especially for traveling, traversing, and hoisting dynamics in (4.4–1), (4.4–2), and (4.4–3), respectively, they can be readily written in transfer function form in Laplace domain knowing the integral relation between position ($x$, $y$, $l$) and velocity ($v_x$, $v_y$, $v_l$), which makes the system to be type one, as follows,

$$x(s) = \frac{1}{s} v_x(s), \tag{4.6–1a}$$

$$v_x(s) = \frac{K_{ex}}{J_{ex}s + B_{ex}} v_{ax}(s) - \frac{1}{J_{ex}s + B_{ex}} (f_{dx} + f_{cfx}), \tag{4.6–1b}$$

$$y(s) = \frac{1}{s} v_y(s), \tag{4.6–2a}$$

$$v_y(s) = \frac{K_{ey}}{J_{ey}s + B_{ey}} v_{ay}(s) - \frac{1}{J_{ey}s + B_{ey}} (f_{dy} + f_{cfy}), \tag{4.6–2b}$$

$$l(s) = \frac{1}{s} v_l(s), \tag{4.6–3a}$$

$$v_l(s) = \frac{K_{el}}{J_{el}s + B_{el}} v_{al}(s) - \frac{1}{J_{el}s + B_{el}} (f_{dl} + f_{cfl}). \tag{4.6–3b}$$

As can be seen, the overall load disturbances ($f_{di} + f_{cfi}$) affect the velocities with the first-order transfer function similar to the one that relates the input voltages to the velocities in (4.6–1b)–(4.6–3b). And, the positions are simply obtained by integrating the velocities as shown in (4.6–1a)–(4.6–3a).

Considering the fact that the control system is eventually implemented on digital processors, it would be beneficial to design the control system directly in discrete-time to be able to deal with quantization errors and sampling time issues. Thus, using the techniques given in [107], the above transfer functions can be converted into discrete-time transfer functions in Z-domain as the following,

$$x(z) = \frac{T_s}{z-1} v_x(z), \tag{4.6–4a}$$

$$v_x(z) = \frac{b_{1x}}{z-a_{1x}} v_{ax}(z) - \frac{b_{d1x}}{z-a_{1x}} (f_{dx}(z) + f_{cfx}(z)), \tag{4.6–4b}$$

$$y(z) = \frac{T_s}{z-1} v_y(z), \tag{4.6–5a}$$

$$v_y(z) = \frac{b_{1y}}{z-a_{1y}} v_{ay}(z) - \frac{b_{d1y}}{z-a_{1y}} (f_{dy}(z) + f_{cfy}(z)), \tag{4.6–5b}$$

$$l(z) = \frac{T_s}{z-1} v_l(z), \tag{4.6–6a}$$

$$v_l(z) = \frac{b_{1l}}{z-a_{1l}} v_{al}(z) - \frac{b_{d1l}}{z-a_{1l}} (f_{dl}(z) + f_{cfl}(z)), \tag{4.6–6b}$$

where ($a_{1x}$, $b_{1x}$, $b_{d1x}$), ($a_{1y}$, $b_{1y}$, $b_{d1y}$), and ($a_{1l}$, $b_{1l}$, $b_{d1l}$), are the discrete-time transfer function coefficients for traveling, traversing and hoisting models, respectively, and they are all considered to have positive values, and $T_s$ is defined as the sampling time. The discrete-time transfer function coefficients are obtained using the zero-order-hold (ZOH) equivalent of the first-order continuous-time transfer function as follows [107],

$$G(s) = \frac{a}{s+a} \xleftrightarrow{Z\text{-}Transform} G_{ZOH}(z) = \frac{1-e^{-aT_s}}{z-e^{-aT_s}} v_l(z),$$

$$\Rightarrow \quad a_{1i} = e^{-\frac{B_{ei}T_s}{J_{ei}}}, \quad b_{1i} = \frac{K_{ei}}{B_{ei}}\left(1-e^{-\frac{B_{ei}T_s}{J_{ei}}}\right), \quad b_{d1i} = \frac{1}{B_{ei}}\left(1-e^{-\frac{B_{ei}T_s}{J_{ei}}}\right), \tag{4.6–7a}$$

$$for \quad i = x, y, l.$$

Separating the position transfer functions ((4.6–1a)–(4.6–3a)) from velocity transfer functions ((4.6–1b)–(4.6–3b)) makes it easier to build discrete-time state space representation of the overhead crane model in three subsystems. First, let us simplify the transfer functions obtained in (4.6–4)–(4.6–6) by multiplying both side of the equations by their denominators as below,

$$zx(z) = x(z) + T_s v_x(z), \tag{4.6–8a}$$

$$zv_x(z) = a_{1x} v_x(z) + b_{1x} v_{ax}(z) - b_{d1x}( f_{dx}(z) + f_{cfx}(z)), \tag{4.6–8b}$$

$$zy(z) - y(z) = T_s v_y(z), \tag{4.6–9a}$$

$$zv_y(z) = a_{1y} v_y(z) + b_{1y} v_{ay}(z) - b_{d1y}( f_{dy}(z) + f_{cfy}(z)), \tag{4.6–9b}$$

$$zl(z) = l(z) + T_s v_l(z), \tag{4.6–10a}$$

$$zv_l(z) = - a_{1l} v_l(z) + b_{1l} v_{al}(z) - b_{d1l}( f_{dl}(z) + f_{cfl}(z)). \tag{4.6–10b}$$

Now, using time shift property of Z-transform, discrete-time equations are given as below,

$$\begin{bmatrix} x(k+1) \\ v_x(k+1) \end{bmatrix} = \begin{bmatrix} 1 & T_s \\ 0 & a_{1x} \end{bmatrix} \begin{bmatrix} x(k) \\ v_x(k) \end{bmatrix} + \begin{bmatrix} 0 \\ b_{1x} \end{bmatrix} v_{ax}(k) + \begin{bmatrix} 0 \\ -b_{d1x} \end{bmatrix} ( f_{dx}(k) + f_{cfx}(k)), \tag{4.6–11}$$

$$\begin{bmatrix} y(k+1) \\ v_y(k+1) \end{bmatrix} = \begin{bmatrix} 1 & T_s \\ 0 & a_{1y} \end{bmatrix} \begin{bmatrix} y(k) \\ v_y(k) \end{bmatrix} + \begin{bmatrix} 0 \\ b_{1y} \end{bmatrix} v_{ay}(k) + \begin{bmatrix} 0 \\ -b_{d1y} \end{bmatrix} ( f_{dy}(k) + f_{cfy}(k)), \tag{4.6–12}$$

$$\begin{bmatrix} l(k+1) \\ v_l(k+1) \end{bmatrix} = \begin{bmatrix} 1 & T_s \\ 0 & a_{1l} \end{bmatrix} \begin{bmatrix} l(k) \\ v_l(k) \end{bmatrix} + \begin{bmatrix} 0 \\ b_{1l} \end{bmatrix} v_{al}(k) + \begin{bmatrix} 0 \\ -b_{d1l} \end{bmatrix} ( f_{dl}(k) + f_{cfl}(k)). \tag{4.6–13}$$

Therefore, by choosing the position and velocity of girder, trolley, and hoisting rope as state vector and the positions as the output vector, the discrete-time state space representation of the overhead crane model is obtained by combining (4.6–11)–(4.6–13) as follows,

$$\begin{aligned} \boldsymbol{x}(k+1) &= A\boldsymbol{x}(k) + B\boldsymbol{u}(k) + W_d \boldsymbol{f}_d(k), \\ \boldsymbol{y}(k) &= C\boldsymbol{x}(k), \end{aligned} \tag{4.6–14}$$

where $\boldsymbol{x}(k)$ is defined as state vector[15]; $\boldsymbol{u}(k)$ is the control input vector; $\boldsymbol{f}_d(k)$ is the vector of input disturbances including coulomb friction forces as well; $\boldsymbol{y}(k)$ is the output vector;

[15] Bold notations are used for vector variables and to avoid confusion between scalar position of trolley in $XY$ plane ($x$, $y$) with state and output vectors i.e., $\boldsymbol{x}(k)$ and $\boldsymbol{y}(k)$, respectively.

$A = BlockDiag\{A_x, A_y, A_l\}$ is the system matrix; $B = BlockDiag\{B_x, B_y, B_l\}$ is the control input matrix; $W_d = BlockDiag\{W_{dx}, W_{dy}, W_{dl}\}$ is the input disturbance matrix, and $C = BlockDiag\{C_x, C_y, C_l\}$ is the output matrix, all given as the following,

$$\begin{gathered} \boldsymbol{x}(k) = [x(k) \;\; v_x(k) \;\; y(k) \;\; v_y(k) \;\; l(k) \;\; v_l(k)]^T, \\ \boldsymbol{u}(k) = [v_{ax}(k) \;\; v_{ay}(k) \;\; v_{al}(k)]^T, \\ \boldsymbol{f}_d(k) = [\,f_{dx}(k) + f_{cfx}(k) \quad f_{dy}(k) + f_{cfy}(k) \quad f_{dl}(k) + f_{cfl}(k)]^T, \\ \boldsymbol{y}(k) = [x(k) \;\; y(k) \;\; l(k)]^T, \end{gathered} \tag{4.6–15}$$

$$\begin{gathered} A = \begin{bmatrix} A_x & \underline{0} & \underline{0} \\ \underline{0} & A_y & \underline{0} \\ \underline{0} & \underline{0} & A_l \end{bmatrix}, \; B = \begin{bmatrix} B_x & \underline{0} & \underline{0} \\ \underline{0} & B_y & \underline{0} \\ \underline{0} & \underline{0} & B_l \end{bmatrix}, \\ W_d = \begin{bmatrix} W_{dx} & \underline{0} & \underline{0} \\ \underline{0} & W_{dy} & \underline{0} \\ \underline{0} & \underline{0} & W_{dl} \end{bmatrix}, \; C = \begin{bmatrix} C_x & \underline{0} & \underline{0} \\ \underline{0} & C_y & \underline{0} \\ \underline{0} & \underline{0} & C_l \end{bmatrix} \end{gathered} \tag{4.6–16}$$

$$A_i = \begin{bmatrix} 1 & T_s \\ 0 & a_{1i} \end{bmatrix}, \; B_i = \begin{bmatrix} 0 \\ b_{1i} \end{bmatrix}, \; W_{di} = \begin{bmatrix} 0 \\ -b_{d1i} \end{bmatrix}, \; C_i = [1 \quad 0], \quad for \quad i = x, y, l, \tag{4.6–17}$$

where $\underline{0}$ is defined as a zero matrix with proper size.

It should be noted that although swing dynamics in (4.4–4) and (4.4–5) are not explicitly incorporated in the discrete-time state space model obtained for overhead crane, their effects are included in the model through disturbances $f_{dx}$, $f_{dy}$ and $f_{dl}$. Later on in Section 5.3, a robust load swing control will be developed based on swing dynamics and then integrated to the overall control systems designed in this thesis for overhead crane control to deal with the effects of load swings. Moreover, combining the decoupled traveling, traversing and hoisting equation obtained from independent joint modeling into state space form enables us to formulate the control system as one MIMO controller in charge of the entire control operation. This is an advantage over traditional independent joint control and computed torque control where a series of SISO controllers are used for each decoupled system. The reason is that if any of the SISO controllers starts malfunctioning during the operation, other controllers could be still working which would lead to faulty control operation, and perhaps total control system failure.

## 4.7 Discussion and Conclusion

In this Chapter, the overhead crane equations of motion in both 3D and 2D forms have been derived, and then the idea of independent joint model is applied to construct a dynamic model with linear-in-parameter form and system nonlinearities as disturbances acting on each actuator. The system identification procedure for determining the unknown parameters of an overhead crane has been developed which includes estimation of the coulomb friction constants and PM DC motors parameters as well. The procedure is simple yet quite effective in terms of the accuracy of the identified model since each actuator is driven separately for traveling, traversing, and hoisting motions as demonstrated by the practical validation results. The idea of independent joint model has made it possible to apply linear RLS technique to estimate the parameters. Furthermore, formulating the obtained independent joint model for the overhead crane in discrete-time state space form is advantageous from practical point of view since the designed control systems can be easily implemented on any digital computer and processor with less complexity regarding sampling time and quantization errors.

# Chapter 5

# Anti-Swing Tracking Control of Overhead Crane

This chapter addresses the procedure of designing the discrete-time control systems for the overhead crane. The main goals in the design of a high-performance overhead crane control system are expressed in Section 5.1. The details of the different parts of the overall control system structure are covered in Section 5.2. The design process of the control system starts with load swing control in Section 5.3 that also includes the design of swing angle observer. Reference trajectory planning is explained in Section 5.4 followed by the design of reference signal generator in Section 5.5. The formulation of the first discrete-time controller which is designed for trajectory tracking purpose is provided in Sections 5.6 containing model predictive control and state observer design. Section 5.7 describes the design of state feedback control and feedforward signal generation as the second discrete-time controller. An alternative method to computed torque control for estimating load disturbances is demonstrated in Section 5.8. The description of the experimental overhead crane setup for which the control system are implemented is given in Section 5.9 along with the extensive test results under different scenarios and trajectory speeds to validate the stability and show the performance of the designed discrete-time control system. Finally, Section 5.10 concludes the chapter with a brief discussion.

## 5.1 Control Objectives and Requirements

In Chapter 1, the overall objective for overhead crane control was stated. This objective is to develop and test advance control algorithms for automatic load transportation using overhead crane with the aim of improving the control performance of the overhead crane. To achieve this goal, the control requirements for high-performance load transportation using overhead crane should be defined. The main requirement, as in many other transportation systems, is to move the load as fast as possible with high accuracy at the final position where the load should be placed to increase efficiency. However, the mechanical structure of the overhead crane makes it difficult and quite challenging for achieving this requirement. Since overhead crane is built to carry heavy loads, the pendulum behavior of the hoisting rope creates load swing which could be considerably large, and potentially dangerous, if it is moved very fast without proper control, especially at the final point. A simple analogy to this situation is a fast moving car with a passenger having seat belt attached to his body rather than the seat. Once the car approaches its destination, the braking action will throw the passenger forward, and then pull him back heavily as a result of a large inertia and jerk. This will jeopardize the life of the passenger as the body moves back and forth with massive momentum. Now image what would happen if a huge load is moved at high speed by an overhead crane without proper control action. The resulting load swings create a significant load force on the overhead crane that could potentially damage the entire overhead crane, if not breaking it down, not to mention the potential danger to the surrounding objects that might be hit by the overhead crane load swings.

As a result, in manual operation of the overhead crane, an expert operator with many hours of training operates the overhead crane motion mostly with the help of a second operator on the ground giving him important hand signals to guide the main operator. Fig. 5.1–1 illustrates a sample of hand signals given by a second operator on the ground as feedback signals to the main operator controlling the overhead crane motion.

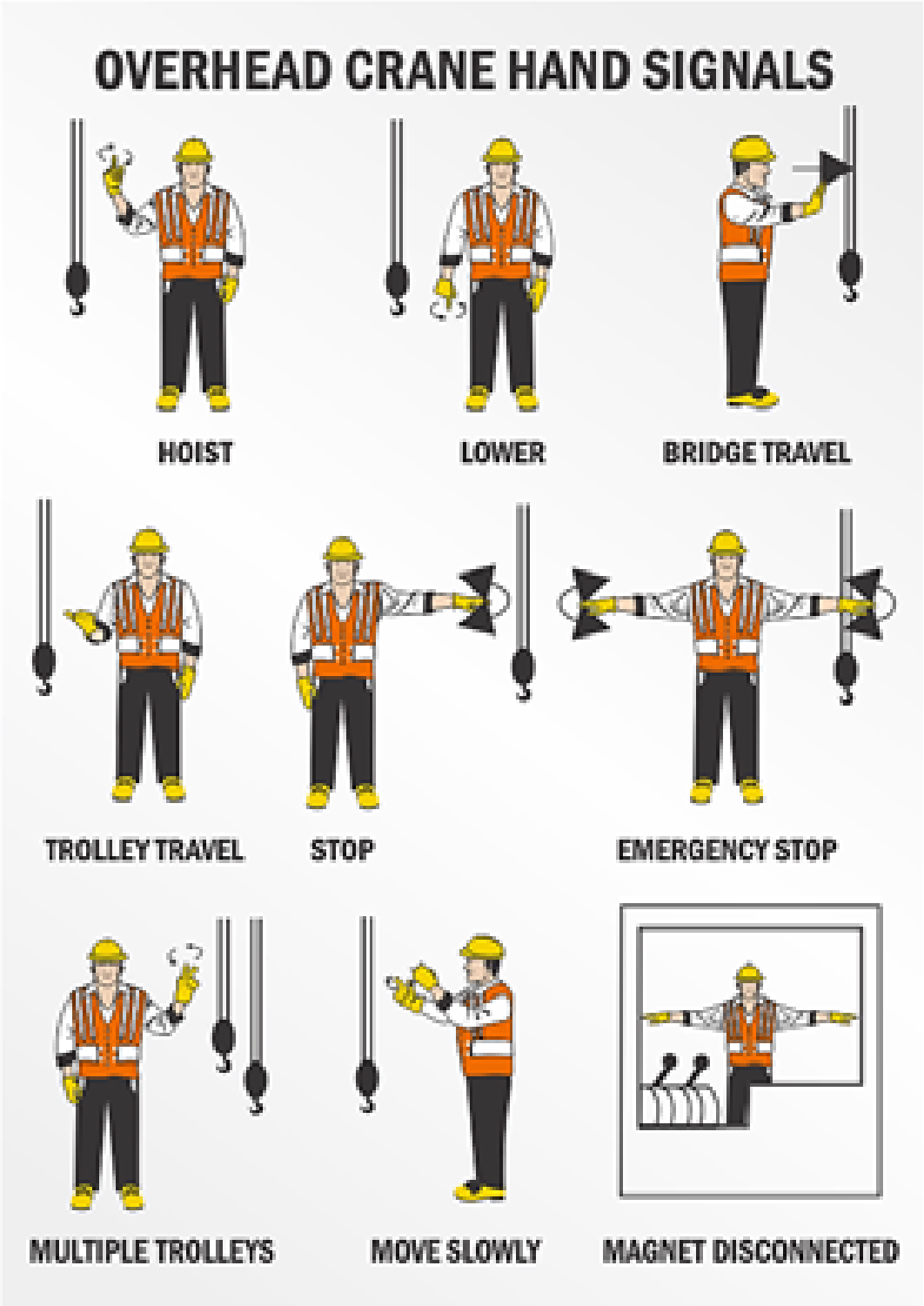

Fig. 5.1–1. The hand signals illustration required in manual control of the overhead crane.

Thus, when it comes to controlling the overhead crane operation automatically, the following problems should be taken into account,

– Overhead crane is a highly nonlinear system which makes it difficult to design the control system only based on its nonlinear dynamics ((4.2–21)–(4.2–25)).
– Overhead crane is classified as an underactuated system since there are no direct control inputs for swing dynamics ((4.2–24), (4.2–25)). This makes it even harder to damp load swings without affecting overhead crane load positioning.
– Overhead crane exhibits non-minimum phase behavior since the load tends to move in the opposite direction of the applied forces. This increases the load forces on the overhead crane which affects the precision of load positioning as well as intensifying load swings
– Load hoisting during crane acceleration intensifies load swing that reduces safety and efficiency for high-speed load transportation (this will be shown in the stability analysis of swing dynamics in Section 5.3). That is why in some manual operation the load is first hoisted up and then it is moved horizontally, and even in some literature, the hoisting rope is assumed to be fixed during the entire operation.

Considering the above-mentioned problems, the control objectives should be modified such that they meet the high-performance requirements for overhead crane operation and resolve the problems stated above. Therefore, the control objectives are narrowed down as the following,

- The control system should be able to deliver high-speed load transportation without compromising the safety of the overhead crane operation.
- The control system should be designed such that it could maintain load swing as minimum as possible and suppress them should they tend to increase (anti-swing control).
- The position of the overhead crane load should be controlled with high accuracy, especially at the final destination (tracking control).
- To improve the time efficiency, the control system should be capable of handling high-speed load hoisting when the overhead crane is accelerated.
- From practical point of view, the designed control system should be able to perform high-performance control in repetitive tasks which is known as repeatability characteristic.
- The control system design should not be very complicated so that it can be easily programmed and implemented on digital computers, industrial controllers and processors such as programmable logic controllers (PLCs).
- The control settings and configurations should be simple enough so that the overhead crane operator can understand how to change or modify them for better control operation.

To achieve the goals and objectives for high-performance control of the overhead crane mentioned above, it is required to develop new control systems to be simple in design yet effective in delivering anti-swing tracking control. In the following sections, the basics of the proposed control systems structure will be presented in further detail.

# 5.2 Control Configuration

## 5.2.1 Independent Joint Control Strategy

In Section 4.4, a new dynamic model was developed for the overhead crane based on the idea that the actuators are considered as the main plant that should be controlled. This idea is a common method used to design control systems in robot manipulator control field known as independent joint control strategy [31], [32]. That is why the model is called independent joint model since each joint of the robot can be controlled separately due to the resulting decoupled model. The nonlinearities of the system are then treated as disturbances acting on each actuator. In this way, the overall control system design becomes less complicated compared to including nonlinear dynamics as part of the control design core. This idea is displayed in Fig. 5.2–1.

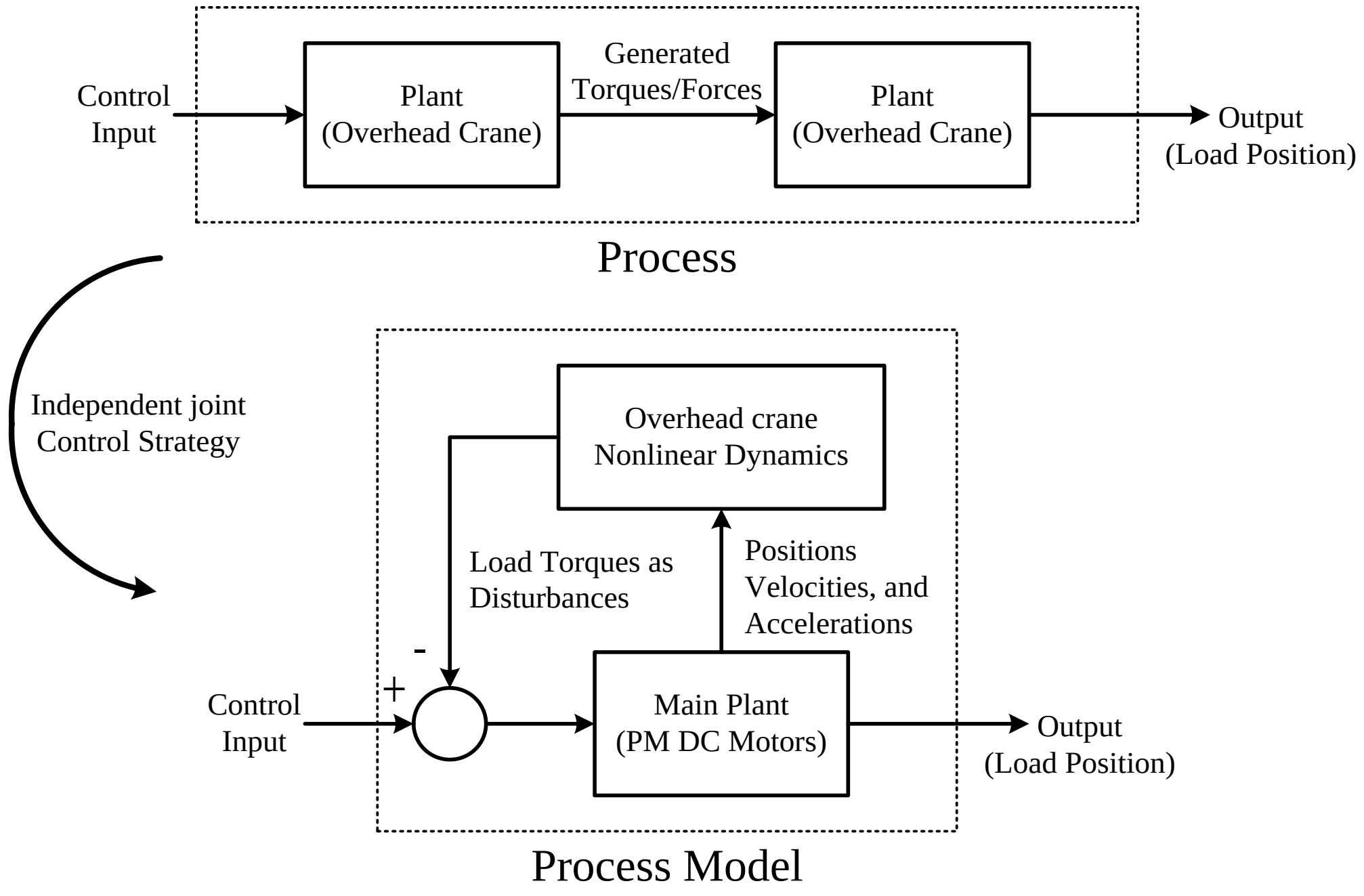

Fig. 5.2–1. Application of independent joint control strategy in controlling the overhead crane.

## 5.2.2 Computed Torque Control

It should be noted that since the nonlinear dynamics of the overhead crane in independent joint control strategy is modeled as disturbances, in high-speed operations the effects of load disturbances could be significant on the overhead crane and it could lead to poor control performance if they are not taken into account in the design. Different nonlinear robust control designs can be used to deal with these nonlinearities, but at a cost of complicating the overall control system design. A simpler yet effective approach is the method known as computed torque control which is common to be used in conjunction with independent joint control strategy in robot manipulator control field [31]. Since the nonlinear dynamics of the robot arm and the desired/reference end-effector trajectories are known in advance, it is common practice to use them to calculate a good estimate of the required torques for moving the robot arm with the desired acceleration, speed and position profiles in the reference trajectories. This method also known as inverse dynamic control method since the computed torques would theoretically drive the robot arm in the same direction, speed, and acceleration as intended for the manipulator when applied as the control inputs considering no uncertainties in the system. However, this method can be used as feedforward control action in addition to the main control system which is designed based on independent joint control strategy to effectively reduce the effect of disturbances caused by nonlinear couplings as illustrated in Fig. 5.2–2.

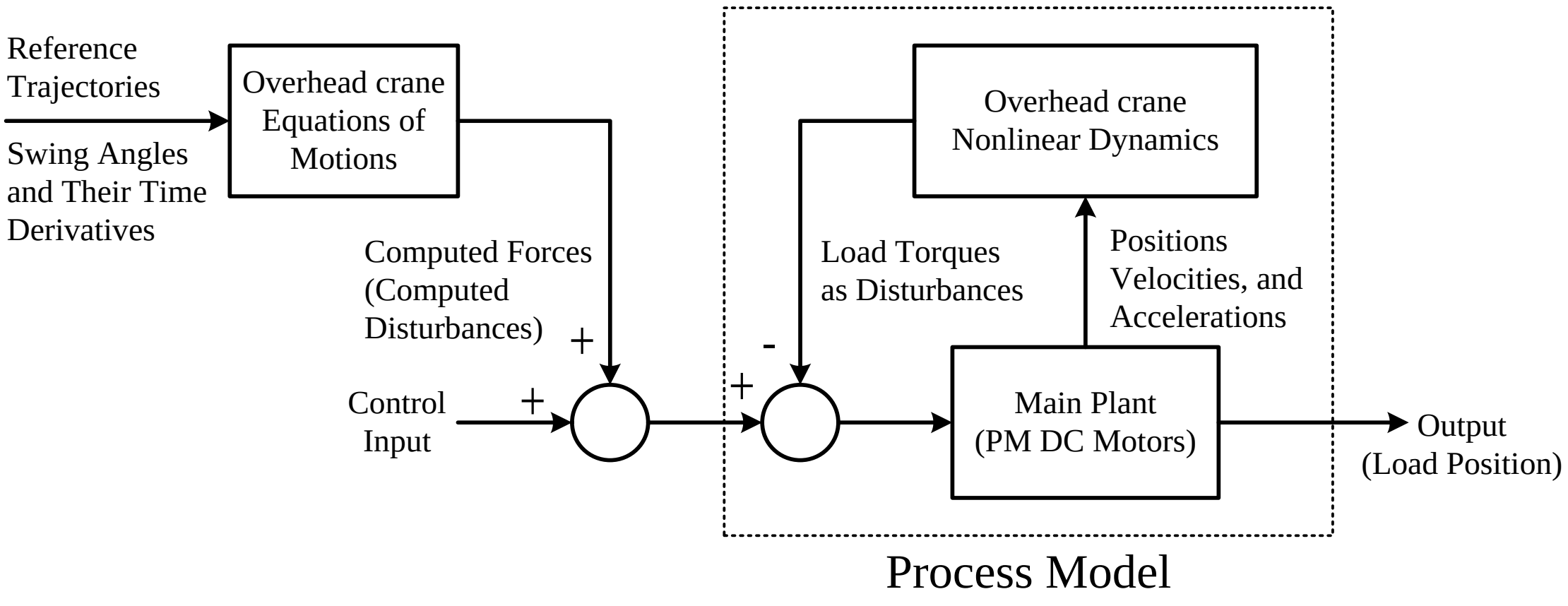

Fig. 5.2–2. Application of computed torque control with independent joint control for the overhead crane.

As can be seen in Fig. 5.2–2, to have a better estimate of load disturbances using computed torque control, it is required that the desired trajectories for traveling, traversing and hoisting motions are designed in advance as reference trajectories which the overhead crane should follow. Unlike set-point control which is used in many previous works in literature where the reference signal is considered to be fixed for traveling, traversing and hoisting motions, the control problem here is to track some pre-designed reference trajectories for smooth and better control performance which is classified as servo control. This will give more control advantages on how to operate the overhead crane more effectively and robustly. Later on in Section 5.4, we will show how trajectory planning can be smartly designed to robustly suppress load swings, especially when the load approaches its final destination at the end of the trajectory.

Moreover, to reflect the effects of load swings in the computation of load torques, the online measurements of swing angles can be used. However, the sensors used for swing angles cannot measure their first and second time-derivatives, which are needed for calculating load disturbances. To deal with this issue, recall that the overhead crane equations of motion were simplified in Section 4.2.2 to remove second time-derivative of swing angles from traveling and traversing dynamics in (4.2–21) and (4.2–22), respectively. Thus, to compute the load torques more accurately, we only need to have access to swing angles and their first time-derivative. This could be done by designing a load swing observer to estimate swing angles and their first time-derivatives which will be explained later in Section 5.3.2.

Furthermore, including the coulomb friction force model as part of load disturbances makes it possible to be used in computed toque control to attenuate the effects of coulomb friction forces which can significantly reduce the accuracy of load positioning at the beginning and the end of the trajectory where the overhead crane operates at low speed.

### 5.2.3 Control System Structure

After obtaining the discrete-time state space model for overhead crane in Section 4.6 and establishing control objectives for high-performance control of overhead crane operation, the overall control system structure to be used for designing our advance control systems is presented in Fig. 5.2–3. As can be seen, the proposed structure for the overhead crane control consists of four main blocks. Reference signal generator is responsible for supplying reference state trajectory profiles for traveling, traversing, and hoisting motions considering the physical limitations of the actuator admissible torque and speed, and overhead crane workspace using a reference dynamic model and reference accelerations as input.

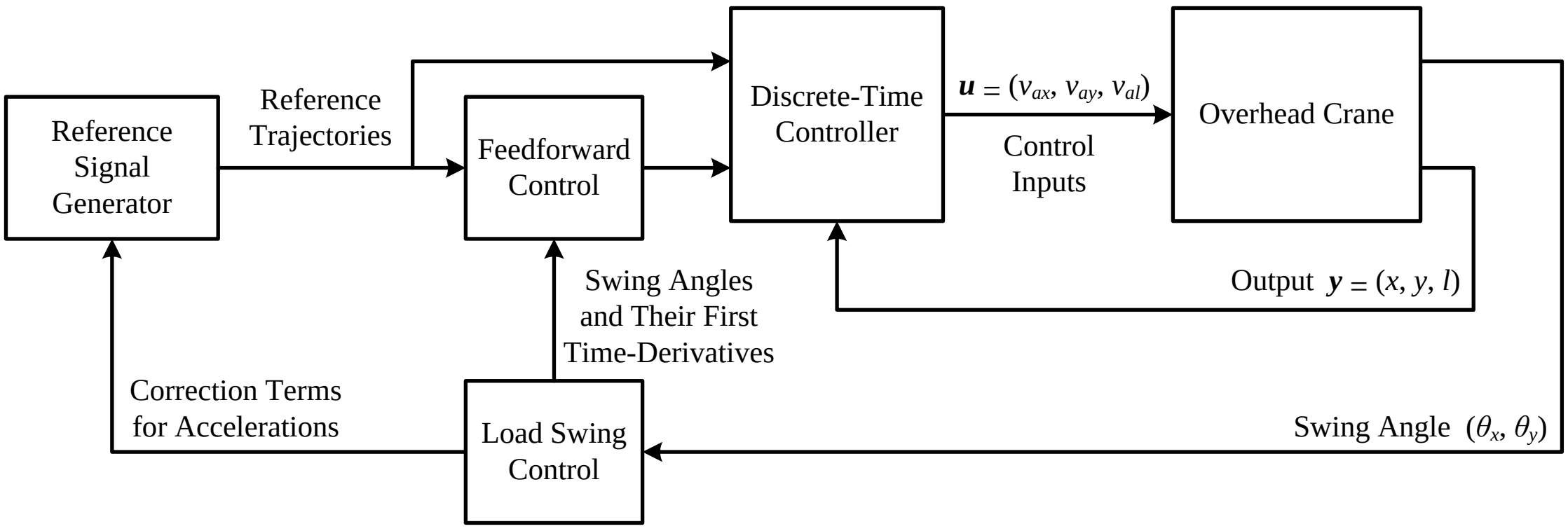

Fig. 5.2–3. The structure of the high-performance discrete-time control system for overhead crane

However, as it will be proven in Section 5.3, the reference accelerations, particularly reference traveling and traversing accelerations, are required to be modified so that load swings can be robustly suppressed. According to swing dynamics in (4.4–4) and (4.4–5), the traveling and traversing accelerations are in fact acting as inputs to the swing dynamics and determining the behavior of the swing angles. Thus, these accelerations can be controlled such that load swings remain bounded in a small range. To do this, some correction terms are added to the reference traveling and traversing accelerations which are generated by the load swing control block to update reference trajectories for traveling and traversing motions. Since modifying the reference accelerations would cause deviation in the reference position and velocity trajectories, a new trajectory planning is developed in Section 5.4 as part of reference signal generation. This trajectory planning allows load swing suppression throughout the trajectory as well as fixing the changes in the reference position and velocity for the traveling and traversing during the final section of the trajectory. Swing angels and their first time-derivatives, which are estimated by a swing angle observer in the load swing control block are needed to generate these correction terms.

The main control algorithm is implemented in discrete-time controller block to generate the final control input voltages. It is designed based on the discrete-time model developed in (4.6–14)–(4.6–17) in conjunction with a feedforward control action which uses the idea of computed torque control to compensate for the effect of nonlinear disturbances and coulomb friction forces and improve the accuracy of trajectory tracking. For the purpose of designing the discrete-time controller, two approaches have been adopted based on their merits, advantages, and integration with other parts of the control system structure to deliver the control objectives on high-performance anti-swing trajectory tracking control for the overhead crane motion as follows,

- Model Predictive Control (MPC)
- Discrete-time State Feedback Control

The main advantage of MPC compared to conventional controllers, as also mentioned before, is that the process constraints can be explicitly taken into account in the controller design, naming input voltage constraints of the PM DC motors and overhead crane workspace limits. Discrete-time nature of MPC also provides easy implementation on digital computers, not to mention, it offers online optimization which is quite useful in real-time control applications. State feedback approach is, on the other hand, simpler in the design and faster in generating the control signals. The feedforward disturbance compensation can be easily incorporated in both control formulations in addition to state observer for estimating positions and velocities from sensor measurements which only provides positions of traveling, traversing, and hoisting rope length.

Moreover, as it will been shown in Section 5.8, a disturbance observer can be designed using the state estimation error signal to smartly estimate the total disturbances acting on each of the actuators and to be used for feedforward signal generation. When using computed torque control to estimate the disturbance forces, the value of the load mass should be known since the load mass is formulated as part of disturbances, and therefore, large uncertainties in the load mass can deteriorate the tracking performance. The significance of disturbance observer is in the fact that it only uses state estimation error to use predict disturbance without the need to know the value of the load mass or the nonlinear structure of overhead crane equations of motion. Unlike the computed torque control which needs the knowledge of the load mass and nonlinear dynamics, disturbance observer makes the control system to be robust against uncertainties in the mass of the overhead crane load and its nonlinear dynamics. In the following sections, the procedures of designing each part of the control system structure described above will be explained in detail.

## 5.3 Load Swing Control

In the discrete-time dynamic model obtained for the overhead crane in (4.6–14)–(4.6–17), the swing dynamics are not apparently part of the equations although they affect the disturbance forces corresponding to each direction of motion. Feedforward control action is able to damp the load swings indirectly to some extent since it is used to compensate for the effects of disturbances (either using the computed torque control or the disturbance observe). The reason is that if swing angles tend to increase, the disturbances on the actuators will increase as well. The control system would then try to calculate control input voltages such that the effects of disturbance are reduced with the help of feedforward action (This will be shown in practical results of the designed control systems in operation). Subsequently, the load swings would remain bounded because of this disturbance compensation.

Nevertheless, to guarantee the suppression of load swings throughout the entire overhead crane control operation, particularly in high-speed motions, a separate load swing control is required. To do this, we first need to understand the behavior of load swing by further examining swing dynamics. Let us begin with swing dynamics for 2D overhead crane for simplicity and then we can extend the proposed load swing control design for 3D overhead crane which has more complicated swing dynamics.

Recall from simplified swing dynamics for 2D overhead crane obtained in (4.4–13),

$$l\ddot{\theta}_x + C_{\theta_x}\ddot{x} + 2\dot{l}\dot{\theta}_x + gS_{\theta_x} = 0. \tag{5.3–1}$$

The above equation can be written in state space form as follows,

$$\dot{\boldsymbol{x}} = f(\boldsymbol{x}) + g(\boldsymbol{x})u, \tag{5.3–2}$$

where $\boldsymbol{x}$ is the state vector, $u$ is the input, and $f$ and $g$ are nonlinear functions given as below,

$$\boldsymbol{x} = \begin{bmatrix} x_1 \\ x_2 \end{bmatrix} = \begin{bmatrix} \theta_x \\ \dot{\theta}_x \end{bmatrix}, \qquad u = \ddot{x}, \qquad f(\boldsymbol{x}) = \begin{bmatrix} \dot{\theta}_x \\ -\dfrac{1}{l}(2\dot{l}\dot{\theta}_x + gS_{\theta_x}) \end{bmatrix}, \qquad g(\boldsymbol{x}) = \begin{bmatrix} 0 \\ -\dfrac{C_{\theta_x}}{l} \end{bmatrix} \tag{5.3–3}$$

As can be seen, the trolley acceleration $\ddot{x}$ is in fact acting as the input to the swing dynamics and determining the behavior of the swing angle. Since the swing dynamics are nonlinear, we have to apply Lyapunov stability analysis and some other nonlinear analysis tools to investigate how the swing dynamics can be stabilized via controlling the trolley acceleration.

### 5.3.1 Passivity-Based Control and $\mathcal{L}_2$ Stability

The nonlinear analysis tools that can help us to design the load swing control are passivity-based control and $\mathcal{L}_2$ stability theorem and their relationship with Lyapunov stability. The idea of passivity-based control, as its name suggests, is to design the control input such that the total energy absorbed by the system over a finite time is greater than the increase in the stored energy in the system. This implies that the system acts like a passive element and dissipate more energy rather than storing it, and thus, it remains stable. $\mathcal{L}_2$ stability, on the other hand, is a method to workout stability of the system in the input-output sense, mainly for the square-integrable input and output signals. This is quite similar to the concept of bounded-input bonded-output stability. Both methods are closely connected to Lyapunov stability as a basic tool to establish them. Before going further, we need to give some definitions (Def.) and lemmas to be able to relate these methods together, and then apply them for designing load swing control and stability proof. More detailed discerptions and theorems for input-output stability and passivity for linear and nonlinear systems can be found in [108].

***Def. 1.*** *The piecewise continuous, square-integrable signal/function* $\boldsymbol{u}$ *in vector form is said to belong to* $\mathcal{L}_2$ *space if its* $\mathcal{L}_2$ *norm* (*denoted by* $\|\boldsymbol{u}\|_{\mathcal{L}_2}$) *is bounded, i.e.,*

$$\boldsymbol{u} \in \mathcal{L}_2 \quad \Leftrightarrow \quad \|\boldsymbol{u}\|_{\mathcal{L}_2} = \sqrt{\int_0^t \boldsymbol{u}^T(t)\boldsymbol{u}(t)dt} < \infty, \qquad \textit{for} \quad t \in [0,\infty) \,. \tag{5.3–4}$$

***Def. 2***. *Consider the following time-invariant nonlinear system with* $\boldsymbol{x} \in R^n$ *as the system states,* $\boldsymbol{u} \in R^p$ *as the input, and* $\boldsymbol{y} \in R^m$ *as the output,*

$$\begin{aligned} \dot{\boldsymbol{x}} &= f(\boldsymbol{x},\boldsymbol{u}), \qquad \boldsymbol{x}(0) = \boldsymbol{x}_0, \\ \boldsymbol{y} &= h(\boldsymbol{x}), \end{aligned} \tag{5.3–5}$$

*where functions* $f(\boldsymbol{x},\boldsymbol{u})$ *and* $h(\boldsymbol{x})$ *are locally Lipschitz with* $f(0,0) = 0$ *and* $h(0) = 0$. *For all* $\boldsymbol{u} \in \mathcal{L}_2$ *the system* (5.3–5) *is finite-gain* $\mathcal{L}_2$ *stable if there exist nonnegative constants* $\gamma$ *and* $\beta$ *such that*

$$\|\boldsymbol{y}\|_{\mathcal{L}_2} \le \gamma \|\boldsymbol{u}\|_{\mathcal{L}_2} + \beta, \tag{5.3–6}$$

*where* $\gamma$ *is known as the upper bound of* $\mathcal{L}_2$ *gain of the system.*

***Def. 3.*** *The system* (5.3–5) *is said to be passive if there exists a continuously differentiable positive semi-definite function* $V(\boldsymbol{x})$ (*called the storage function*) *such that*

$$\boldsymbol{u}^T \boldsymbol{y} \ge \dot{V} = \frac{\partial V}{\partial \boldsymbol{x}} f(\boldsymbol{x},\boldsymbol{u}). \tag{5.3–7}$$

*Moreover, it is said to be output strictly passive if*

$$\boldsymbol{u}^T \boldsymbol{y} \ge \dot{V} + \boldsymbol{y}^T \varphi(\boldsymbol{y}) \quad \text{with} \quad \boldsymbol{y}^T \varphi(\boldsymbol{y}) > 0 \quad \text{and} \quad \forall\ \boldsymbol{y} \ne 0, \tag{5.3–8}$$

*and in both cases, the inequality should hold for all* $(\boldsymbol{x}, \boldsymbol{u})$.

***Def. 4***. *The system* (5.3–5) *is said to be zero-state observable if no solution of zero-input response of the system, i.e.,* $\dot{\boldsymbol{x}} = f(\boldsymbol{x})$, *can stay identically in* $S = \{\boldsymbol{x} \in R^n, \boldsymbol{u} = 0 \mid h(\boldsymbol{x}) = 0\}$, *other than trivial solution* $\boldsymbol{x}(t) \equiv 0$. *In other words, when* $\boldsymbol{u} = 0$, *for all solution of* $\dot{\boldsymbol{x}} = f(\boldsymbol{x})$,

$$\text{if} \quad \boldsymbol{y}(t) \equiv 0 \quad \Leftrightarrow \quad \boldsymbol{x}(t) \equiv 0. \tag{5.3–9}$$

It is interesting to see that the storage function $V(\boldsymbol{x})$ in Def. 3. has the same concept as the Lyapunov function candidate. Therefore, Lyapunov stability tools can be used to establish the connection between $\mathcal{L}_2$ stability and passivity for nonlinear systems represented by state space models.

***Lemma 1.*** *Consider the system* (5.3–5), *the equilibrium point of* $\dot{\boldsymbol{x}} = f(\boldsymbol{x})$ *is asymptotically stable if the system is output strictly passive and zero-state observable. Furthermore, if the storage function is radially unbounded, the equilibrium point will be globally asymptotically stable.*

***Lemma 2****. If the system* (5.3–5) *is output strictly passive with* $\boldsymbol{u}^T\boldsymbol{y} \geq \dot{V} + \delta\boldsymbol{y}^T\boldsymbol{y}$ *for some* $\delta > 0$, *then it is finite-gain* $\mathcal{L}_2$ *stable and its* $\mathcal{L}_2$ *gain is less than or equal to* $1/\delta$ *if* $\boldsymbol{u} \in \mathcal{L}_2$ *since it can be proven that*

$$\|\boldsymbol{y}\|_{\mathcal{L}_2} \leq \frac{1}{\delta}\|\boldsymbol{u}\|_{\mathcal{L}_2} + \sqrt{\frac{2}{\delta}V(\boldsymbol{x}(0))}\,. \tag{5.3–10}$$

Now, we can investigate the stability of swing dynamics using the above definitions and lemmas. Let us first begin with 2D Overhead crane swing dynamics given in (5.3–1). The following positive definite Lyapunov function candidate/storage function is considered,

$$V_{\theta_x} = \frac{1}{2} l\dot{\theta}_x^2 + g(1 - C_{\theta_x}), \tag{5.3–11}$$

which is also radially unbounded since $(\theta_x, \dot{\theta}_x) \rightarrow \infty$ , $V_{\theta_x} \rightarrow \infty$. The first time-derivative of $V_{\theta_x}$ is obtained by replacing $l\ddot{\theta}_x$ form (5.3–1) as follows,

$$\dot{V}_{\theta_x} = -1.5\dot{l}\dot{\theta}_x^2 - C_{\theta_x}\dot{\theta}_x\ddot{x}. \tag{5.3–12}$$

Knowing the fact that $|C_{\theta_x}| \leq 1$, the upper bound of (5.3–12) is given as below,

$$\dot{V}_{\theta_x} \leq -1.5\dot{l}\dot{\theta}_x^2 + \dot{\theta}_x\ddot{x}, \tag{5.3–13}$$

and then it can be rearrange as follows,

$$\ddot{x}\dot{\theta}_x \geq \dot{V}_{\theta_x} + 1.5\dot{l}\dot{\theta}_x^2. \tag{5.3–14}$$

It is interesting to see that (5.3–14) has the same form as an output strictly passive system as in (5.3–8) if $\dot{\theta}_x$ is chosen as the output with $\varphi(\boldsymbol{y}) = 1.5\dot{l}\dot{\theta}_x$ and $\ddot{x}$ as the input (Def. 3.). In addition, the swing dynamics in (5.3–1) is zero-state observable because when $\ddot{x} = 0$, if $\dot{\theta}_x \equiv 0$ (which is both output and system state as in (5.3–3)), then $\theta_x \equiv 0$ (Def. 4). However, the hoisting velocity is not always positive and so is $\varphi(\boldsymbol{y})$. Based on the generalized coordinates described in Section 4.2.1, when the load is lifted up hoisting velocity is negative $\dot{l} < 0$, and when it is hoisted down the velocity is positive $\dot{l} > 0$. Therefore, when the load is hoisted up during acceleration or deceleration of the trolley, the load swing tends to increase and the system becomes unstable because swing dynamics is no longer output strictly passive. Whereas, when it is hoisted down the load swing is asymptotically going to zero if trolley moves with constant velocity ($\ddot{x}$= 0) according to Lemma 1., and if trolley is accelerated or decelerated while load is hoisted down, the swing dynamics is finite-gain $\mathcal{L}_2$ stable (Lemma 2.).

Thus, it is important to have a control action on swing dynamics such that the stability of load swing is guaranteed for any form of overhead crane motion without any restriction on when to hoist the load or accelerate the trolley as it would reduce time efficiency. This would also enable the control system to handle high-speed load hoisting during acceleration as one of the control objectives. To achieve this goal, trolley acceleration should be manipulated such that it makes the second term in (5.3–14) always positive, and therefore, (5.3–14) attains output strictly passive form, and subsequently, finite-gain $\mathcal{L}_2$ stable. This is possible by using the tracking controller. As discussed in control system structure in Section 5.2.3, the discrete-time controller is designed to calculate control input voltages such that the positon and velocity of trolley and hoisting rope length can track some reference trajectories generated from reference accelerations ($a_{xref}$, $a_{lref}$). According to the principal of Kinematics in mechanics [31], [32], to move an object from one point to another point following a specific position and velocity trajectories within a finite period, the acceleration of the object should be a function of position and velocity profiles. Therefore, it is a true assumption that when the position and velocity of the trolley follow the reference traveling trajectories by the discrete-time controller, i.e., $x \to x_{ref}$ and $v_x \to v_{xref}$, the trolley acceleration will ultimately follow the reference traveling acceleration $a_{xref}$ designed to generate $x_{ref}$ and $v_{xref}$, in the reference signal generator block, i.e., $\ddot{x} \to a_{xref}$. Otherwise, the overhead crane would never reach the final destination the way it was designed to. Therefore, we can consider reference traveling acceleration $a_{xref}$ as the input to the swing dynamics and then write (5.3–12) as below,

$$\dot{V}_{\theta_x} = -1.5\dot{l}\dot{\theta}_x^2 - C_{\theta_x}\dot{\theta}_x a_{xref}. \tag{5.3–15}$$

Now, to stabilize the load swing, reference traveling acceleration is modified by adding a correction term to $a_{xref}$ as follows [109], [110],

$$u_{cx} = a_{xref} + k_{\theta_x}\frac{\dot{\theta}_x}{C_{\theta_x}}, \tag{5.3–16}$$

where $k_{\theta_x}$ is defined as the swing control gain and $u_{cx}$ is defined as the traveling acceleration command signal. By replacing $a_{xref}$ with $u_{cx}$ in (5.3–15) we have

$$\dot{V}_{\theta_x} = -1.5\dot{l}\dot{\theta}_x^2 - C_{\theta_x}\dot{\theta}_x\left(a_{xref} + k_{\theta_x}\frac{\dot{\theta}_x}{C_{\theta_x}}\right), \tag{5.3–17}$$

and rearranging the terms in the above equation to find the upper bound of $\dot{V}_{\theta_x}$ results in the following, knowing that $|C_{\theta_x}| \leq 1$,

$$a_{xref}\dot{\theta}_x \geq \dot{V}_{\theta_x} + (1.5\dot{l} + k_{\theta_x})\dot{\theta}_x^2. \tag{5.3–18}$$

It can be seen from (5.3–18) that by choosing $a_{xref} \in \mathcal{L}_2$ as the input, $\dot{\theta}_x$ as the output, and $V_{\theta_x}$ as the storage function with $\varphi(\boldsymbol{y}) = (1.5\dot{l} + k_{\theta_x})\dot{\theta}_x$, swing dynamics will be output strictly passive if swing control gain is chosen as

$$k_{\theta_x} \geq 1.5|\dot{l}|_{\max}, \tag{5.3–19}$$

where $|\dot{l}|_{\max}$ is the maximum accessible hoisting velocity. Subsequently, the swing dynamics becomes finite-gain $\mathcal{L}_2$ stable (Lemma 2) with $\mathcal{L}_2$ gain less than or equal to $1/(k_{\theta_x} + 1.5|\dot{l}|_{\max})$. It should be noted that swing control gain guarantees that $\varphi(\boldsymbol{y})$ is always positive and the higher $k_{\theta_x}$ the smaller $\dot{\theta}_x$. Furthermore, since swing angle has sinusoidal behavior, if $\mathcal{L}_2$ gain of $\dot{\theta}_x$ is bounded, its integration over a fixed period also has bounded $\mathcal{L}_2$ gain, meaning that $\| \theta_x \|_{\mathcal{L}_2} \leq c_1 < \infty$ for a positive constant $c_1$. Therefore, (5.3–16) is considered and the load swing control law for 2D overhead crane.

**Remark**: As it is shown, to suppress load swings during load transportation, some sort of damping action should be applied on load swings, knowing that there is no direct control input for swing dynamics in the overhead crane. Therefore, by adding the correction term to $a_{xref}$, an indirect swing damping force is exerted on swing dynamics through traveling acceleration to decrease load swing. This would act like a virtual friction force on swing angle since the correction term is a function of swing angle velocity $\dot{\theta}_x$. Moreover, this proposed load swing control can be easily implemented in discrete-time since the reference traveling acceleration can be updated at each sampling time using the discrete-time values of $(\theta_x, \dot{\theta}_x)$.

Now that we proved how to stabilize load swings in 2D overhead crane, the same approach can be extended for 3D overhead crane by adding a correction term to the traveling and traversing reference accelerations, $\ddot{x}$ and $\ddot{y}$, respectively, in a matrix form. Let us first recall the simplified swing dynamics for 3D overhead crane obtained in (4.4–4) and (4.4–5) as below,

$$lC_{\theta_y}\ddot{\theta}_x + C_{\theta_x}\ddot{x} + 2C_{\theta_y}\dot{l}\dot{\theta}_x - 2lS_{\theta_y}\dot{\theta}_x\dot{\theta}_y + gS_{\theta_x} = 0, \tag{5.3–20}$$

$$l\ddot{\theta}_y + C_{\theta_y}\ddot{y} - S_{\theta_x}S_{\theta_y}\ddot{x} + 2\dot{l}\dot{\theta}_y + lC_{\theta_y}S_{\theta_y}\dot{\theta}_x^2 + gC_{\theta_x}S_{\theta_y} = 0. \tag{5.3–21}$$

These equations can be written in matrix form as follows,

$$M_\theta\ddot{\boldsymbol{\theta}} + C_\theta\dot{\boldsymbol{\theta}} + G_\theta + H_\theta\boldsymbol{a}_{xy} = 0, \tag{5.3–22}$$

where

$$M_\theta = \begin{bmatrix} lC_{\theta_y}^2 & 0 \\ 0 & l \end{bmatrix}, \quad \boldsymbol{\theta} = \begin{bmatrix} \theta_x \\ \theta_y \end{bmatrix}, \quad C_\theta = \begin{bmatrix} 2C_{\theta_y}^2\dot{l} - 2lS_{\theta_y}C_{\theta_y}\dot{\theta}_y & 0 \\ lC_{\theta_y}S_{\theta_y}\dot{\theta}_x & 2\dot{l} \end{bmatrix},$$
$$G_\theta = \begin{bmatrix} gS_{\theta_x}C_{\theta_y} \\ gC_{\theta_x}S_{\theta_y} \end{bmatrix}, \quad H_\theta = \begin{bmatrix} C_{\theta_x}C_{\theta_y} & 0 \\ -S_{\theta_x}S_{\theta_y} & C_{\theta_y} \end{bmatrix}, \quad \boldsymbol{a}_{xy} = \begin{bmatrix} \ddot{x} \\ \ddot{y} \end{bmatrix} \tag{5.3–23}$$

The positive definite radially unbounded storage function is then defined as the following,

$$V_\theta = \frac{1}{2}\dot{\theta}^T M_\theta \dot{\theta} + g(1 - C_{\theta_x} C_{\theta_y}). \qquad (5.3\text{–}24)$$

The first time-derivative of the storage function $V_\theta$ is obtained by replacing $M_\theta \ddot{\theta}$ from (5.3–22) as below,

$$\dot{V}_\theta = \dot{\theta}^T (\frac{1}{2}\dot{M}_\theta - C_\theta)\dot{\theta} - \dot{\theta}^T H_\theta \alpha_{xy}. \qquad (5.3\text{–}25)$$

Using (5.3–23), $(0.5\dot{M}_\theta - C_\theta)$ can be simplified as follows,

$$\begin{aligned}\frac{1}{2}\dot{M}_\theta - C_\theta &= \begin{bmatrix} \frac{1}{2}\dot{l}C_{\theta_y}^2 - lS_{\theta_y}C_{\theta_y}\dot{\theta}_y & 0 \\ 0 & \frac{1}{2}\dot{l} \end{bmatrix} - \begin{bmatrix} 2\dot{l}C_{\theta_y}^2 - 2lS_{\theta_y}C_{\theta_y}\dot{\theta}_y & 0 \\ lC_{\theta_y}S_{\theta_y}\dot{\theta}_x & 2\dot{l} \end{bmatrix} \\ &= \begin{bmatrix} -\frac{3}{2}\dot{l}C_{\theta_y}^2 + lS_{\theta_y}C_{\theta_y}\dot{\theta}_y & 0 \\ -lC_{\theta_y}S_{\theta_y}\dot{\theta}_x & -\frac{3}{2}\dot{l} \end{bmatrix}\end{aligned} \qquad (5.3\text{–}26)$$

By expanding $\dot{\theta}^T(0.5\dot{M}_\theta - C_\theta)\dot{\theta}$ and then rearranging it in matrix form we have

$$\begin{aligned}\dot{\theta}^T(\frac{1}{2}\dot{M}_\theta - C_\theta)\dot{\theta} &= \begin{bmatrix}\dot{\theta}_x & \dot{\theta}_y\end{bmatrix} \begin{bmatrix} (-\frac{3}{2}\dot{l}C_{\theta_y}^2 + lS_{\theta_y}C_{\theta_y}\dot{\theta}_y)\dot{\theta}_x \\ (-lC_{\theta_y}S_{\theta_y}\dot{\theta}_x)\dot{\theta}_x - \frac{3}{2}\dot{l}\dot{\theta}_y \end{bmatrix} \\ &= -\frac{3}{2}\dot{l}C_{\theta_y}^2\dot{\theta}_x^2 - \frac{3}{2}\dot{l}\dot{\theta}_y^2 = -\frac{3}{2}\dot{\theta}^T \underbrace{\begin{bmatrix} \dot{l}C_{\theta_y}^2 & 0 \\ 0 & \dot{l} \end{bmatrix}}_{\hat{M}_\theta} \dot{\theta},\end{aligned} \qquad (5.3\text{–}27)$$

and finally, $\dot{V}_\theta$ is obtained as follows,

$$\dot{V}_\theta = -\frac{3}{2}\dot{\boldsymbol{\theta}}^T M'_\theta \dot{\boldsymbol{\theta}} - \dot{\boldsymbol{\theta}}^T H_\theta \boldsymbol{a}_{xy}. \tag{5.3–28}$$

Now, as it was explained for 2D overhead crane, the reference traveling and traversing accelerations in the vector form $\boldsymbol{a}_{xy_ref} = [a_{xref} \;\; a_{yref}]^T$ are considered as the input and hence being replaced with $\boldsymbol{a}$ in (5.3–28) as below,

$$\dot{V}_\theta = -\frac{3}{2}\dot{\boldsymbol{\theta}}^T M'_\theta \dot{\boldsymbol{\theta}} - \dot{\boldsymbol{\theta}}^T H_\theta \boldsymbol{a}_{xy_ref}. \tag{5.3–29}$$

Thus, to stabilize swing dynamics, $\boldsymbol{a}_{xy_ref}$ is modified by adding a correction term as follows,

$$\boldsymbol{u}_{c_xy} = \boldsymbol{a}_{xy_ref} + K_\theta H_\theta^{-1} \dot{\boldsymbol{\theta}}, \tag{5.3–30}$$

where $K_\theta = Diag\{k_{\theta_x}, k_{\theta_y}\}$ is the swing control gain and $\boldsymbol{u}_{c_xy}$ is defined as the trolley acceleration command signal in $XY$ plane[16] $\boldsymbol{u}_{c_xy} = [u_{cx} \;\; u_{cy}]^T$. It should be noted that $H_\theta$ is invertible for all $|\theta_x| \neq \pi/2$ and $|\theta_y| \neq \pi/2$ since using (5.3–23) we have

$$H_\theta^{-1} = \frac{1}{\det(H_\theta)}\begin{bmatrix} C_{\theta_y} & 0 \\ S_{\theta_x}S_{\theta_y} & C_{\theta_x}C_{\theta_y} \end{bmatrix} = \frac{1}{C_{\theta_x}C_{\theta_y}^2}\begin{bmatrix} C_{\theta_y} & 0 \\ S_{\theta_x}S_{\theta_y} & C_{\theta_x}C_{\theta_y} \end{bmatrix},$$

$$\det(H_\theta) = 0 \quad \Rightarrow \quad C_{\theta_x}C_{\theta_y}^2 = 0 \quad \Rightarrow \quad \begin{cases} |\theta_x| = \dfrac{\pi}{2}, \\ |\theta_y| = \dfrac{\pi}{2}. \end{cases} \tag{5.3–31}$$

By replacing $\boldsymbol{a}_{xy_ref}$ with $\boldsymbol{u}_{c_xy}$ in (5.3–29), $\dot{V}_\theta$ is given as the following,

$$\dot{V}_\theta = -\frac{3}{2}\dot{\boldsymbol{\theta}}^T M'_\theta \dot{\boldsymbol{\theta}} - \dot{\boldsymbol{\theta}}^T H_\theta (\boldsymbol{a}_{ref} + K_\theta H_\theta^{-1} \dot{\boldsymbol{\theta}}). \tag{5.3–32}$$

[16] Trolley position in 3D overhead crane means both traveling and traversing positions in $XY$ plane unless 2D overhead crane is considered which trolley position means traveling position.

Rearranging the terms in the above equation to find the upper bound of $\dot{V}_\theta$ results in the following, knowing that $\| H_\theta \|_1 \leq 1$,[17]

$$\boldsymbol{a}_{xy_ref}^T \dot{\boldsymbol{\theta}} \geq \dot{V}_\theta + \dot{\boldsymbol{\theta}}^T \left(\frac{3}{2} M'_\theta + K_\theta\right)\dot{\boldsymbol{\theta}}. \tag{5.3–33}$$

It can be seen from (5.3–33) that by choosing $\boldsymbol{a}_{xy_ref} \in \mathcal{L}_2$ as the input, $\dot{\boldsymbol{\theta}}$ as the output, and $V_\theta$ as the storage function with $\varphi(\boldsymbol{y}) = (1.5M'_\theta + K_\theta)\dot{\boldsymbol{\theta}}$, the swing dynamics for the 3D overhead crane will be output strictly passive if $\dot{\boldsymbol{\theta}}^T(1.5M'_\theta + K_\theta)\dot{\boldsymbol{\theta}} > 0$. This requires that swing control gains for traveling and traversing are chosen as follows,

$$\{ k_{\theta_x}, k_{\theta_y} \} \geq 1.5 |\dot{l}|_{\max}. \tag{5.3–34}$$

Subsequently, the swing dynamics becomes finite-gain $\mathcal{L}_2$ stable (Lemma 2) with $\mathcal{L}_2$ gain less than or equal to $1 / (\max\{k_{\theta_x}, k_{\theta_y}\} + 1.5|\dot{l}|_{\max})$. Furthermore, similar to the statement for 2D overhead carne, having $\| \dot{\boldsymbol{\theta}} \|_{\mathcal{L}_2} \leq c_1 < \infty$, its integration over a fixed period also has bounded $\mathcal{L}_2$ gain since swing angles have sinusoidal behavior, meaning that $\| \boldsymbol{\theta} \|_{\mathcal{L}_2} \leq c_2 < \infty$ for a positive constants $c_1$ and $c_2$. Therefore, (5.3–30) is considered and the load swing control law for 3D overhead crane.

## 5.3.2 Swing Angle Observer

As we showed in the previous section, to stabilize the load swing in the sense of finite-gain $\mathcal{L}_2$ stability, a correction term should be added to the reference trolley accelerations, which also stabilizes the equilibrium point in the absence of the trolley accelerations (in constant-velocity motion). This term is a function of swing angles and their speeds ((5.3–16) for 2D and (5.3–30) for 3D overhead crane) which acts as a state feedback. However, only swing angles measurements are available. Therefore, swing angles speed should be estimated using a reliable method. High-gain observers are the best choice for this task since they can asymptotically estimate the states using output measurements in nonlinear systems whose dynamics can be written as the sum of a

[17] $\| H_\theta \|_1$ is the $p$-norm of $H_\theta$ for $p = 1$ defined as follows,

$$\max_{1 \leq j \leq n} \sum_{i=1}^{m} \left| h_{\theta_{ij}} \right| = \max_{\forall\, |\theta_x|, |\theta_y| < \frac{\pi}{2}} \{ (C_{\theta_x} C_{\theta_y} - S_{\theta_x} S_{\theta_y}), C_{\theta_y} \} = \max_{\forall\, |\theta_x|, |\theta_y| < \frac{\pi}{2}} \{ C_{\theta_x + \theta_y}, C_{\theta_y} \} = 1.$$

linear part plus a nonlinear perturbation part. This is possible based on the *separation principle* that allows us to separate the overall design into two tasks: designing the stabilizing state feedback controller, and then obtaining the equivalent output feedback controller by replacing the states by their estimates provided by the high-gain observer [108][18]. To design the high-gain observer for load swing control, the following lemma is utilized.

***Lemma 3.*** *Consider the following time-invariant nonlinear square system,*

$$\begin{aligned} \dot{\boldsymbol{x}} &= A\boldsymbol{x} + Bg(\boldsymbol{x},\boldsymbol{u}), \qquad \boldsymbol{x}(0) = \boldsymbol{x}_0, \\ \boldsymbol{y} &= C\boldsymbol{x}, \end{aligned} \tag{5.3–35}$$

*where* $\boldsymbol{x} \in R^{2n}$ *is the state vector,* $\boldsymbol{u} \in R^n$ *is the input,* $\boldsymbol{y} \in R^n$ *is the output, and* $g(\boldsymbol{x},\boldsymbol{u})$ *is a locally Lipschitz function with* $g(0,0) = 0$*. The* $2n\times 2n$ *matrix A, the* $2n\times n$ *matrix B, and the* $n\times 2n$ *matrix C are given by*

$$A = \begin{bmatrix} A_1 & \underline{0} & \cdots & \underline{0} \\ \underline{0} & A_2 & \ddots & \vdots \\ \vdots & \ddots & \ddots & \underline{0} \\ \underline{0} & \cdots & \underline{0} & A_n \end{bmatrix} \quad B = \begin{bmatrix} B_1 & \underline{0} & \cdots & \underline{0} \\ \underline{0} & B_2 & \ddots & \vdots \\ \vdots & \ddots & \ddots & \underline{0} \\ \underline{0} & \cdots & \underline{0} & B_n \end{bmatrix} \quad C = \begin{bmatrix} C_1 & \underline{0} & \cdots & \underline{0} \\ \underline{0} & C_2 & \ddots & \vdots \\ \vdots & \ddots & \ddots & \underline{0} \\ \underline{0} & \cdots & \underline{0} & C_n \end{bmatrix}$$
$$A_i = \begin{bmatrix} 0 & 1 \\ 0 & 0 \end{bmatrix} \qquad B_i = \begin{bmatrix} 0 \\ 1 \end{bmatrix} \qquad C_i = \begin{bmatrix} 1 & 0 \end{bmatrix} \qquad for \quad i = 1, 2, \ldots, n, \tag{5.3–36}$$

*where* $\underline{0}$ *is a zero matrix with proper size. The stabilizing output feedback control law* $\boldsymbol{u} = g(\hat{\boldsymbol{x}})$ *is obtained by using the estimates of the state vector* $\hat{\boldsymbol{x}}$ *generated by the high-gain observer as follows,*

$$\dot{\hat{\boldsymbol{x}}} = A\hat{\boldsymbol{x}} + B\hat{g}(\hat{\boldsymbol{x}},\boldsymbol{u}) + L(\boldsymbol{y} - C\hat{\boldsymbol{x}}), \tag{5.3–37}$$

*with* $\hat{g}(\boldsymbol{x},\boldsymbol{u})$ *as the nominal model of* $g(\boldsymbol{x},\boldsymbol{u})$ *required to be locally Lipschitz function with* $\hat{g}(0,0) = 0$*, and L as the observer gain chosen as*

[18] Practical application of high-gain observer in induction motor and some mechanical systems can be found in [111]–[113].

$$L = \begin{bmatrix} L_1 & \underline{0} & \cdots & \underline{0} \\ \underline{0} & L_2 & \ddots & \vdots \\ \vdots & \ddots & \ddots & \underline{0} \\ \underline{0} & \cdots & \underline{0} & L_n \end{bmatrix}, \qquad L_i = \begin{bmatrix} \delta_{1i}/\varepsilon \\ \delta_{2i}/\varepsilon^2 \end{bmatrix}, \qquad for \quad i = 1, 2, \ldots, n, \tag{5.3–38}$$

*where* $\varepsilon \ll 1$ *is a positive constant and the positive constants* $\delta_{1i}$ *and* $\delta_{2i}$ *are chosen such that the roots of* $s^2 + \delta_{1i}\, s + \delta_{2i} = 0$ *are located in open left-hand plane for all* $i = 1, 2, \ldots, n$. *This high gain observer guarantees that for any given* $\mu$, *there exists* $\varepsilon_1 > 0$, *dependent on* $\mu$, *such that for every* $0 < \varepsilon < \varepsilon_1$, *the state estimation error is bounded by* $\mu$ *starting at* $\boldsymbol{x}(0)$, *i.e.,*

$$\| \boldsymbol{x}(t) - \hat{\boldsymbol{x}}(t) \| \le \mu, \qquad \forall\, t \ge 0. \tag{5.3–39}$$

One of the main results of the above lemma is that the smaller the value of $\varepsilon$, the higher the robustness of the observer against uncertainties in the nominal model $\hat{g}$. However, realizing that the high-gain observer is basically an approximate differentiator, particularly when the nominal function is chosen to be zero, we can see that measurement noise and unmodeled high-frequency sensor dynamics will put a practical limit on how small $\varepsilon$ could be. Also, when the nominal model is not good enough, the observer could work better if the nominal function is chosen to be zero which makes it to be linear [108]. Thus, we can use the linear high gain observer for estimating swing angles and their speeds, and then discretize it to be compatible with the discrete-time nature of the overall control system design for the overhead crane. The following continuous-time model for estimating swing angles and their speed is considered,

$$\begin{aligned} \dot{\boldsymbol{x}}_q(t) &= A_{cq}\boldsymbol{x}_q(t), \\ \boldsymbol{y}_q(t) &= C_{cq}\boldsymbol{x}_q(t), \end{aligned} \tag{5.3–40}$$

$$\begin{aligned} &A_{cq} = \begin{bmatrix} A_{cq_x} & \underline{0} \\ \underline{0} & A_{cq_y} \end{bmatrix}, \quad C_{cq} = \begin{bmatrix} C_{cq_x} & \underline{0} \\ \underline{0} & C_{cq_y} \end{bmatrix}, \\ &A_{cq_i} = \begin{bmatrix} 0 & 1 \\ 0 & 0 \end{bmatrix}, \quad C_{cq_i} = [1 \quad 0], \qquad for \quad i = x, y, \end{aligned} \tag{5.3–41}$$

where $\boldsymbol{x}_\theta = [\theta_x \;\; \dot{\theta}_x \;\; \theta_y \;\; \dot{\theta}_y]^T$ is the state vector containing traveling and traversing swing angles and their speeds, and the matrices $A_{c\theta}$ and $C_{c\theta}$ are continuous-time model matrices with $\underline{0}$ being a zero matrix with proper size. The discrete-time linear high-gain observer is then obtained as follows [109],

$$\hat{\boldsymbol{x}}_\theta(k+1) = A_\theta \hat{\boldsymbol{x}}_\theta(k) + L_\theta(\boldsymbol{y}_\theta(k) - C_\theta \hat{\boldsymbol{x}}_\theta(k)), \tag{5.3–42}$$

$$A_\theta = e^{A_{c\theta}T_s} = \begin{bmatrix} A_{\theta_x} & \underline{0} \\ \underline{0} & A_{\theta_y} \end{bmatrix}, \quad L_\theta = \begin{bmatrix} L_{\theta_x} & \underline{0} \\ \underline{0} & L_{\theta_y} \end{bmatrix}, \quad C_\theta = C_{c\theta},$$
$$A_{\theta_i} = \begin{bmatrix} 1 & T_s \\ 0 & 1 \end{bmatrix}, \quad L_{\theta_i} = \begin{bmatrix} l_{1\theta_i} \\ l_{2\theta_i} \end{bmatrix} \quad for \quad i = x, y, \tag{5.3–43}$$

where $L_\theta$ is the swing angle observer gain with $l_{1\theta_i}$ and $l_{2\theta_i}$ to be chosen as

$$l_{1\theta_i} = \frac{\delta_{1i} + 2}{\varepsilon}, \quad l_{2\theta_i} = \frac{\varepsilon^2 + \delta_{1i}\varepsilon + \delta_{2i}}{\varepsilon T_s} \quad for \quad i = x, y, \tag{5.3–44}$$

for a positive constant $\varepsilon \ll 1$ similar to (5.3–38) except that the positive constants $\delta_{1i}$ and $\delta_{2i}$ are chosen such that the roots of $z^2 + \delta_{1i} z + \delta_{2i} = 0$ are located inside the unit circle for $i = x, y$. In Section 5.9.3, the performance of the swing angle observer will be shown under practical results obtained from several tests conducted on the designed control systems for the overhead crane system.

## 5.4 Trajectory Planning

For the case of the overhead crane, it is desired to plan a trajectory from an initial point $\boldsymbol{q}_{ref}(t_0) = (x_{ref}(t_0), y_{ref}(t_0), l_{ref}(t_0))$ to a final point $\boldsymbol{q}_{ref}(t_f) = (x_{ref}(t_f), y_{ref}(t_f), l_{ref}(t_f))$ as fast as possible with a minimum load swing during the operation within the period of $t_0 < t < t_f$. The trajectory planning is subject to constraints on the maximum permissible velocity, acceleration or toque, workspace, and the amount of time for moving the load. In practical applications, the desired trajectory is divided into three zones: accelerating zone, constant-velocity zone, and decelerating zones. The overhead crane is initially accelerated to a normal velocity. The load can be hoisted up in this zone if it was not hoisted fully before accelerating to increase time efficiency. This process allows a certain level of load swing until the normal velocity is reached and the load swings are damped since load hoisting during acceleration creates load swing. Then, in the constant-velocity zone, the overhead crane is moved at the normal velocity. Finally, in the decelerating zone, the overhead crane is decelerated to a complete stop. The load can be hoisted down after decelerating is finished or during deceleration if necessary. This process also allows a certain level of load swing until a full stop is achieved. This type of motion is known as typical anti-swing trajectory, especially if the load is hoisted down during decelerating zone, which is mostly performed manually by the expert operator and it is not time efficient. The reason is that according to stability analysis of swing dynamics described in previous section, swing dynamics is naturally output strictly passive if the load is hoisted down during acceleration or deceleration since hoisting velocity is positive, and therefore, load swings will be suppressed. We are going to take advantage of this property in planning the reference trajectories when load swing control is applied.

However, to have high speed load transportation, the load should be lifted up during accelerating zone and hoisted down during decelerating zone with high speed without pause between zones for load swings to reduce. The load swing tends to increase in the accelerating zone and depending on the hoisting speed it would be large (negative hoisting velocity). Then, in constant-velocity zone, the load swings remain unchanged since traveling and traversing accelerations and hoisting velocity are zero, and consequently, there will be no change in the rate of the load swing energy ($\dot{V}_\theta = 0$ in (5.3–28)). This situation makes it difficult for the operator to damp the load swings during the first two zones and could pose a real danger to the operation. Therefore, they normally try to avoid high speed load hoisting during accelerating zone at a cost of lower time efficiency. That is why a load swing damping action is needed if the overhead carne is operated automatically under high speed load transportation to maintain the safety as well as meeting the high-performance control requirements.

A suitable choice for generating the typical trajectories described above is the combination of polynomial functions known as linear segments with parabolic blends or LSBP for short [32]. The LSPB trajectory is designed such that the velocity is initially ramped up to its desired value (normally between 70% and 80% of the maximum velocity), and then ramped down when it approaches the goal position. The equations for an LSPB trajectory and with the constraints on velocity, acceleration and time are given as follows with the associated trajectory profiles shown in Fig. 5.4–1.

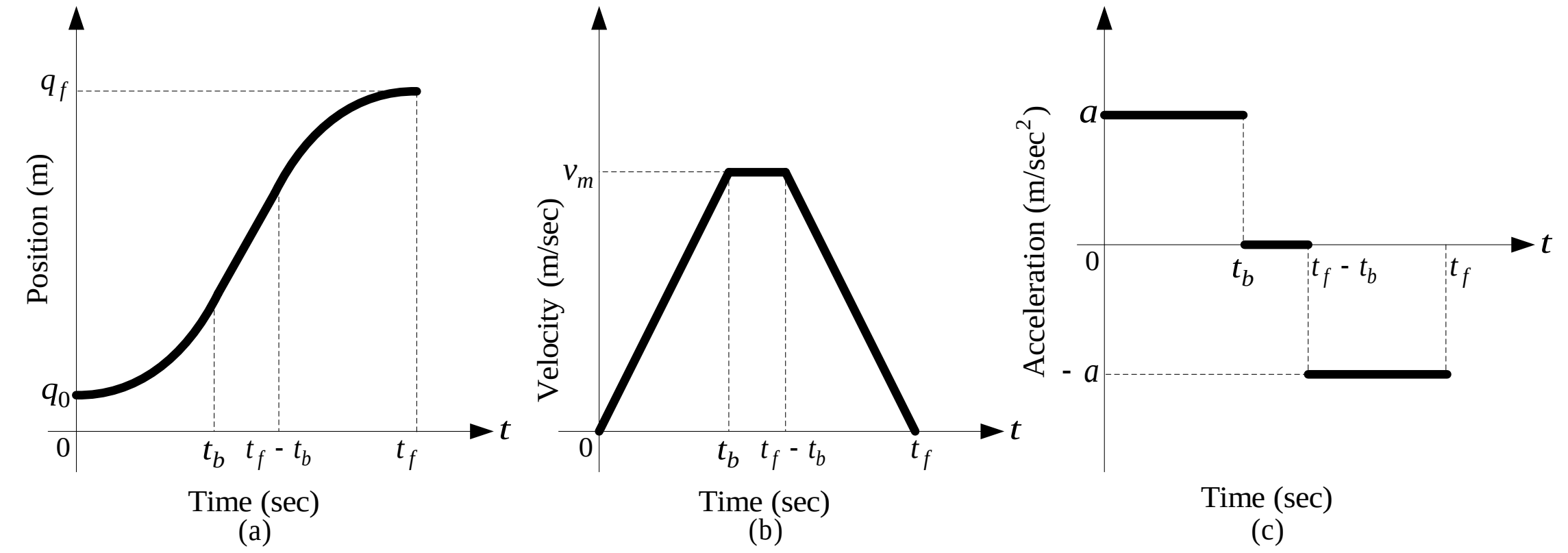

Fig. 5.4–1. LSPB trajectory. (a) Position profile, (b) Velocity profile, (c) Acceleration profile.

$$q(t)=\begin{cases} q_0+\dfrac{a}{2}t^2 & 0\le t<t_b \\ (q_0-\dfrac{a}{2}t_b^2)+v_m t & t_b\le t<t_f-t_b, \\ (q_f-\dfrac{a}{2}t_f^2)+at_f t-\dfrac{a}{2}t^2 & t_f-t_b\le t<t_f \end{cases} \qquad \begin{cases} v_m=at_b, \\ q_f-q_0=v_m(t_f-t_b), \\ t_f>2t_b. \end{cases} \tag{5.4–1}$$

where $q_0$ is the initial point ($q(t_0)$) for each direction of motion), $q_f$ is the final point ($q(t_f)$), $a$ is the acceleration, $v_m$ is the normal/maximum velocity, and $t_b$ is the time when the trajectory reaches the normal velocity.

If the constant-velocity section is removed from the profile, the trajectory is called *minimum-time trajectory* since it allows finding the fastest trajectory between $q_0$ and $q_f$ with a given constant acceleration in a symmetric way. The optimum solution for this approach is usually achieved with the acceleration at its maximum admissible. Thus, the equations for minimum-time trajectory is given as below,

$$q(t)=\begin{cases} q_0+\dfrac{a}{2}t^2 & 0\le t<\dfrac{t_f}{2} \\ (q_f-\dfrac{a}{2}t_f^2)+at_f t-\dfrac{a}{2}t^2 & \dfrac{t_f}{2}\le t<t_f \end{cases}, \qquad \begin{cases} v_m=a\dfrac{t_f}{2}, \\ q_f-q_0=v_m\dfrac{t_f}{2}. \end{cases} \tag{5.4–2}$$

The LSPB trajectory corresponds to that of the practical trajectory for overhead crane with three zones including accelerating zone ($0 \le t \le t_b$), constant-velocity zone ($t_b \le t \le t_f - t_b$), and decelerating zone ($t_f - t_b \le t \le t_f$). Therefore, the reference trajectories for traveling and traversing motions are designed using LSBP trajectory and for hoisting motion, minimum-time trajectory is used for accelerating and decelerating zones as illustrated in Fig. 5.4–2.

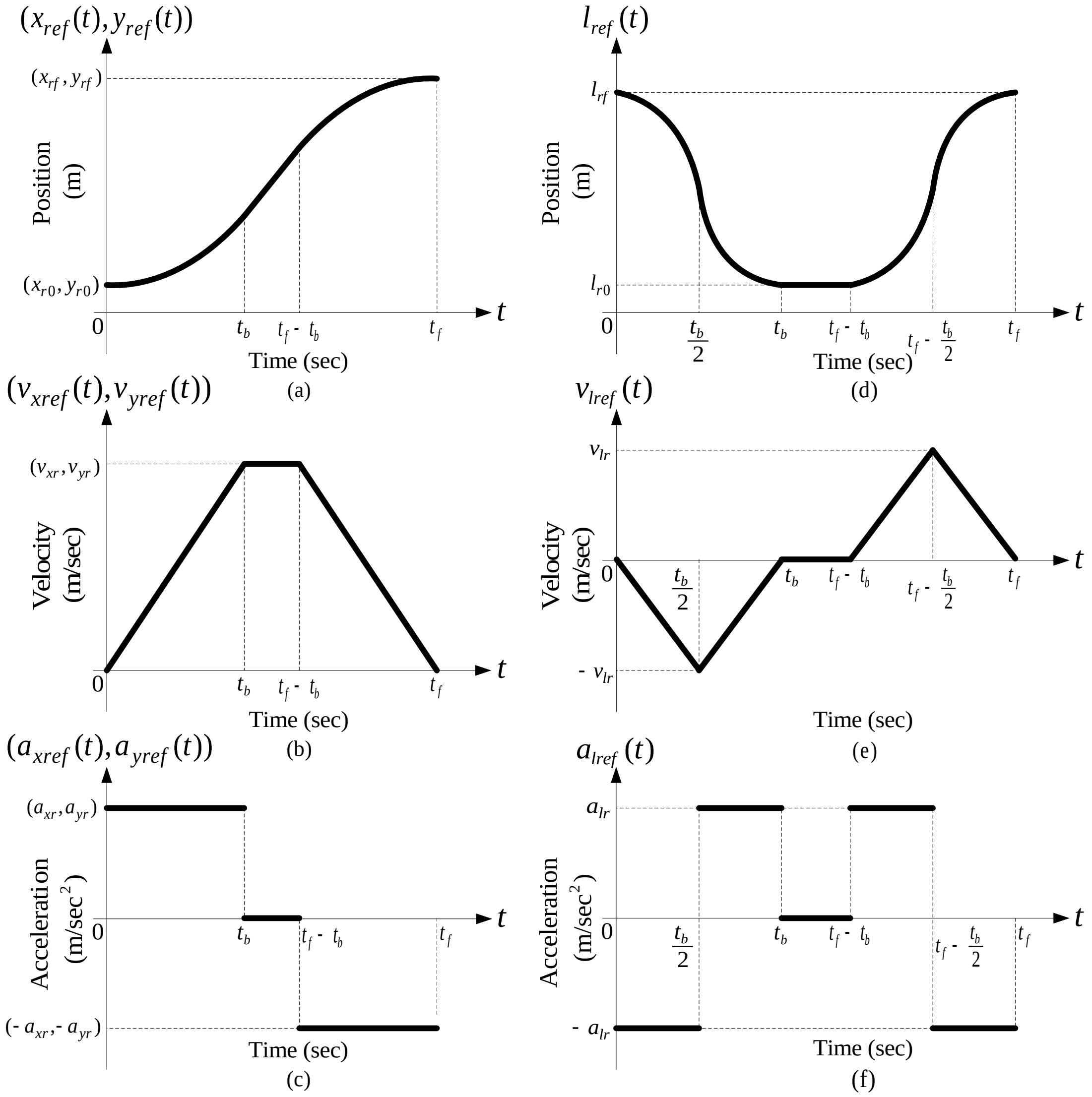

Fig. 5.4–2. LSPB trajectories for overhead crane. (a) Position, (b) velocity, and (c) acceleration profiles for traveling and traversing motions. (d) Position, (e) velocity, and (f) acceleration profiles for hoisting motion.

As we explained in Section 5.3, the load swing control is designed such that by modifying the reference traveling and traversing accelerations, the suppression of load swings during the overhead crane operation will be guaranteed. However, one complication of this approach is the deviation of the reference position and velocity of the trolley in $X$ and $Y$ directions, ($x_{ref}$ , $y_{ref}$) and ($v_{xref}$ , $v_{yref}$), respectively, since they are generated by the reference signal generator block, which is designed based on discrete-time double integrator as a reference model. Therefore, using the modified accelerations in $\boldsymbol{u}_{c_xy}$ as the input to the reference signal generator rather than the original ones in $\boldsymbol{a}_{xy_ref}$ will generate the reference trajectories which are deviated from the original desired ones. This amount of deviation created in reference trajectories will depend on the amount of initial load swing and the speed of swing damping imposed by swing control gain $K_\theta$. This deviation will result in the control system to track the deviated reference trajectories and end up reaching to the final point with probably huge position error. In order to solve this problem, we can take advantage of the natural swing damping property in decelerating zone with the load being hoisted down. That means, the load swing control will be active during accelerating and constant-velocity zone to suppress load swing. Then, to resolve the deviation in reference trajectories due to load swing control, we design a plan to recalculate the amount of velocity and acceleration for traveling and traversing such that the reference trajectories return to the original final point during the decelerating zone with no load swing control ($K_\theta = 0$). In this way, not only any remaining load swing will be damped due to natural load swing stability in decelerating zone, but also, the deviation in reference trajectories will be fixed and the control system will be able to bring the overhead crane to the original final point designed in the first place. This procedure is summarized in the following steps for both traveling and traversing reference trajectories. Fig. 5.4–3 also demonstrates this procedure for reference traveling trajectory as an example.

**Step1:** Find the correction velocity ($v_{xrc}$, $v_{yrc}$) needed to move the trolley from its deviated reference position at the end of constant-velocity zone ($x_{rd}$, $y_{rd}$) to the final designed point ($x_{rf}$, $y_{rf}$) within decelerating time ($t_b$ seconds) in parabolic form, i.e., $v_{xrc} = 2(x_{rf} - x_{rd})/t_b$ and $v_{yrc} = 2(y_{rf} - y_{rd})/t_b$.

**Step2:** Find the correction reference trolley acceleration ($a_{xrc}$, $a_{yrc}$) required for linear velocity from ($v_{xrc}$, $v_{yrc}$) to zero and set it in $\boldsymbol{u}_{c_xy}$ with $K_\theta = 0$, i.e., $u_{cx} = a_{xrc} = (v_{xrc}/t_b)$ and $u_{cy} = a_{yrc} = (v_{yrc}/t_b)$ at time $t_f - t_b$.

**Step3:** Set the initial conditions of the traveling and traversing reference models in reference signal generator block to $[x_{rd}\ \ v_{xrc}]^T$ and $[y_{rd}\ \ v_{yrc}]^T$ at time $t_f - t_b$.

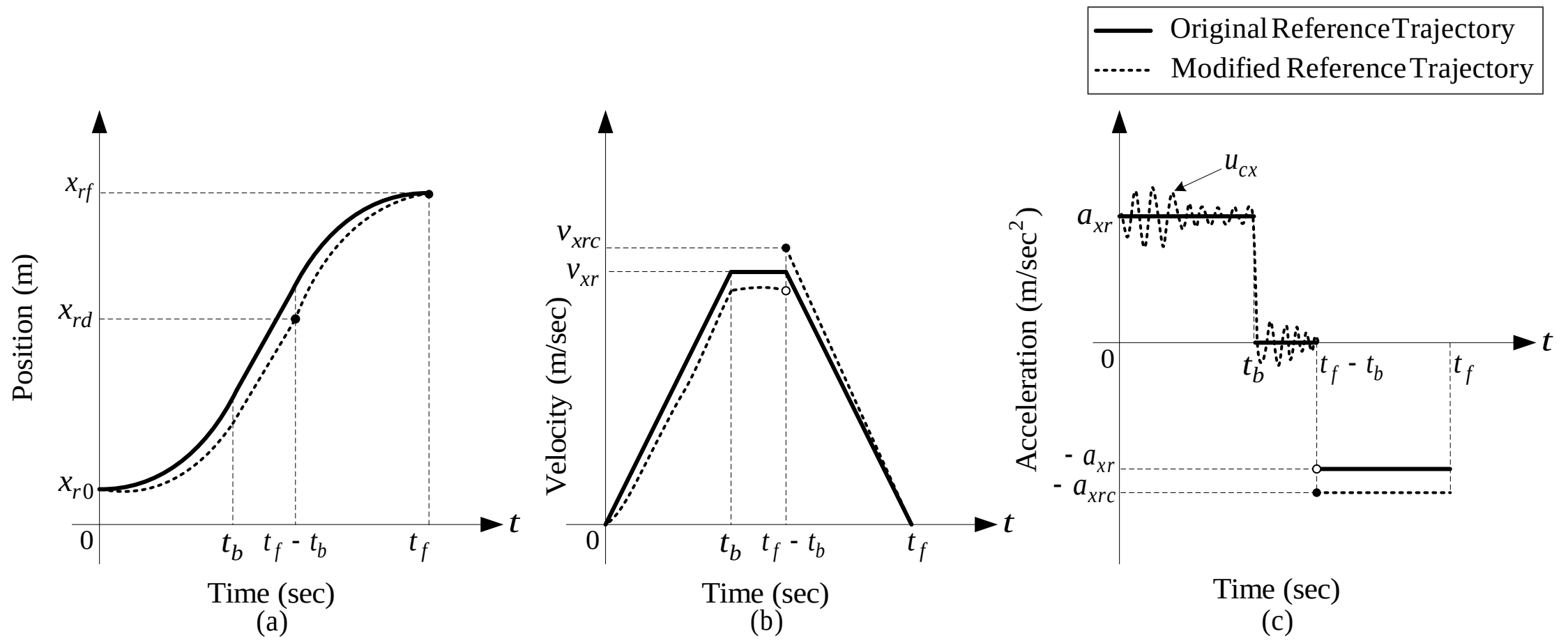

Fig. 5.4–3. Comparison between original and modified reference traveling trajectory for the case where the load swig control causes the reference position to fall back. (a) Position profile, (b) Velocity profile, (c) Acceleration profile.

The correction steps explained above for fixing the deviation in reference traveling and traversing trajectories caused by load swing control would be sufficient if the amount of deviation in position is not significant [109]. Large deviation in reference trajectories could happen because of some unexpected disturbances acting directly on load swing, like a sudden strong wind blow, that intensifies load swings. As a result, the load swing control would need to change reference accelerations significantly to suppress them which could lead to over-expected deviation in reference trolley position. Thus, to replan the reference traveling and traversing positions in decelerating zone so that they reach to the original final destination within the predesigned time, the correcting velocities may exceed the maximum permissible velocity of the actuators for either of the traveling or traversing motions (if the updated trajectories fall behind the original ones). In that case, the velocities can be set to their maximum values to protect the actuators, but the reference trolley position will not get to the original final destination within the original decelerating time and we will still have position error at the end of the trajectory. In addition, it is possible that either of the correction accelerations become greater than the maximum admissible acceleration generated by the actuators that can further complicate the situation. Therefore, there is no other way except than increasing the decelerating time to allow the correction being conducted within permitted velocity and acceleration range. Moreover, if the decelerating time increases, the hoisting trajectory should be adapted to the new decelerating zone.

Even though it is unlikely that such an incident happens, especially when the overhead crane is operated indoors, as a precaution, there should be an automatic procedure to specify the optimum decelerating time extension considering all the constraints on the LSBP trajectory. This will guarantee that the load could be transported as fast as possible with robust load swing suppression and without compromising the violating the constraints on the maximum permissible velocity and acceleration of the traveling and traversing actuators. The following flowchart illustrates how replanning of the decelerating zone is performed, particularly if decelerating time extension should happen due to the violation of velocity or acceleration constraints.

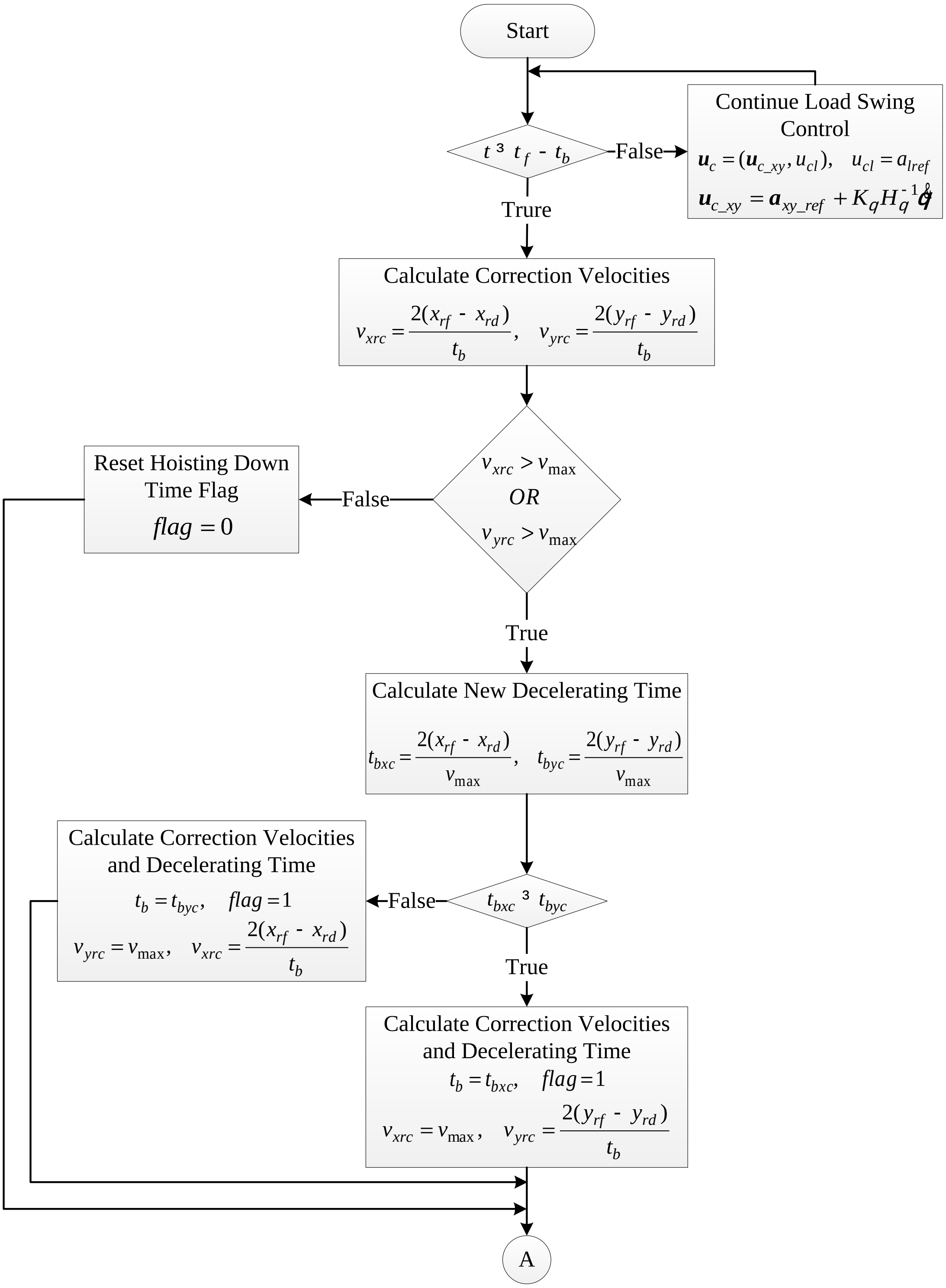
Start
Continue Load Swing Control
$\boldsymbol{u}_c = (\boldsymbol{u}_{c_xy}, u_{cl}), \quad u_{cl} = a_{lref}$
$\boldsymbol{u}_{c_xy} = \boldsymbol{a}_{xy_ref} + K_q H_q^{-1} \dot{\boldsymbol{q}}$
$t \geq t_f - t_b$
False
Trure
Calculate Correction Velocities
$v_{xrc} = \frac{2(x_{rf} - x_{rd})}{t_b}, \quad v_{yrc} = \frac{2(y_{rf} - y_{rd})}{t_b}$
Reset Hoisting Down Time Flag
$flag = 0$
False
$v_{xrc} > v_{max}$ OR $v_{yrc} > v_{max}$
True
Calculate New Decelerating Time
$t_{bxc} = \frac{2(x_{rf} - x_{rd})}{v_{max}}, \quad t_{byc} = \frac{2(y_{rf} - y_{rd})}{v_{max}}$
Calculate Correction Velocities and Decelerating Time
$t_b = t_{byc}, \quad flag = 1$
$v_{yrc} = v_{max}, \quad v_{xrc} = \frac{2(x_{rf} - x_{rd})}{t_b}$
False
$t_{bxc} \geq t_{byc}$
True
Calculate Correction Velocities and Decelerating Time
$t_b = t_{bxc}, \quad flag = 1$
$v_{xrc} = v_{max}, \quad v_{yrc} = \frac{2(y_{rf} - y_{rd})}{t_b}$
A

(a)

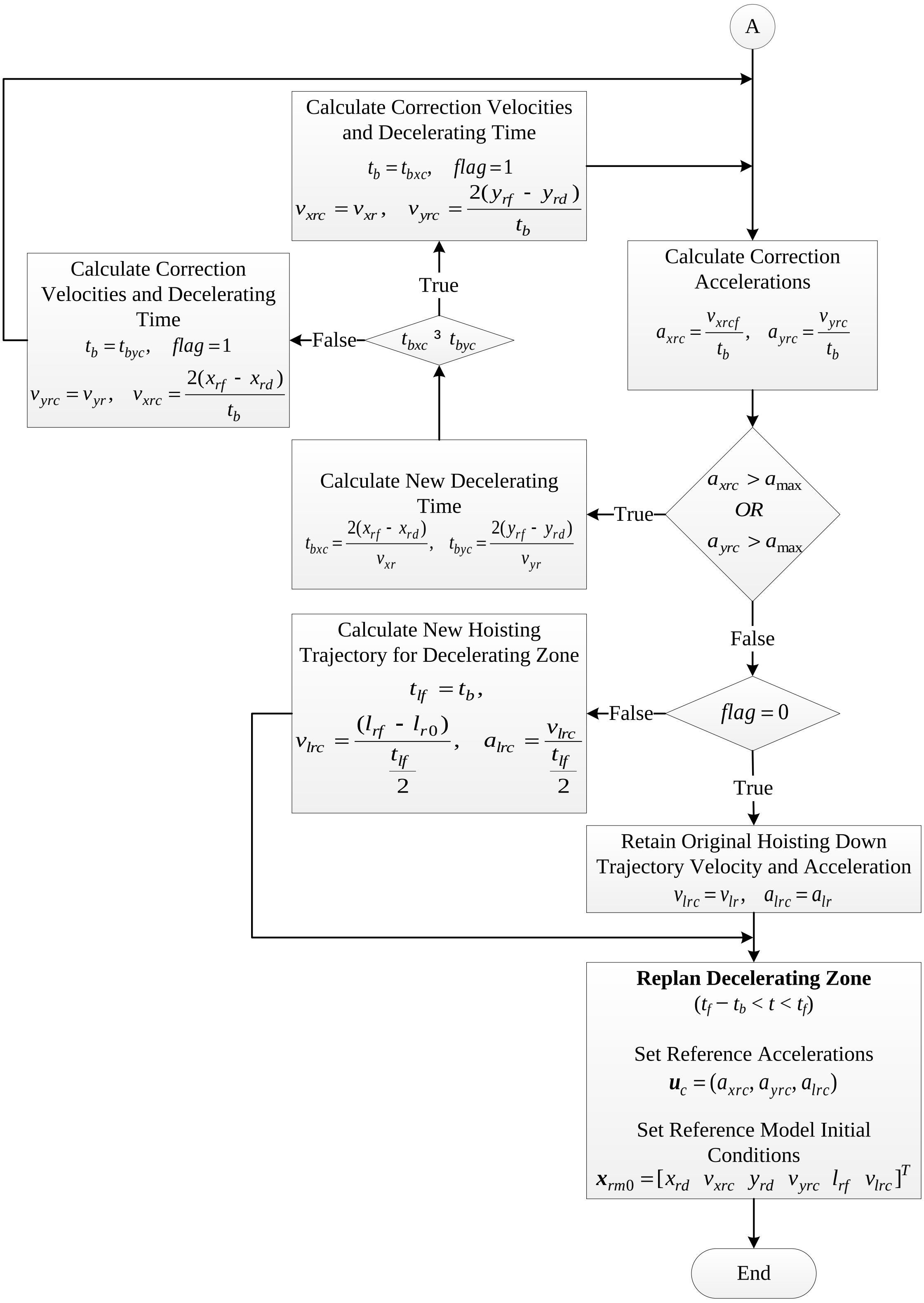

(b)

Fig. 5.4–4. The proposed flowchart for replanning the decelerating zone. (a) Calculation of correction velocities and initial check for violation of maximum permissible velocity, (b) Calculation of correction acceleration, checking for violation of maximum admissible acceleration, recalculation of correction velocities, and recalculation of hoisting velocity and acceleration if decelerating time has extended.

The decelerating zone replanning procedure shown in Fig. 5.4–4 is similar to those three steps mentioned earlier. The load swing control is active until reaching decelerating zone at $t = t_f - t_b$ (It should be mentioned though that all the times are chosen to be an integer multiplier of sampling time for better accuracy, i.e., $t = kT_s = t_f - t_b$). $\boldsymbol{u}_c = [\boldsymbol{u}_{c_xy}\ u_{cl}]^T = [u_{cx}\ u_{cy}\ a_{lref}]^T$ is the command signal for the reference model in the reference signal generator block responsible for generating reference state trajectories $\boldsymbol{x}_{rm} = [\boldsymbol{x}_{rmx}\ \ \boldsymbol{x}_{rmy}\ \ \boldsymbol{x}_{rml}]^T = [x_{ref}\ \ v_{xref}\ \ y_{ref}\ \ v_{yref}\ \ l_{ref}\ \ v_{lref}]^T$. It contains modified reference traveling and traversing accelerations ($\boldsymbol{u}_{c_xy}$ in (5.3–30)) plus the reference hoisting acceleration $a_{lref}$ (the details of the reference signal generator are given in Section 5.5). Once the decelerating zone is reached, the correction velocities are calculated according to Step 1, and then they are checked against the maximum permissible velocity $v_{\max}$ using logic operator OR (it is assumed that maximum permissible velocity is the same for both traveling and traversing motions). If both of the correction velocities are less than or equal to $v_{\max}$, the result is false and the flowchart goes to calculate correction accelerations in Fig. 5.4–4(b) and the hoisting down time flag is set to zero showing that up to this stage there is no need for changing reference hoisting trajectory. If any of them is greater than $v_{\max}$, extended decelerating times are calculated using $v_{\max}$ for both traveling and traversing motions $t_{bxc}$ and $t_{bxc}$, respectively. Between these two times, the greater one determines the new decelerating time as it shows which one requires more time to fix the deviation using the maximum permissible velocity. After checking that, the new correction velocities are calculated based on the new decelerating time, and the hoisting down flag is set to one showing that the hoisting trajectory needs to be updated as shown in Fig. 5.4–4(a) .

Next, correction accelerations should be calculated as expressed in Step 2. Even if the decelerating time has extended in previous step, we still need to make sure that with new velocities and decelerating time the required acceleration will not exceed the maximum admissible acceleration $a_{\max}$ for both traveling and traversing. If the correction accelerations are less than $a_{\max}$, then it is just needed to check whether the decelerating time has extended in the previous step or not by checking the hoisting down time flag. If the flag is zero, it means that neither the correction velocities nor the corresponding correction accelerations were needed to be recalculated using a longer decelerating time. Otherwise, with flag equal one, the reference hoisting velocity and acceleration should be updated based on the new decelerating time, and then the reference trajectories for decelerating zone is replanned based on Step 3.

However, if either of the correction accelerations turns out to become greater than $a_{\max}$, we use a different approach since calculating new decelerating time using $a_{\max}$ and following similar procedure done for velocities would complicated the situation and there is chance that the procedure could not converge to the end of the flowchart. Therefore, regardless of whether the decelerating time has extended in the previous step, if one of the correction accelerations is greater than $a_{\max}$, the correction velocities are set back to their original normal velocities that had been designed for the reference traveling and traversing trajectories in the first place, i.e., $v_{rx}$ and $v_{ry}$, respectively. The decelerating time for fixing the deviations is then recalculated using $v_{rx}$ and $v_{ry}$ which is longer than both the original decelerating time and the one calculated in the previous step (if that occurred). In this way, the new correction accelerations are guaranteed to be less than $a_{\max}$, even less than the original accelerations, at a cost of lengthening the decelerating time long enough to make sure none of the velocities and accelerations would violate their maximum values. Therefore, the second check of the correction accelerations would always pass to the hoisting down time flag checking, and ultimately, the whole procedure depicted in the flowchart in Fig. 5.4–4 will end in maximum 15 cycles when decelerating zone is reached after constant-velocity zone. Thus, we have to make sure that the processor frequency of the main controller responsible for executing control program lines is faster than $1/(15\times T_s)$. This will guarantee that replanning of the decelerating zone will be successfully finalized before the next sampling time after velocity zone at $t_f - t_b$.

In a nut shell, the proposed trajectory planning in designed in conjunction with load swing control that allows a slight deviation in the reference traveling and traversing positions to be able to suppress load swing during accelerating and constant-velocity zones. Then, just before decelerating zone, load swing control will be shut down to let the deviation being fixed by replanning the trajectories, and the remaining load swings will be damped due to natural damping property of the swing dynamics when the load is hoisted down. In the worst-case scenario, if fixing the deviation in the reference position trajectories cannot be occurred within the original decelerating time as one of the correction velocities or accelerations exceeds their maximum values, we have to slow down the deceleration of the overhead crane load. This can be done by finding the optimum decelerating time extension so that the control system can still be operated in a safe range of velocities and accelerations to protect the actuators from being damaged, without significantly reducing the time efficiency of the overall operation.

## 5.5 Reference Signal Generator

As we explained earlier, the reference trajectories for traveling, traversing and hoisting positions and velocities are generated via discrete-time integration of the accelerations. The trolley accelerations ($a_{xref}$ and $a_{yref}$) are modified by the load swing control to robustly suppress load swings following the design procedure for trajectory planning described in Section 5.4. We can now define a reference model that uses the modified accelerations as the input to generate reference trajectories as follows,

$$\begin{aligned} \boldsymbol{x}_{rm}(k+1) &= A_m \boldsymbol{x}_{rm}(k) + B_m \boldsymbol{u}_c(k), \\ \boldsymbol{y}_{ref}(k) &= C_m \boldsymbol{x}_{rm}(k), \end{aligned} \qquad (5.5\text{–}1)$$

where $\boldsymbol{u}_c(k)$ = $[u_{cx}\,(k)\ u_{cy}\,(k)\ u_{cl}\,(k)]^T$ is the command signal for the reference model containing the modified reference traveling and traversing accelerations $\boldsymbol{u}_{c_xy}$ in (5.3–30) and reference hoisting acceleration $u_{cl}\,(k)$ = $a_{lref}\,(k)$ (This will be updated as weel should the decelerating time is extended); $\boldsymbol{x}_{rm}(k)$ = $[\boldsymbol{x}_{rmx}(k)\ \ \boldsymbol{x}_{rmy}(k)\ \ \boldsymbol{x}_{rml}(k)]^T$ = $[x_{ref}\,(k)$ $v_{xref}\,(k)\ \ y_{ref}\,(k)\ \ v_{yref}\,(k)\ \ l_{ref}\,(k)\ \ v_{lref}\,(k)]^T$ is the reference state trajectories; $\boldsymbol{y}_{ref}\,(k)$ = $[x_{ref}$ $(k)\ y_{ref}\,(k)\ l_{ref}\,(k)]^T$ is reference output response; $A_m$ = $BlockDiag\{A_{mx}, A_{my}, A_{ml}\}$; $B_m$ = $BlockDiag\{B_{mx}, B_{my}, B_{ml}\}$, and $C_m$ = $BlockDiag\{C_{mx}, C_{my}, C_{ml}\}$ are system matrix, input matrix, and output matrix for reference model, respectively, with inner matrices given as follows,

$$A_{mi} = \begin{bmatrix} 1 & T_s \\ 0 & 1 \end{bmatrix}, \quad B_{mi} = \begin{bmatrix} 0 \\ T_s \end{bmatrix}, \quad C_{mi} = \begin{bmatrix} 1 & 0 \end{bmatrix}, \quad for \quad i = x, y, l. \qquad (5.5\text{–}2)$$

To have a better understanding on how reference trajectories are generated in conjunction with the trajectory planning and load swing control, the internal block diagram of reference signal generator is depicted in Fig. 5.5–1. As can be seen, the original reference accelerations $\boldsymbol{a}_{ref}$ are initially modified by the correction term generated by the load swing control. Then, they are sent to the decelerating zone replanning block to check when the constant-velocity zone is finished. Once reaching the decelerating zone, the correction velocities and accelerations are calculated by measuring the amount of deviation in the reference traveling and traversing positions caused by the load swing control. The modified accelerations in the command signal $\boldsymbol{u}_c$ are then fed to the reference model to generate reference state trajectories $\boldsymbol{x}_{rm}$ and reference output $\boldsymbol{y}_{ref}$. It should be mentioned that in both accelerating and constant-velocity zones if the modified accelerations, and subsequently the resulting reference velocities become greater than their maximum values, i.e., $a_{\max}$ and $v_{\max}$, respectively, the reference signal generator block would replace their generated values with their maximum ones knowing that in the decelerating zone any possible deviation in trolley trajectory will be fixed.

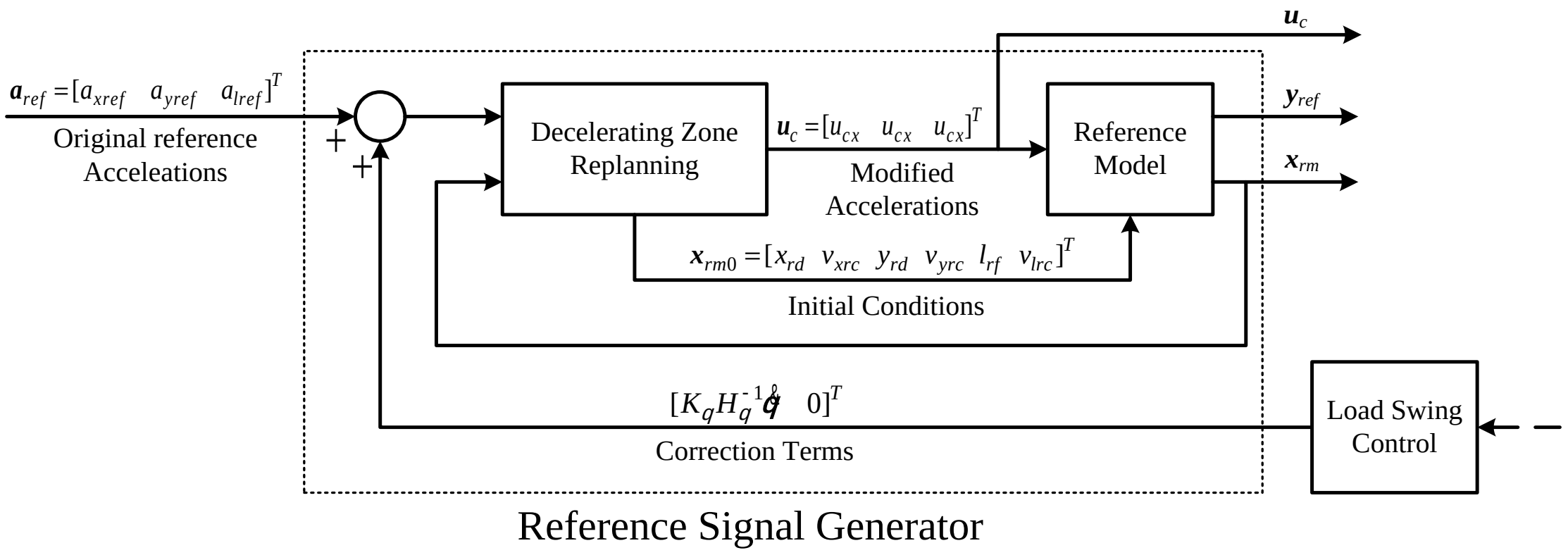

Fig. 5.5–1. The reference generator signal block.

## 5.6 MPC Formulation for Overhead Crane

As we mentioned in Section 5.1, to meet the control objectives for high-performance anti-swing tracking control of overhead crane, two approaches are utilized to design the discrete-time controller. In this section, we use MPC as the discrete-time controller for the overhead crane due its discrete-time nature, constraint handling, easy implementation, and its capability in compensating disturbances trough integration with feedforward control based on its formulation in Section 3.5. Let us recall the discrete-time state-space model we obtained for overhead crane in Section 4.6 given as below,

$$\begin{aligned} \boldsymbol{x}(k+1) &= A\boldsymbol{x}(k) + B\boldsymbol{u}(k) + W_d \boldsymbol{f}_d(k), \\ \boldsymbol{y}(k) &= C\boldsymbol{x}(k), \end{aligned} \tag{5.6–1}$$

$$\begin{aligned} \boldsymbol{x}(k) &= [x(k) \;\; v_x(k) \;\; y(k) \;\; v_y(k) \;\; l(k) \;\; v_l(k)]^T, \\ \boldsymbol{u}(k) &= [v_{ax}(k) \;\; v_{ay}(k) \;\; v_{al}(k)]^T, \\ \boldsymbol{f}_d(k) &= [f_{dx}(k) + f_{cfx}(k) \quad f_{dy}(k) + f_{cfy}(k) \quad f_{dl}(k) + f_{cfl}(k)]^T, \\ \boldsymbol{y}(k) &= [x(k) \;\; y(k) \;\; l(k)]^T, \end{aligned} \tag{5.6–2}$$

$$A = \begin{bmatrix} A_x & \underline{0} & \underline{0} \\ \underline{0} & A_y & \underline{0} \\ \underline{0} & \underline{0} & A_l \end{bmatrix} \; B = \begin{bmatrix} B_x & \underline{0} & \underline{0} \\ \underline{0} & B_y & \underline{0} \\ \underline{0} & \underline{0} & B_l \end{bmatrix}$$
$$W_d = \begin{bmatrix} W_{dx} & \underline{0} & \underline{0} \\ \underline{0} & W_{dy} & \underline{0} \\ \underline{0} & \underline{0} & W_{dl} \end{bmatrix} \; C = \begin{bmatrix} C_x & \underline{0} & \underline{0} \\ \underline{0} & C_y & \underline{0} \\ \underline{0} & \underline{0} & C_l \end{bmatrix} \tag{5.6–3}$$

$$A_i = \begin{bmatrix} 1 & T_s \\ 0 & a_{1i} \end{bmatrix}, \; B_i = \begin{bmatrix} 0 \\ b_{1i} \end{bmatrix}, \; W_{di} = \begin{bmatrix} 0 \\ -b_{d1i} \end{bmatrix}, \; C_i = [1 \quad 0], \quad for \quad i = x, y, l, \tag{5.6–4}$$

and the cost function defined in Section 3.5.2 for MPC that penalizes trajectory tracking error and control input changes in (3.5–2) subject to system equations (5.6–1), control input, and output constraints as follows,

$$V(k) = \sum_{i=1}^{H_p} \left\| \hat{\boldsymbol{y}}(k+i\,|\,k) - \boldsymbol{y}_{ref}(k+i\,|\,k) \right\|_{Q(i)}^2 + \sum_{i=0}^{H_u-1} \left\| \Delta\hat{\boldsymbol{u}}(k+i\,|\,k) \right\|_{R(i)}^2, \tag{5.6–5}$$

$$\begin{aligned} \boldsymbol{y}_{min} \le \boldsymbol{y}(k) \le \boldsymbol{y}_{max} \qquad & k = 1, 2, \ldots, H_p, \\ \boldsymbol{u}_{min} \le \boldsymbol{u}(k) \le \boldsymbol{u}_{max} \qquad & k = 0, 1, \ldots, H_u - 1, \end{aligned} \tag{5.6–6}$$

where $\boldsymbol{y}_{ref}(k)$ is the reference trajectory generated by the reference signal generator as shown in Fig. 5.5–1; $\boldsymbol{y}_{min} = [x_{min} \;\; y_{min} \;\; l_{min}]^T$ and $\boldsymbol{y}_{max} = [x_{max} \;\; y_{max} \;\; l_{max}]^T$ are the vectors of lower and upper bounds of crane workspace, respectively; $\boldsymbol{u}_{min} = [v_{ax_min} \;\; v_{ay_min} \;\; v_{al_min}]^T$ and $\boldsymbol{u}_{max} = [v_{ax_max} \;\; v_{ay_max} \;\; v_{al_max}]^T$ are the vectors of lower and upper bounds of DC motor voltages, respectively;

Now, the cost function given in (5.6–5) can be written in a matrix form as below,

$$
\begin{aligned}
V(k) &= \| Y(k) - Y_{ref}(k) \|_Q^2 + \| \Delta U(k) \|_R^2 \\
&= (Y(k) - Y_{ref}(k))^T Q (Y(k) - Y_{ref}(k)) + \Delta U(k)^T R \Delta U(k),
\end{aligned}
\tag{5.6–7}
$$

where $Y(k) = [\hat{\boldsymbol{y}}(k+1|\ k)\ \ \hat{\boldsymbol{y}}(k+2|\ k)\ \ \ldots\ \ \hat{\boldsymbol{y}}(k+H_p\ |\ k)]^T$ is the vector of output predictions; $Y_{ref}(k) = [\boldsymbol{y}_{ref}(k+1|\ k)\ \ \boldsymbol{y}_{ref}(k+2|\ k)\ \ \ldots\ \ \boldsymbol{y}_{ref}(k+H_p\ |\ k)]^T$ is the vector of future values for reference trajectories; $\Delta U(k) = [\Delta\hat{\boldsymbol{u}}(k\ |\ k)\ \Delta\hat{\boldsymbol{u}}(k+1|\ k)\ \ldots\ \Delta\hat{\boldsymbol{u}}(k+H_u - 1|\ k)]^T$ is the vector of control input changes; $Q = BlockDiag\{Q(1), Q(2), \ldots Q(H_p)\}$, and $R = BlockDiag\{R(0), R(1), \ldots R(H_u-1)\}$ are block-diagonal matrices containing 3×3 square diagonal weighting matrices for tracking error $Q(k) = Diag\{q_x, q_y, q_l\}$, and control input changes $R(k) = Diag\{r_x, r_y, r_l\}$, respectively. To solve the optimization problem defined in (5.6–7), the output predictions should be obtained for the prediction horizon $H_p$ knowing that $\boldsymbol{u}(k-1)$ and $\boldsymbol{x}(k)$ are available at time $k$. The overhead crane dynamic model given in (5.6–1) can be used to calculate $\hat{\boldsymbol{y}}(k+i\ |\ k)$ as follows,

$$
\hat{\boldsymbol{y}}(k+1|k) = C\hat{\boldsymbol{x}}(k+1|k) = CA\boldsymbol{x}(k) + CB\hat{\boldsymbol{u}}(k|k) + CW_d\hat{\boldsymbol{f}}_d(k|k),
\tag{5.6–8}
$$

$$
\begin{aligned}
\hat{\boldsymbol{y}}(k+2|k) &= CA^2\boldsymbol{x}(k) + CAB\hat{\boldsymbol{u}}(k|k) + CAW_d\hat{\boldsymbol{f}}_d(k|k) + CB\hat{\boldsymbol{u}}(k+1|k) \\
&\quad + CW_d\hat{\boldsymbol{f}}_d(k+1|k),
\end{aligned}
\tag{5.6–9}
$$

$$\vdots$$

$$
\begin{aligned}
\hat{\boldsymbol{y}}(k+H_p|k) &= CA^{H_p}\boldsymbol{x}(k) + CA^{H_p-1}B\hat{\boldsymbol{u}}(k|k) + \ldots + CB\hat{\boldsymbol{u}}(k+H_p-1|k) \\
&\quad + CA^{H_p-1}W_d\hat{\boldsymbol{f}}_d(k|k) + \ldots + CW_d\hat{\boldsymbol{f}}_d(k+H_p-1|k),
\end{aligned}
\tag{5.6–10}
$$

where $\hat{\boldsymbol{f}}_d(k+i\ |\ k)$ is the prediction of load disturbances at $k+i$ made at time step $k$ using computed torque control as explained in Section 5.2.2. It should be mentioned that these computed disturbances are generated using (4.4–7)–(4.4–10), reference trajectories $\boldsymbol{x}_{rm}$, the modified accelerations set in command signal $\boldsymbol{u}_c$, and swing angels and their first time-derivatives, $\boldsymbol{\theta}$ and $\dot{\boldsymbol{\theta}}$, respectively, provided by the swing angle observer. It is assumed that the future values of disturbances are constant and equal to those calculated at time step $k$, i.e., $\hat{\boldsymbol{f}}_d(k+i\ |\ k) = \hat{\boldsymbol{f}}_d(k\ |\ k)$ for $0 \le i \le H_p - 1$. Although this is a common practice [89], reference trajectories are subject to change due to load swing control in our case, which prevents having the predictions of $\hat{\boldsymbol{f}}_d$ for some steps ahead. Unlike the traditional computed torque control, that uses desired values of trajectories, the actual swing angles and their first time-derivatives (coming from swing angle observer) are used in the computations of disturbances to have more accurate estimation of the load disturbances. That is also why we simplified the overhead crane equations of motion to remove second time-derivative from the equations so that by using only swing angles and their first time-derivatives, the load disturbances can be estimated as explained in Section 5.2.2.

Since the cost function will be minimized against the control input changes $\Delta\boldsymbol{u}(k)$ rather than control input $\boldsymbol{u}(k)$, we have to obtain output prediction in (5.6–8)–(5.6–10)in terms of $\Delta\hat{\boldsymbol{u}}(k)$. Also, it is assumed that the control input will remain constant after control horizon $H_c$. Therefore, knowing that $\Delta\hat{\boldsymbol{u}}(k+i \mid k) = \hat{\boldsymbol{u}}(k+i \mid k) - \hat{\boldsymbol{u}}(k+i-1 \mid k)$ and for $H_u \le i \le H_p - 1$ we have $\Delta\hat{\boldsymbol{u}}(k+i \mid k) = 0$ or $\hat{\boldsymbol{u}}(k+i \mid k) = \hat{\boldsymbol{u}}(k+H_u-1 \mid k)$, (5.6–8)–(5.6–10) are rewritten as follows,

$$\hat{\boldsymbol{y}}(k+1|k) = CA\boldsymbol{x}(k) + CB\Delta\hat{\boldsymbol{u}}(k|k) + CB\boldsymbol{u}(k-1) + CW_d\hat{\boldsymbol{f}}_d(k|k), \tag{5.6–11}$$

$$\begin{aligned}\hat{\boldsymbol{y}}(k+2|k) = {} & CA^2\boldsymbol{x}(k) + C(AB+B)\Delta\hat{\boldsymbol{u}}(k|k) + CB\Delta\hat{\boldsymbol{u}}(k+1|k) + C(AB+B)\boldsymbol{u}(k-1)\\ & + CAW_d\hat{\boldsymbol{f}}_d(k|k) + CW_d\hat{\boldsymbol{f}}_d(k+1|k),\end{aligned} \tag{5.6–12}$$

$$\vdots$$

$$\begin{aligned}\hat{\boldsymbol{y}}(k+H_u|k) = {} & CA^{H_u}\boldsymbol{x}(k) + C(A^{H_u-1}B + \ldots + AB + B)\Delta\hat{\boldsymbol{u}}(k|k)\\ & + \ldots + CB\Delta\hat{\boldsymbol{u}}(k+H_u-1|k) + C(A^{H_u-1}B + \ldots + AB + B)\boldsymbol{u}(k-1)\\ & + CA^{H_u-1}W_d\hat{\boldsymbol{f}}_d(k|k) + \ldots + CW_d\hat{\boldsymbol{f}}_d(k+H_u-1|k),\end{aligned} \tag{5.6–13}$$

$$\begin{aligned}\hat{\boldsymbol{y}}(k+H_u+1|k) = {} & CA^{H_u+1}\boldsymbol{x}(k) + C(A^{H_u}B + \ldots + AB + B)\Delta\hat{\boldsymbol{u}}(k|k)\\ & + \ldots + C(AB+B)\Delta\hat{\boldsymbol{u}}(k+H_u-1|k) + C(A^{H_u}B + \ldots + AB + B)\boldsymbol{u}(k-1)\\ & + CA^{H_u}W_d\hat{\boldsymbol{f}}_d(k|k) + \ldots + CW_d\hat{\boldsymbol{f}}_d(k+H_u|k),\end{aligned} \tag{5.6–14}$$

$$\vdots$$

$$\begin{aligned}\hat{\boldsymbol{y}}(k+H_p|k) = {} & CA^{H_p}\boldsymbol{x}(k) + C(A^{H_p-1}B + \ldots + AB + B)\Delta\hat{\boldsymbol{u}}(k|k)\\ & + \ldots + C(A^{H_p-H_u}B + \ldots + AB + B)\Delta\hat{\boldsymbol{u}}(k+H_u-1|k)\\ & + C(A^{H_p-1}B + \ldots + AB + B)\boldsymbol{u}(k-1)\\ & + CA^{H_p-1}W_d\hat{\boldsymbol{f}}_d(k|k) + \ldots + CW_d\hat{\boldsymbol{f}}_d(k+H_p-1|k).\end{aligned} \tag{5.6–15}$$

These predictions can be written in matrix form as the following,

$$Y(k) = \Psi \boldsymbol{x}(k) + \Gamma \boldsymbol{u}(k-1) + \Theta \Delta U(k) + \Lambda F_d(k), \tag{5.6–16}$$

where $F_d(k) = [\hat{\boldsymbol{f}}_d(k \mid k) \;\; \hat{\boldsymbol{f}}_d(k+1 \mid k) \; \ldots \; \hat{\boldsymbol{f}}_d(k+H_p-1 \mid k)]^T$, and matrices $\Psi$, $\Gamma$, $\Theta$, and $\Lambda$ are obtained using (5.6–11)–(5.6–15) as below,

$$\Psi = \begin{bmatrix} CA \\ CA^2 \\ \vdots \\ CA^{H_p} \end{bmatrix}, \tag{5.6–17}$$

$$\Gamma = \begin{bmatrix} CB \\ CB + CAB \\ CB + CAB + CA^2B \\ \vdots \\ C\sum_{i=0}^{H_p-1} A^i B \end{bmatrix}, \tag{5.6–18}$$

$$\Lambda = \begin{bmatrix} CW_d & \underline{0} & \cdots & \underline{0} \\ CAW_d & CW_d & \ddots & \vdots \\ \vdots & \vdots & \ddots & \underline{0} \\ CA^{H_p-1}W_d & CA^{H_p-2}W_d & \cdots & CW_d \end{bmatrix}, \tag{5.6–19}$$

$$\Theta = \begin{bmatrix} CB & \underline{0} & \cdots & \underline{0} \\ CB + CAB & CB & \ddots & \vdots \\ CB + CAB + CA^2B & CB + CAB & \ddots & \underline{0} \\ \vdots & & \ddots & CB \\ & \ddots & & CB + CAB \\ \vdots & & \ddots & \vdots \\ C\sum_{i=0}^{H_p-1} A^i B & \cdots & \cdots & C\sum_{i=0}^{H_p-H_u} A^i B \end{bmatrix} \tag{5.6–20}$$

As mentioned before, the system constraints given in (5.6–6) should also be translated into linear inequalities in terms of $\Delta\hat{\boldsymbol{u}}(k+i \mid k)$. It should be noted that the system constraints should hold for the entire prediction and control horizon. Let us first find the control input constrains, i.e., $\boldsymbol{u}_{min} \le \hat{\boldsymbol{u}}(k \mid k) \le \boldsymbol{u}_{max}$ and considering the fact that $\hat{\boldsymbol{u}}(k \mid k) = \Delta\hat{\boldsymbol{u}}(k \mid k) + \boldsymbol{u}(k-1)$, Thus we have,

$$\boldsymbol{u}_{min} \le \hat{\boldsymbol{u}}(k\,|\,k) \le \boldsymbol{u}_{max} \quad \Rightarrow \quad \boldsymbol{u}_{min} - \boldsymbol{u}(k-1) \le \Delta\hat{\boldsymbol{u}}(k\,|\,k) \le \boldsymbol{u}_{max} - \boldsymbol{u}(k-1),$$
$$\Rightarrow \begin{array}{l} \Delta\hat{\boldsymbol{u}}(k\,|\,k) \le \boldsymbol{u}_{max} - \boldsymbol{u}(k-1) \\ \Delta\hat{\boldsymbol{u}}(k\,|\,k) \ge \boldsymbol{u}_{min} - \boldsymbol{u}(k-1) \end{array}. \tag{5.6–21}$$

This inequality can be written into two separate inequalities if the lower bound of (5.6–21) is inverted, i.e., $-1\times(\boldsymbol{u}_{min} - \boldsymbol{u}(k-1) \le \hat{\boldsymbol{u}}(k\,|\,k))$, as below,

$$\begin{array}{r} \Delta\hat{\boldsymbol{u}}(k\,|\,k) \le \boldsymbol{u}_{max} - \boldsymbol{u}(k-1), \\ -\Delta\hat{\boldsymbol{u}}(k\,|\,k) \le -\boldsymbol{u}_{min} + \boldsymbol{u}(k-1). \end{array} \tag{5.6–22}$$

Similarly, $\hat{\boldsymbol{u}}(k+1\,|\,k) = \Delta\hat{\boldsymbol{u}}(k+1|\,k) + \hat{\boldsymbol{u}}(k\,|\,k) = \Delta\hat{\boldsymbol{u}}(k+1|\,k) + \Delta\hat{\boldsymbol{u}}(k\,|\,k) + \boldsymbol{u}(k-1)$, which results in the following inequalities,

$$\begin{array}{r} \Delta\hat{\boldsymbol{u}}(k+1|k) + \Delta\hat{\boldsymbol{u}}(k\,|\,k) \le \boldsymbol{u}_{max} - \boldsymbol{u}(k-1), \\ -\Delta\hat{\boldsymbol{u}}(k+1|k) - \Delta\hat{\boldsymbol{u}}(k\,|\,k) \le -\boldsymbol{u}_{min} + \boldsymbol{u}(k-1). \end{array} \tag{5.6–23}$$

Repeating this for $\hat{\boldsymbol{u}}(k+i\,|\,k)$ up to $i = H_u - 1$ leads to the following control input constraints written in terms of control input changes,

$$\Omega_1 \Delta U(k) \le U_m + I_1 \boldsymbol{u}(k-1), \tag{5.6–24}$$

where the $6H_u\times 3H_u$ matrix $\Omega_1$, the $6H_u\times 1$ vector $U_m$, and $6H_u\times 3$ matrix $I_1$ are given by

$$\Omega_1 = \begin{bmatrix} I_{3\times 3} & 0_{3\times 3} & \cdots & 0_{3\times 3} \\ I_{3\times 3} & I_{3\times 3} & \ddots & \vdots \\ \vdots & \vdots & \ddots & 0_{3\times 3} \\ I_{3\times 3} & I_{3\times 3} & \cdots & I_{3\times 3} \\ -I_{3\times 3} & 0_{3\times 3} & \cdots & 0_{3\times 3} \\ -I_{3\times 3} & -I_{3\times 3} & \ddots & \vdots \\ \vdots & \vdots & \ddots & 0_{3\times 3} \\ -I_{3\times 3} & -I_{3\times 3} & \cdots & -I_{3\times 3} \end{bmatrix}, \tag{5.6–25}$$

$$U_m = \begin{bmatrix} \boldsymbol{u}_{max} \\ \boldsymbol{u}_{max} \\ \vdots \\ \boldsymbol{u}_{max} \\ -\boldsymbol{u}_{min} \\ -\boldsymbol{u}_{min} \\ \vdots \\ -\boldsymbol{u}_{min} \end{bmatrix} \qquad I_1 = \begin{bmatrix} -I_{3\times 3} \\ -I_{3\times 3} \\ \vdots \\ -I_{3\times 3} \\ I_{3\times 3} \\ I_{3\times 3} \\ \vdots \\ I_{3\times 3} \end{bmatrix} \tag{5.6–26}$$

with $I_{3\times3}$ as the identity matrix. The output constraints given in (5.6–6) can also be separated into two inequalities in the same way as in (5.6–21) and extended over the prediction horizon, and then written in matrix form with the $6H_p \times 3H_p$ matrix $\Omega_2$ and the $6H_p \times 1$ vector $Y_m$ as below,

$$\Omega_2 Y(k) \le Y_m, \tag{5.6–27}$$

$$\Omega_2 = \begin{bmatrix} I_{3\times3} & 0_{3\times3} & \cdots & 0_{3\times3} \\ 0_{3\times3} & I_{3\times3} & \ddots & \vdots \\ \vdots & \vdots & \ddots & 0_{3\times3} \\ 0_{3\times3} & 0_{3\times3} & \cdots & I_{3\times3} \\ -I_{3\times3} & 0_{3\times3} & \cdots & 0_{3\times3} \\ 0_{3\times3} & -I_{3\times3} & \ddots & \vdots \\ \vdots & \vdots & \ddots & 0_{3\times3} \\ 0_{3\times3} & 0_{3\times3} & \cdots & -I_{3\times3} \end{bmatrix} \qquad Y_m = \begin{bmatrix} \boldsymbol{y}_{max} \\ \boldsymbol{y}_{max} \\ \vdots \\ \boldsymbol{y}_{max} \\ -\boldsymbol{y}_{min} \\ -\boldsymbol{y}_{min} \\ \vdots \\ -\boldsymbol{y}_{min} \end{bmatrix} \tag{5.6–28}$$

Now, by using (5.6–16) which relates predicted outputs $Y(k)$ to the future control input changes $\Delta U(k)$, both the inequalities obtained in (5.6–24) and (5.6–27) can be combined in one set of linear constraints on control input changes as follows,

$$\Omega \Delta U(k) \le \omega, \tag{5.6–29}$$

where

$$\Omega = \begin{bmatrix} \Omega_1 \\ \Omega_2 \Theta \end{bmatrix}, \qquad \omega = \begin{bmatrix} U_m + I_1 \boldsymbol{u}(k-1) \\ Y_m - \Omega_2 \left( \Psi \boldsymbol{x}(k) + \Gamma \boldsymbol{u}(k-1) + \Lambda F_d(k) \right) \end{bmatrix} \tag{5.6–30}$$

Having all the constraints written in terms of control input changes, the optimal control input voltages ($\boldsymbol{u}(k)_{opt}$) is obtained based on the receding horizon strategy in the sense of MPC for overhead crane [114]. This is conducted by discrete integration of the first element of optimal control input changes ($\Delta\boldsymbol{u}(k)_{opt}$), and then applying it to the plant as the control input at time step $k$ as follows,

$$\Delta U(k)_{opt} = \arg \min_{\substack{\Delta U(k) \\ W\Delta U(k) \le w}} \left\| Y(k) - Y_r(k) \right\|_Q^2 + \left\| \Delta U(k) \right\|_R^2, \qquad (5.6\text{–}31)$$
$$\Delta U(k)_{opt} = [\Delta\boldsymbol{u}(k)_{opt}\ \Delta\boldsymbol{u}(k+1)_{opt} \ldots \Delta\boldsymbol{u}(k+H_u-1)_{opt}]^T,$$

$$\begin{aligned} \boldsymbol{u}(k)_{opt} &= \Delta\boldsymbol{u}(k)_{opt} + \boldsymbol{u}(k-1), \\ &= [v_{ax}(k)_{opt} \quad v_{ay}(k)_{opt} \quad v_{al}(k)_{opt}]^T. \end{aligned} \qquad (5.6\text{–}32)$$

### 5.6.1 State Observer Design

Since only the position of the trolley in $XY$ plane and the hoisting rope length are available as outputs, i.e., $\boldsymbol{y} = [x \ \ y \ \ l]^T$, to make the predictions, the current values of the state vector $\boldsymbol{x}(k)$ is needed. As mentioned earlier, a state observer can be designed to estimate the system state at each sampling time as a replacement for $\boldsymbol{x}(k)$ based on the *separation principle* to be used in MPC for predictions and optimization, as long as the dynamics of the observer are stable and also faster than the system dynamics to avoid undesirable delays in the control system. Therefore, to estimate state variables $\hat{\boldsymbol{x}}(k)$, the following dynamic observer is considered [89], [107],

$$\begin{aligned} \hat{\boldsymbol{x}}(k|k) &= \hat{\boldsymbol{x}}(k|k-1) + L'(\boldsymbol{y}(k) - C\hat{\boldsymbol{x}}(k|k-1)), \\ \hat{\boldsymbol{x}}(k+1|k) &= A\hat{\boldsymbol{x}}(k|k) + B\boldsymbol{u}(k|k-1) + W_d \hat{\boldsymbol{f}}_d(k|k), \end{aligned} \qquad (5.6\text{–}33)$$

which can be simplified by eliminating $\hat{\boldsymbol{x}}(k \mid k)$ as below,

$$\hat{\boldsymbol{x}}(k+1|k) = (A - LC)\hat{\boldsymbol{x}}(k|k-1) + B\boldsymbol{u}(k|k-1) + W_d \hat{\boldsymbol{f}}_d(k|k) + L\boldsymbol{y}(k), \qquad (5.6\text{–}34)$$

where the 6×3 matrix $L = AL' = BlockDiag\{L_x, L_y, L_l\}$ is the observer gain with $L_i = [l_{1i}\ l_{2i}]^T$ for $i = x, y, l$. If the state estimation error is defined as $\hat{\boldsymbol{e}}(k) = \boldsymbol{x}(k) - \hat{\boldsymbol{x}}(k \mid k-1)$, then using system model in (5.6–1) we have

$$\hat{\boldsymbol{e}}(k+1) = (A - LC)\hat{\boldsymbol{e}}(k) + W_d(\boldsymbol{f}_d(k) - \hat{\boldsymbol{f}}_d(k)), \tag{5.6–35}$$

which shows that if the pair ($A$, $C$) is observable, then there exist an observer gain $L$ such that the eigenvalues of the observer, i.e., eig($A - LC$) are placed inside unit circle. Furthermore, the state estimation error will be uniformly bounded if the amount of uncertainty is known and finite, i.e., $\|\hat{\boldsymbol{e}}(k)\| \le c_1 < \infty$ if $\|\boldsymbol{f}_d(k) - \hat{\boldsymbol{f}}_d(k)\| \le c_2 < \infty$ for small positive constants $c_1$ and $c_2$ [107]. It should be noted that using computed torque control guarantees that the model uncertainties is small enough to have both bounded state estimation error and also feasible optimization problem.

Finally, the overall control system for overhead crane using MPC as its discrete-time controller is illustrated in Fig. 5.6–1. As can be seen, the proposed control system is compatible with the general control system structure expressed in Section 5.2.

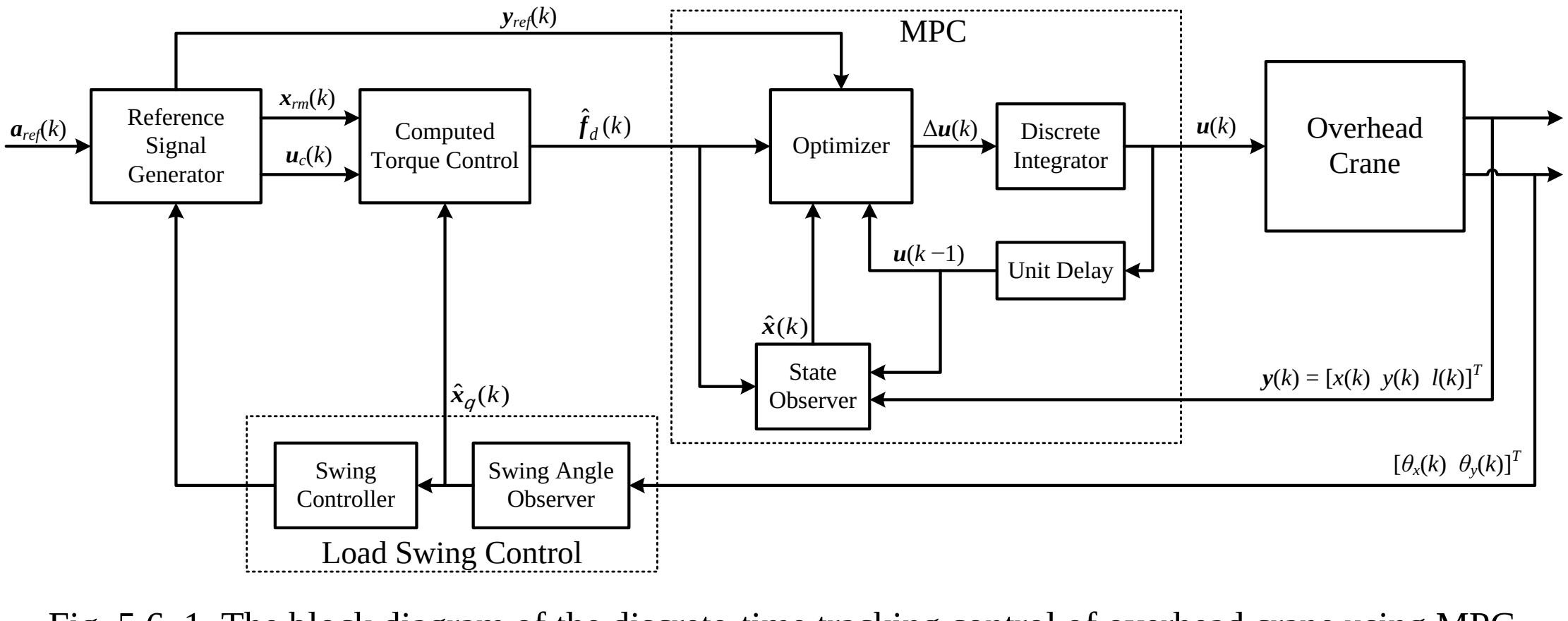

Fig. 5.6–1. The block diagram of the discrete-time tracking control of overhead crane using MPC.

## 5.7 State feedback Control

Although state feedback control is commonly used for regulation problems with the aim to drive the states of the system to the equilibrium point, it can be formulated such that it provides the servo control properties, in which the objective is to make the states and the outputs of the system respond to reference signals in a specified way, as well as regulation. This can be solved by using the two-degree-of-freedom (2DOF) control structure which contains a feedback part and a feedforward part [107]. In the case of controlling overhead crane as we established in this thesis, the control configurations described in Section 5.2 allows us to apply state feedback approach for the purpose of tracking the reference traveling, traversing and hoisting positions. Similar to MPC design in the previous section, the discrete-time controller can be designed using the state feedback approach and then it can be completed by the feedforward control to deliver the task of following the reference trajectories as well as rejecting the effect of load disturbances caused by the overhead crane nonlinearities.

Therefore, discrete-time control law for generating control input voltages in overhead crane control system using the state feedback approach is given as follows [109], [110], [115],

$$\begin{aligned} \boldsymbol{u}(k) &= \boldsymbol{u}_{fb}(k) + \boldsymbol{u}_{ff}(k) \\ &= K(\boldsymbol{x}_{rm}(k) - \hat{\boldsymbol{x}}(k)) + \boldsymbol{u}_{ff}(k), \end{aligned} \tag{5.7–1}$$

where $\boldsymbol{u}_{fb}(k) = K(\boldsymbol{x}_{rm}(k) - \hat{\boldsymbol{x}}(k))$ is the feedback signal aiming to reduce the error between the reference state trajectories generated by the reference model and the system states; the 3×6 matrix $K = BlockDiag\{K_x, K_y, K_l\}$ is the feedback gain with $K_i = [k_{1i}\ \ k_{2i}]$ for $i = x, y, l.$, and $\boldsymbol{u}_{ff}(k) = [u_{ffx}(k)\ u_{ffx}(k)\ u_{ffl}(k)]^T$ is the feedforward signal that gives the desired output when applied to the open-loop system. Similar to MPC, it is assumed that state measurements are not available and hence, the estimation of system states $\hat{\boldsymbol{x}}(k)$ can be used in the control law by using the similar state observer designed in Section 5.6.1.

### 5.7.1 Feedforward Signal Generation

It is interesting to see that how load disturbances can be compensated using feedforward action in the context of state feedback. Based on the definition of the feedforward signal, $\boldsymbol{u}_{ff}$ would ideally produce desired output response if there was no feedback control action. In other word, $\boldsymbol{u}_{ff}$ is supposed to be the ideal motor voltages that can generate the driving forces such that the overhead crane can move identically to what is designed in reference trajectories. Recall that in computed torque control, the reference trajectories are used to calculate the ideal forces for moving the crane based on those trajectory profiles using inverse dynamic technique as explained in Section 5.2.2. The same concept can be used here to obtain the ideal motor voltages by using discrete-time dynamic model derived in (4.6–11)–(4.6–13) for overhead crane that relates motor voltages to the overhead crane velocities and the reference velocity trajectories as follows,

$$b_{1i} u_{ffi}(k) = v_{iref}(k+1) - a_{1i} v_{iref}(k) + b_{d1i} \hat{f}_{di}(k), \quad for \quad i = x, y, l, \tag{5.7–2}$$

where $\hat{f}_{di}$ is the computed load disturbances generated by the computed torque control (and later on by disturbance observer) corresponds to each direction of motion. In addition, the reference velocities are generated by discrete integration of the reference accelerations using the reference model in reference signal generator block as described in Section 5.5. That means we have $v_{iref}(k + 1) = v_{iref}(k) + T_s u_{ci}(k)$ which helps us to generate feedforward signal by using the command signal $\boldsymbol{u}_c$ and the reference model as follows,

$$\boldsymbol{u}_{ff}(k) = \Phi_{ff}\, \boldsymbol{x}_{rm}(k) + \Gamma_{ff}\, \boldsymbol{u}_c(k) + \Lambda_{ff}\, \hat{\boldsymbol{f}}_d(k), \tag{5.7–3}$$

where $\Phi_{ff} = BlockDiag\{\Phi_{ffx}, \Phi_{ffy}, \Phi_{ffl}\}$; $\Gamma_{ff} = Diag\{\gamma_{ffx}, \gamma_{ffy}, \gamma_{ffl}\}$; $\Lambda_{ff} = Diag\{\lambda_{ffx}, \lambda_{ffy}, \lambda_{ffl}\}$, with inner matrices given as below,

$$\Phi_{ffi} = \begin{bmatrix} 0 & \dfrac{1 - a_{1i}}{b_{1i}} \end{bmatrix}, \ \gamma_{ffi} = \frac{T_s}{b_{1i}}, \ \lambda_{ffi} = \frac{b_{d1i}}{b_{1i}}, \quad for \quad i = x, y, l. \tag{5.7–4}$$

It can be seen from (5.7–3) and (5.7–4) that the overhead crane nonlinear effects can be compensated through feedforward signal at each sampling time due to having computed disturbances $\hat{\boldsymbol{f}}_d(k)$ as part of $\boldsymbol{u}_{ff}(k)$.

## 5.7.2 Tracking Error Dynamics

In order to show how the discrete-time servo control law given in (5.7–1) with the feedforward signal obtained in (5.7–3) can provide both stability in trajectory tracking and disturbance rejection, we need to construct the tracking error equation. Let us define $\boldsymbol{e}(k) = \boldsymbol{x}_{rm}(k) - \boldsymbol{x}(k)$ as the tracking error. Thus, by subtracting the reference model equation in (5.5–1) from the overhead crane model in (5.6–1), the tracking error equation is given as follows,

$$\boldsymbol{e}(k+1) = A_m \boldsymbol{x}_{rm}(k) - A\boldsymbol{x}(k) + B_m \boldsymbol{u}_c(k) - B\boldsymbol{u}(k) - W_d \boldsymbol{f}_d(k). \tag{5.7–5}$$

Now, by substituting servo control input $\boldsymbol{u}(k)$ and feedforward signal $\boldsymbol{u}_{ff}(k)$ from (5.7–1) and (5.7–3), respectively, into (5.7–5), the tracking error equation is obtained as the following,

$$\begin{aligned} \boldsymbol{e}(k+1) = A\boldsymbol{x}(k) - (A_m - B\Phi_{ff})\boldsymbol{x}_{rm}(k) - BK\boldsymbol{e}(k) + BK\hat{\boldsymbol{e}}(k) \\ - (B_m - B\Gamma_{ff})\boldsymbol{u}_c(k) + B\Lambda_{ff}\hat{\boldsymbol{f}}_d(k) + W_d \boldsymbol{f}_d(k), \end{aligned} \tag{5.7–6}$$

where $\hat{\boldsymbol{e}}(k) = \boldsymbol{x}(k) - \hat{\boldsymbol{x}}(k)$ is the state estimation error. Recall from inner matrices for system model in (5.6–4), reference model in (5.5–2), and feedforward signal in (5.7–4) and the fact all the matrices all in block-diagonal form. The tracking error equation in (5.7–6) can be simplified since

$$\begin{aligned} A_{mi} - B_i\Phi_{ffi} &= \begin{bmatrix} 1 & T_s \\ 0 & 1 \end{bmatrix} - \begin{bmatrix} 0 \\ b_{1i} \end{bmatrix}\begin{bmatrix} 0 & \dfrac{1-a_{1i}}{b_{1i}} \end{bmatrix} \\ &= \begin{bmatrix} 1 & T_s \\ 0 & a_{1i} \end{bmatrix} = A_i, \\ B_{mi} - B_i\gamma_{ffi} &= \begin{bmatrix} 0 \\ T_s \end{bmatrix} - \begin{bmatrix} 0 \\ b_{1i} \end{bmatrix}\frac{T_s}{b_{1i}} = \begin{bmatrix} 0 \\ 0 \end{bmatrix}, \\ B_i\lambda_{ffi} &= \begin{bmatrix} 0 \\ b_{1i} \end{bmatrix}\frac{b_{d1i}}{b_{1i}} = \begin{bmatrix} 0 \\ b_{d1i} \end{bmatrix} \\ &= -W_{di}, \qquad \qquad for \quad i = x, y, l, \end{aligned} \tag{5.7–7}$$

and therefore $A_m - B\Phi_{ff} = A$, $B_m = B\Gamma_{ff}$, and $B\Lambda_{ff} = -W_d$, which leads to the following tracking error equation,

$$\boldsymbol{e}(k+1) = (A - BK)\boldsymbol{e}(k) + BK\hat{\boldsymbol{e}}(k) + W_d(\boldsymbol{f}_d(k) - \hat{\boldsymbol{f}}_d(k)). \tag{5.7–8}$$

As can be seen, the tracking error depends on the state estimation error and the amount of uncertainties in the system. Thus, we can write the augmented error equation by combining (5.7–8) with the state estimation error given in (5.6–35) to obtain the tracking error dynamics for the overhead crane control using state feedback as follows [109],

$$\begin{bmatrix} \boldsymbol{e}(k+1) \\ \hat{\boldsymbol{e}}(k+1) \end{bmatrix} = \begin{bmatrix} A - BK & BK \\ \underline{0} & A - LC \end{bmatrix} \begin{bmatrix} \boldsymbol{e}(k) \\ \hat{\boldsymbol{e}}(k) \end{bmatrix} + \begin{bmatrix} W_d \\ W_d \end{bmatrix} (\boldsymbol{f}_d(k) - \hat{\boldsymbol{f}}_d(k)), \tag{5.7–9}$$

where $\underline{0}$ is a $6 \times 6$ zero matrix. It can be seen from (5.7–9) that the model uncertainty $(\boldsymbol{f}_d - \hat{\boldsymbol{f}}_d)$ acts as the input to the tracking error dynamics. By using computed torque control to calculate $\hat{\boldsymbol{f}}_d$, it is guaranteed that $\|\boldsymbol{f}_d(k) - \hat{\boldsymbol{f}}_d(k)\| \le c_1 < \infty$ for a small positive constant $c_2$. Thus, choosing feedback gain $K$ and observer gain $L$ such that $(A - BK)$ and $(A - LC)$ are stable, tracking error and state estimation error are proven to be uniformly bounded [107], i.e., $\|\boldsymbol{e}(k)\| \le \varepsilon_1 < \infty$ and $\|\hat{\boldsymbol{e}}(k)\| \le \varepsilon_2 < \infty$ for small positive constants $\varepsilon_1$, and $\varepsilon_2$.

It should be noted that feedback signal $\boldsymbol{u}_{fb}(k)$ in discrete-time state feedback control law in (5.7–1) is defined by the error between the reference state trajectories and the estimate of system states. Thus, by defining $\boldsymbol{e}_c(k) = \boldsymbol{x}_{rm}(k) - \hat{\boldsymbol{x}}(k)$ as the controller error, the dynamic equation of controller error can be found similar to tracking error dynamics as follows,

$$\boldsymbol{e}_c(k+1) = A_m\boldsymbol{x}_{rm}(k) - A\boldsymbol{x}(k) + B_m\boldsymbol{u}_c(k) - B\boldsymbol{u}(k) - W_d\hat{\boldsymbol{f}}_d(k) - L(\boldsymbol{y} - C\hat{\boldsymbol{x}}), \tag{5.7–10}$$

which can be simplified by substituting servo control input $\boldsymbol{u}(k)$ and feedforward signal $\boldsymbol{u}_{ff}(k)$ from (5.7–1) and (5.7–3), respectively, adding and subtracting $C\boldsymbol{x}_{rm}(k)$ in the last term of (5.7–10), and using (5.7–7) as below,

$$\boldsymbol{e}_c(k+1) = (A - BK - LC)\boldsymbol{e}_c(k) + LC\boldsymbol{e}(k). \tag{5.7–11}$$

It can be seen from the obtained controller error dynamics in (5.7–11) that in addition to stability condition on ($A$ – $BK$) and ($A$ – $LC$) for uniformly boundedness of tracking error $\boldsymbol{e}(k)$, feedback gain matrix $K$ and observer gain matrix $L$ should also be chosen such that the matrix ($A$ – $BK$ – $LC$) have all its eigenvalues inside unit circle. This will guarantee to have controller error uniformly bounded, i.e., $\|\boldsymbol{e}_c(k)\| \leq \varepsilon_3 < \infty$ for a small positive constant $\varepsilon_3$.

These three stability conditions imply the two-degree-of-freedom (2DOF) nature of the proposed discrete-time servo control system for overhead crane. Moreover, the results obtained above indicate that the state feedback approach can deliver the control objectives for high-performance control of the overhead crane in conjunction with load swing control and reference signal generator. The overall control system block diagram for the overhead crane using state feedback approach in its discrete-time controller is illustrated in Fig. 5.7–1.

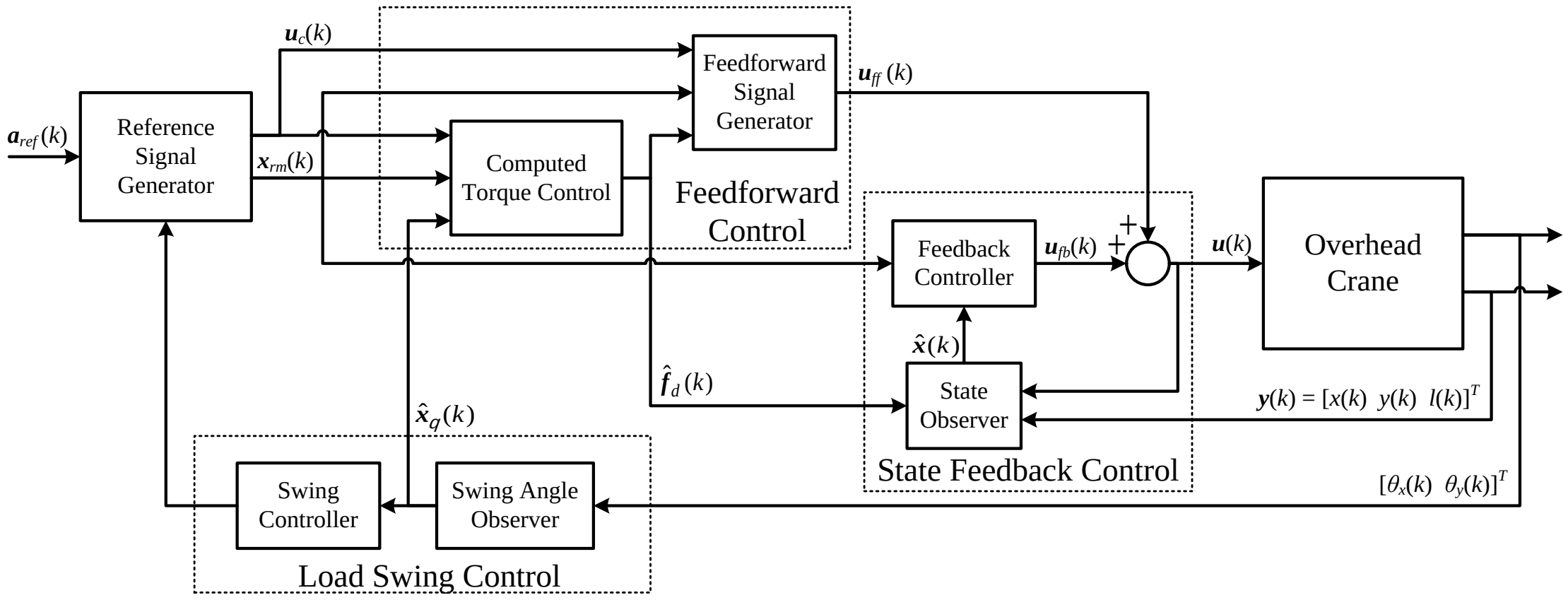

Fig. 5.7–1. The block diagram of the discrete-time tracking control of overhead crane using state feedback.

## 5.8 Disturbance Observer

In both MPC and state feedback control for the overhead crane, computed torque control is applied to generate the computed disturbances so that feedforward control can compensate the effect of load disturbances in overhead crane. The only downside of using computed torque control is that the nonlinear dynamics of the overhead crane should be known in advance. This causes disturbance rejection to depend on the parameters of the overhead crane that are changing during the operation. Except the mass of the overhead crane load *m*, the rest of the parameters remain unchanged during the operation and they are included as part of the discreet-time model proposed for the overhead crane with high estimated accuracy as shown in Section 4.5. As we explained in independent joint modeling approach, all nonlinearities are considered as disturbances and the overhead crane load *m* is part of the computed disturbances $\hat{\boldsymbol{f}}_d(k)$ (recall (4.4–7)–(4.4–9)). The problem here is that overhead crane load can have different values in each operation, which is quite normal, and to have better disturbance compensation, the value of *m* should be known in advance to compute $\hat{\boldsymbol{f}}_d(k)$. In this section, we introduce a method to estimate the amount of computed disturbance without the need to have the knowledge of the nonlinear dynamics of the overhead carne and the value of *m*. The equation to obtain the computed disturbance $\hat{\boldsymbol{f}}_d(k)$ is given as follows [115],

$$\hat{\boldsymbol{f}}_d(k+1) = \hat{\boldsymbol{f}}_d(k) + L_w(\boldsymbol{y}(k) - C\hat{\boldsymbol{x}}(k)), \tag{5.8–1}$$

where $L_w = Diag\{l_{wx}, l_{wy}, l_{wl}\}$ is disturbance observer gain; $\boldsymbol{y}(k)$ is the system output, $C$ is the output matrix, and $\hat{\boldsymbol{x}}(k)$ is the estimate of system states.

To show how the proposed disturbance observer is capable of estimating the real disturbance $\boldsymbol{f}_d$, by only using state estimation error, let us write (5.8–1) in *Z*-domain as below,

$$\begin{aligned}\hat{\boldsymbol{f}}_d(z) &= (zI - I)^{-1} L_w C\hat{\boldsymbol{e}}(z),\\ &= H_w(z) C\hat{\boldsymbol{e}}(z),\end{aligned} \tag{5.8–2}$$

where $\hat{\boldsymbol{e}}(z)$ is the state estimation error and $H_w(z) = (zI - I)^{-1}L_w$ is a 3×3 diagonal transfer function matrix, with $I_{3\times 3}$ as the identity matrix, given by

$$H_w(z) = \begin{bmatrix} \frac{l_{wx}}{z-1} & 0 & 0 \\ 0 & \frac{l_{wy}}{z-1} & 0 \\ 0 & 0 & \frac{l_{wl}}{z-1} \end{bmatrix} \quad (5.8\text{–}3)$$

Recall from (5.6–35) that state estimation error is obtained as follows,

$$\hat{\boldsymbol{e}}(k+1) = (A - LC)\hat{\boldsymbol{e}}(k) + W_d(\boldsymbol{f}_d(k) - \hat{\boldsymbol{f}}_d(k)), \quad (5.8\text{–}4)$$

and its transformation into *Z*-domain results in the following,

$$\hat{\boldsymbol{e}}(z) = (zI - (A - LC))^{-1} + W_d(\boldsymbol{f}_d(z) - \hat{\boldsymbol{f}}_d(z)), \quad (5.8\text{–}5)$$

with $I_{6\times6}$ as the identity matrix. Thus, by substituting (5.8–5) into (5.8–2), $\hat{\boldsymbol{f}}_d(z)$ is obtained as below,

$$\hat{\boldsymbol{f}}_d(z) = H_w(z)C(zI - (A - LC))^{-1}W_d(\boldsymbol{f}_d(z) - \hat{\boldsymbol{f}}_d(z)). \quad (5.8\text{–}6)$$

It is interesting to see that (5.8–6) is in fact a decoupled unit feedback close-loop system with actual load disturbances $\boldsymbol{f}_d$ as the input and the computed ones $\hat{\boldsymbol{f}}_d$ as the output knowing that all the matrices are in block-diagonal form. Therefore, by defining $G_w(z) = C(zI - (A - LC))^{-1}W_d$, the close-loop transfer function matrix for estimating load disturbance is given as follows with its block diagram demonstrated in Fig. 5.8–1.

$$\hat{\boldsymbol{f}}_d(z) = ((I + H_w(z)G_w(z))^{-1}H_w(z)G_w(z))\boldsymbol{f}_d(z). \quad (5.8\text{–}7)$$

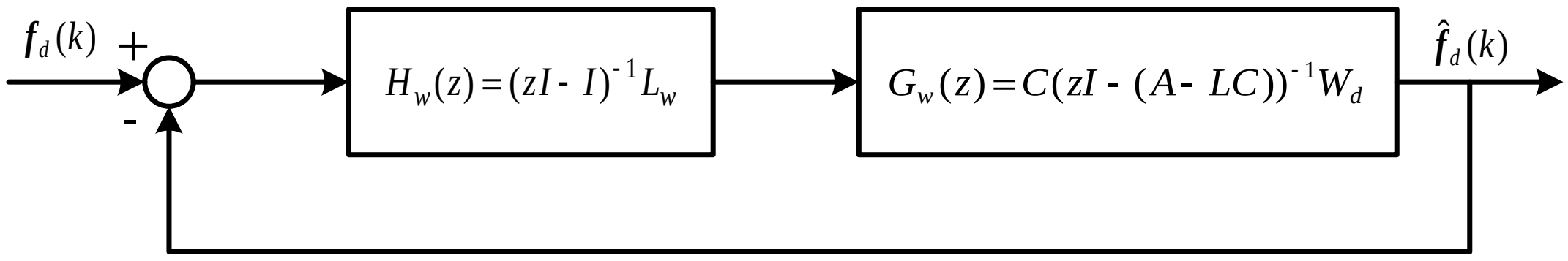

Fig. 5.8–1. The block diagram of the disturbance observer.

**Remark:** An important outcome of using the disturbance observer is that the control system will perform robustly against variations in the overhead crane load mass *m*. This is possible since both the discrete-time model used to design the controller and the estimation of load disturbances are independent from load mass *m* as any changes in *m* would be reflected in $\boldsymbol{f}_d$, and subsequently the disturbance observer would follow those changes in its output $\hat{\boldsymbol{f}}_d$ if the closed-loop system is stable. Moreover, the proposed disturbance observer is a 2DOF system since the stability of the closed-loop system depends on both the disturbance observer gain $L_w$ and the state observer gain $L$. Thus, we have to make sure that after determining the proper values for state observer $L$, disturbance observer gain $L_w$ is chosen such that the close-loop eigenvalues are located inside the unit circle.

It is interesting to see that due to the decupled nature of the disturbance observer structure, the load disturbances can be identified for each direction of motion separately, and hence, the disturbance observer gain $L_w$. Recall from the discrete-time state-space model (5.6–1)–(5.6–4) and the fact that all the matrices are in block-diagonal form. This makes the open-loop transfer function matrix $H_w(z)G_w(z)$ have block-diagonal form as well, i.e., $H_w(z)G_w(z) = BlockDiag\{H_{wx}(z)G_{wx}(z), H_{wy}(z)G_{wy}(z), H_{wl}(z)G_{wl}(z)\}$. Therefore, we can find the open-loop disturbance observer transfer function for each direction of motion, and then determine how to choose $L_w$ to stabilize the disturbance observer. Let us first work out with each decoupled open-loop transfer function as follows,

$$\begin{aligned} &H_{wi}(z)G_{wi}(z) = \frac{l_{wx}}{z-1} C_i (zI - (A_i - L_i C_i))^{-1} W_{di}, \\ &A_i = \begin{bmatrix} 1 & T_s \\ 0 & a_{1i} \end{bmatrix}, \; L_i = \begin{bmatrix} l_{1i} \\ l_{2i} \end{bmatrix}, \; W_{di} = \begin{bmatrix} 0 \\ -b_{d1i} \end{bmatrix}, \; C_i = \begin{bmatrix} 1 & 0 \end{bmatrix}, \quad for \quad i = x, y, l. \end{aligned} \tag{5.8–8}$$

By substituting $A_i$, $L_i$, $W_{di}$, and $C_i$ into the equation for $H_{wi}(z)G_{wi}(z)$ in (5.8–8) we have

$$\begin{aligned} H_{wi}(z)G_{wi}(z) &= \frac{l_{wi}}{z-1} \begin{bmatrix} 1 & 0 \end{bmatrix} \begin{bmatrix} z-1+l_{1i} & -T_s \\ l_{2i} & z-a_{1i} \end{bmatrix}^{-1} \begin{bmatrix} 0 \\ -b_{d1i} \end{bmatrix} \\ &= \frac{-l_{wi} b_{d1i} T_s}{(z-1)(z-1+l_{1i})(z-a_{1i}) + l_{2i} T_s (z-1)}, \quad for \quad i = x, y, l. \end{aligned} \tag{5.8–9}$$

Now, the closed-loop transfer function for each direction of motion in disturbance observer is obtained as below,

$$\frac{\hat{f}_{di}(z)}{f_{di}(z)} = \frac{H_{wi}(z)G_{wi}(z)}{1+H_{wi}(z)G_{wi}(z)}$$

$$= \frac{-l_{wi}b_{d1i}T_s}{z^3+(l_{1i}-2-a_{1i})z^2+(1-l_{1i}(1+a_{1i})+2a_{1i}+l_{2i}T_s)z+a_{1i}(l_{1i}-1)-T_s(l_{2i}+l_{wi}b_{d1i})}, \quad \text{for} \quad i=x,y,l. \qquad (5.8–10)$$

Once the state observer gain is determined, we can use pole placement technique to find each $l_{wi}$ such that the roots of the denominator in (5.8–10) are located inside unit circle. It should be noted that the disturbance observer gain $L_W$ should be negative for all its elements to cancel the negative phase of the transfer function (5.8–9). Otherwise, we end up with positive feedback which destabilizes the disturbance estimation. Therefore, having a stable ($A - LC$) matric and choose $L_W$ properly to make (5.8–10) stable for traveling, traversing and hoisting motions, it is guaranteed that $\|\boldsymbol{f}_d(k) - \hat{\boldsymbol{f}}_d(k)\| \leq c_1 < \infty$ for a small positive constant $c_1$. This also leads to stable state estimation (as mentioned in Section 5.6.1) and load disturbance compensation through feedforward control in both MPC and state feedback control for the overhead crane.

It should be mentioned that some might argue that conventional integrator would have similar effect on reducing disturbances, specifically when having slow-motion trajectory that results in almost constant disturbances. However, this argument may not be thoroughly true as it adds another eigenvalue at $z = 1$ in addition to the one already in the system which increases the chance of instability and hardens the pole-placement approach to assign suitable closed-loop eigenvalues. Not to mention that the proposed disturbance observer could estimate quite accurately the amount of nonlinear disturbances regardless of the speed of the trajectory as already proved in this section.

## 5.9 Practical Results

In this section, the designed discrete-time control systems for overhead crane in this thesis are implemented on a laboratory-sized overhead crane setup. Several tests have been carried out to not only evaluate the performance of the proposed control systems, but also to demonstrate their capability in delivering high-performance anti-swing control in accordance with the control objectives including high precision in load positioning for fast motions as well as suppressing load swings. The control systems are examined under different scenarios in both 2D and 3D overhead crane cases. Two different reference trajectories are designed, one slow and one fast, to compare the performance of control systems in handling high-speed load transportation. These trajectories are designed for multiple repetitions to test repeatability starting from an initial point, going to the destination and coming back to the starting point. In addition, the overhead crane load, which is supposed to be transferred following the reference trajectories, is assumed to have two different masses, one light ($m$ = 0.4 kg) and one heavy ($m$ = 0.8 kg) to investigate how disturbance observer will be able to estimate overall disturbances without knowing the mass of the load.

To further indicate the ability of the discrete-time control system in both load swing damping and robustly tracking the reference trajectories, three different scenarios are considered in running the experiments. In the first scenario (Scenario I), the control system is operated with no load swing control (LSC) and no feedforward control (FFC), (i.e., LSC = Off, FFC = Off, which means $K_q = \underline{0}$, $\hat{\boldsymbol{f}}_d(k) = \underline{0}$ in MPC, and $\boldsymbol{u}_{ff}(k) = \underline{0}$ in state feedback control). Trajectory tracking is then conducted with feedforward compensation in the second scenario (Scenario II) but still with no load swing control (FFC = On, LSC = Off). Finally, both load swing control and feedforward control are active during the third scenario (Scenario III: LSC = On, FFC = On). These experiments will show how each part of the control system contributes in controlling the overhead crane for automatic load transportation. Due to extensive number of experiments, only the results of the 3D overhead carne control are included in this section. For the results and tests on the 2D overhead crane, please see [109], [110], and [115].

### 5.9.1 Experimental Overhead Crane Setup

The laboratory-sized overhead crane setup used in this research is manufactured by INTECO Limited [116], and it is shown in Fig. 5.9–1.This setup is driven by three 24-volt PM DC motors. The measurements are made by five identical position encoders with the resolution of 4069 pulses per rotation which provide the measurements for traveling and traversing positions, hoisting rope length, and swing angles in *X* and *Y* directions. The setup is equipped with RT-DAC/PCI9030 multipurpose digital I/O board connected to a power interface board and installed on a personal computer (Intel® Core2Due 3.00GHz CPU with 3GB RAM). This setup works with the sampling time $T_s$ = 0.01 seconds and all functions of the board are accessible from a Toolbox provided by the manufacturer that operates in MATLAB® software and SIMULINK® environment. The real-time codes of the constructed control systems were created by the Real-Time Workshop toolbox of MATLAB.

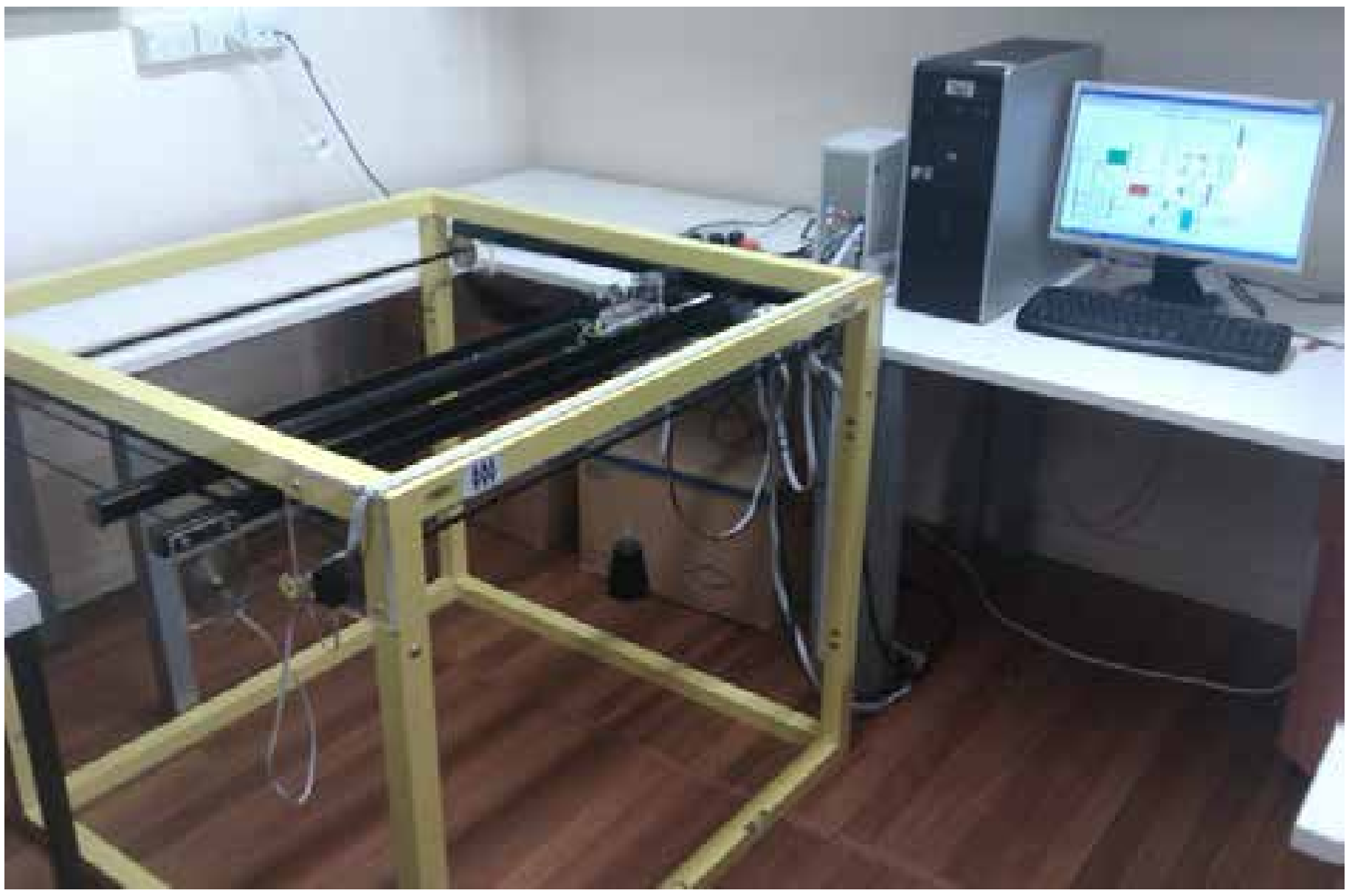

Fig. 5.9–1. The experimental overhead crane setup used in this thesis.

### 5.9.2 Reference Trajectory

As it is mentioned earlier, the reference trajectories are designed with two different speeds in term of transporting the overhead crane load. They designed using the LSPB form for traveling and traversing motions, as illustrated in Fig. 5.9–2(a), and minimum-time trajectory for hoisting motion as shown in Fig. 5.9–2(b) in accordance with trajectory planning explained in Section 5.4. The reference trajectory parameters for both slow and fast trajectories are provided in Table 5.9–1 for traveling and traversing motions, and in Table 5.9–2 for hoisting motion. They include the reference accelerations for accelerating and decelerating zones ($a_{xr}$, $a_{yr}$, $a_{lr}$), normal velocities for constant-velocity zone ($v_{xr}$, $v_{yr}$, $v_{lr}$), the starting and finishing points, i.e., $\boldsymbol{q}_{ref}(t_0) = (x_{r0}, y_{r0}, l_{r0})$ and $\boldsymbol{q}_{ref}(t_f) = (x_{rf}, y_{rf}, l_{rf})$, respectively, and the zone timings ($t_b$, $t_f$). The values of these parameters are determined taking into account the maximum admissible torque, velocity and load capacity of the overhead crane setup and the PM DC motors that are provided by the manufacture in the setup datasheet. The maximum permissible velocity and the maximum allowable acceleration for traveling and traversing are given as $v_{\max}$ = 0.3 m/sec and $a_{\max}$ = 0.2 m/sec$^2$. The repetitions of reference trajectories for traveling, traversing, and hoisting motions are displayed in Fig. 5.9–3(a) and Fig. 5.9–3(b) for slow trajectories and fast trajectories, respectively. In addition, the desired path that the overhead crane load should follow in 3D space is displayed in Fig. 5.9–4 for both slow and fast trajectories. Due to the constraints of the overhead crane setup, in the fast trajectory, the hoisting length is less than the slow trajectory since we wanted to have the same amount of movement for traveling and traversing in both fast and slow trajectories.

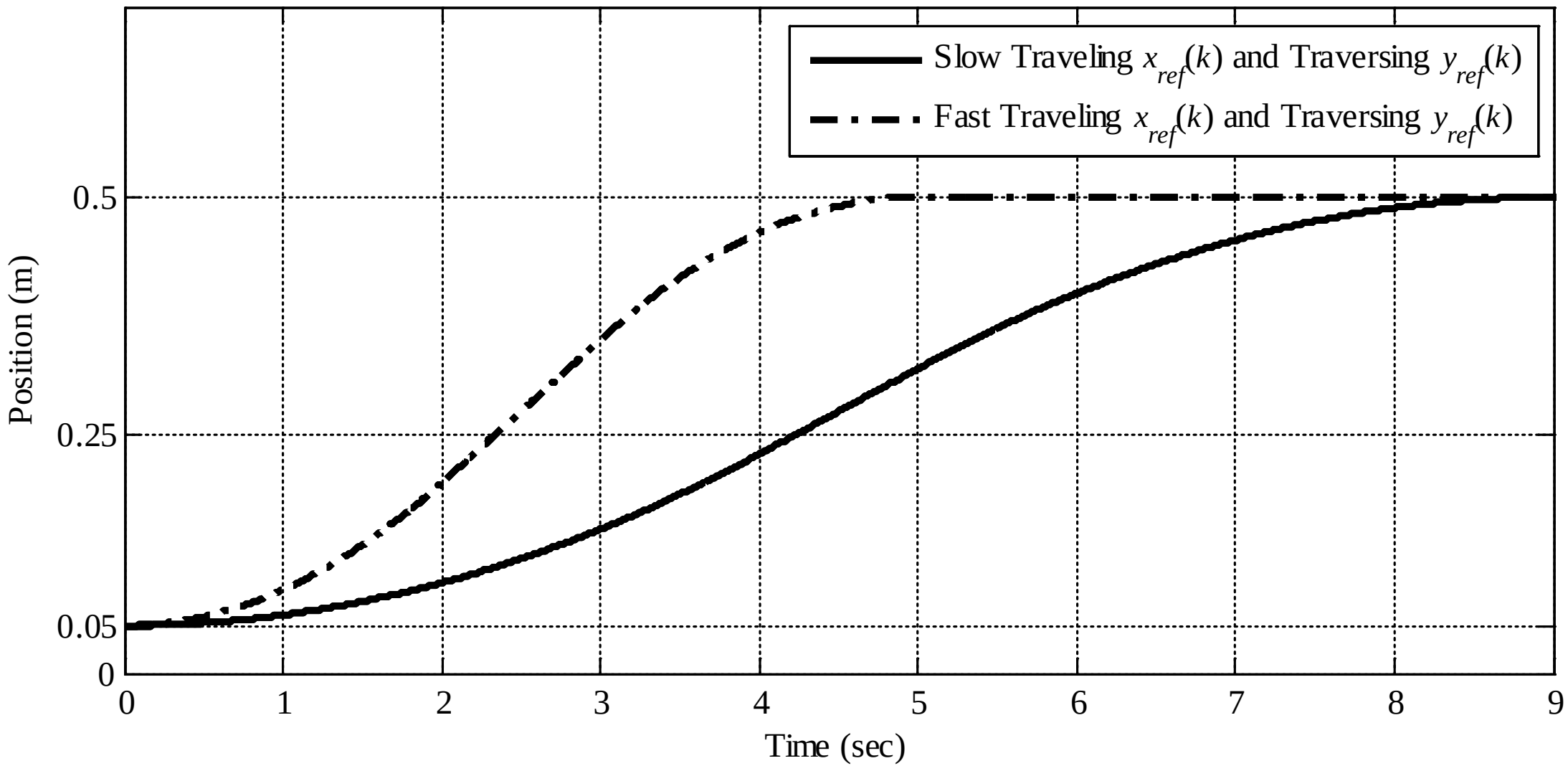

(a)

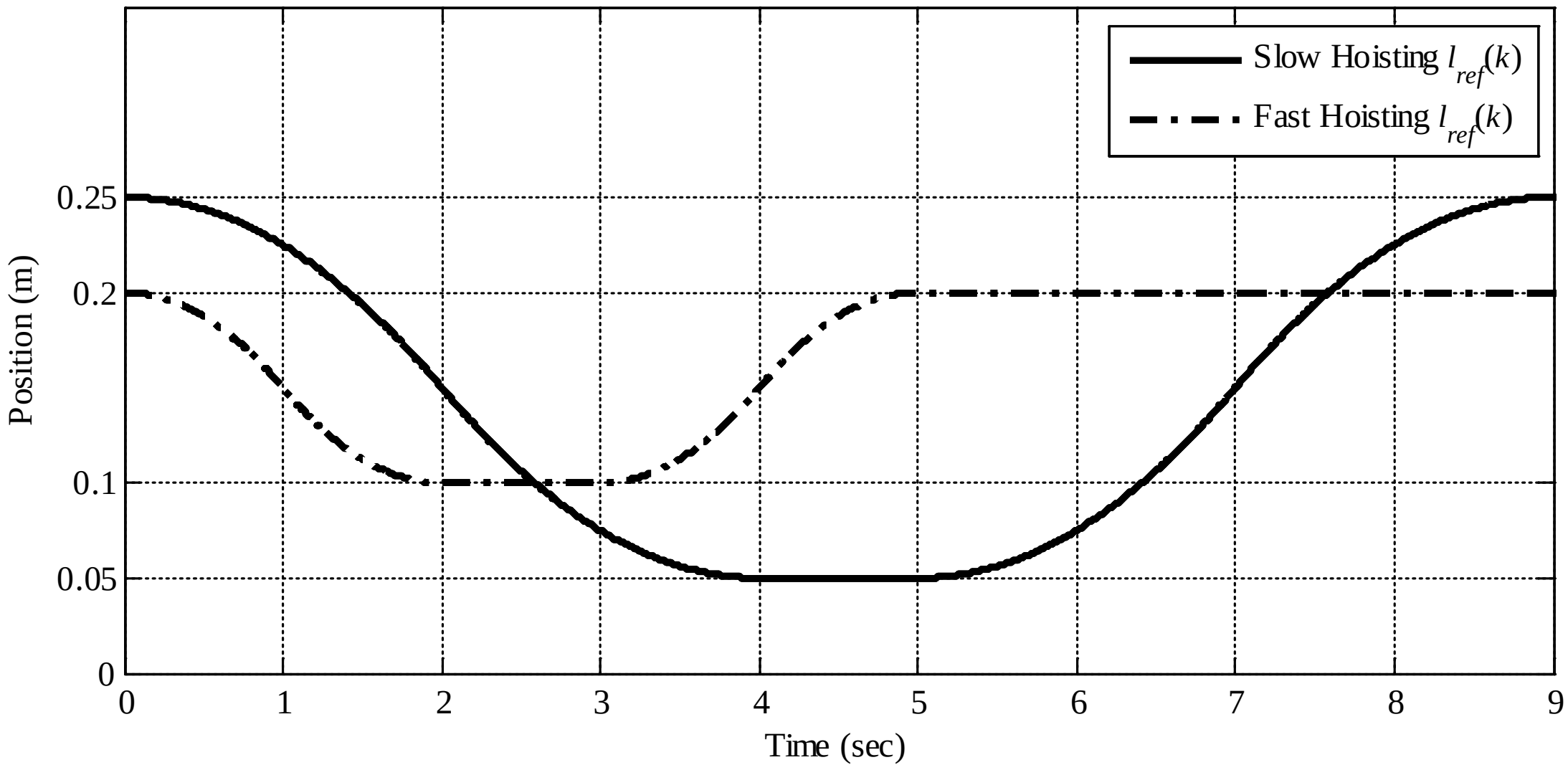

(b)

Fig. 5.9–2. Reference trajectories. (a) Slow and fast trajectories for traveling and traversing motions, (b) Slow and fast trajectories for hoisting motion.

Table 5.9–1. Reference Trajectory Parameters for Traveling and Traversing Motions

| *Parameters* | $(a_{xr}, a_{yr})$ (m/sec$^2$) | $(v_{xr}, v_{yr})$ (m/sec) | $(x_{r0}, y_{r0})$ (m) | $(x_{rf}, y_{rf})$ (m) | $t_b$ (sec) | $t_f$ (sec) |
|---|---|---|---|---|---|---|
| Slow Traveling/Traversing | 22.5e−3 | 9e–2 | 5e–2 | 50e–2 | 4 | 9 |
| Fast Traveling/Traversing | 75e−3 | 15e–2 | 5e–2 | 50e–2 | 2 | 5 |

Table 5.9–2. Reference Trajectory Parameters for Hoisting Down/Up Motion

| *Parameters* | $a_{lr}$ (m/sec$^2$) | $v_{lr}$ (m/sec) | $l_{r0}$ (m) | $l_{rf}$ (m) | $t_f$ (sec) |
|---|---|---|---|---|---|
| Slow Hoisting | 50e–3 | 10e–2 | 25e–2 | 5e–2 | 4 |
| Fast Hoisting | 100e–3 | 10e–2 | 2e–2 | 10e–2 | 2 |

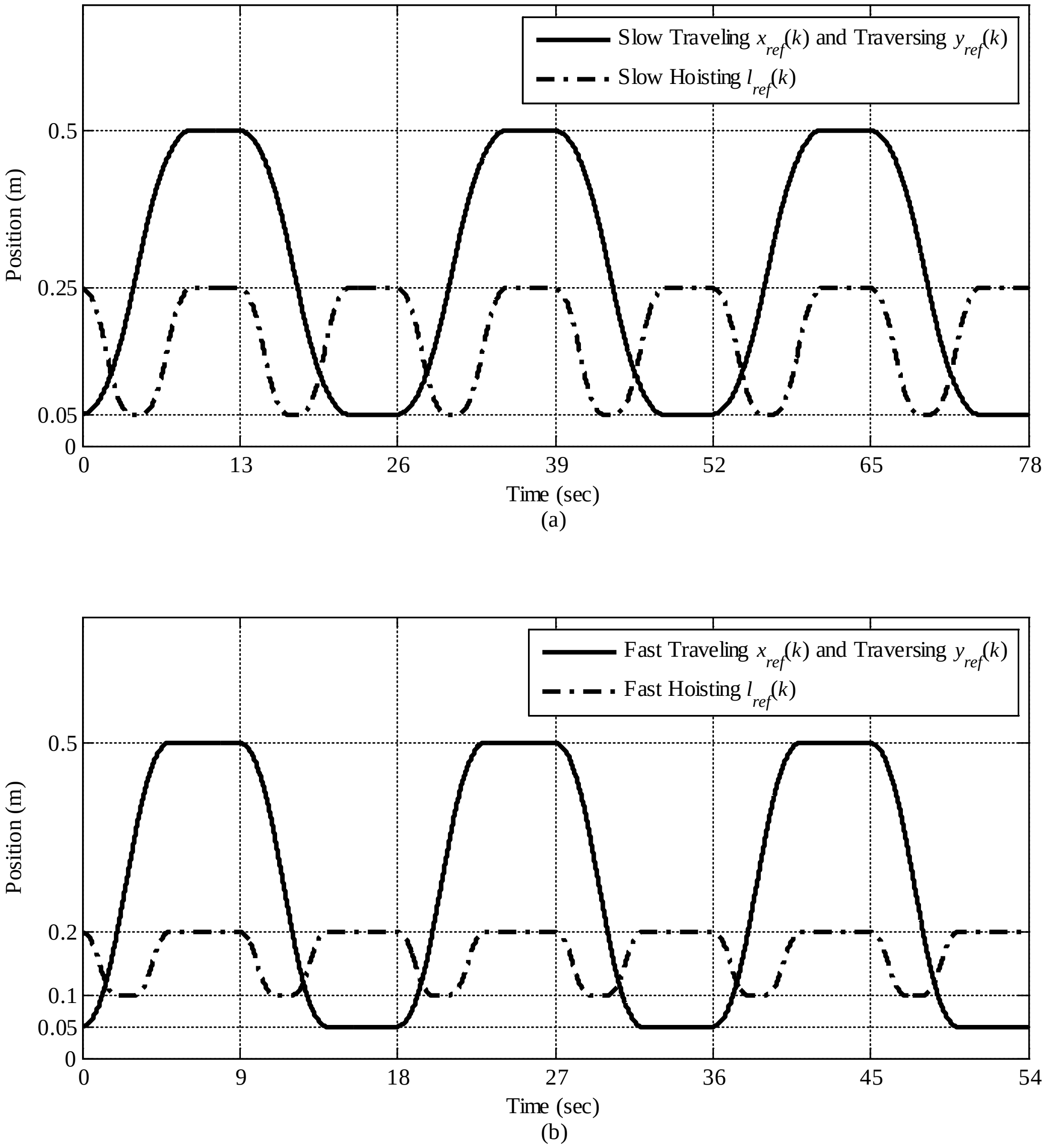

Fig. 5.9–3. Reference trajectories in repetition. (a) Slow trajectory for traveling, traversing, and hoisting motions, (b) Fast trajectory for traveling, traversing, and hoisting motions.

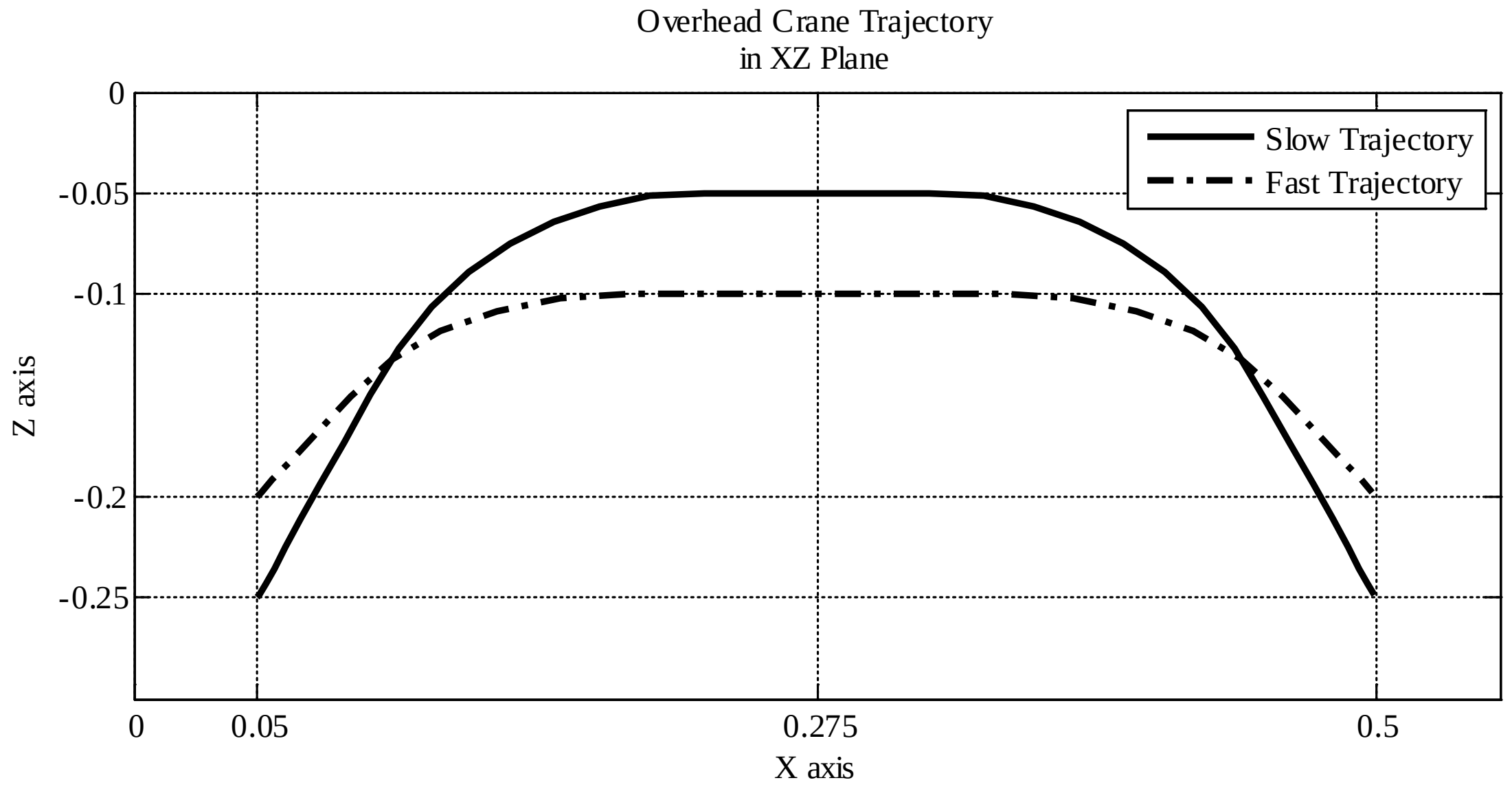

(a)

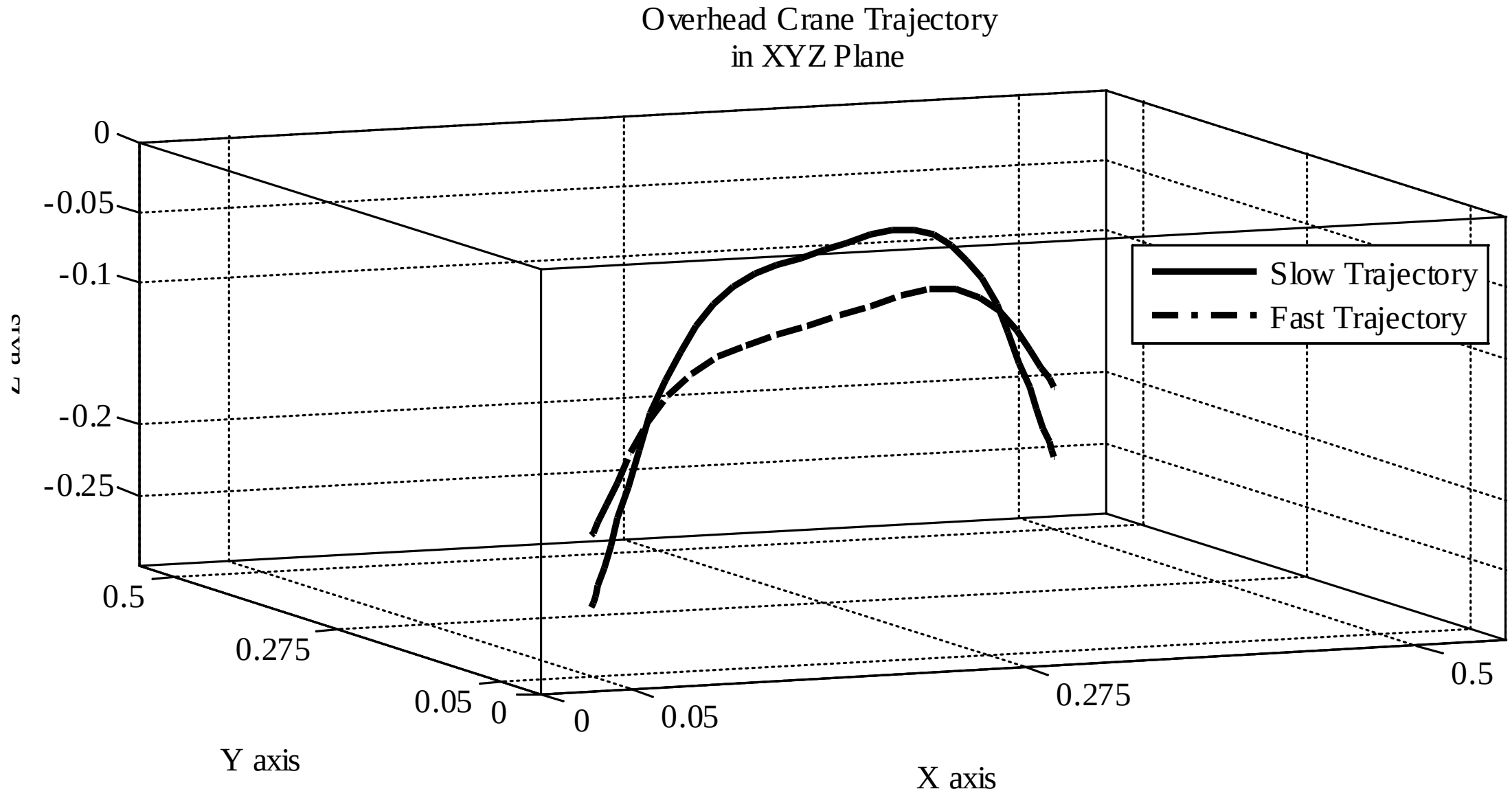

(b)

Fig. 5.9–4. The desired path of the overhead crane load for slow and fast trajectories. (a) 2D space view of the trajectories, (b) 3D space view of the trajectories.

It should be noted that the actual reference trajectories that the control system would follow are the modified ones generated in real-time by the reference signal generator using the correction terms generated by the load swing control throughout the control operation. Therefore, for evaluating the precision of load position the original reference trajectories shown in Fig. 5.9–3(a) and Fig. 5.9–3(b) will be considered not the modified ones which is compatible with the definition of real tracking error as we defined in previous sections of this chapter.

### 5.9.3 Test Results Validation

As mentioned earlier, the designed control systems are implemented and tested on the overhead crane setup and the validation results are provided in this section. The parameters of the load swing control block are given in Table 4.5–1. The values for swing control gain and swing angle observer gain shown in Table 4.5–1 are used in all the experiments for consistency of the results. Table 5.9–4 contains the parameters for MPC in discrete-time controller, and finally the parameters for state feedback control are provided in Table 5.9–5, which also contains the disturbance observer gains.

Table 5.9–3. Load Swing Control Parameters

| *Parameters* | Swing Control Gain $K_\theta$ ($k_{\theta_x}$, $k_{\theta_y}$) | Swing Angle Observer Gain $L_\theta$ ($L_{\theta_x}$, $L_{\theta_y}$) |
|---|---|---|
| Traveling Swig Angle $\theta_x$ | 17e−2 | $[1\ \ 25]^T$ |
| Traversing Swig Angle $\theta_y$ | 17e−2 | $[1\ \ 25]^T$ |

Table 5.9–4. Model Predictive Control Parameters

| *Parameters* | Tracking Error Weighting Matrix $Q(k)$ | Control Input Weighting Matrix $R(k)$ | Control Input Constraints $\boldsymbol{u}_{min}$ (V) | $\boldsymbol{u}_{max}$ (V) | Output Constraints $\boldsymbol{y}_{min}$ (m) | $\boldsymbol{y}_{max}$ (m) | Prediction Horizon $H_p$ | Control Horizon $H_u$ |
|---|---|---|---|---|---|---|---|---|
| Traveling | 50e+2 | 1e–3 | –24 | 24 | 0 | 6e–1 | | |
| Traversing | 50e+2 | 1e–3 | –24 | 24 | 0 | 6e–1 | 20 | 3 |
| Hoisting | 50e+2 | 1e–3 | –24 | 24 | 1e–3 | 6e–1 | | |

Table 5.9–5. State Feedback Control and Disturbance Observer Parameters

| *Parameters* | Feedback Gain $K$ ($K_x$, $K_y$, $K_l$) | Observer Gain $L$ ($L_x$, $L_y$, $L_l$) | Disturbance Observer Gain $L_w$ ($l_{wx}$, $l_{wy}$, $l_{wl}$) |
|---|---|---|---|
| Traveling | [12.9e+2 1.1e+2] | $[42.9e{-}2 \quad 26.5e{-}2]^T$ | –10e–2 |
| Traversing | [25.9e+2 1.2e+2] | $[41.5e{-}2 \quad 27.7e{-}2]^T$ | –10e–2 |
| Hoisting | [38.4e+2 1.2e+2] | $[43.5e{-}2 \quad 29.7e{-}2]^T$ | –50e–2 |

Let us first discuss the results obtained from implementing the discrete-time control system with MPC as its discrete-time controller which are shown in Fig. 5.9–5– Fig. 5.9–15. The comparison between the reference trajectories and the actual ones for both slow and fast trajectories are pictured in Fig. 5.9–5 and Fig. 5.9–6, respectively, when both load swing control and feedforward control are active (Scenario III) with $m$ = 0.8kg. It can be seen that throughout all the repetitions of the trajectories, the control system can successfully track the reference trajectories with high performance in both slow and fast motions. The control input voltages are illustrated in Fig. 5.9–7 with their maximum values bounded within the nominal voltage range of the PM DC motors, i.e., ±24V. It is interesting to see that for fast trajectory, the traveling and traversing voltages in Fig. 5.9–7(d) and Fig. 5.9–7(e) are higher than their counterparts for slow trajectories in Fig. 5.9–7(a) and (b). Whereas, the hoisting voltages as shown in Fig. 5.9–7(c) and Fig. 5.9–7(f) are quite similar since the hoisting distance in fast trajectory is less than the one in slow trajectory.

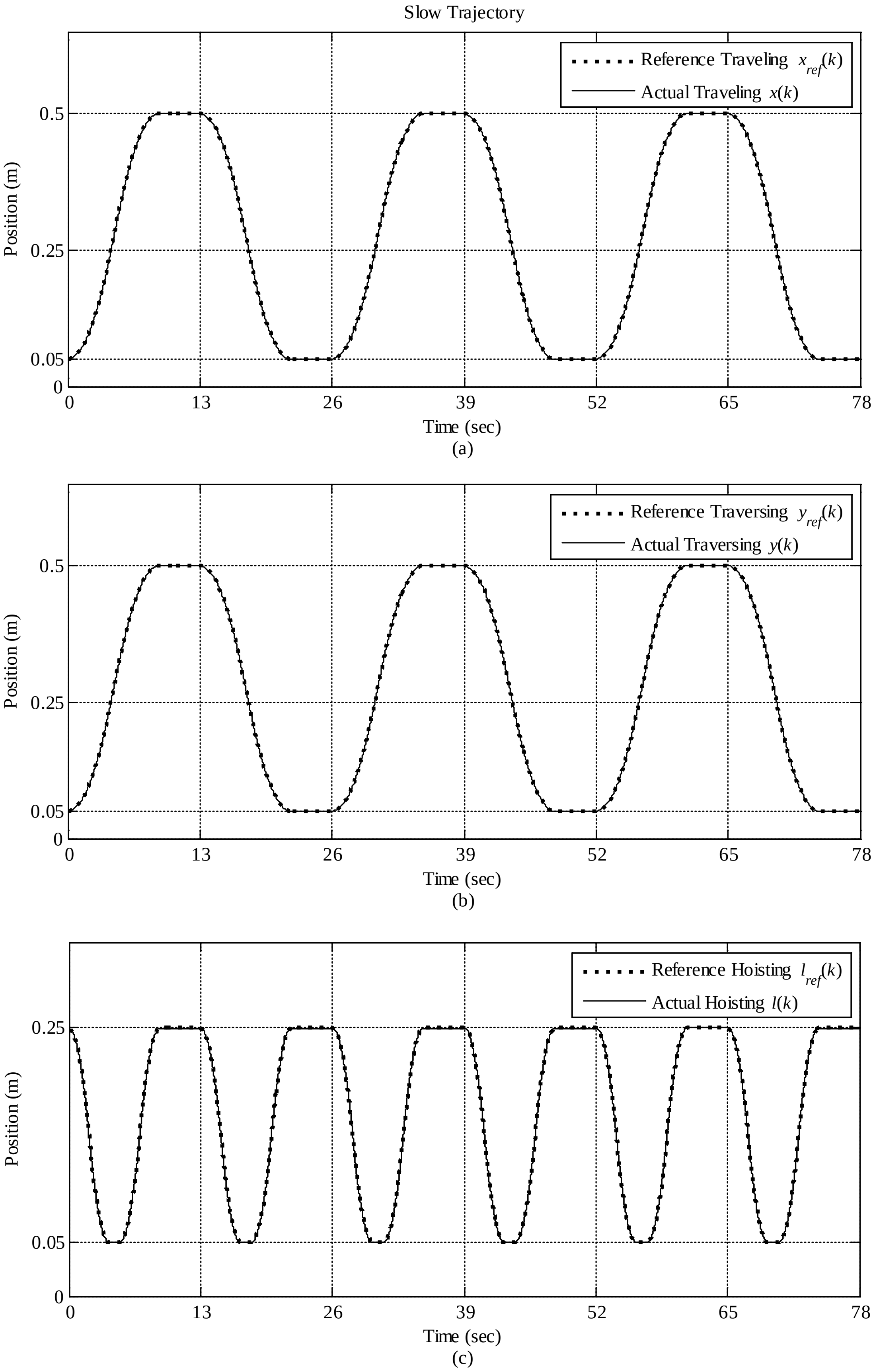

Fig. 5.9–5. Comparison between reference and actual trajectories using MPC for slow trajectory. (a) Traveling trajectory, (b) Traversing trajectory, (c) Hoisting trajectory.

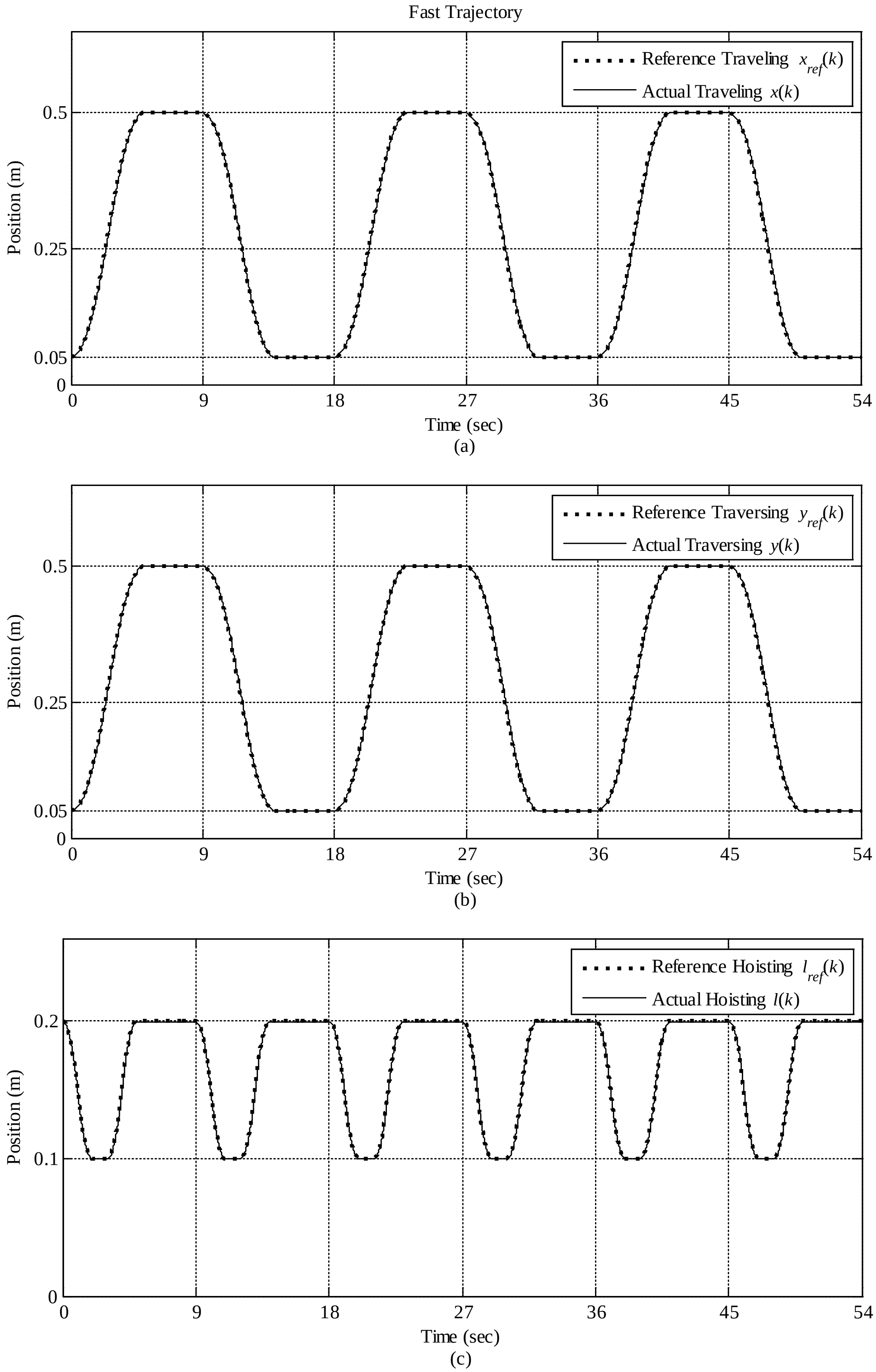

Fig. 5.9–6. Comparison between reference and actual trajectories using MPC for fast trajectory. (a) Traveling trajectory, (b) Traversing trajectory, (c) Hoisting trajectory.

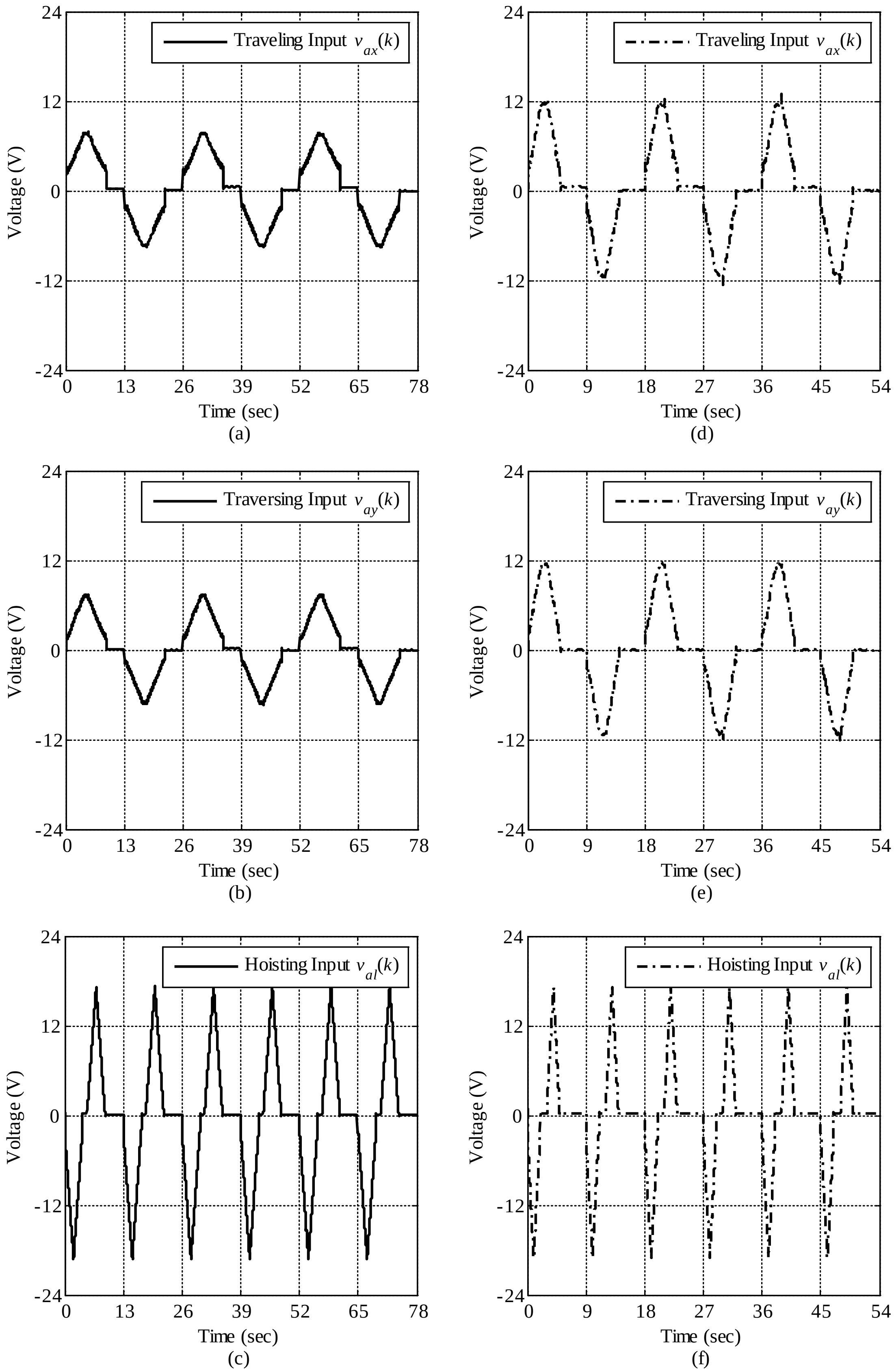

Fig. 5.9–7. Control input voltages generated by MPC. (a) Traveling input voltage, (b) Traversing input voltage, and (c) Hoisting input voltage for slow trajectory; (d) Traveling input voltage, (e) Traversing input voltage, and (f) Hoisting input voltage for fast trajectory.

To show the performance of the swing angle observer, the comparison between the estimates of swing angles and their actual measured ones along with their first time-derivatives are plotted in Fig. 5.9–8 and Fig. 5.9–9 in slow and fast motion controls, respectively. The results are provided for the first transition of the trajectories indicating the stable operation of the designed swing angle observer. It can be seen though that in fast trajectory, the first time-derivative of the swing angles are estimated with better accuracy and less oscillations in Fig. 5.9–9(b) and (d) compared to the slow trajectory ones in Fig. 5.9–8(b) and (d) since the load swing are larger in fast trajectory with lower frequency.

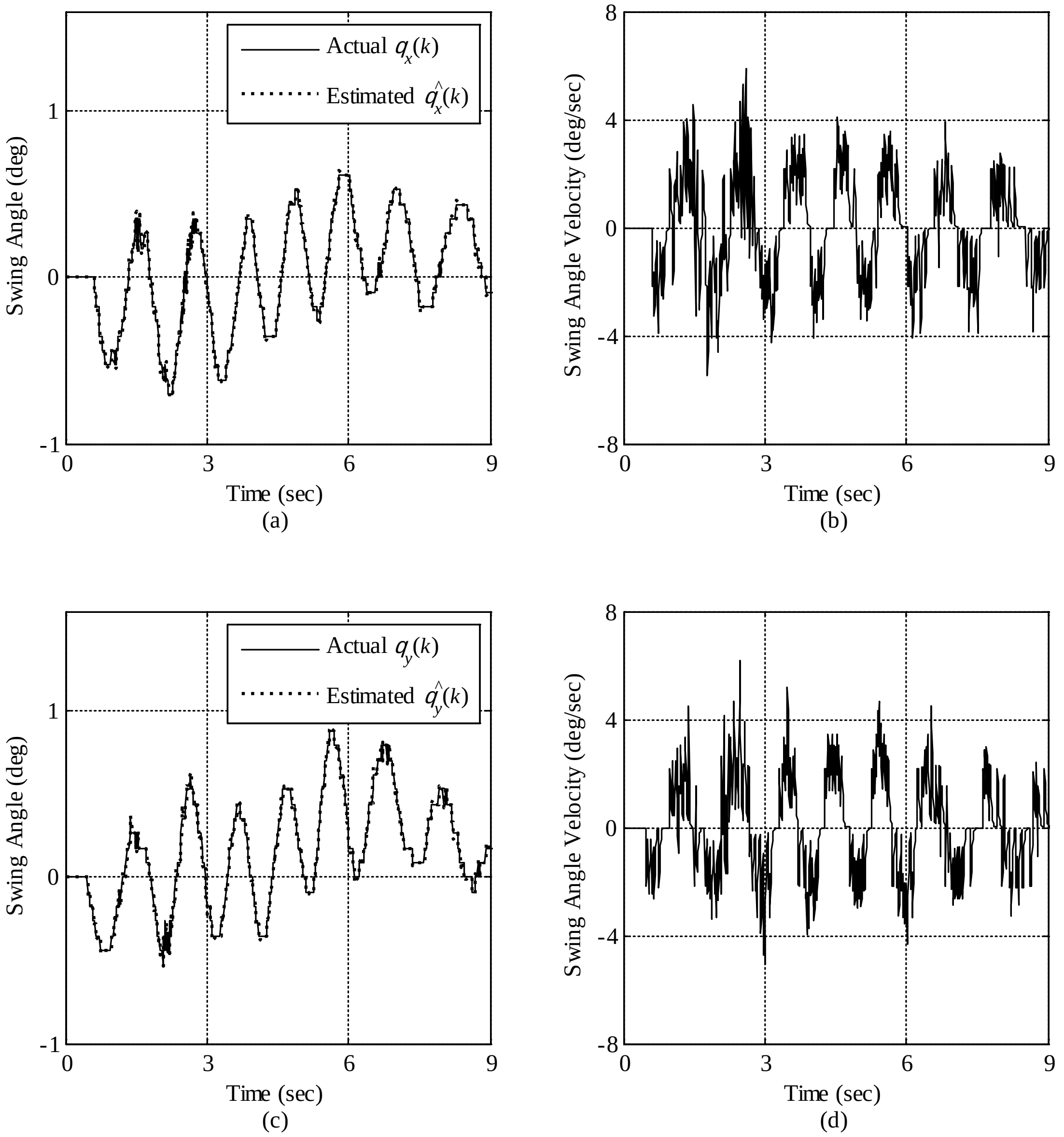

Fig. 5.9–8. Load swing estimation for slow trajectory in one transition $t_f$ = 9sec. (a) Comparison of estimated and actual swing angles, (b) Estimated first time-derivative of swing angles.

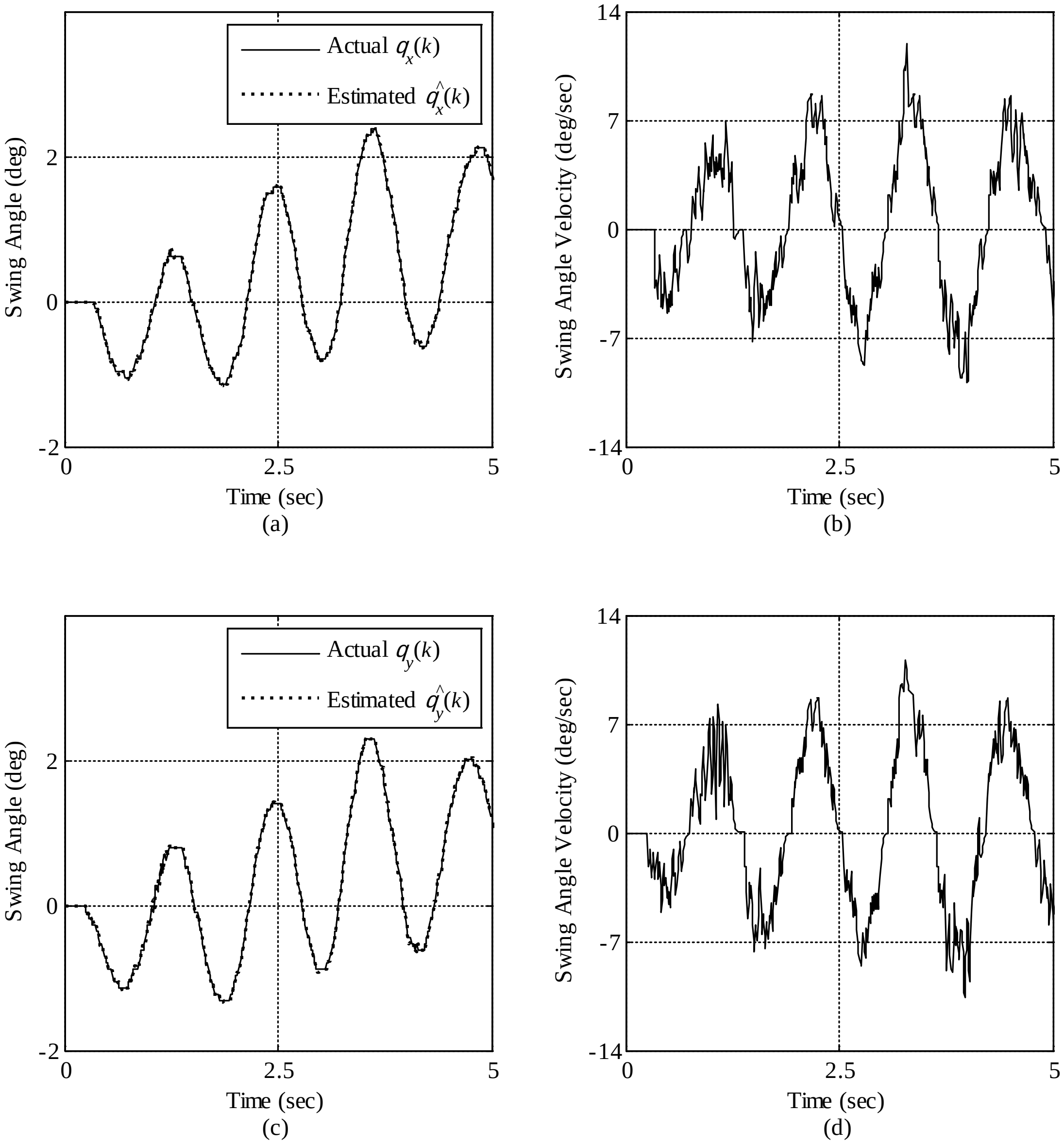

Fig. 5.9–9. Load swing estimation for fast trajectory in one transition $t_f$ = 5sec. (a) Comparison of estimated and actual swing angles, (b) Estimated first time-derivative of swing angles.

As we mentioned earlier, each designed control system for overhead crane is tested under three scenarios for both slow and fast trajectory to investigate the contribution of the main parts of the control system to the overall performance of the control operation in addition to the effects of slow and fast motions. The measurements of load swings under these scenarios are provided in Fig. 5.9–10 for slow trajectory and in Fig. 5.9–11 for fast trajectory. As can be seen, load swings are much higher in the first scenario where no swing control and feedforward control are in action. However, due to the natural property of the LSBP trajectory in damping load swings during decelerating zone, load swings in slow trajectory are close to zero at the end of each transition which is shown in Fig. 5.9–10 (a) and (b) for both swing angles $\theta_x$ and $\theta_y$. Whereas, in fast trajectory in Fig. 5.9–11(a) and (b), they tend to increase significantly over time which is what is expected in Scenario I with no proper damping action. It is interesting to see that in the second scenario, the overall magnitude of swing angles are dropped in both trajectories in Fig. 5.9–10 (c) and (d) and Fig. 5.9–11(c) and (d) compared to the first scenario. The reason is that in the absence of load swig control, the use of feedforward control in Scenario II can lead to reduction in load swings indirectly as it tries to compensate the increase in load disturbances caused by the elevation of load swings as we also explained in Section 5.3.

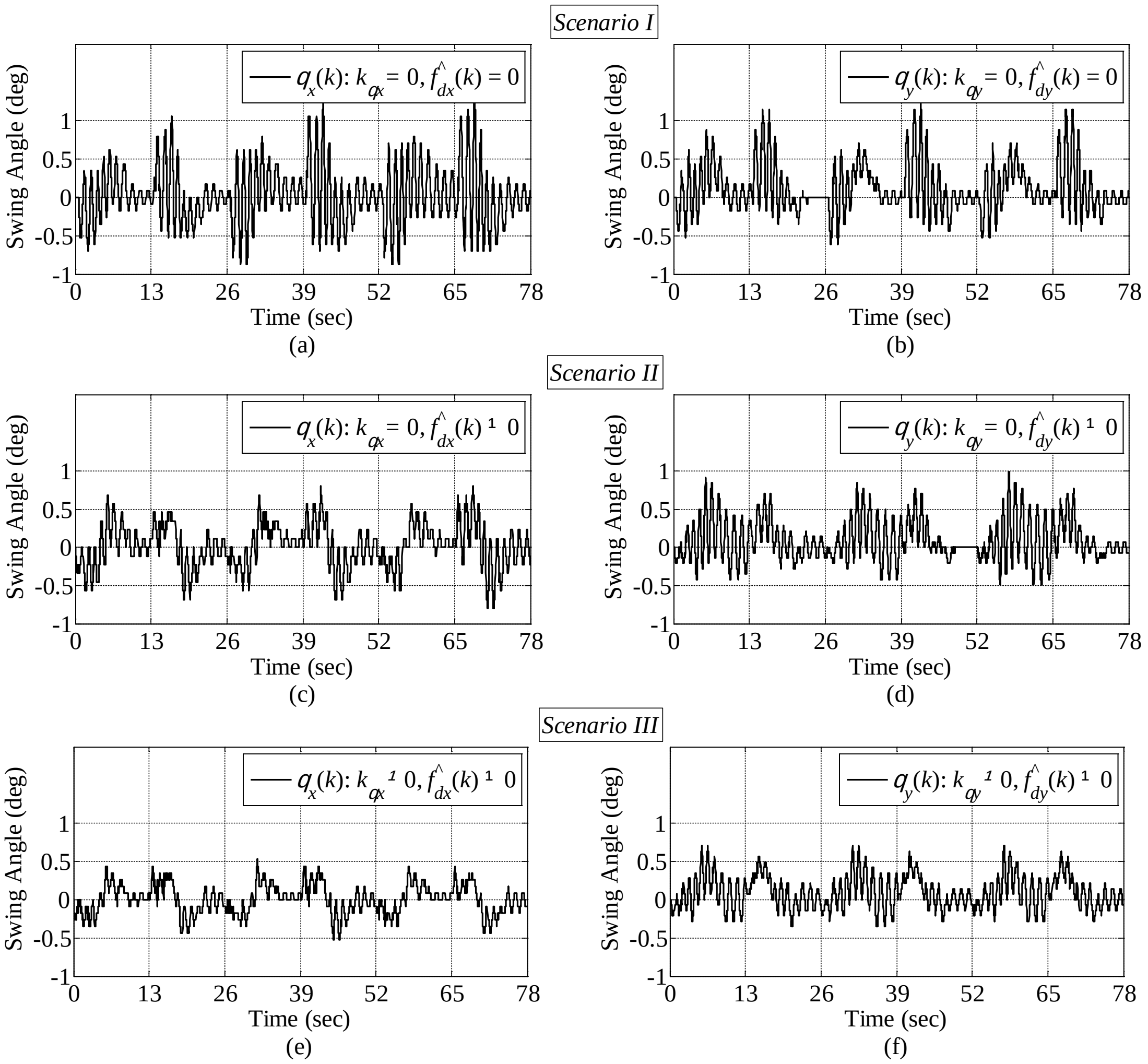

Fig. 5.9–10. Swing angle measurements with MPC for slow trajectory. (a) Traveling swing angle and (b) traversing swing angle in Scenario I: Load swing control and feedforward control are off, (c) Traveling swing angle and (d) traversing swing angle in Scenario II: Load swing control is off and feedforward control is on, (e) Traveling swing angle and (f) traversing swing angle in Scenario III: Both load swing control and feedforward control are on.

In Scenario III, load swings are noticeably suppressed, particularly in fast trajectory in Fig. 5.9–11(e) and (f), as a result of having load swing control besides feedforward control. Load swing damping in slow trajectory as shown in Fig. 5.9–10 (e) and (f) may not be significant by activating swing control which is obvious due to much slower velocity in moving the load. Nevertheless, the combination of load swing control and feedforward control can prove to robustly suppress load swings even for fast load transportation and keep them around ±2 degrees. We also examined the effect of increasing the swing control gain $K_\theta$ in fast trajectory to see how further it can reduce load swing as illustrated in Fig. 5.9–11(g) and (h). As we also showed in the proof of load swing stability in Section 5.3.1, as long as swing control gain is chosen to be greater than 1.5 times the maximum hoisting speed in the operation ($1.5|\dot{l}|_{\max} \equiv 1.5v_{l\max}$), the upper bound of load swings will be bounded and increasing the swing control gain may not have a considerable impact on load swing. Since the hoisting rope length follows the reference hoisting trajectory, the maximum hoisting speed would reach the normal velocity in the minimum-time trajectory designed for hoisting motion in both slow and fast trajectories (Fig. 5.4–2) with the value given in Table 5.9–2 ($v_{lr}$ = 0.1 m/sec) as discussed in Section 5.4. That is why swing control gain was chosen as $K_\theta = Diag\{0.17, 0.17\}$ to guarantee the stability condition obtained for stability of load swing in (5.3–34) in Section 5.3.1.

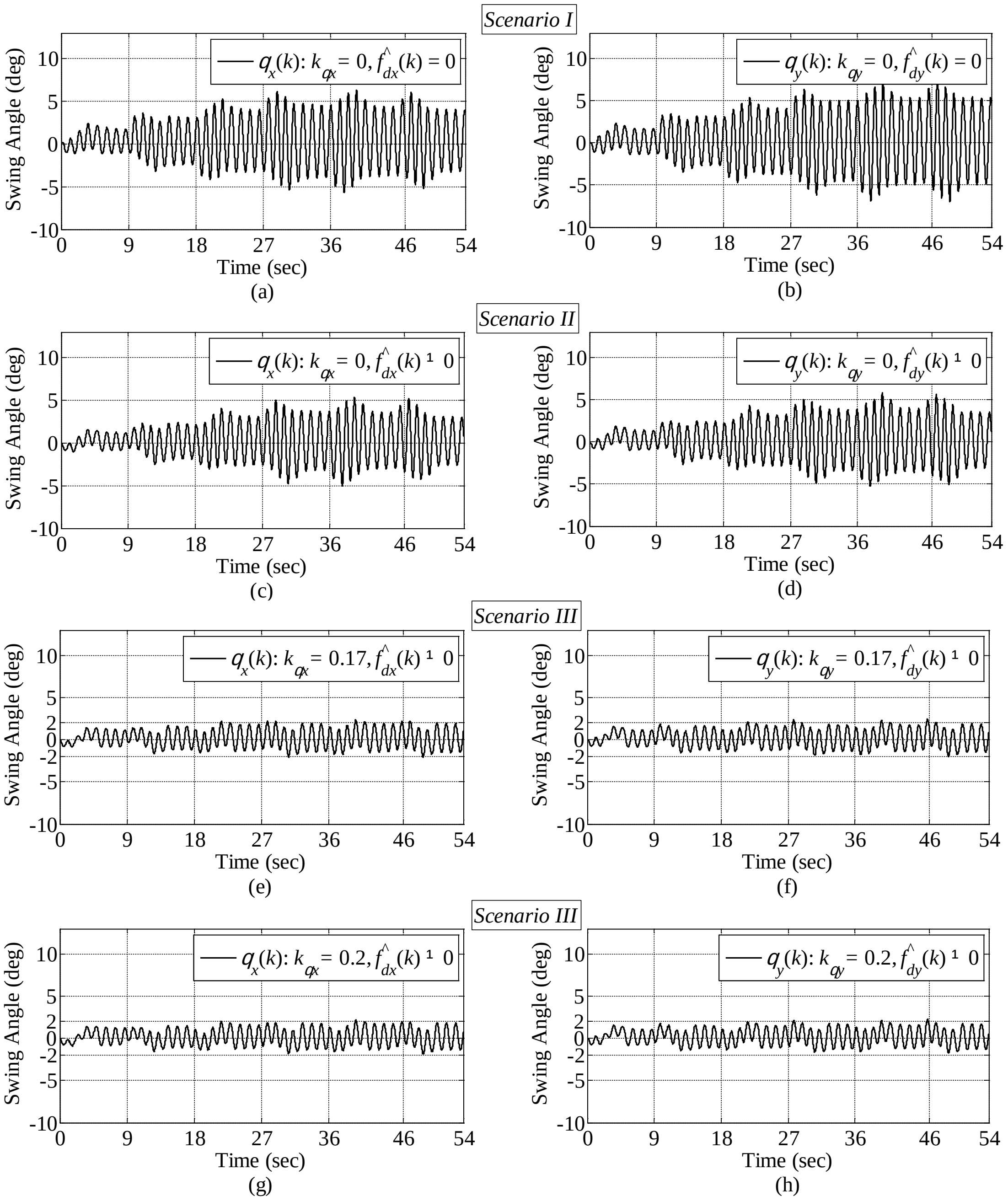

Fig. 5.9–11. Swing angle measurements with MPC for fast trajectory. (a) Traveling swing angle and (b) traversing swing angle in Scenario I: Load swing control and feedforward control are off, (c) Traveling swing angle and (d) traversing swing angle in Scenario II: Load swing control is off and feedforward control is on, (e) Traveling swing angle and (f) traversing swing angle in Scenario III: Both load swing control and feedforward control are on, (g) Traveling swing angle and (h) traversing swing angle in Scenario III: Both load swing control and feedforward control are on but with different swing control gain, i.e., $K_{\theta} = Diag\{0.2, 0.2\}$.

Furthermore, the position tracking error results for each direction of motion under the aforementioned scenarios with MPC as the discrete-time control are provided in Fig. 5.9–12 for slow trajectory and in Fig. 5.9–13 for fast trajectory. It can be clearly seen that the overall accuracy in following the reference trajectories is significantly improved by about 50 percent when feedforward control is active in Scenario II for both slow and fast motions as shown in Fig. 5.9–12(b) and Fig. 5.9–13(b) compared to Scenario I results in Fig. 5.9–12(a) and Fig. 5.9–13(a) with no proper disturbance compensation. When load swig control is added to the control system in Scenario III in conjunction with feedforward control, the temporary jumps in tracking error is evident as we expected in Fig. 5.9–12(c) and Fig. 5.9–13(c). According to what we explained in trajectory planning, when load swing control is applied, the traveling and traversing accelerations are modified such that the load swing can be stabilized in the sense of $\mathcal{L}_2$ stability. This creates a deviation in the reference position and velocity trajectories and that is why during accelerating and constant-velocity zones we have higher tracking error. However, the decelerating replanning procedure is designed to correct the reference trajectories, which then allows the control system to bring back the position of the load to its original final position at the end of each trajectory. As can be seen in Fig. 5.9–12(c) and Fig. 5.9–13(c), the control system could successfully reduce the tracking error at the end of each transition with the precision of around ±1 millimeter indicating the high performance of the control system in following the reference trajectories. It should also be reminded that the control system works on the error between the modified reference trajectories and the measured ones that is certainly less than the actual tracking error we displayed in Fig. 5.9–12 and Fig. 5.9–13 due to the deviation in the modified reference trajectories.

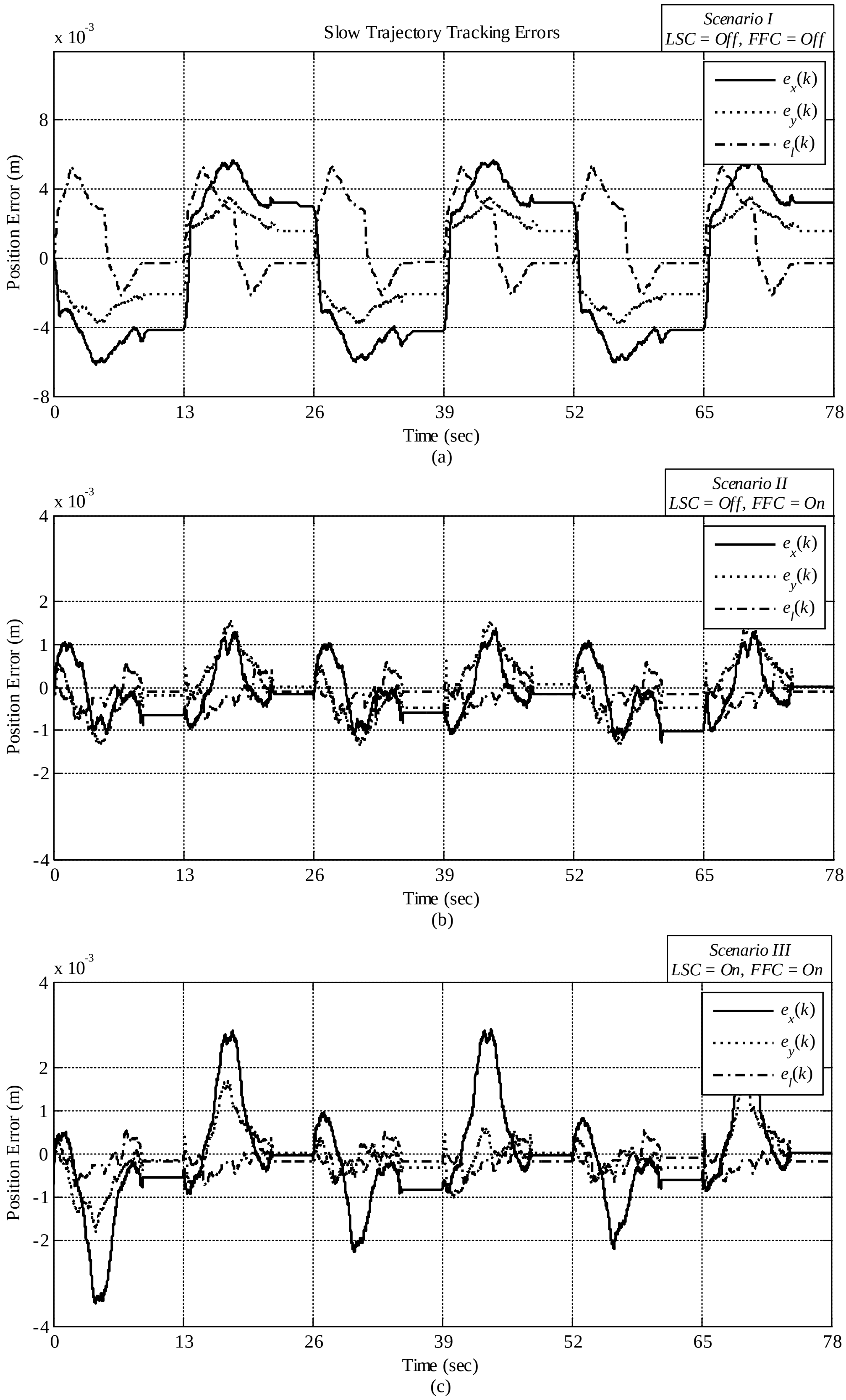

Fig. 5.9–12. Position tracking error with MPC for traveling $e_x(k)$, traversing $e_y(k)$, and hoisting $e_l(k)$ motions in slow trajectory. (a) Scenario I: Load swing control (LSC) and feedforward control (FFC) are off, (b) Scenario II: Load swing control is off and feedforward control is on, (c) Scenario III: Both load swing control and feedforward control are on.

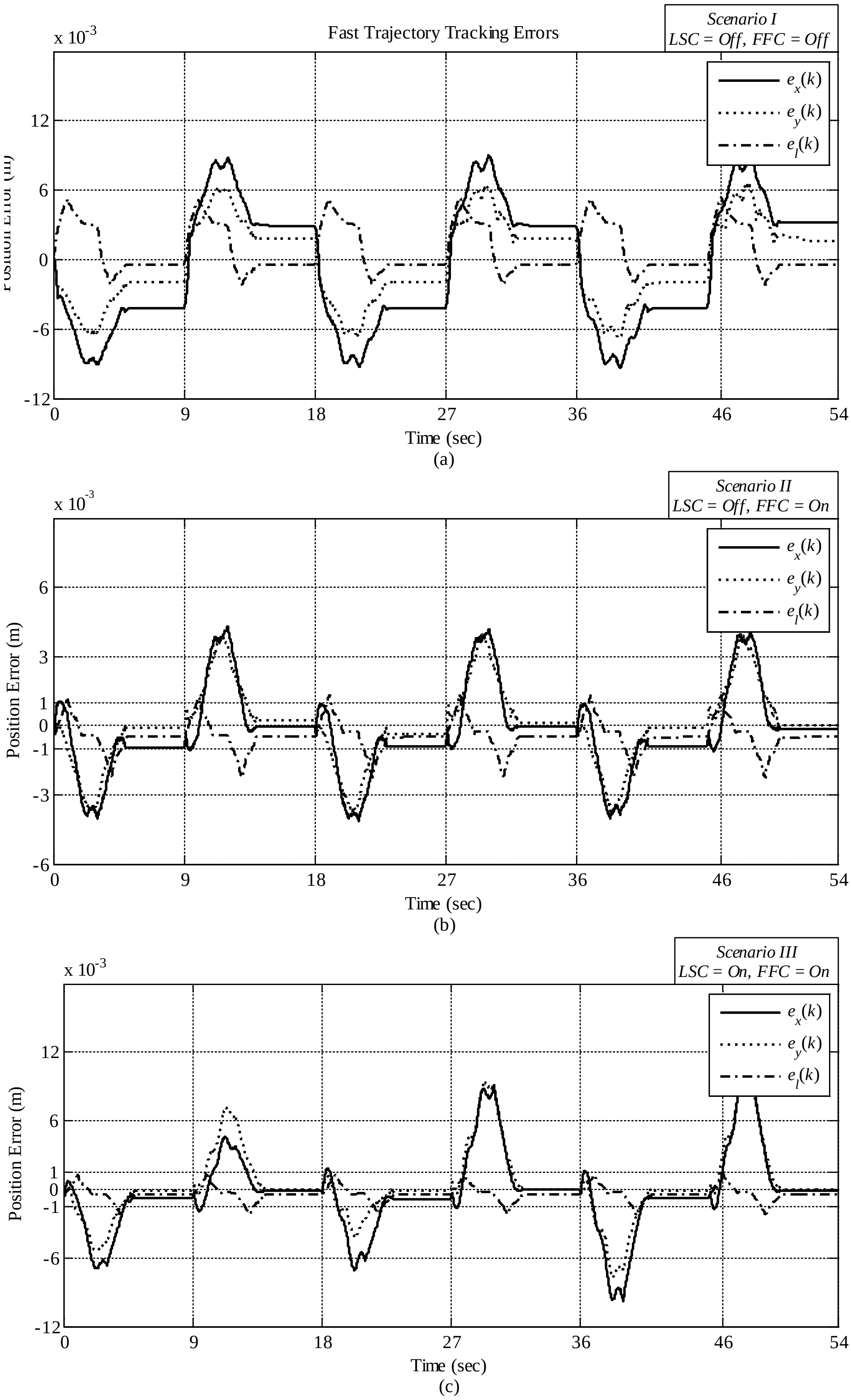

Fig. 5.9–13. Position tracking error with MPC for traveling $e_x(k)$, traversing $e_y(k)$, and hoisting $e_l(k)$ motions in fast trajectory. (a) Scenario I: Load swing control (LSC) and feedforward control (FFC) are off, (b) Scenario II: Load swing control is off and feedforward control is on, (c) Scenario III: Both load swing control and feedforward control are on.

It would be interesting to see how the reference signal generator provides the modified reference trajectories which are updated through adding the correction term to the reference trolley accelerations generated by the load swing control. As an example, Fig. 5.9–14 and Fig. 5.9–15 shows the comparison between the original reference trajectories and the modified ones in the third repetition of the fast trajectory for traveling and traversing motions, respectively, when the overhead crane was controlled under the third scenario with MPC. The deviation in the reference position trajectory, as shown in Fig. 5.9–14(a) and Fig. 5.9–15(a), may not be very visible due to the scale of figures. However, Fig. 5.9–14(c) and Fig. 5.9–15(c) clearly depict how the modified reference accelerations differ from the original ones, and consequently affects the reference velocity profiles as illustrated in Fig. 5.9–14(b) and Fig. 5.9–15(b). Moreover, it can be seen that at the end of the constant-velocity zone at time 39 seconds, the decelerating replanning procedure calculates the correction velocities ($v_{rxc}$, $v_{ryc}$) in Fig. 5.9–14(b) and Fig. 5.9–15(b) and the correction accelerations ($a_{rxc}$, $a_{ryc}$) for traveling and traversing motions, respectively, such that the reference position profiles can approach to the original final values at the end of the decelerating zone. Therefore, it is guaranteed that the load will be located at the intended final destination knowing that the discrete-time control system can successfully track the reference trajectories for traveling, traversing, and hoisting motions.

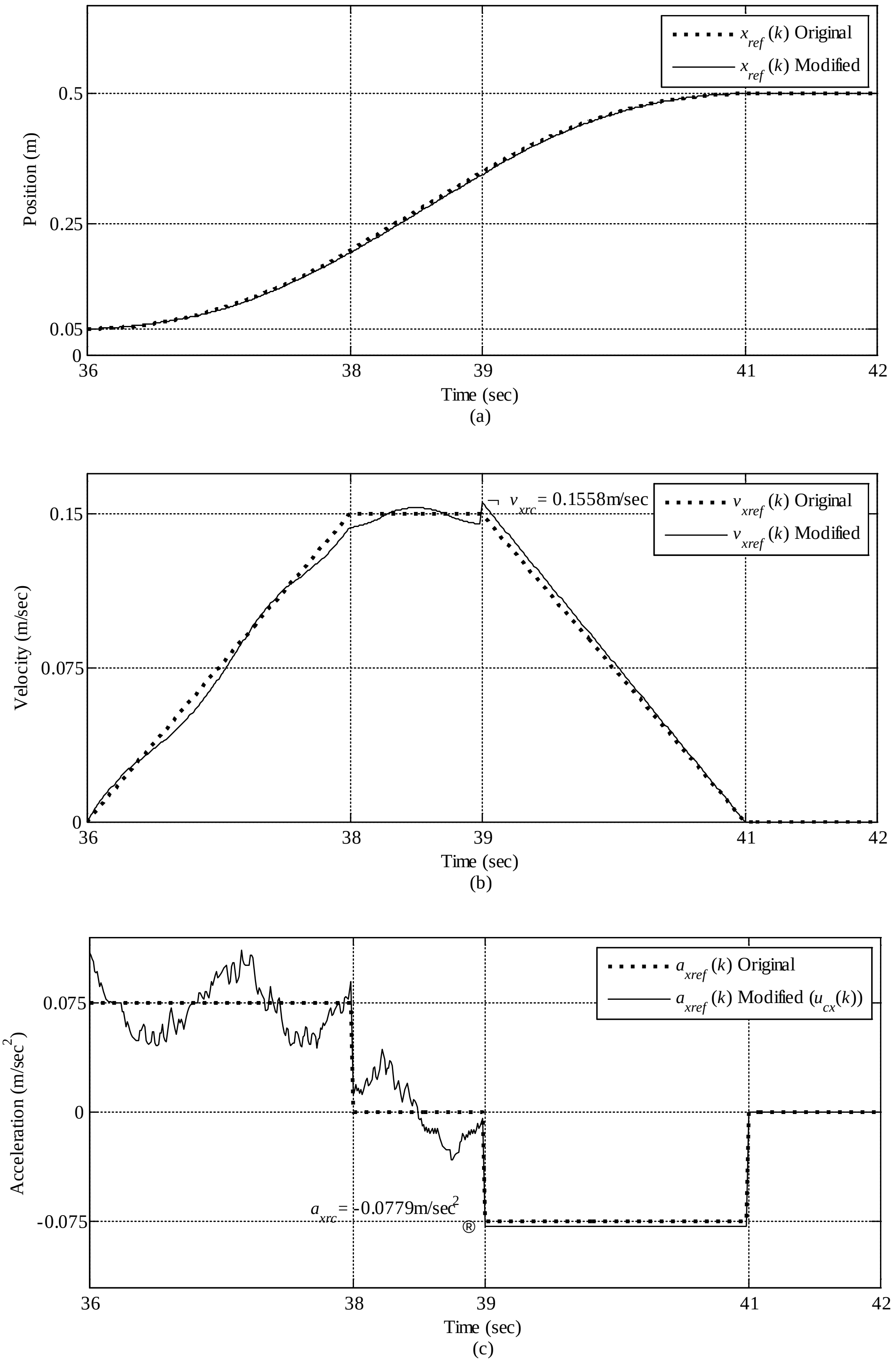

Fig. 5.9–14. Comparison between original and modified reference trajectories in the third repetition of the fast trajectory for traveling motion. (a) Reference position profiles, (b) Reference velocity profiles, (a) Reference acceleration profiles.

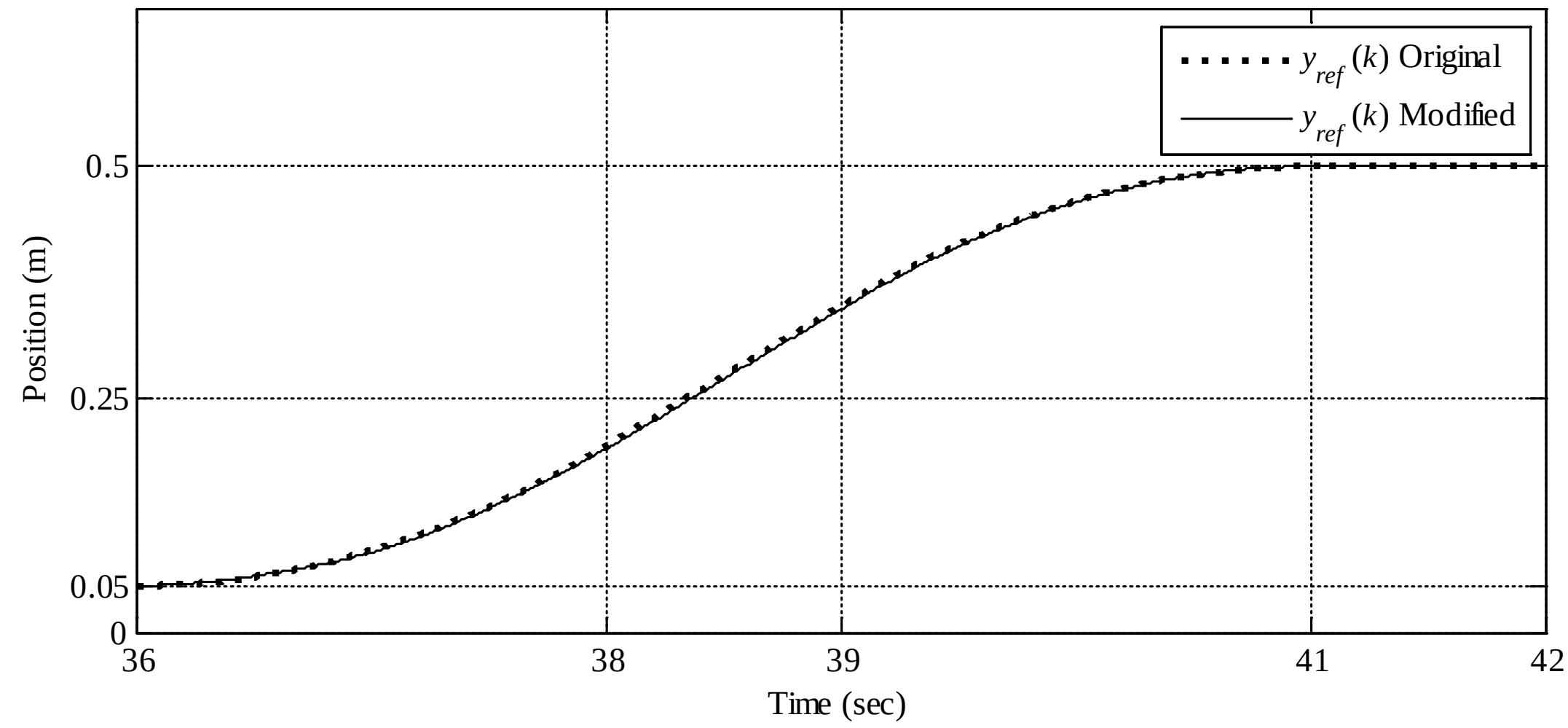

(a)

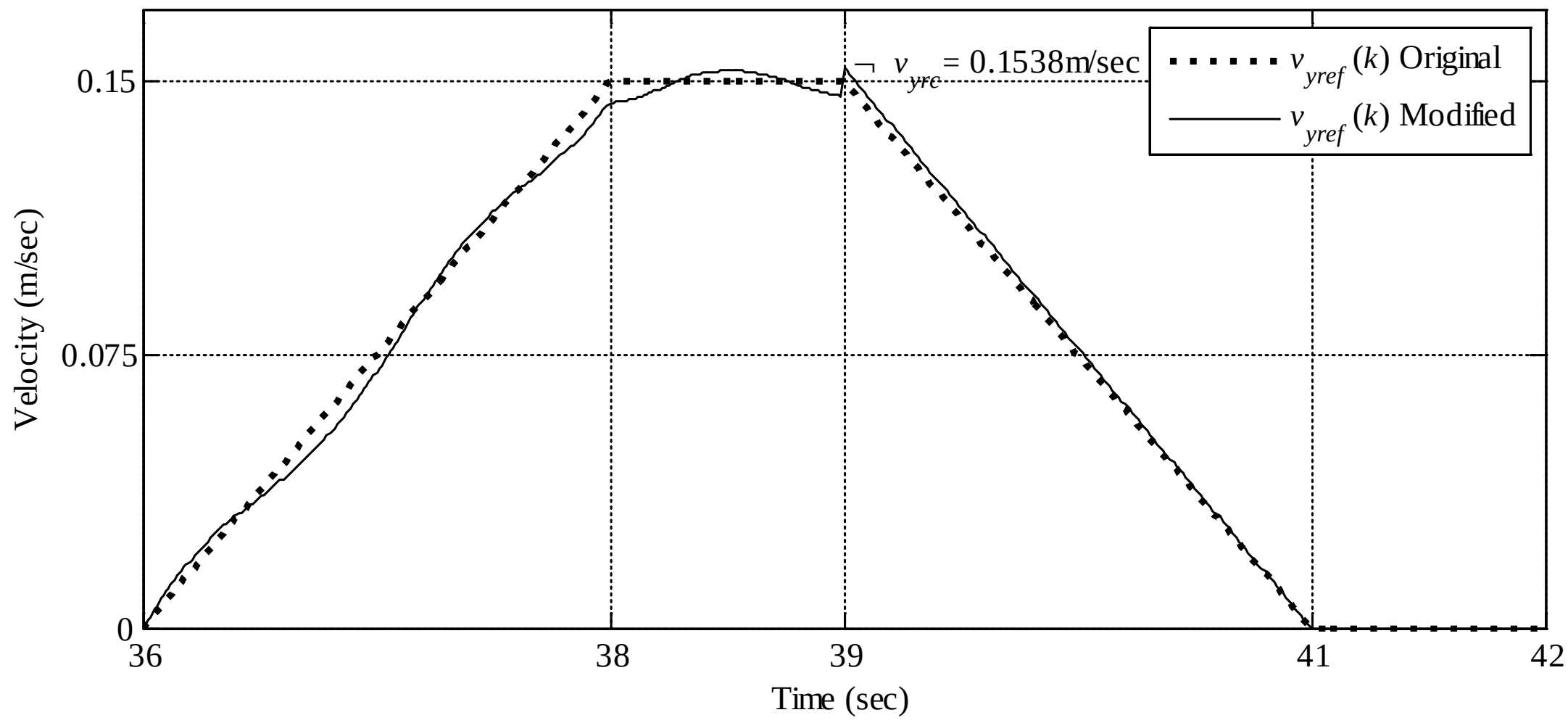

(b)

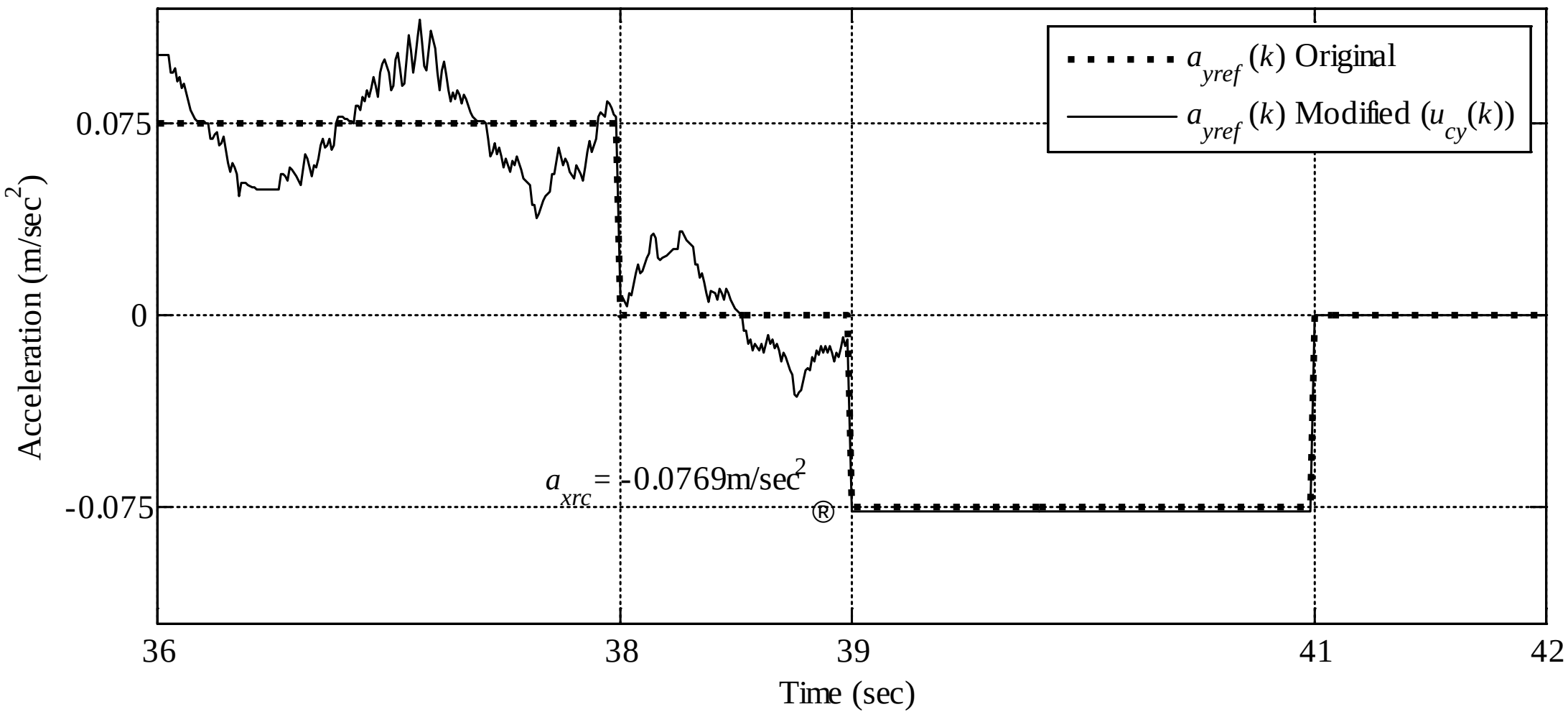

(c)

Fig. 5.9–15. Comparison between original and modified reference trajectories in the third repetition of the fast trajectory for traversing motion. (a) Reference position profiles, (b) Reference velocity profiles, (a) Reference acceleration profiles.

Now, let us discuss the results of the discrete-time control system on the 3D overhead crane with state feedback control as its discrete-time controller. As we mentioned before, the three test scenarios under which the overhead crane should be controlled are conducted with the addition of using disturbance observer as well with two different load masses. The trajectory tracking results are displayed in Fig. 5.9–16 for slow trajectory and in Fig. 5.9–17 for fast trajectory, both under Scenario III where the load swing control and feedforward control are active with crane load mass $m$ = 0.8kg. The comparison between the actual trajectories and the reference ones (original reference trajectories) clearly indicates that the designed discrete-time control system with state feedback control can follow the reference trajectories with high performance and accuracy even for multiple repetitions similar to the results obtained with MPC. The control input voltages are also shown in Fig. 5.9–18 for both trajectories under the third scenario as mentioned above. It can be seen that all input voltages for PM DC motors are maintained within the nominal voltage range of ±24V. The similarity between the slow and fast hoisting voltages in Fig. 5.9–18(c) and (d) comes from the fact that the hoisting speed was chosen to be the same in both slow and fast trajectories to demonstrate that the proposed control system can handle high speed load hoisting as it was also shown in MPC results before in Fig. 5.9–7.

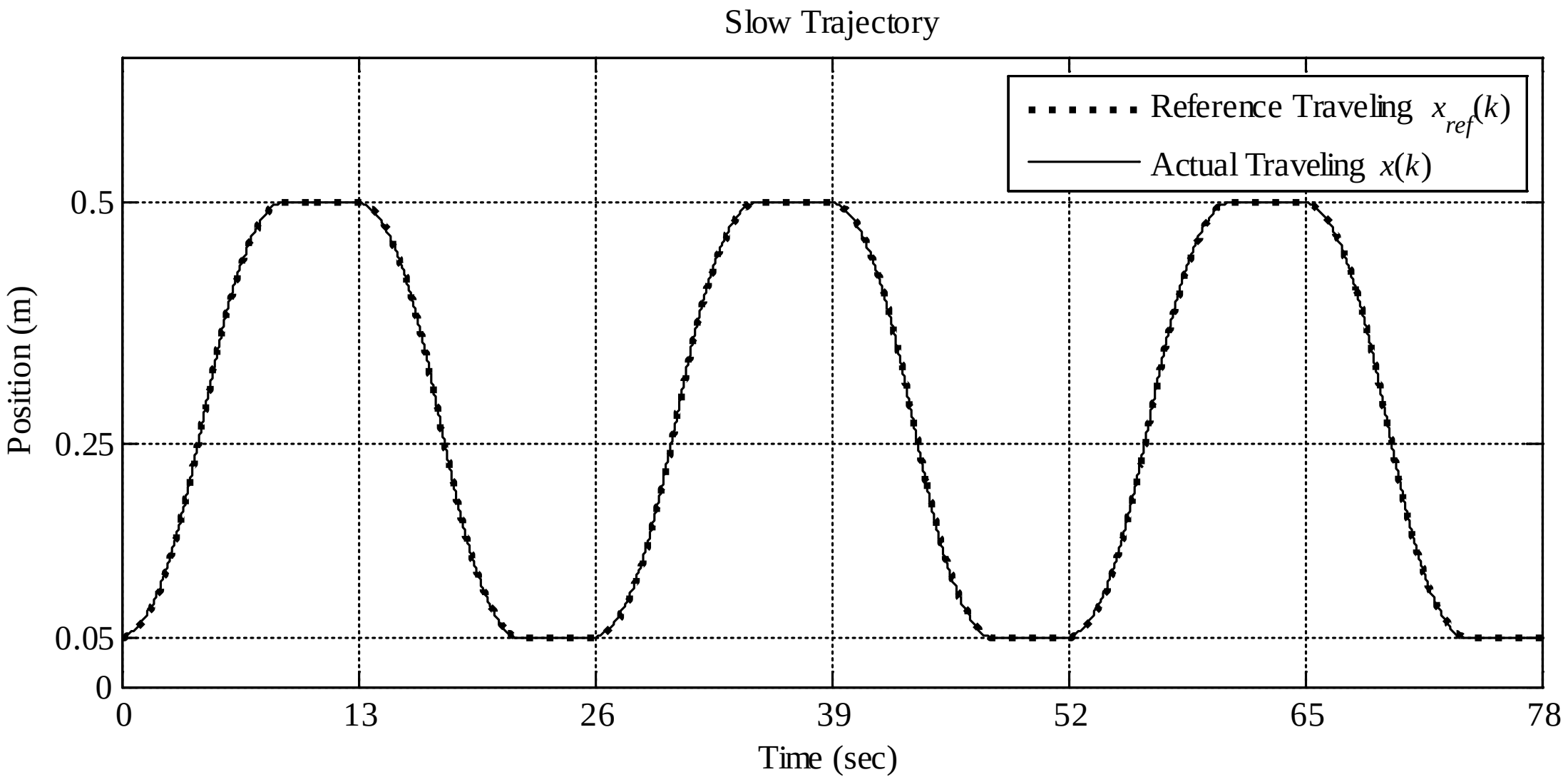

(a)

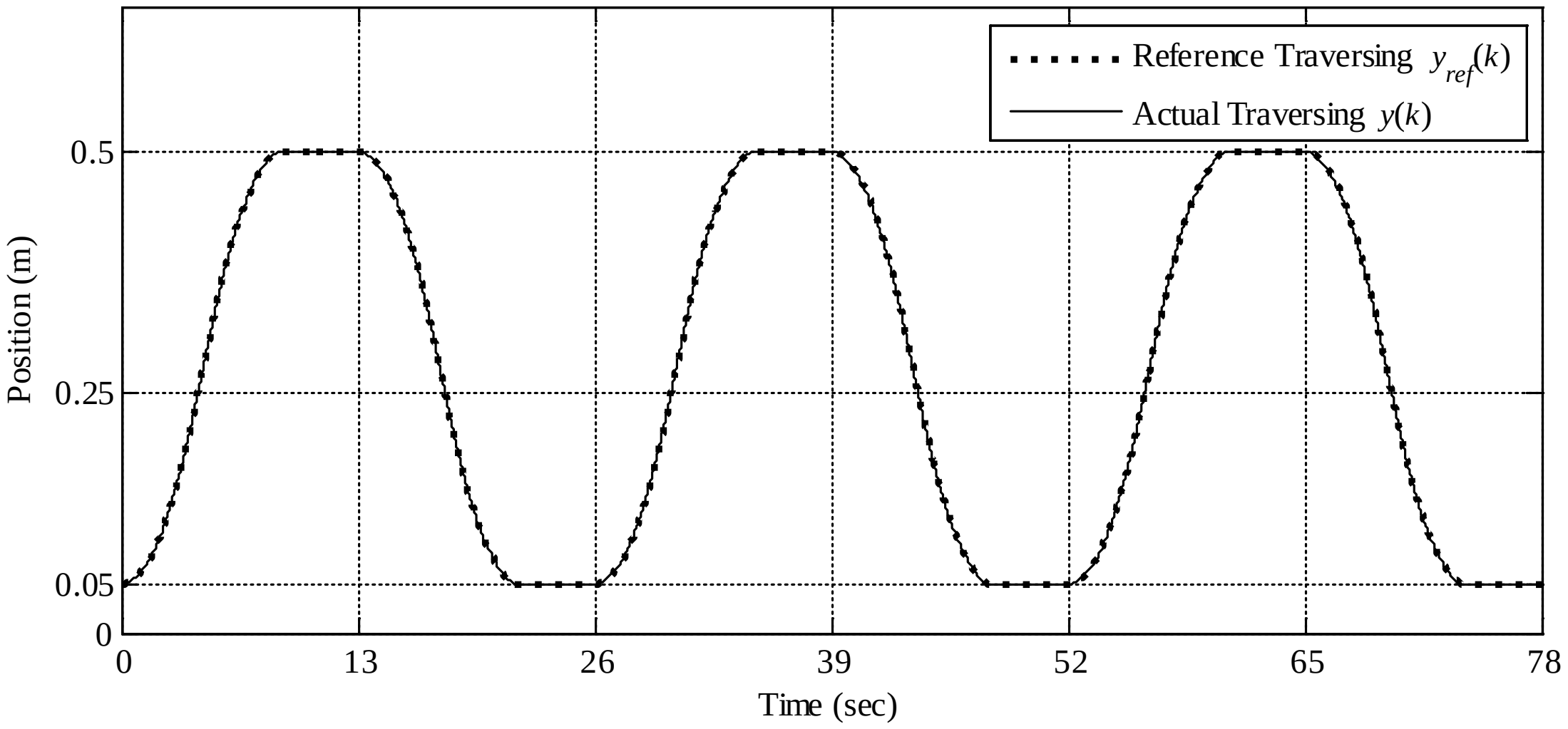

(b)

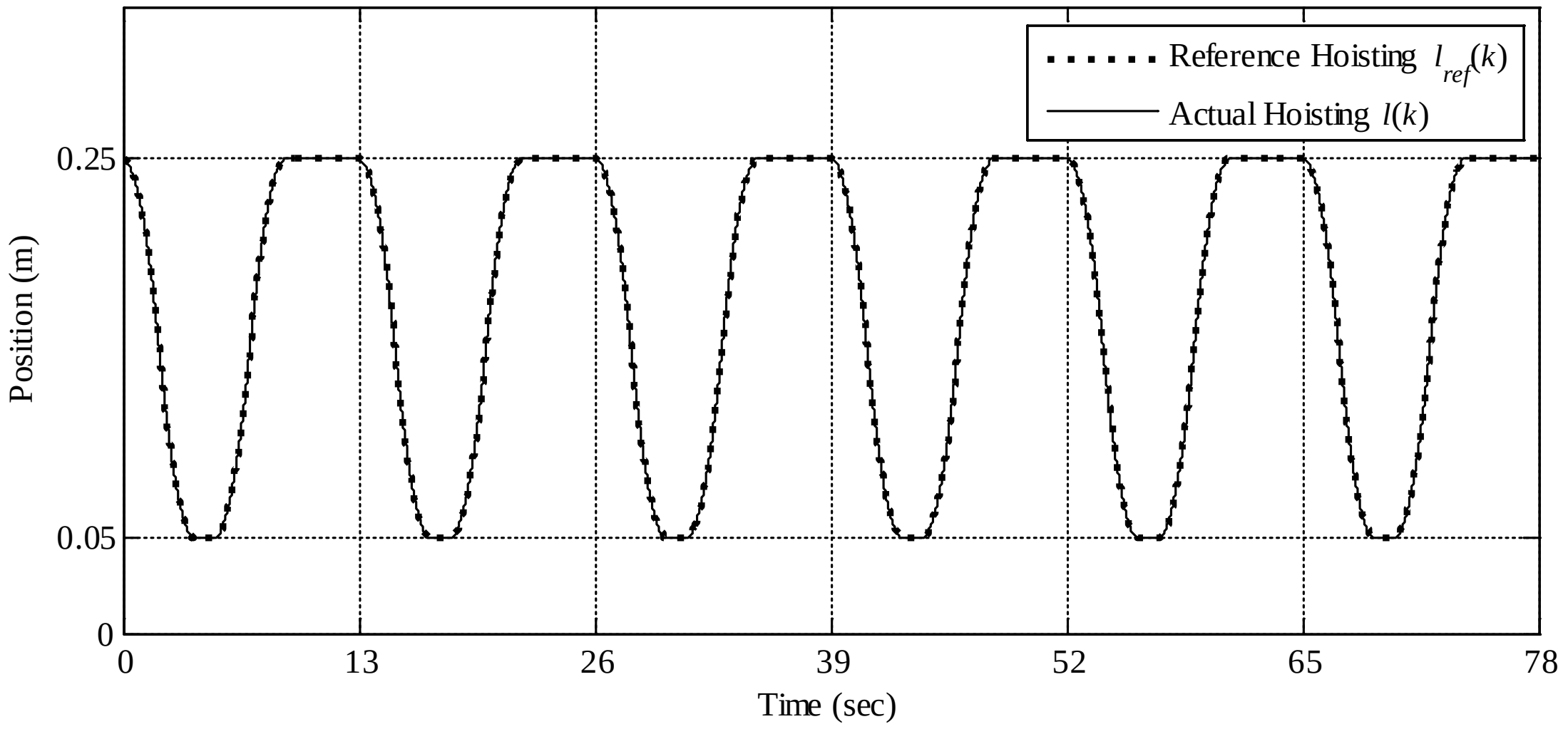

(c)

Fig. 5.9–16. Comparison between reference and actual trajectories using state feedback control for slow trajectory. (a) Traveling trajectory, (b) Traversing trajectory, (c) Hoisting trajectory.

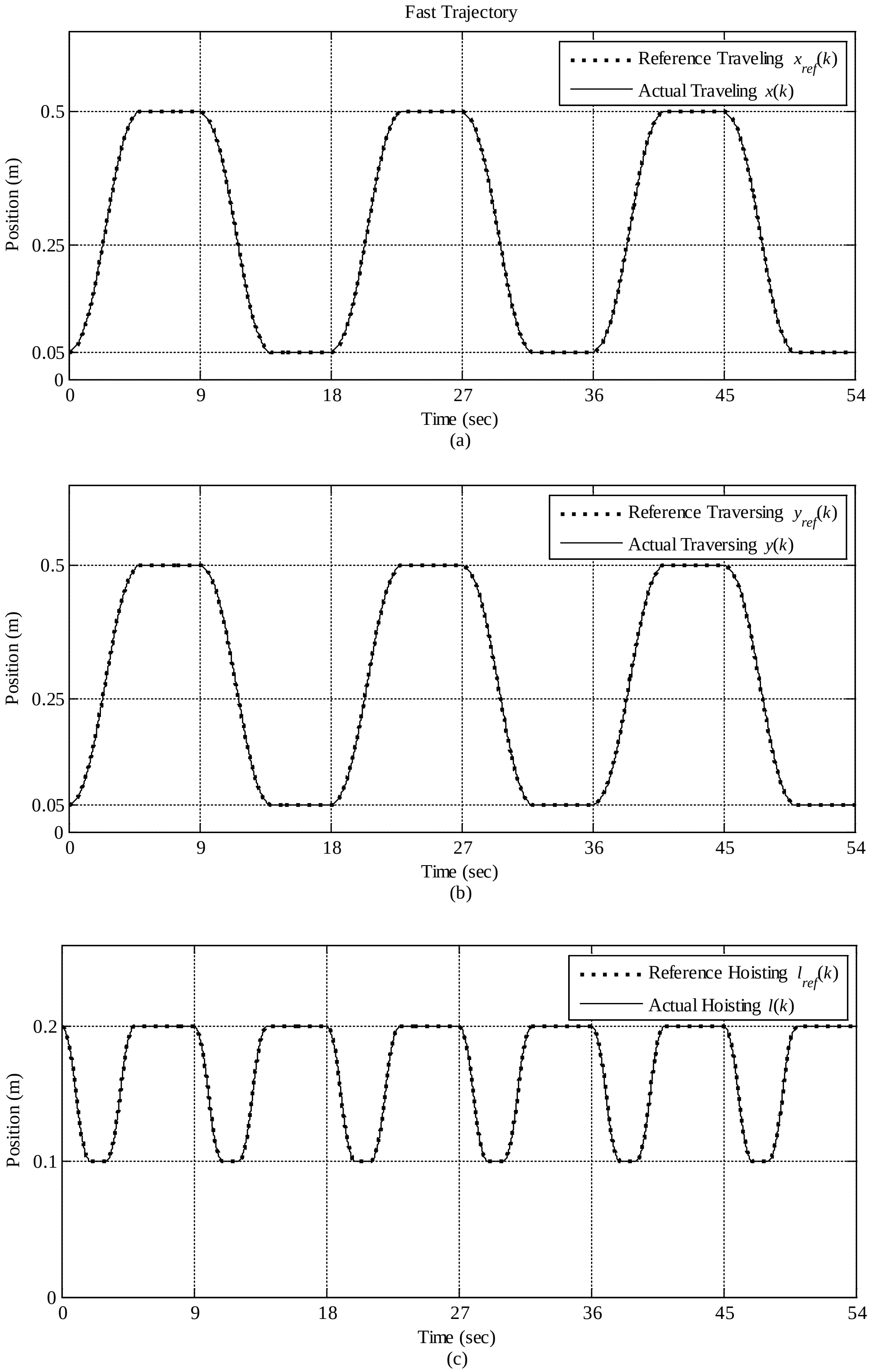

Fig. 5.9–17. Comparison between reference and actual trajectories using state feedback control for fast trajectory. (a) Traveling trajectory, (b) Traversing trajectory, (c) Hoisting trajectory.

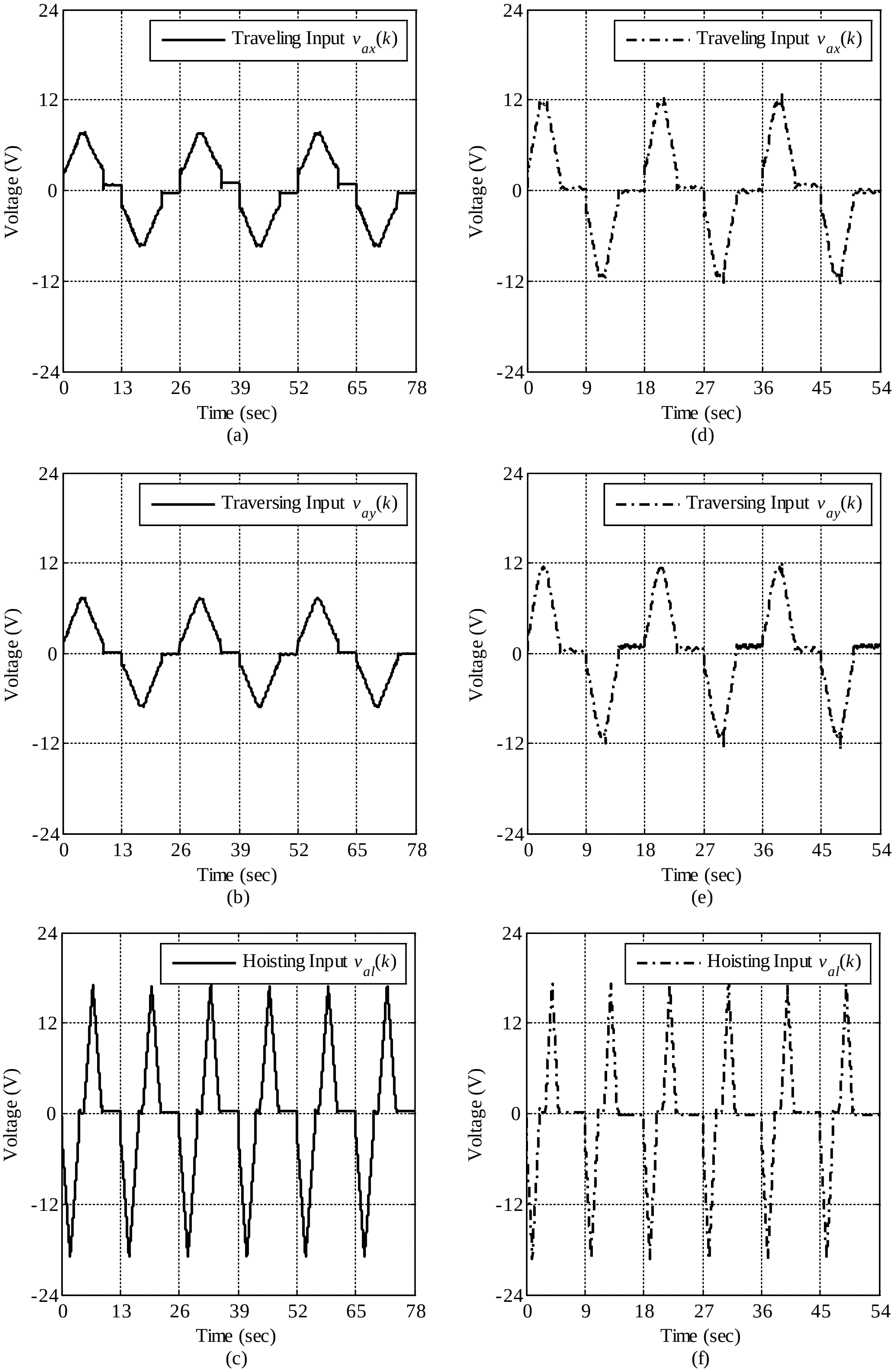

Fig. 5.9–18. Control input voltages generated by state feedback control. (a) Traveling input voltage, (b) Traversing input voltage, and (c) Hoisting input voltage for slow trajectory; (d) Traveling input voltage, (e) Traversing input voltage, and (f) Hoisting input voltage for fast trajectory.

The measurements of load swings under the three scenarios for both slow and fast trajectories are given in Fig. 5.9–19 and Fig. 5.9–20, respectively. Similar to the results obtained with MPC, it can be seen that the largest load swings happens when no load swing control and feedforward control are active in Scenario I for both slow and fast trajectories in Fig. 5.9–19(a) and Fig. 5.9–20(a). The effect of adding feedforward control to the state feedback control in Scenario II can be seen in Fig. 5.9–19(b) and Fig. 5.9–20(b) which reduces the overall amount of load swing throughout the operation to some extent as we expected. However, with both load swing control and feedforward control in action, the amplitude of swing angles is significantly declined in the third scenario. This drop in load swing is more obvious in fast trajectory as can be seen in Fig. 5.9–20(c) with around 60 percent reduction from maximum magnitude of 5 degrees in Scenario I and II to about 2 degrees in Scenario III. A higher swing control gain has also tested like the previous experiment with MPC to see whether the swing angle will decrease further or not as illustrated in Fig. 5.9–20(d) but the effect is not considerable. These results along with those obtained with MPC show that our load swing control could robustly suppress load swings as we have also proved in the stability analysis of load swing given in Section 5.3.

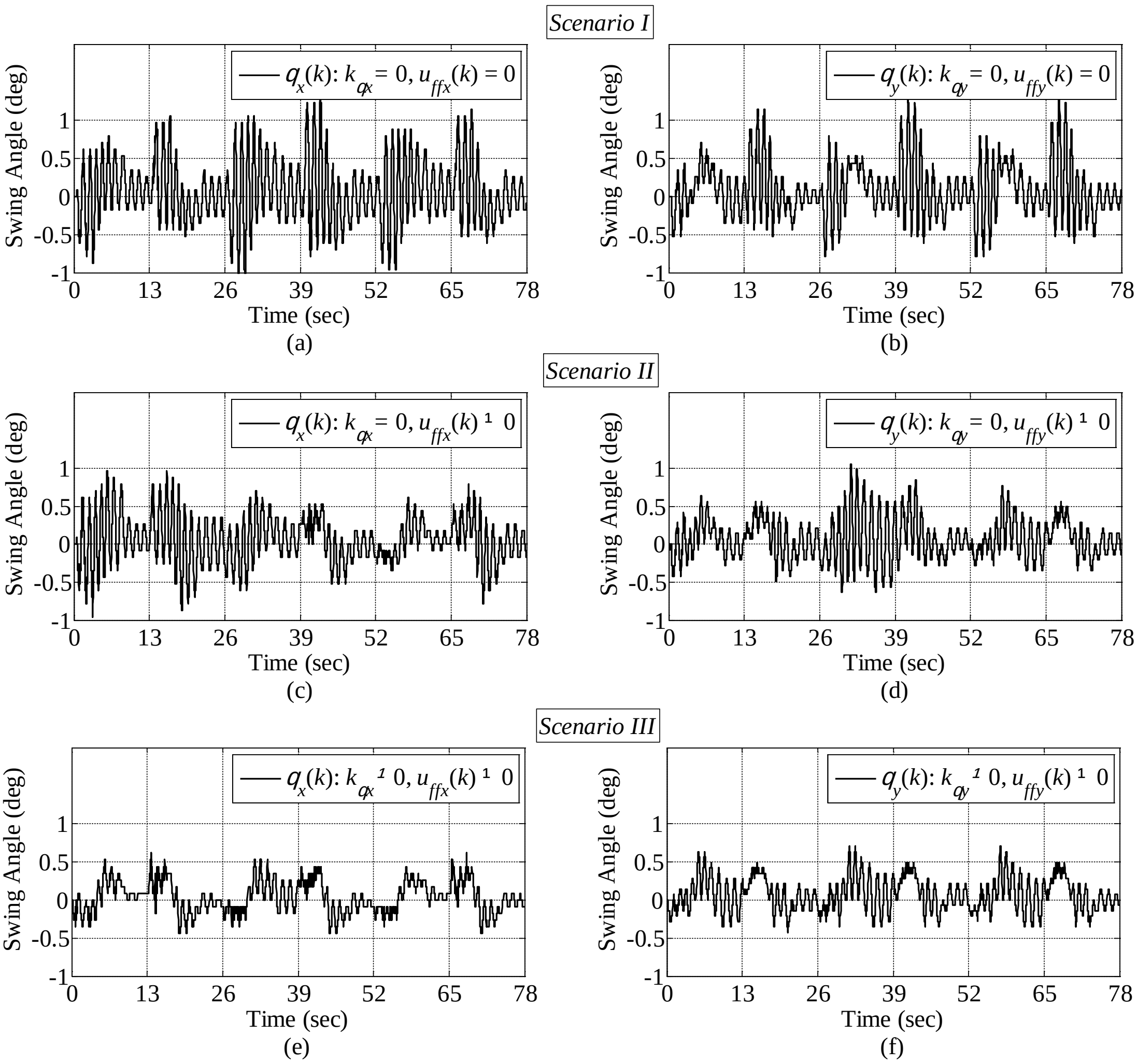

Fig. 5.9–19. Swing angle measurements with state feedback control for slow trajectory. (a) Traveling swing angle and (b) traversing swing angle in Scenario I: Load swing control and feedforward control are off, (c) Traveling swing angle and (d) traversing swing angle in Scenario II: Load swing control is off and feedforward control is on, (e) Traveling swing angle and (f) traversing swing angle in Scenario III: Both load swing control and feedforward control are on.

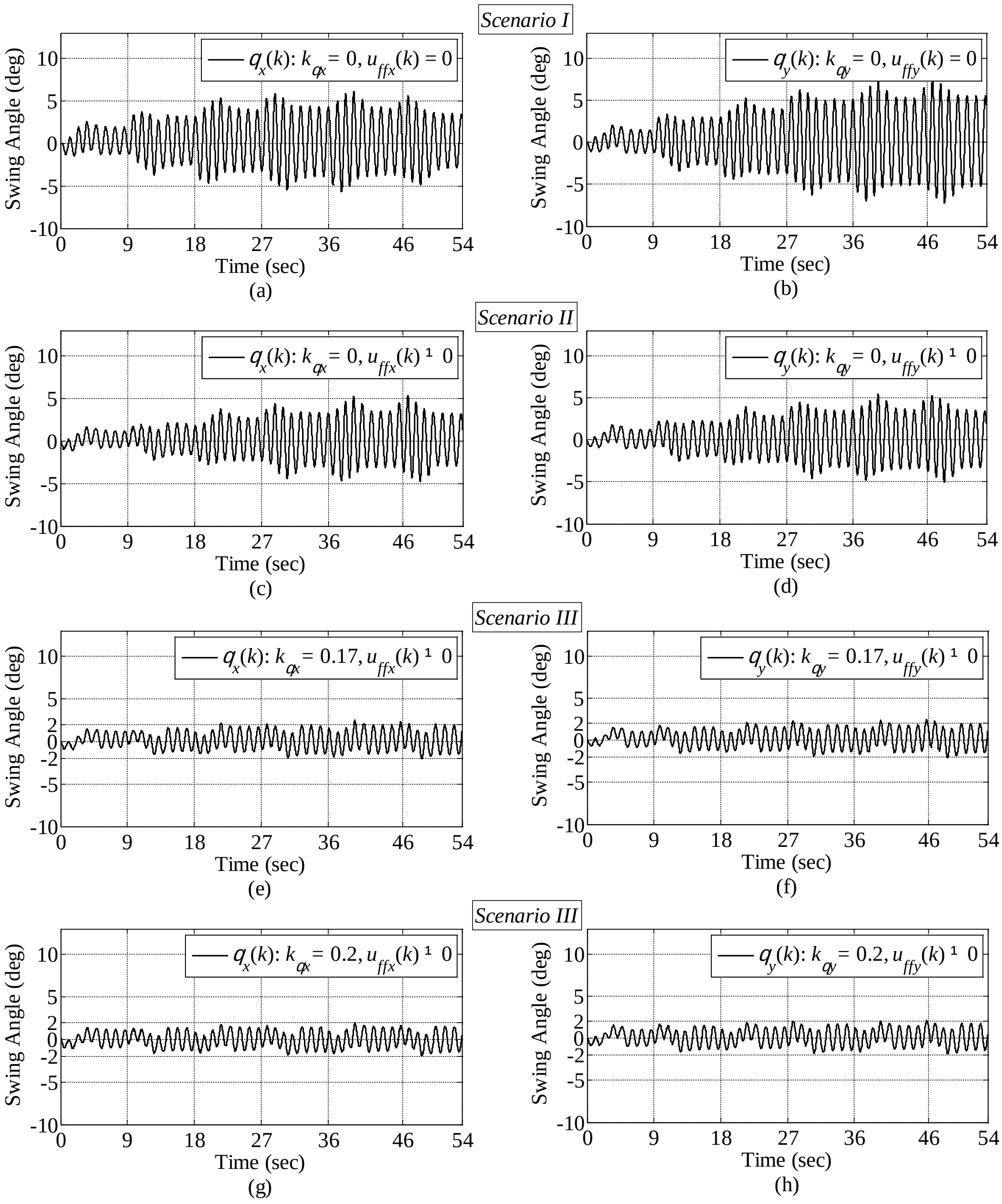

Fig. 5.9–20. Swing angle measurements with state feedback control for fast trajectory. (a) Traveling swing angle and (b) traversing swing angle in Scenario I: Load swing control and feedforward control are off, (c) Traveling swing angle and (d) traversing swing angle in Scenario II: Load swing control is off and feedforward control is on, (e) Traveling swing angle and (f) traversing swing angle in Scenario III: Both load swing control and feedforward control are on, (g) Traveling swing angle and (h) traversing swing angle in Scenario III: Both load swing control and feedforward control are on but with different swing control gain, i.e., $K_\theta = Diag\{0.2, 0.2\}$.

The estimates of system states are demonstrated in Fig. 5.9–21 and Fig. 5.9–22 for slow and fast trajectories, respectively. As can be seen, the state observer can provide good estimates of positions as shown in Fig. 5.9–21(a), (c), (e) and in Fig. 5.9–22(a), (c), (e). However, the maximum value of velocity estimates in Fig. 5.9–21(b), (d), (f) and Fig. 5.9–22(b), (d), (f) could not reach to the expected normal velocity of the reference trajectories. The reason is that the state observer gains corresponding to the traveling, traversing and hoisting velocities, i.e., ($\hat{v}_x(k)$, $\hat{v}_y(k)$, $\hat{v}_l(k)$), cannot be chosen very large since they can excite high frequency modes of the PM DC motors and destabilize the system. Therefore, it is just needed to design the observer gain such that the state estimation error is bounded with small range as discussed in Section 5.6.1.

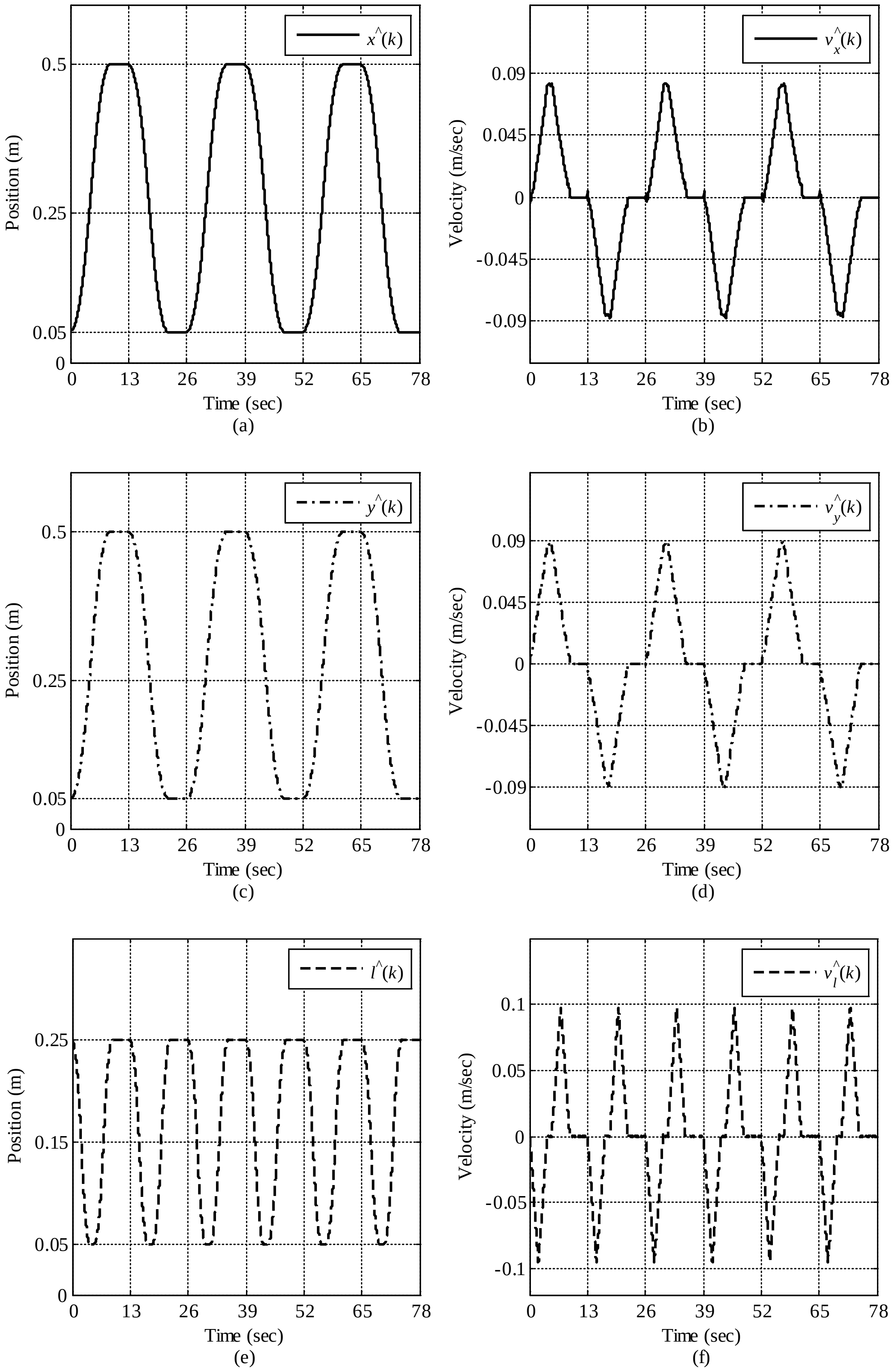

Fig. 5.9–21. The estimate of system states generated by the state observer in slow trajectory with state feedback control. (a) Traveling position estimate, (b) Traveling velocity estimate, (c) Traversing position estimate, (d) Traversing velocity estimate, (e) Hoisting position estimate, (f) Hoisting velocity estimate.

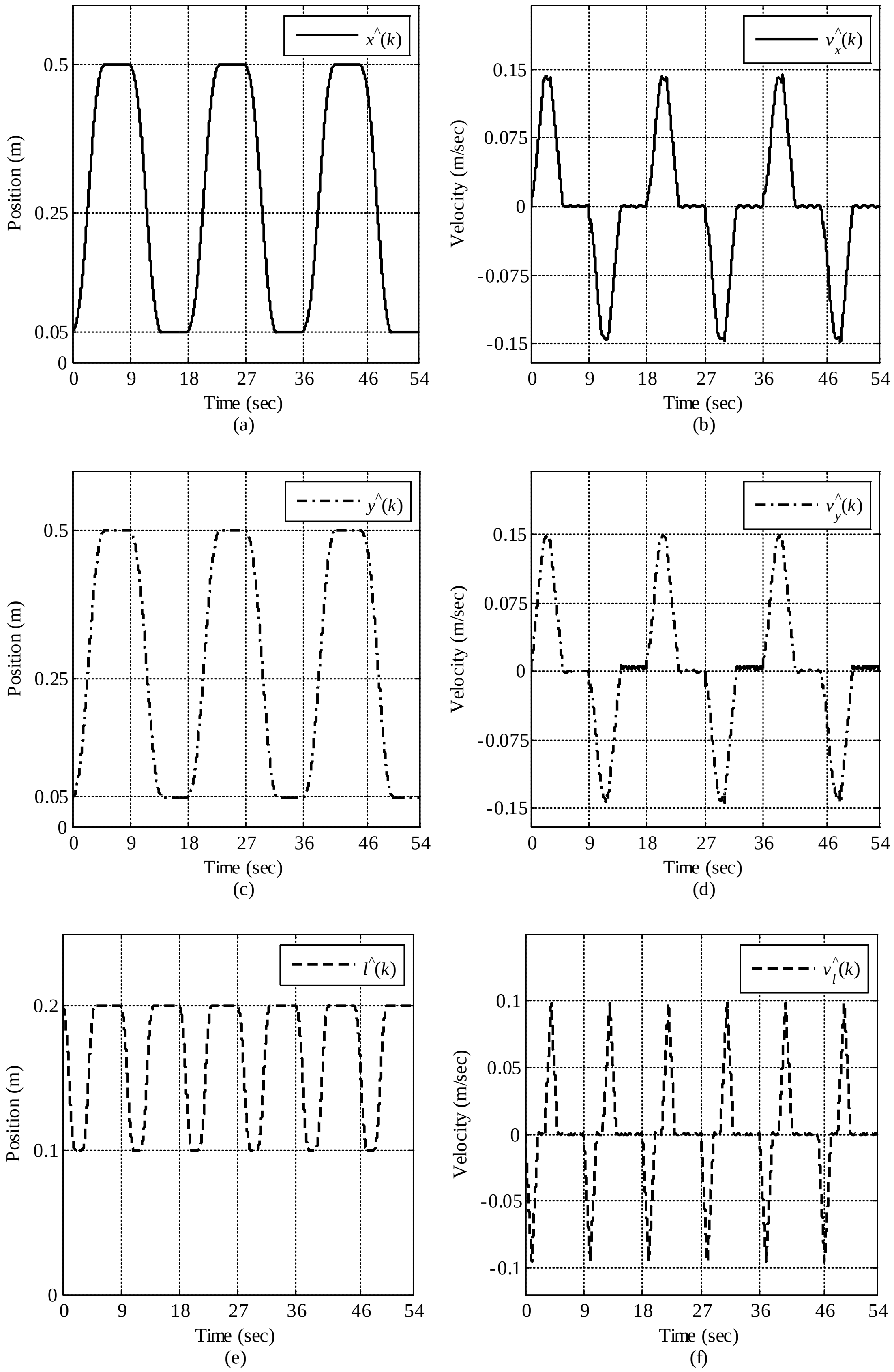

Fig. 5.9–22. The estimate of system states generated by the state observer in fast trajectory with state feedback control. (a) Traveling position estimate, (b) Traveling velocity estimate, (c) Traversing position estimate, (d) Traversing velocity estimate, (e) Hoisting position estimate, (f) Hoisting velocity estimate.

To demonstrate the performance of the overhead crane control operation in following the reference trajectories using state feedback control more accurately, the trajectory tracking error for traveling, traversing, and hoisting motions under the aforementioned three scenarios are provided in Fig. 5.9–23 and Fig. 5.9–24 for slow and fast trajectories, respectively. As expected, the position of traveling, traversing and hoisting has the highest error in Scenario I with no load swing control, and specifically, no feedforward control to compensate for load disturbances as shown in Fig. 5.9–23(a) in slow trajectory, and more noticeable in Fig. 5.9–24(a) in fast trajectory. Since the disturbances are intensified when the overhead crane moves with high speed, the tracking error is much higher without any compensation measure. In the second and third Scenarios where feedforward control is active, the performance of load positioning is improved considerably, particularly at the end of each transition with the tracking error less than ±1 millimeter for both slow and fast trajectory. However, due to using load swing control in Scenario III, the tracking error grows more during accelerating and constant-velocity zones in each transition of the trajectories compared to Scenario II as can be seen in Fig. 5.9–24(b) and (c) for fast trajectory. For slow trajectory, however, the deterioration of tracking error is not significant between the second and third scenarios, as shown in Fig. 5.9–23(a) and (b) due to lower speed of the overhead crane motion. Nevertheless, the combination of load swing control and feedforward control creates a trade-off between suppressing load swings and maintaining a low tracking error to provide high-performance control operation. As a major objective, it is important to be able to get the overhead crane load to the final destination with a high accuracy as well as keeping load swings as small as possible, and that is what our proposed discrete-time control system can successfully deliver.

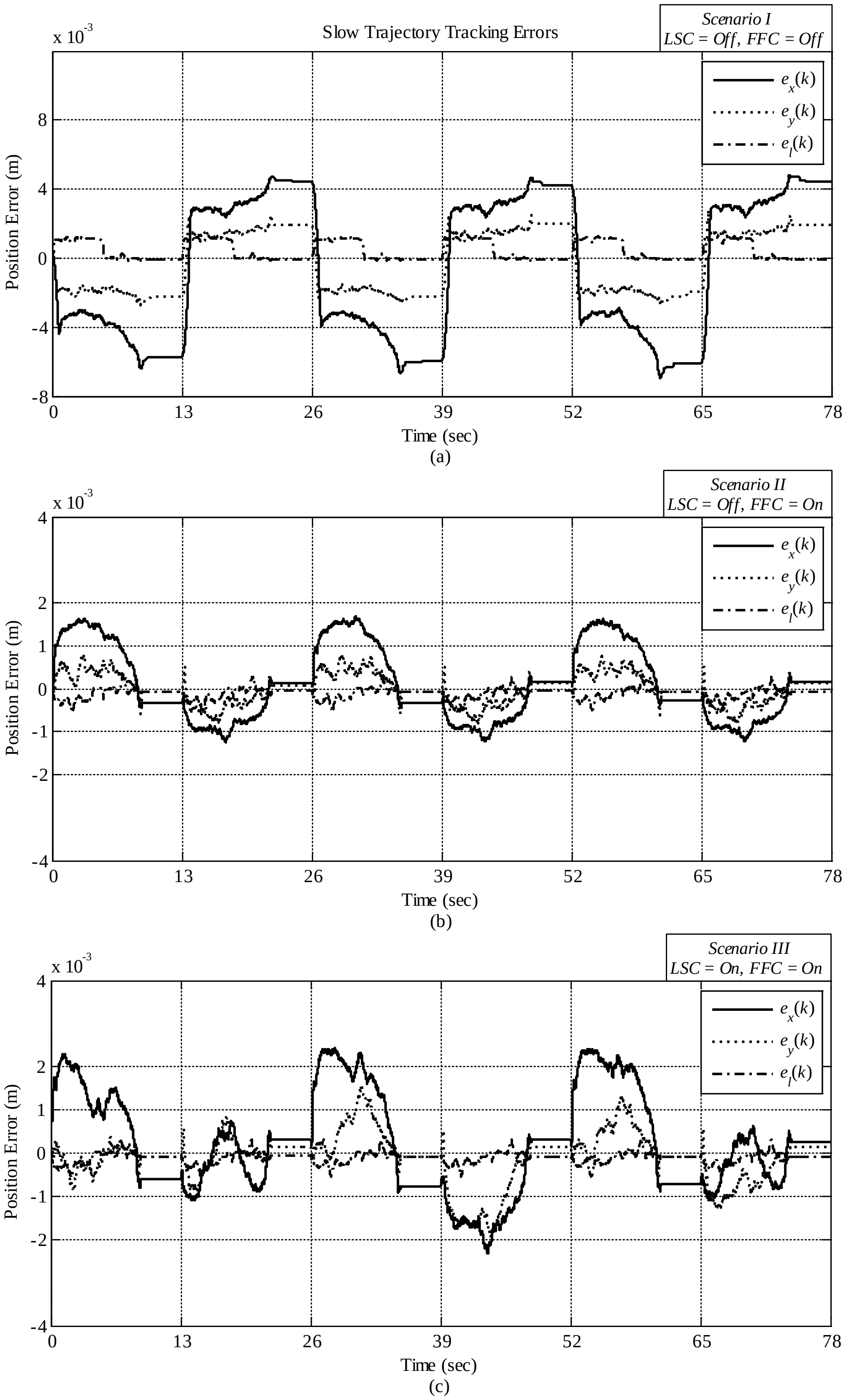

Fig. 5.9–23. Position tracking error with state feedback control for traveling $e_x(k)$, traversing $e_y(k)$, and hoisting $e_l(k)$ motions in slow trajectory. (a) Scenario I: Load swing control (LSC) and feedforward control (FFC) are off, (b) Scenario II: Load swing control is off and feedforward control is on, (c) Scenario III: Both load swing control and feedforward control are on.

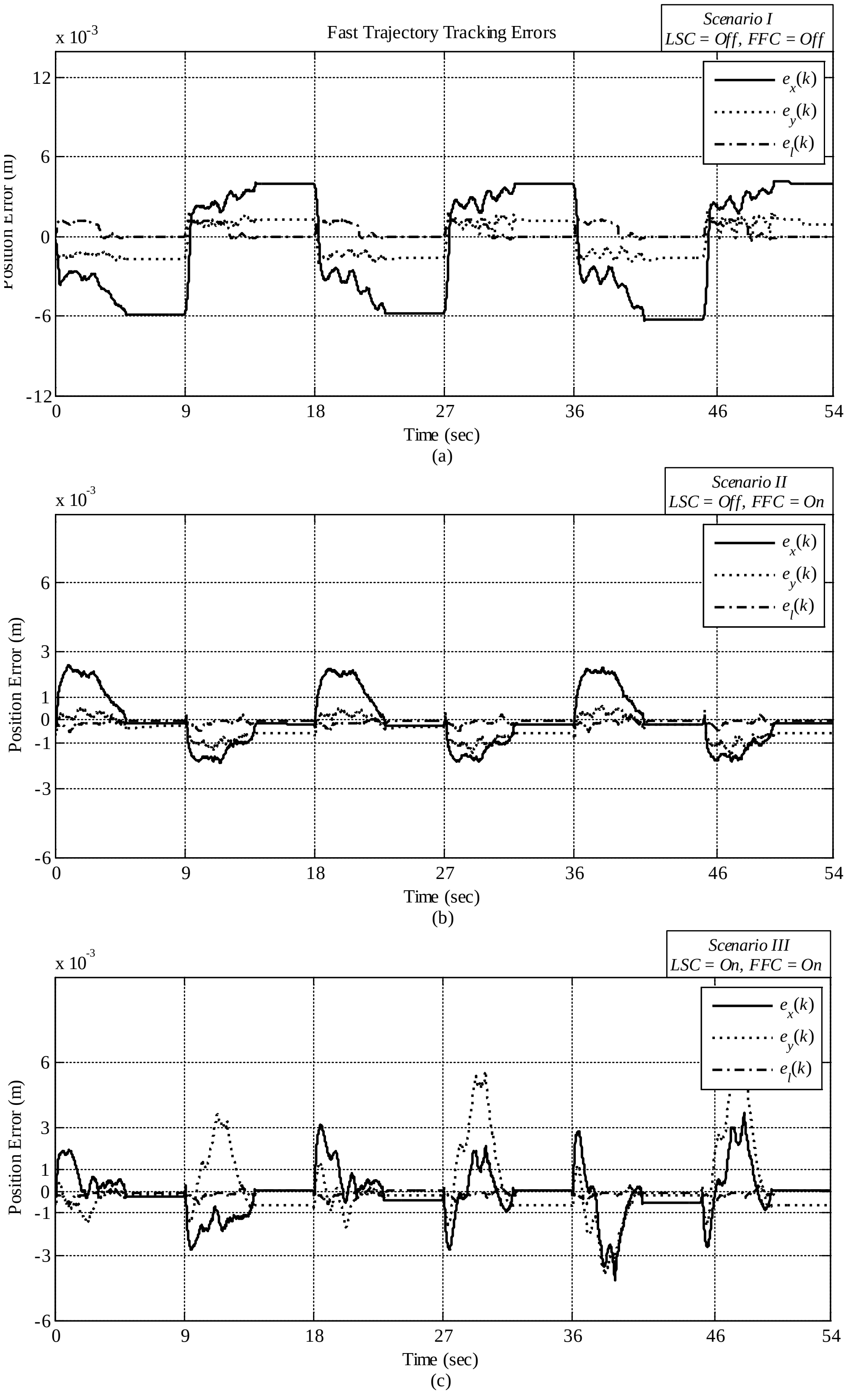

Fig. 5.9–24. Position tracking error with state feedback control for traveling $e_x(k)$, traversing $e_y(k)$, and hoisting $e_l(k)$ motions in fast trajectory. (a) Scenario I: Load swing control (LSC) and feedforward control (FFC) are off, (b) Scenario II: Load swing control is off and feedforward control is on, (c) Scenario III: Both load swing control and feedforward control are on.

We have developed a method to estimate the load disturbances using a disturbance observer as a replacement for load disturbances generated via inverse dynamic method in feedforward control as explained in Section 5.8. To evaluate the performance of disturbance observer, it is implemented in the discrete-time control system with state feedback control to generate the load disturbances for feedforward control. As we mentioned earlier, inverse dynamic method requires to have the knowledge of the system dynamics and parameters, particularly the value of the load mass, to generate the estimates of load disturbances. Since the mass of the load can vary significantly during the performance control operation, the control system will be vulnerable to the uncertainties of in the load mass. The significance of the disturbance observer is that it only uses the estate estimation error to calculate the load disturbance that makes it robust against any uncertainty in the load mass. Moreover, coulomb friction forces as a major component of the disturbances may not be consistence over time due to lubrication of bearing and some other mechanical issues, and hence, the estimated friction constants may be inaccurate to be used in inverse dynamic for load disturbance computation. The utilization of disturbance observer brings the benefit of robustness against these uncertainties which can definitely improve the performance of trajectory tracking and disturbance compensation. Thus, we have examined the performance of disturbance observer by comparing its generated load disturbance with those calculated via traditional computed torque control using inverse dynamic method with two different load masses: $m$ = 0.4kg and $m$ = 0.8kg. These tests have been conducted under the third scenario for slow and fast trajectories where both load swing control and feedforward control are active.

The results are pictured in Fig. 5.9–25 and Fig. 5.9–26. As can be seen, load disturbances generated by disturbance observer is more dynamic whereas inverse dynamics method generates fairly step-form signals and both change sign with respect to the direction of motion of traveling, traversing and hoisting. This is a clear indication of the coulomb friction effect in the load disturbance that is more dominant compared to the effects of nonlinear dynamic of the overhead crane. The main reason for this is the use of independent joint model which considers the effects of reduction gearbox in the connection of actuators to the moving parts of the overhead crane that can effectively reduce the amount of load torque on the PM DC motors. When $m$ = 0.4kg, there is not a significant increase in the maximum value of the estimated load disturbances between slow trajectory in Fig. 5.9–25(a) and (b) and fast trajectory in Fig. 5.9–25(c) and (d).

However, moving a heavier load with higher speed would create bigger load disturbance as illustrated in Fig. 5.9–26(c) and (d) compare to the same load mass but with lower speed motion in in Fig. 5.9–26(a) and (b). It is interesting to see that the disturbance observer can estimate the amount of load disturbances with more dynamically compared to the computed torque control without the need to know the values of the load mass and coulomb friction constants in advance.

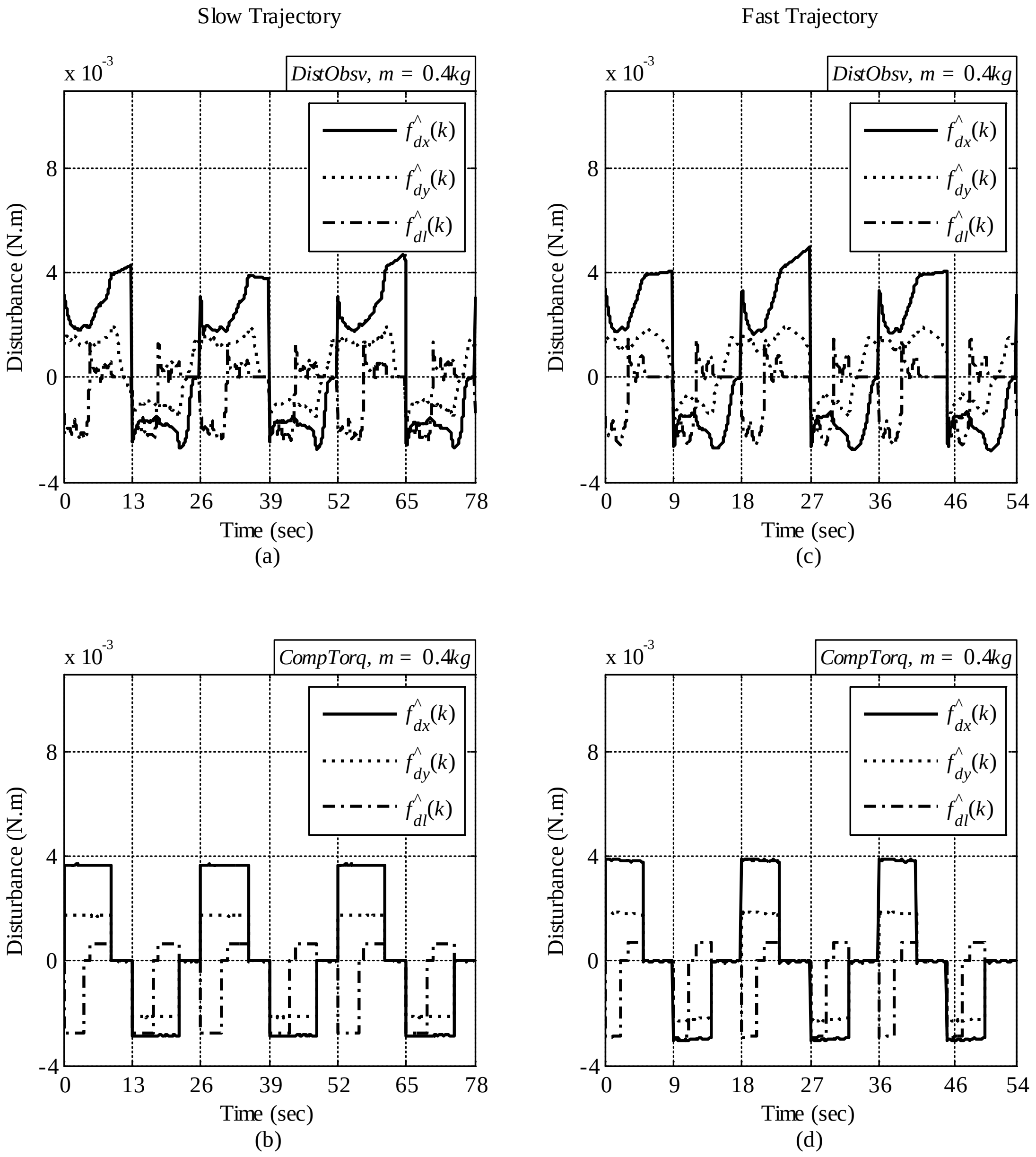

Fig. 5.9–25. Computed load disturbances to be used in feedforward control with *m* = 0.4kg. (a) Load disturbances generated via disturbance observer in slow trajectory, (b) Load disturbances generated via traditional computed torque control in slow trajectory, (c) Load disturbances generated via disturbance observer in fast trajectory, (d) Load disturbances generated via traditional computed torque control in fast trajectory.

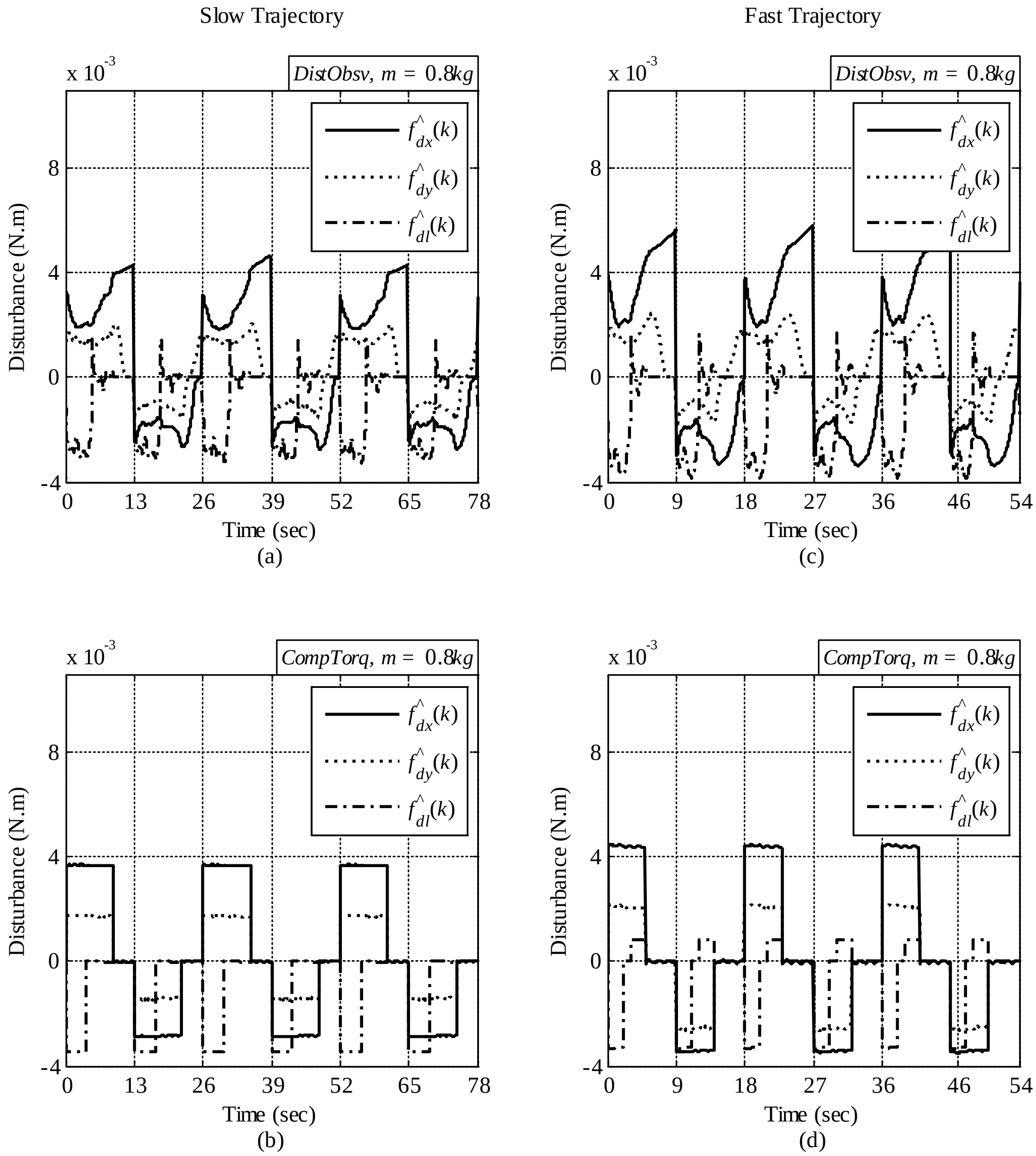

Fig. 5.9–26. Computed load disturbances to be used in feedforward control with $m$ = 0.8kg. (a) Load disturbances generated via disturbance observer in slow trajectory, (b) Load disturbances generated via traditional computed torque control in slow trajectory, (c) Load disturbances generated via disturbance observer in fast trajectory, (d) Load disturbances generated via traditional computed torque control in fast trajectory.

After scrutinizing the performance of the designed discrete-time control systems for overhead crane under several experiments and tests separately, it is time to have a combined judgment for an ultimate evaluation on the performances of these control systems. To do this, let us recall that the actual position trajectory of the overhead crane load in the workspace is a 3D path which starts from the initial location of the load and ends at the final destination. The load position $\boldsymbol{q}_m = (x_m, y_m, z_m)$ at each moment with respect to the reference coordinates mounted on one corner of the overhead crane framework (Fig. 4.2–1) is obtained as he following (also given in (4.2–1)),

$$\begin{aligned} x_m &= x + lS_{q_x}C_{q_y}, \\ y_m &= y + lS_{q_y}, \\ z_m &= -lC_{q_x}C_{q_y}. \end{aligned} \tag{5.9–1}$$

The actual position trajectories of the overhead crane load in the 3D workspace under the third scenario for fast trajectory with all the repetitions are shown in Fig. 5.9–27, Fig. 5.9–28, and Fig. 5.9–29 (with $m$ = 0.8kg). The first figure is the results of the control system with MPC as its discrete-time controller (Fig. 5.9–27). The response of the control system with state feedback control as the discrete-time controller and traditional computed torque control as the source of feedforward signal generator is depicted in Fig. 5.9–28. And the last one showing the load position trajectory with state feedback control and disturbance observer for generating feedforward signal (Fig. 5.9–29). As can be seen, in all the transitions of the overhead crane load, our discrete-time control system can deliver a high-performance load transportation with as minimum load swing as possible with high precision in load positioning.

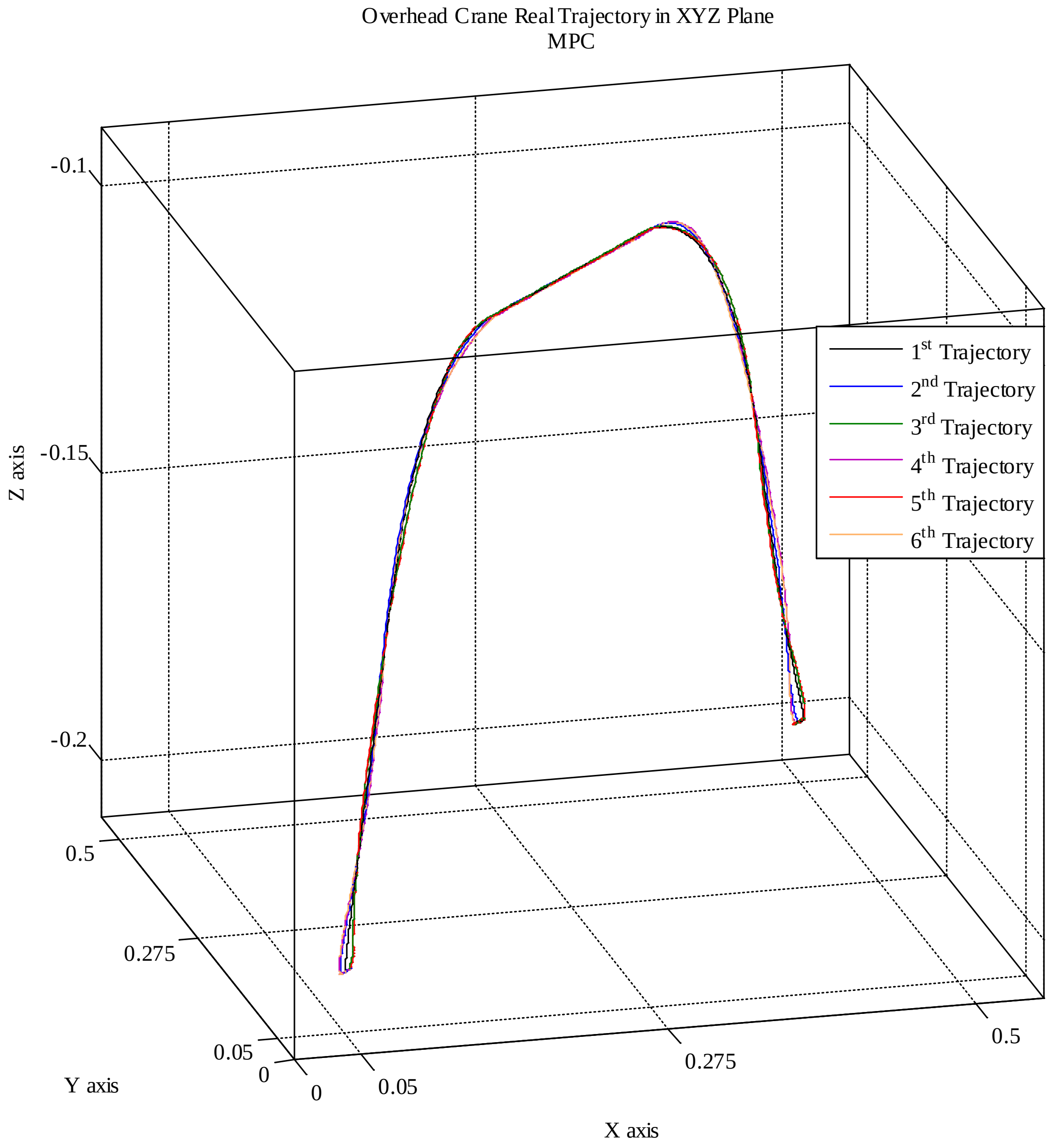

Fig. 5.9–27. Actual trajectory of the overhead crane load in 3D workspace with MPC as the discrete-time controller and traditional computed torque control for feedforward control.

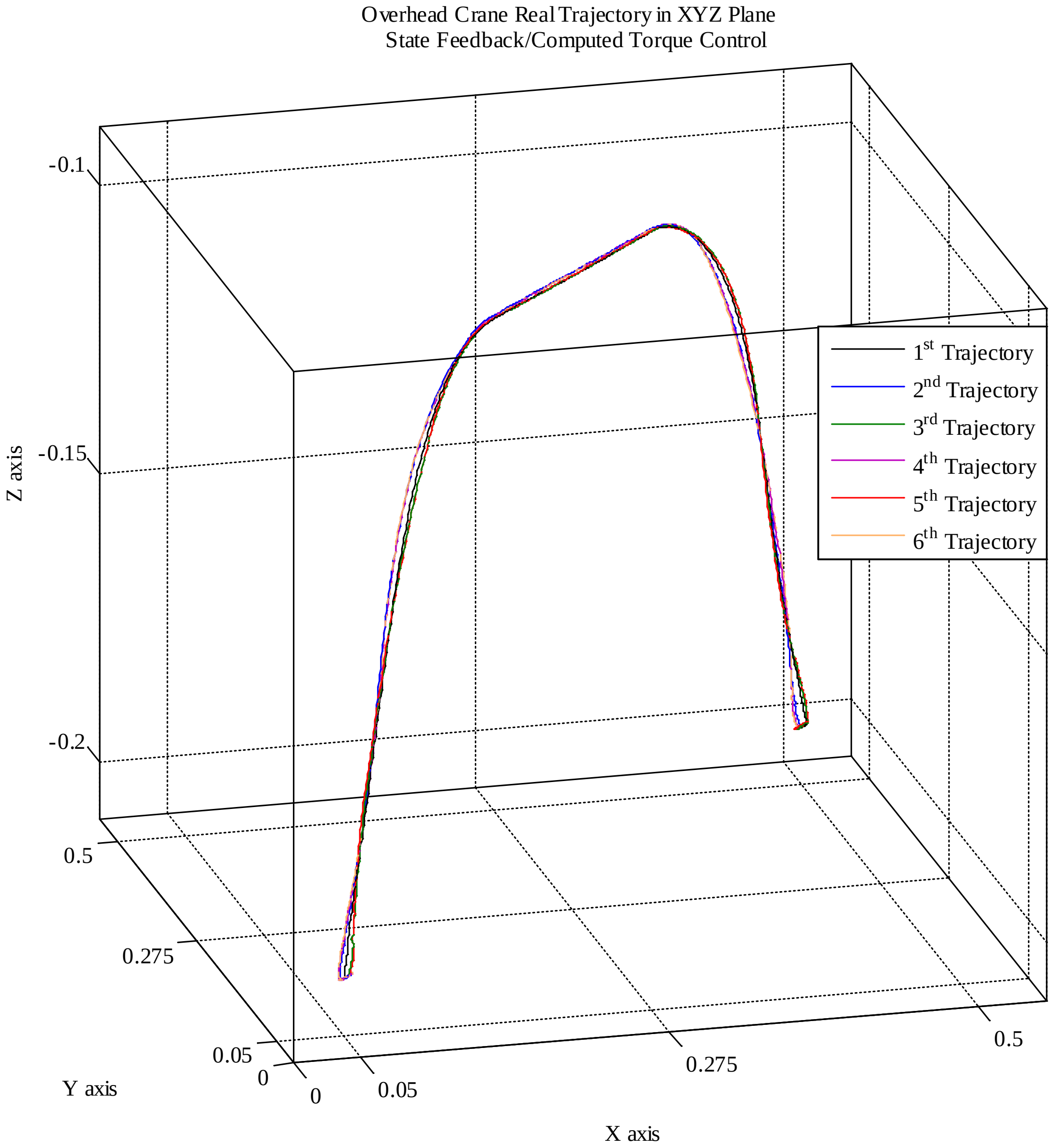

Fig. 5.9–28. Actual trajectory of the overhead crane load in 3D workspace with state feedback control as the discrete-time controller and traditional computed torque control for feedforward control.

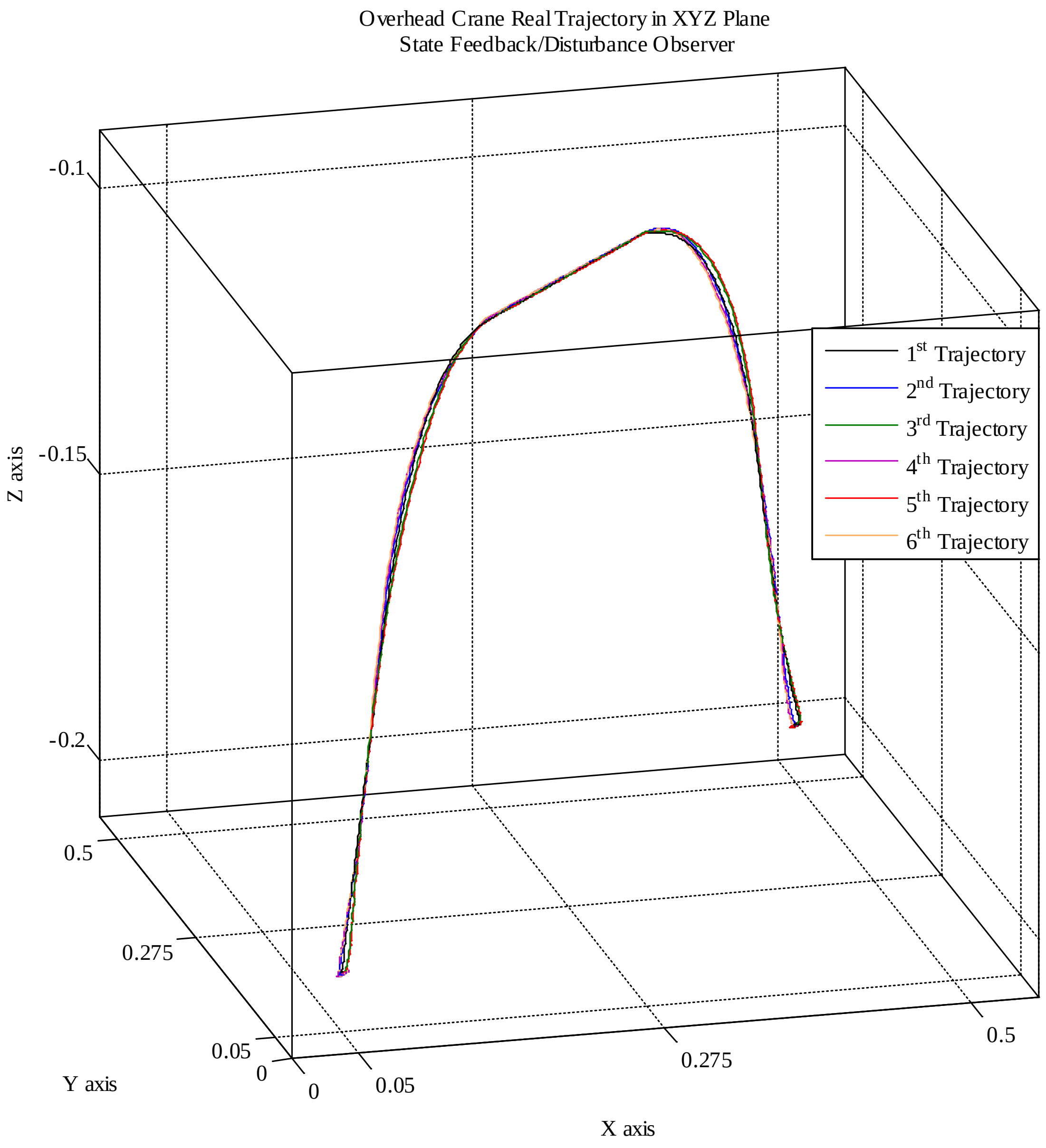

Fig. 5.9–29. Actual trajectory of the overhead crane load in 3D workspace with state feedback control as the discrete-time controller and disturbance observer for feedforward control.

Furthermore, the comparison between the performance of our discrete-time control systems in terms of the accuracy of the actual position of the load in the workspace can be further investigated by plotting the difference between the actual location of the load at each time step, $\boldsymbol{q}_m(k)$, and the reference location, i.e., $\boldsymbol{q}_{ref}(k) = (x_{ref}(k), y_{ref}(k), z_{ref}(k))$ as previously shown in Fig. 5.9–4 ($z_{ref}(k) = -l_{ref}(k)$). This difference is in fact the physical distance between two points at each time step that can be calculated using the second norm or Euclidean norm of the actual load position error, i.e., $E_q = \| \boldsymbol{q}_{ref}(k) - \boldsymbol{q}_m(k) \|_2$ , defined as follows,

$$\begin{aligned} E_q(k) &= \| \boldsymbol{q}_{ref}(k) - \boldsymbol{q}_{ref}(k) \|_2 \\ &= \sqrt{(x_{ref}(k) - x_m(k))^2 + (y_{ref}(k) - y_m(k))^2 + (z_{ref}(k) - z_m(k))^2}, \qquad (5.9\text{–}2) \\ & \qquad \textit{for} \quad k = 0, 1, 2, \ldots . \end{aligned}$$

The comparison results are illustrated in Fig. 5.9–30 and Fig. 5.9–31 for slow and fast trajectory, respectively under the third scenario with $m$ = 0.8kg. The effects of load swings in the actual position of the load can be clearly seen in the form of the oscillation in the distance error. The performance of discrete-time control systems in slow trajectory is quite similar as can be seen in Fig. 5.9–30(a), (b), and (c) which average distance error of about of ±2 millimeters. When the overhead crane is operated under fast trajectory, the distance error increases by about 60 percent which is completely normal due to much larger load swings (from ±0.5 degree in slow trajectory to ±2 degrees in fast trajectory) and higher disturbances. However, the performance of the control system with MPC as shown in Fig. 5.9–31(a) seems to be inferior with some jumps up to 15 millimeters compare to the other two. The reason is that MPC calculates the optimum control input by penalizing the deviation in the actual system output and the reference trajectory which is defined as the position of traveling, traversing and hoisting. Whereas, in state feedback control, both position and velocity errors are used in the control law directly which acceptably should perform better in terms of the accuracy of load positioning. Interestingly, the best performance in fast trajectory is obtained when disturbance observer is used to estimate load disturbance for feedforward control in Fig. 5.9–31(c) compare to Fig. 5.9–31(b) where traditional computed torque control is used. As we showed in the comparison results of the load disturbance estimation, we expected to get better performance using disturbance observer performs.

Nevertheless, all the discrete-time control systems have proven to be able to deliver high-performance control for overhead crane operation.

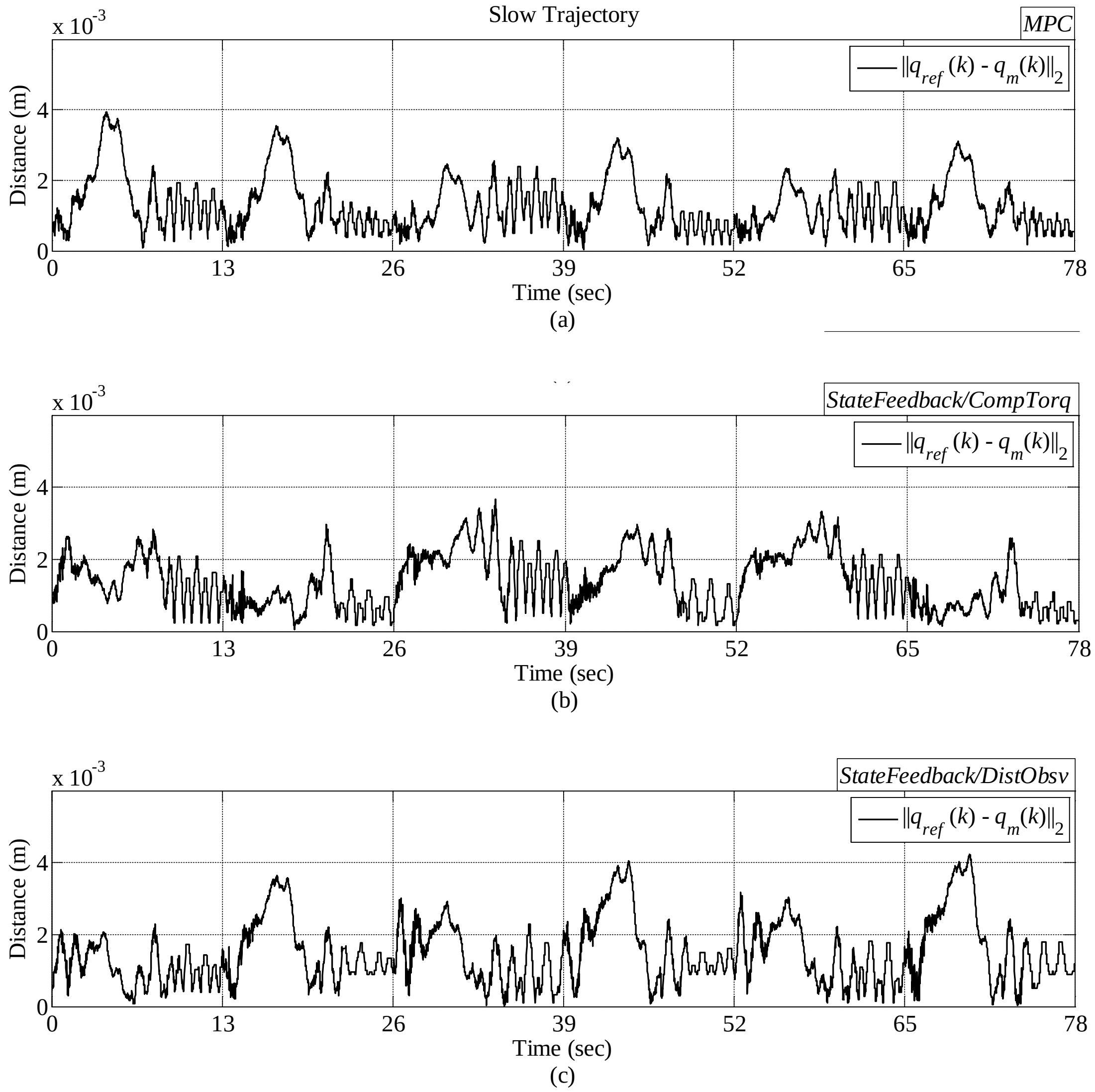

Fig. 5.9–30. The physical distance error between the actual position of the overhead crane load and the reference position trajectory under Scenario III with $m$ = 0.8kg in slow trajectory.

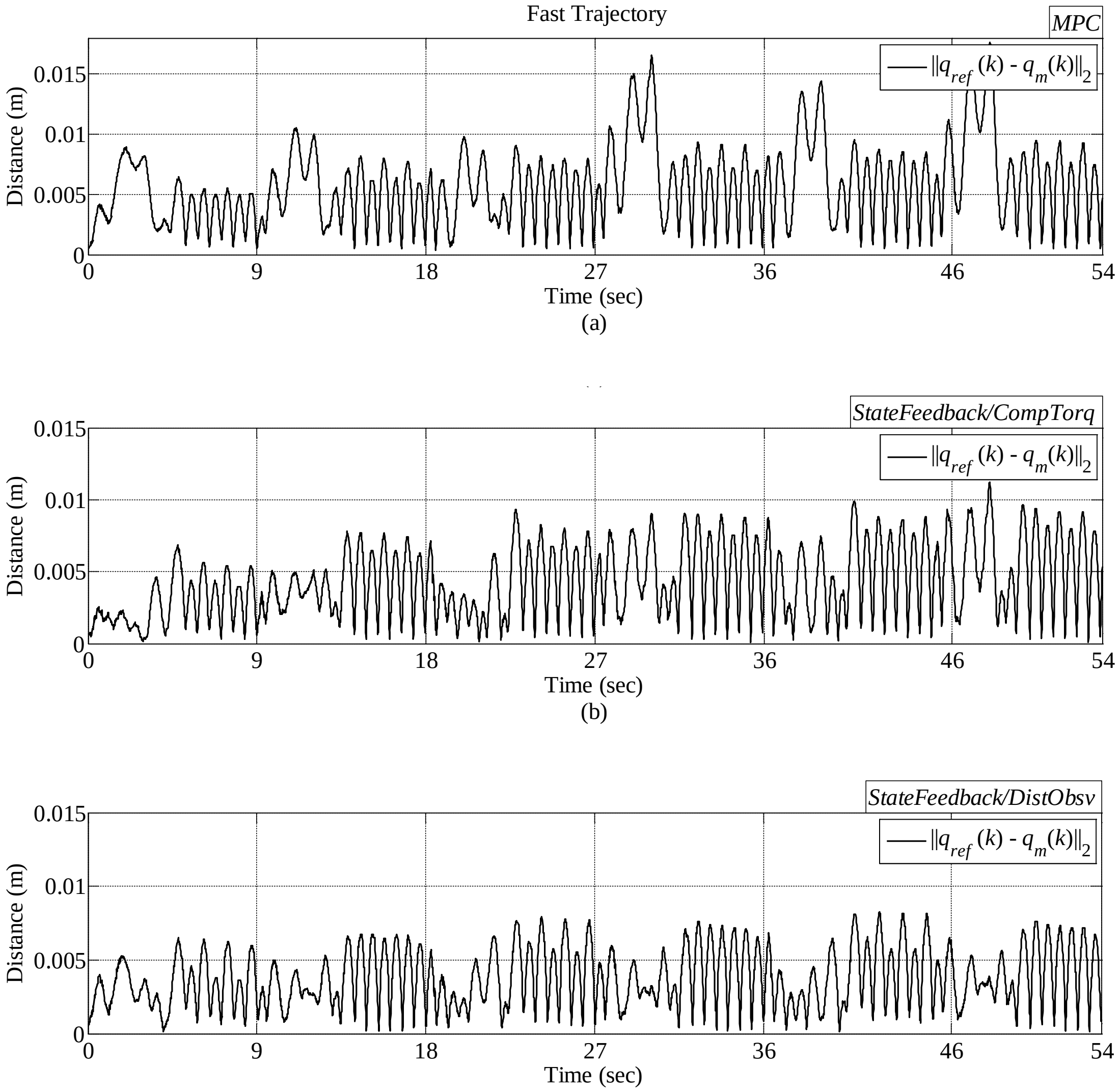

Fig. 5.9–31. The physical distance error between the actual position of the overhead crane load and the reference position trajectory under Scenario III with $m$ = 0.8kg in fast trajectory.

Finally, as an example, an interactive flash video file is attached in Fig. 5.9–32 demonstrating the actual experiment on 2D overhead crane for discrete-time state feedback with computed torque control being run under Scenarios I and III with fast trajectory as comparison side by side, which was reported in [109] (Adobe Flash Player software is required to play the video). This provides a real and visual sense on the high performance of the proposed control system in not only tracking the reference trajectory but also suppressing load swing robustly in high-speed overhead crane motion. The video can also be watched on YouTube website via the link provided in [117].

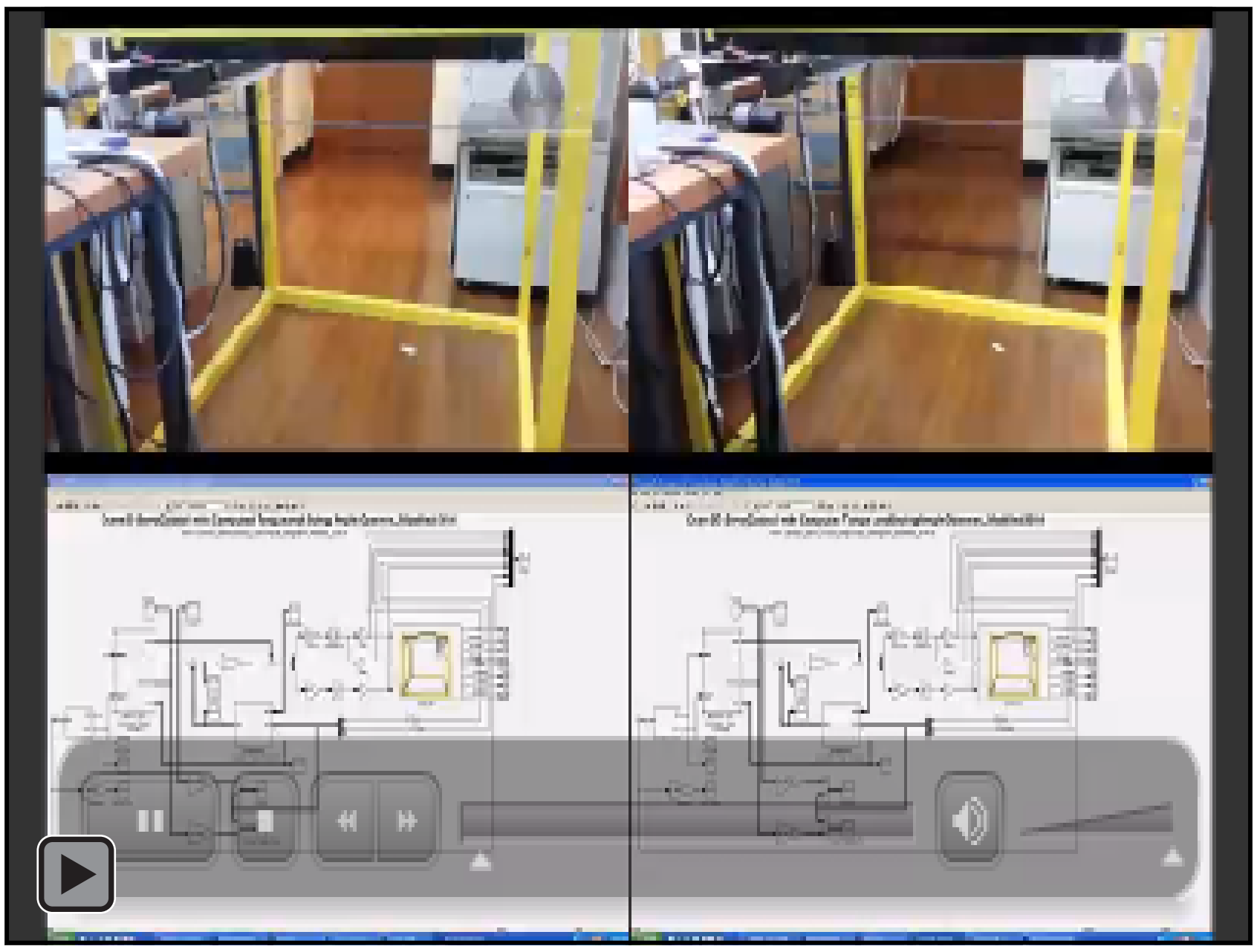

Fig. 5.9–32. Flash video of the real-tile experiment on 2D overhead crane with state feedback approach with computed torque control under Scenarios I and III.

The video above contains four sections. The two sections on the right showing the real-time control operation of 2D overhead crane in top-right corner and real-time graphs representing load swing and trajectory tracking for traveling and hoisting on bottom-right corner, which is run with discrete-time state feedback control without feedforward and load swing controls (Scenario I) in fast trajectory. On the left side, the real-time experiment and graphs similar to the right side of the video are attached demonstrating control operation using discrete-time state feedback under Scenario III with fast trajectory (feedforward control and load swing control are On). The side-by-side comparison clearly is the indication of the superiority and capability of the proposed discrete-time control system in practice.

## 5.10 Discussion and Conclusion

The core of the discrete-time control system design for the overhead crane was provided in this chapter. The fundamental control objectives and requirements were elaborated which, in brief, expresses that a high-performance overhead crane control operation requires that the load can be transported as fast as possible with high accuracy in load positioning with as minimum load swing as possible. In addition, the control system design and settings should not be very complicated and difficult to understand for the operator. Based on these objectives, the configuration of the proposed discrete-time control system was described which is mainly inspired by the idea of independent joint control for simplifying the overall design procedure, and computed torque control as a mean to deal with nonlinear disturbances caused by coupling effects between the actuator and mechanical dynamics of the overhead crane. Therefore, the structure of the proposed control system was proposed to have four main parts.

Load swing control was designed based on passivity based control and $\mathcal{L}_2$ stability theorem to suppress load swing by modifying the reference traveling and traversing accelerations such that load swing are maintained bounded for the entire control operation. Reference signal generator designed to provide reference trajectories according to the proposed trajectory planning which allows smooth collaboration between reference trajectory generation and load swing damping. Two discrete-time tracking controllers based on MPC and state feedback control approach were then designed to generate the required control inputs for following the reference trajectories with minimum tracking error. These discrete-time controllers were integrated with feedforward control to cope with nonlinear disturbances using computed torque control, and disturbance observer as an alternative for estimating load disturbances without the need to know the value of systems parameters especially load mass. In addition to the analytical stability proof, several experiments were carried out on a laboratory-sized overhead crane to demonstrate the ability of the designed control systems to deliver the control objectives established in this chapter. The tests were designed for three different scenarios with two trajectory speeds to access the capability of the control systems when operating with and without load swing control and feedforward control. Swing angle measurements and trajectory tracking errors were compared with each other for MPC and state feedback control. These practical results indicated that the proposed discrete-time control systems are able to control the overhead crane with high performance and less complexity compared to the existing methods in the literature.

# Chapter 6

# Conclusion

Control theory has made substantial progress since 1955. Only some of this theory, however, has made its way into existing computer-controlled systems, even though feasibility studies have indicated that significant improvements can be made. For example, model predictive control and adaptive control are some of the theoretical areas that are being applied in the industry today due to their discrete-time nature and easy implementation in digital computers. In addition, the practical limitations of the process can be included into these control techniques which make them attractive and applicable. Designing the control system in discrete-time for implementation in digital computers has substantial advantages. Many of the difficulties with continuous-time controllers can be avoided. For example, the problems associated with sampling time choice, quantization errors, and approximation of calculus operators for solving ordinary differential equations. Logic statements and nonlinear functions can also be included in the discrete-time control law easily. These advantages of discrete-time control systems could bring more efficiency and benefit to the final product from the industry point of view. Therefore, the intention of the work presented in this thesis was to study and highlight the capability of discrete-time control systems for industrial applications. In particular, we chose two processes from different fields of industry and developed advanced control methods to increase their performance in real-time operation. The presented work was separated into two main parts in this thesis.

## 6.1 Concluding Remarks on Part I

The first process considered in this thesis was the wind power integrated with a BESS in a grid-connected mode. A novel discrete-time control system was designed with the aim of increasing financial benefits for the wind power generation through the sale of power in the electricity market in time shifting application using a BESS, as well as improving the controllability of wind power dispatch. Due to discrete-time nature of power dispatch process and electricity pricing in the Australian NEM, a simple discrete-time model was considered for dispatching wind power and charging/discharging the BESS. According to time shifting application that allows selling more power at peak times and storing extra wind power at off-peaks for taking advantage of electricity price variations in the market, the control scheme was designed in three parts.

The first part is a decision-making system to generate the proper reference power signal based on electricity price and time of the day just before each 5-minute dispatch using fuzzy logic. The second part is the heart of the control system which is the discrete-time controller designed using MPC. The task of this discrete-time controller is to calculate the amount of charging-discharging power for the BESS to enable the control system not only to follow the reference power signal but also deal with the constraints on the BESS energy capacity and rated power, i.e., avoiding overcharging and depletion of the BESS. The last part is a fuzzy controller that uses the current condition of the BESS to smartly update the reference power signal for facilitating the task of the discrete-time controller. The combination of these parts to form the discrete-time control system is a new insight to the problem of controlling wind power dispatch with BESS in the electricity market which could provide higher earnings for the wind farm in the long term. The effectiveness of the proposed discrete-time control system was verified by the simulation results based on different scenarios of selling wind power to the grid using the actual wind power and electricity data obtained from the AEMO database. A key performance index was used for earning comparison of the power sale in different scenarios. However, an in-depth economic analysis has to be performed in order to precisely assess the proposed dispatch control scheme from the profit maximization point of view.

### 6.1.1 Future Works

The proposed dispatch control scheme is not limited to wind power and can be developed for other intermittent energy sources like large-scale solar electricity generation systems. Moreover, the problem of BESS sizing and its effects on long term costs and benefits for the wind farm using the proposed discrete-time control system is still open for more research. Furthermore, as the battery in the “time shifting” application operates in a nonstandard conditions including partial state-of-charge cycling and different times between full charging, lifetime prediction is a difficult task to do, which is an essential factor in verifying economic viability and lifecycle cost study. Since finding mathematical models for BESS lifecycle involves many technical parameters that may not be possible to measures mentioned in [99], fuzzy logic systems can be used to model such complicated processes. The basics of fuzzy modeling are similar to the material presented in Section 3.3, however, more details can be found in [85]. Any BESS lifetime model can be combined with the discrete-time model obtained for wind power dispatch with BESS in Chapter 2 and then incorporated into the cost function of the MPC as one of the parameters for optimal operation of the BESS. Thus, finding some model for the BESS lifecycle and include it in the optimization problem is an interesting future line of research. A simpler approach to BESS lifecycle estimation is the one known as *rainflow cycle counting* [118], which was initially proposed for material fatigue [119]. In this way, the counted charging/discharging cycles can be compared with the nominal one provided by the manufacturer to determine the remaining lifetime of the BESS. Thus, finding a model for the BESS lifecycle and including it in the optimization problem is an interesting future line of research.

## 6.2 Concluding Remarks on Part II

Overhead crane system was the second process studied in this thesis. Although considerable amount of research has been carried out on controlling the overhead crane motion over the past couple of decades, very few have tried to consider practicality and compatibility of their designs for real-time industry application. In addition, many of them used complicated control algorithms that might not be favorable as a substitute to an expert human operator. Thus, a new discrete-time control system was developed for the overhead crane to be able to deliver high-performance control for automatic load transportation including high-speed operation with high accuracy in load positioning and minimum load swing. As an underactuated system with a highly nonlinear dynamics, it is quite challenging to design a high-performance control system with less complexity. However, using independent joint control strategy, a new dynamic model was derived for the overhead crane in discrete-time in which the process actuators, mainly PM DC motors, are considered as the main plant and the nonlinearities are treated as load disturbances on each actuator. The proposed independent joint model enabled us to determine the primary physical parameters of the overhead crane with a simple approach yet quite effective with high accuracy in terms of parameter identification. Moreover, the effects of coulomb friction forces was considered in the model as part of disturbances, since they are one of the major elements in reducing the load position accuracy, and their parameters were identified along with other physical parameters. A new discrete-time control system was then designed using the resulting discrete-time model consisting of four main parts.

A reference signal generator was designed as the first part to provide reference trajectory tracking which complies with the typical anti-swing trajectory used by an expert human operator in practice. A trajectory planning was also developed and incorporated with reference signal generation to ensure the practical restrictions on the actuators and the overhead crane workspace. The second part was the new load swing control that uses a high-gain observer for providing the estimates of swing angle and their first-time derivative. To suppress load swing robustly during the entire control operation, the reference traveling and traversing accelerations are modified by the load swing control, which indirectly exerts a damping force to the swing dynamics. A feedforward control was designed as the third part of the control system using the idea of computed torque control and a new disturbance observer to compensate for the nonlinear load disturbances, which significantly improves the accuracy of trajectory tracking. The forth and main part of the control system is the discrete-time controller, which was designed using MPC and state feedback control to calculate the control input voltages for the driving motors such that the control system can track reference trajectories with high performance.

The proposed discrete-time control systems are quite simple in terms of design procedure. In addition, they are easy to implement due to their discrete-time nature, and easy to understand by the operator compared to the existing controllers in the literature, without affecting the control performance as one of the main contributions of his work. Furthermore, they can deliver the control objectives and requirements for high-performance operation including high-speed load hoisting during accelerating zone, which is vastly ignored in the literature. The performance of the proposed discrete-time control systems were verified by an extensive number of tests and experiments on both 2D (traveling and hoisting) and 3D overhead cranes under different scenarios and operation speeds using a practical overhead crane setup. The results indicated the high performance of the control operation in both precision of load positioning and minimizing load swings for high-efficient automatic load transportation. Particularly, the proposed disturbance observer provided high robustness against variation in the overhead crane load mass since the load mass is included as part of load disturbances in the proposed discrete-time model.

### 6.2.1 Future Works

To further improve the performance of the proposed discrete-time control system with MPC for overhead crane, it is suggested that the proposed disturbance observer, described in Section 5.8, and state observer in Section 5.6.1 are combined in the MPC formulations to make a better predictions of the output, rather than having fixed disturbance values for the entire prediction horizon. This would compensate for the exclusion of traveling, traversing, and hoisting velocities in the output definition that led to lower performance compared with the state feedback approach as explained at the end of Section 5.9.3. The state feedback approach for discrete-time controller presented in Section 5.7 can also be improved by applying $H_\infty$ methods to obtain the discrete-time controller gain $K$ in a robust optimal sense [120], [121]. Moreover, linear matrix inequality (LMI) approach can be used to incorporated system constrains into the state feedback control to solve $H_\infty$ problem [122].

Motion-planning schemes for overhead cranes have recently gained attraction among researchers in this area, which is aimed to find a minimum-time trajectory between the initial and final location of the load as well as following the reference trajectory and damping load swing in an open-loop control form [65]–[70]. However, most of them assumed constant hoisting rope in a 2D overhead crane structure, which transform the problem to a pendulum attached to a cart rather than the actual overhead crane. They also seem to be very complicated which is not a positive point from operators' point of view. Our proposed trajectory planning described in Section 5.4 is more realistic and simple in terms of being applicable for generic 3D overhead cranes with load hoisting capability. However, more improvements can be made to the proposed trajectory planning to make it as an independent open-loop control system for overhead crane with less complexity.

The other interesting future topic in this area, which is closely related to motion planning, is the problem of obstacle avoidance during each zone of the trajectory and how to update the reference trajectories such that it does not create undesired load swings, which has rarely been worked on. Moreover, the proposed approach on applying independent joint control strategy in both modeling and controlling the overhead crane can be extended to other underactuated systems with linear actuators.